\documentclass[12pt,document,nofootinbib,superscriptaddress,onecolumn,preprintnumbers,balancelastpage]{revtex4}
\pdfoutput=1
\usepackage{latexsym}
\usepackage{amssymb}
\usepackage{epsfig,amsmath,graphics}
\usepackage{listings}
\usepackage{color}
\usepackage{dcolumn}
\usepackage{slashed}
\usepackage{cancel}
\usepackage{comment,latexsym} 
\usepackage{bm}
\usepackage{verbatim}
\usepackage{tabularx}
\usepackage{dcolumn}
\definecolor{Mahogany}{rgb}{0.62,0.24,0.15}
\definecolor{colorLink}{rgb}{0.6,0,0}
\definecolor{colorCite}{rgb}{0,.6,0}
\definecolor{colorURL}{rgb}{0,0.6,0.0}
\definecolor{colorTC}{rgb}{.2,.7,.2}
\usepackage[colorlinks=true,linktocpage=true,linkcolor=black,citecolor=colorCite,urlcolor=colorURL]{hyperref}
\usepackage{setspace}
\usepackage{xspace}
\usepackage{multirow}
\usepackage{titlesec}
\usepackage[bbgreekl]{mathbbol}
\usepackage{booktabs}
\usepackage{makecell}
\usepackage{array}
\usepackage[export]{adjustbox}
\usepackage{cleveref}
\usepackage[normalem]{ulem}
\usepackage{etoolbox}
\usepackage{mathtools}
\usepackage{relsize,exscale}
\usepackage{tensor}
\usepackage{enumitem}

\usepackage[usenames,dvipsnames]{xcolor}
\usepackage[breakable, theorems, skins]{tcolorbox}
\tcbset{enhanced}

\DeclareRobustCommand{\mybox}[2][gray!14]{%
\begin{tcolorbox}[   
        breakable,
        left=0pt,
        right=0pt,
        top=2pt,
        bottom=4pt,
        colback=#1,
        colframe=#1,
        width=\dimexpr\textwidth\relax, 
        enlarge left by=0mm,
        boxsep=5pt,
        arc=5pt,outer arc=5pt,
        ]
        #2
\end{tcolorbox}
}

\definecolor{colorTech}{rgb}{.6,.2,.55}
\newcounter{scenario}

\renewcommand{\thescenario}{\Alph{scenario}}
\newcommand{\scenario}[1]{%
   \refstepcounter{scenario}%
   \subsection*{\color{colorTech}{Primer~\thescenario.~~#1}}%
}%

\renewcommand{\thesection}{\arabic{section}}

\preto\section{%
  \ifnum\value{section}=0\addtocontents{toc}{\vspace{-5pt}}\fi
}
\preto\section{%
  \ifnum\value{section}=1\addtocontents{toc}{\vspace{-5pt}}\fi
}
\preto\section{%
  \ifnum\value{section}=2\addtocontents{toc}{\vspace{-5pt}}\fi
}
\preto\section{%
  \ifnum\value{section}=3\addtocontents{toc}{\vspace{-5pt}}\fi
}
\preto\section{%
  \ifnum\value{section}=4\addtocontents{toc}{\vspace{-5pt}}\fi
}
\preto\section{%
 \ifnum\value{section}=5\addtocontents{toc}{\vspace{-5pt}}\fi
}
\preto\section{%
  \ifnum\value{section}=6\addtocontents{toc}{\vspace{-5pt}}\fi
}
\preto\section{%
  \ifnum\value{section}=7\addtocontents{toc}{\vspace{-5pt}}\fi
}
\preto\section{%
  \ifnum\value{section}=8\addtocontents{toc}{\vspace{-5pt}}\fi
}

\crefname{table}{Table}{Tables}
\crefname{equation}{Eq.}{Eqs.}
\crefname{appendix}{App.}{Apps.}
\crefname{section}{Sec.}{Secs.}
\crefname{figure}{Fig.}{Figs.}
\crefname{scenario}{\color{colorTech}{\textbf{Primer}}}{\color{colorTech}{\textbf{Primer}}}
\creflabelformat{scenario}{{\color{colorTech}{{\bf #1}}}}

\numberwithin{equation}{section}

\usepackage{enumitem}
\setlist[itemize]{leftmargin=*}

\newcolumntype{L}[1]{>{\raggedright\let\newline\\\arraybackslash\hspace{0pt}}m{#1}}
\newcolumntype{C}[1]{>{\centering\let\newline\\\arraybackslash\hspace{0pt}}m{#1}}
\newcolumntype{R}[1]{>{\raggedleft\let\newline\\\arraybackslash\hspace{0pt}}m{#1}}

\def\be{\begin{equation}}
\def\ee{\end{equation}}
\newcommand{\beq}{\begin{equation}}
\newcommand{\eeq}{\end{equation}}

\newcommand{\lsim}{\!\mathrel{\hbox{\rlap{\lower.55ex \hbox{$\sim$}} \kern-.34em \raise.4ex \hbox{$<$}}}}
\newcommand{\gsim}{\!\mathrel{\hbox{\rlap{\lower.55ex \hbox{$\sim$}} \kern-.34em \raise.4ex \hbox{$>$}}}}
\newcommand{\vev}[1]{ \left\langle {#1} \right\rangle }

\newcommand{\D}{\text{d}}
\newcommand{\muT}{\tilde{\mu}}

\newcommand{\nbsb}{\bar{n}\cdot \bar{\sigma}}

\newcommand{\nsb}{n\cdot \bar{\sigma}} 
\newcommand{\nbs}{\bar{n}\cdot \sigma} 
\newcommand{\ns}{n\cdot \sigma}

\newcommand{\nbp}{\bar{n}\cdot \partial}

\newcommand{\bs}{\bar{\sigma}} 

\newcommand{\ac}{a_c}
\newcommand{\acb}{a_{\bar{c}}}
\newcommand{\as}{a_{us}}
\newcommand{\overbar}[1]{\mkern 1.5mu\overline{\mkern-1.5mu#1\mkern-1.5mu}\mkern 1.5mu}
\newcommand{\asb}{a_{\overbar{us}}}

\newcommand{\FT}{\textsc{Full Theory}}

\newcommand{\Primer}{{\color{colorTech}{\textbf{Primer}}}}
\newcommand{\Primers}{{\color{colorTech}{\textbf{Primers}}}}

\newcommand{\s}{\hspace{0.8pt}}

\expandafter\def\expandafter\normalsize\expandafter{%
    \normalsize
    \setlength\abovedisplayskip{8pt}
    \setlength\belowdisplayskip{8pt}
    \setlength\abovedisplayshortskip{8pt}
    \setlength\belowdisplayshortskip{8pt}
}

\usepackage{feynmp-auto}
\titleformat{\section}{\center\normalfont\fontsize{14}{15}\bfseries}{\thesection.}{1em}{}
\titleformat{\subsubsection}{\center\normalfont\fontsize{12}{15}\scshape}{\thesubsubsection.}{1em}{}

\begin{document}
$\quad$
\vskip 65 pt

\title{
\Large As Scales Become Separated:\\[-10pt]
Lectures on Effective Field Theory
} 

\author{
Timothy Cohen\\
\emph{\small Institute for Fundamental Science \\[-8pt]
Department of Physics \\[-8pt]
University of Oregon \\[-8pt]
 Eugene, OR, 97403}
}

\begin{abstract}
\vskip 30 pt
\begin{center}
{\bf Abstract}
\end{center} 
\vskip -30 pt
$\quad$
\begin{spacing}{1.05}\noindent
These lectures aim to provide a pedagogical introduction to the philosophical underpinnings and technical features of Effective Field Theory (EFT).  Improving control of $S$-matrix elements in the presence of a large hierarchy of physical scales $m \ll M$ is emphasized.  Utilizing $\lambda \sim m/M$ as a power counting expansion parameter, we show how matching an ultraviolet (UV) model onto an EFT makes manifest the notion of separating scales.   Renormalization Group (RG) techniques are used to run the EFT couplings from the UV to the infrared (IR), thereby summing large logarithms that would otherwise reduce the efficacy of perturbation theory.  A variety of scalar field theory based toy examples are worked out in detail.  An approach to consistently evolving a coupling across a heavy particle mass threshold is demonstrated.  Applying the same method to the scalar mass term forces us to confront the hierarchy problem.  The summation of a logarithm that lacks explicit dependence on an RG scale is performed.  After reviewing the physics of IR divergences, we build a scalar toy version of Soft Collinear Effective Theory (SCET), highlighting many subtle aspects of these constructions.  We show how SCET can be used to sum the soft and collinear IR Sudakov double logarithms that often appear for processes involving external interacting light-like particles.  We conclude with the generalization of SCET to theories of gauge bosons coupled to charged fermions.  These lectures were presented at TASI 2018.   
\end{spacing}
\end{abstract}

\maketitle
\newpage
\pagebreak
\renewcommand{\baselinestretch}{0.89}\footnotesize
\tableofcontents
\renewcommand{\baselinestretch}{0.9}\normalsize
\pagebreak
\begin{spacing}{1.3}
\section*{0.~~Some Personal Remarks}
\addcontentsline{toc}{section}{\hspace{-4.2pt} 0\hspace{0.5pt}.\hspace{4.5pt}Some Personal Remarks}
\noindent Dear reader,\\
\indent I could not have been more thrilled to receive the invitation from Tracy Slatyer and Tilman Plehn to lecture at TASI 2018 on Effective Field Theory.   This initiated the exhilarating process of figuring out what I could possibly bring to this infinite\footnote{Pun intended.} subject.  I decided to focus on the techniques, with a central theme of scale separation -- these lectures provide a how-to guide for systematically improving your perturbative expansion in the presence of a hierarchy of scales.  We are going to do a lot of loop integration, we are going to worry about matching corrections, we will take finite integrals and split them into many infinite parts, and we will dimensionally regulate ourselves out of a number of sticky situations.

A goal of these lectures is that you will deepen your understanding of what we mean by phrases like ``separation of scales,'' ``exponentiation of large logs,'' ``mapping the IR to the UV,'' and so on.  Another purpose is to (further) convince you that \emph{field theory works}.  No matter what shenanigans the Feynman diagram approach might pull to obscure the physics, the principle of decoupling and the implications for renormalization group evolution remain.  As such, a patient student can learn to expose and sum large logarithms, thereby mastering the beautiful art of renormalization group improved perturbation theory.  Using a number of concrete examples, we will first explore how to sum UV logs.  Then we will move to SCET, which will give us a framework to sum the notorious Sudakov IR double log.

I chose to approach my charge by stripping away the trappings of real theories; essentially all of my examples are built out of scalar toys.  This makes the Feynman rules trivial, allowing us to focus our attention on simple one-loop integrals.  However, my cardinal offense is that I make no connections to experiment.  I hope you will find this approach enlightening and educational, but if not feel free to request a refund from Tracy and Tilman :-)\s.

I have leaned heavily on my community throughout the process of conceiving, calculating, and writing --  I encourage you to read my extensive acknowledgements.  Any deep insights in what follows are likely due to one of my brilliant colleagues, while any misconceptions or mistakes are mine alone.   Finally, if you come across anything buggy, please send me a message so I can fix it in a future version.  Thanks for going on this journey with me!

\begin{center}
\vspace{7pt}
\includegraphics[width=0.3\textwidth, valign=c]{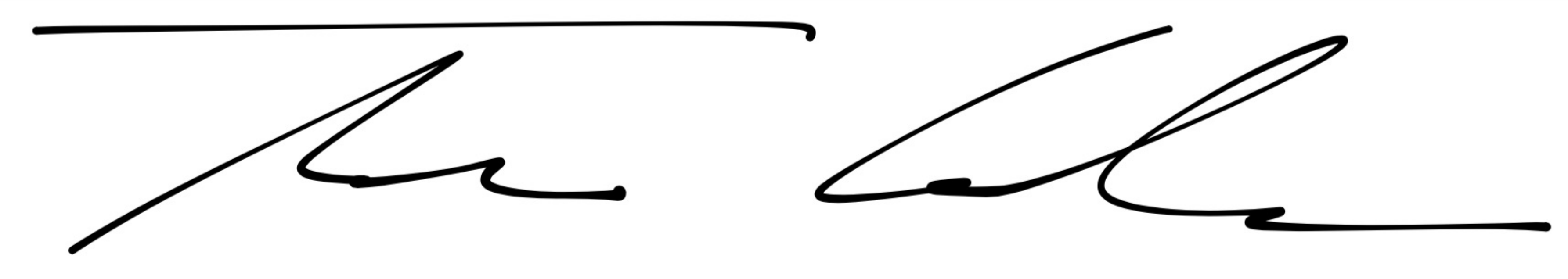}
\vspace{7pt}
\end{center}

\noindent \textbf{\#tl;dr} This project took over my life for awhile, and I'm excited to share it with you. 

\newpage

\section{The View from the Deep IR}
We begin with the big picture.  The purpose of these lectures is to systematically explore quantum field theory\footnote{A straw person might posit that a quantum field theory must be well defined in the UV, while an EFT comes with an associated scale signifying the need for a UV completion.  Specifically, EFTs are often viewed as an expansion that allows for higher dimension operators, implying non-renormalizability.  We do not find this distinction useful here, and as such treat quantum field theory and EFT as identical concepts in these lectures.} in the presence of a large separation of physical scales $m \ll M$.  Feynman diagrams yield an expansion in a coupling(s) $\alpha$.  In the limit $m \ll M$, it is typical that this $\alpha$ expansion will generate terms which go as $\alpha^r \log^s(m/M)$, for some integers $r$ and $s$ (where $s$ is at most $2\s r$ for relativistic four dimensional field theories).  Then there exist regions of parameter space where these logs can become large enough to significantly reduce the quality of the convergence of perturbation theory.  The approach taken here for finding an improved perturbative expansion relies on matching an ultraviolet (UV) ``\s\FT\s''~model onto a judicious choice of Effective Field Theory (EFT) description of the physical system of interest.  This allows one to break apart the large logarithms into single scale pieces and to sum them using renormalization group equations (RGEs).\footnote{\textbf{Disclaimer:} Through these lectures, we have chosen to use the phrase ``summing logarithms'' to describe what is accomplished by integrating the RGEs, in contrast with the common choice to call this ``resumming.''}  This connects the UV multi-scale theory to an infrared (IR) theory that is only aware of the dynamics associated with $m$.  Along the way to systematically setting up the technology that achieves this goal, we will encounter general lessons relevant to many applications within particle physics, condensed matter, cosmology, and more. 

There are two main goals of the EFT program: improving a given UV theory (calculational) and characterizing a physical system systematically (conceptual).  Our focus here will be on the computational aspects of EFTs.  However, it is worth emphasizing that central to the conceptual side of the EFT philosophy is that essentially any system whose dynamics are due to the small fluctuations of some underlying modes can be described as an EFT (see \hyperlink{sec:EFTZoo}{Appendix A} for more details and references).   Physicists are all indoctrinated from birth with the notion that physics is essentially dimensional analysis and the judicious application of Taylor expansions.  The beauty of EFTs is that they make these concepts extremely plain through a notion of ``power counting.''  Power counting allows one to generalize the idea of dimensional analysis by hierarchically organizing the allowed operators, while simultaneously providing a small parameter which can be used to Taylor expand away higher order corrections.  The relation to dimensional analysis should be familiar from previous experience organizing a relativistic Lagrangian in terms of the mass dimension of the operators, which also provides a polynomial expansion of the observables.  Here we will generalize this expansion to loop integrals, by relying on an awesome technique known as the method of regions.  Furthermore, EFTs are shockingly smart in that they both predict how many free parameters are required to compute to a desired accuracy, and that they tend to show signs of inconsistencies when pushed outside their regime of validity.  

In these lectures, we will devote our attention to an amazing aspect of EFTs, namely their ability to model physics across a huge range of physical scales through the application of renormalization group (RG) techniques.\footnote{For emphasis, we repeat that the immensely important conceptual aspects of EFTs and the application to modeling real physical systems are not the focus here.  Specifically, the reader should be aware that EFTs are not just for summing logarithms, \emph{e.g.} Fermi theory, heavy quark effective theory, etc.} We assume the reader is familiar with integrating a set of RGEs to evolve a \FT~description to the IR scale that is characteristic of the physical process of interest, followed by an extraction of observables.  This is known as RG improved perturbation theory.  However, if one encounters a multi-scale problem, large logarithms can in general remain.  The RG approach can be extended by invoking the concept of matching between the \FT~and an appropriate EFT that captures the dynamics of the propagating IR degrees of freedom.  Matching will allow us to systematically model more difficult multi-scale problems using one or more simpler single scale theories.  In other words, we will learn to  interpret the EFT as the single scale IR field theory, whose Lagrangian is constrained by whatever symmetries remain, and whose operator structure is organized as an expansion in terms of the power counting parameter.  This allows one to Taylor expand away as much complexity as possible, while retaining the essential features of the system.  

This multi-scale problem is central to these lectures, so we should explain what we are referring to in a bit more detail.  The Feynman diagram expansion at loop level introduces a spurious unphysical RG scale $\mu$.  Then one can derive a set of RGEs by requiring that observables are independent of $\mu$, which allow one to maintain control of these parameters when evolving them to different scales.  However, when dealing with a multi-scale theory, the Feynman diagram expansion can yield logarithms that do not explicitly depend on $\mu$, and so it is no longer obvious how to derive the appropriate RGEs.  For example, if one performs final state cuts, these parameters can appear inside of logarithms, which can in principle be large.  This is where EFTs, augmented with matching and running, come in to save the day by peeling apart these problematic logs into multiple single-scale contributions.  Furthermore, the action of separating scales inside the logarithm requires the introduction of a new dimensionful parameter.  It is this parameter that can be identified with an RG scale $\mu$, allowing one to simply apply the standard techniques to sum the now separated logarithms.  Furthermore, this approach fails unless the appropriate EFT model of the IR dynamics has been identified, thereby providing physical insight into what degrees of freedom persist to the IR.
 
One non-trivial aspect of the power-counting expansion procedure is the fact that perturbation theory is plagued with infinities, and that the appropriate regulation of these infinities yields logarithms.  In fact, one can prove that the Feynman diagram expansion can only yield powers and logarithms of the power counting parameter~\cite{Slavnov:1973wz}.  Powers are easy to interpret, and they tend to play well with the order of limits issue inherent when choosing to Taylor expand an integrand versus an integrated result.  Physically, this can be traced to the fact that power corrections that emerge from the loop expansion are localized at a scale.  On the other hand, logarithmic corrections arise when an integral receives contributions across all scales.  This feature of logarithmic corrections is straightforward to see by evaluating a simple quadratically divergent integral using a cutoff regulator in the UV:
\begin{align}
i\int \frac{\D^4 \ell}{(2\s\pi)^4} \frac{1}{\ell^2 - m^2} & = \frac{1}{8\s \pi^2} \int_{0}^{\Lambda_\text{UV}} \D \ell\,\frac{\ell^3}{\ell^2+m^2}\notag\\[8pt]
& = \frac{1}{16\s \pi^2}  \left(\Lambda_\text{UV}^2 - m^2 \log\left[\frac{m^2+\Lambda^2_\text{UV}}{m^2}\right]\right)\,,
\label{eq:preTaylorExpand}
\end{align}
where we Wick rotated and integrated over angles in the second step, and this integral is IR finite so we can take the lower limit of integration to zero.  Next, we imagine that we are interested in taking $\Lambda_\text{UV} \gg m$, and so you might think that we can Taylor expand the integrand assuming small $m$, which yields
\begin{align}
&\frac{1}{8\s\pi^2}\int_{0}^{\Lambda_\text{UV}} \D \ell \,\ell^3\,\frac{1}{\ell^2} = \frac{1}{16\s \pi^2}\, \Lambda_\text{UV}^2\notag\\[10pt]
&\frac{1}{8\s\pi^2}\int_{\Lambda_\text{IR}}^{\Lambda_\text{UV}} \D \ell \,\ell^3\left[- \frac{m^2}{\ell^4} + \frac{m^4}{\ell^6}+ \cdots \right] \notag\\[10pt]
&\hspace{50pt}= \frac{1}{16\s \pi^2}\left[ m^2 \log\frac{\Lambda_\text{UV}^2}{\Lambda_\text{IR}^2} +m^4\left( \frac{1}{\Lambda_\text{IR}^2}-\frac{1}{\Lambda_\text{UV}^2}\right) + \cdots \right]\,.
\label{eq:postTaylorExpand}
\end{align}
where now the second (and higher order) integrals are IR divergent, so we introduce an IR cutoff regulator $ \Lambda_\text{IR} \ll m$.  So we see that the analytic properties in the IR for our expanded integral do not track that of the full integral.\footnote{If we did this kind of manipulation using dimensional regularization, we would see that the expanded approach yielded only scaleless integrals which vanish, making the contrast between the two ``theories'' even more stark.  A detailed discussion of scaleless integrals and dimensional regularization is provided in \cref{sec:dimreg} below.}  This mismatch of the IR divergences is a characteristic of working with the wrong EFT description, and we will go to great lengths to ensure that the IR of our \FT~and EFT agree.\footnote{Another feature of this example is that the logarithmic dependence on the physical scale $m$ is naively lost by using this incorrect approach to expanding in $\ell/m$.  However, one could recover the full integral by performing this expansion to all orders and then resumming the integrated results into a logarithm.  Then the limit $\Lambda_\text{IR} \rightarrow 0$ would be finite and we would recover \cref{eq:preTaylorExpand}.}

As we will emphasize many times in what follows, matching and running provides a methodical approach that will allow us to Taylor expand before integration.  Additionally, we can augment this procedure using an interpretation in terms of an EFT operator expansion organized by the power counting parameter.  Imagine that the \FT~includes two dimensionful scales $m \ll M$, so that our power counting parameter is $\lambda \sim m/M$.  Then one of the key aspects of matching between the two theories is that we want to construct a description defined just below the scale $M$  such that the EFT parameters are fully determined by the UV \FT~to arbitrary precision, and are furthermore insensitive to the details of the IR.  This implies that we must match between the \FT~and the EFT in such a way that the description of the physics at $M$ is an analytic function of $m$ (for example, the EFT coefficients in the UV cannot depend on $\log m^2/\mu^2$).  This is critical since we need to be able to take these light parameters to zero without causing any troublesome (non-analytic) behavior in the UV.  Otherwise, the UV would know about the IR, \emph{i.e.}, a breakdown of decoupling.  If we want a local field theory description, we must judiciously apply the methods detailed in these lectures to separate scales and maintain perturbative control of our EFT.  The magic happens if we can find the right EFT description (and can be strong enough to keep track of all the various factors of two that appear). Decoupling will save the day and yield a local improved perturbative expansion.

These lectures are organized around two settings.  The first relies on a variety of simple relativistic scalar field theories, with a light field of mass $m$ and a heavy field of mass $M$.  Depending on the details of the interactions, a variety of interesting physical effects emerge.  After discussing the systematics of constructing EFTs and matching at tree level, we will show how to account for the loop-level decoupling of heavy particle contributions to running couplings, and we will see how the hierarchy problem appears in models with a large ratio of scales (specifically we will clarify how to think about this issue from an EFT point of view using dimensional regularization).  We expect that much of what is presented in this first part will be familiar to the reader.  We will then move to the cornerstone of this section, which is to show how we can separate scales that result from a loop with both a heavy and a light propagator (see \cref{eq:SepScalesHLlog}).  Another benefit of first working within this more familiar setting is that it provides us with the opportunity to frame setting up EFTs in a language that easily generalizes.  

The second main topic is to explore an EFT that captures the soft and collinear divergence structure of relativistic quantum field theory.  When there are light-like external states with restrictions on the external particle phase space, these divergences can manifest as the generic phenomenon of the Sudakov double logarithm.  We will focus on setting up a scalar theory that realizes a large Sudakov double log in the IR, and then will use the techniques of soft collinear effective theory (SCET) to separate the scales inside this double log and sum it.  Learning SCET is useful due to its practical relevance to a wide variety of applications to real world processes (see the discussion in \cref{sec:Conc} below).  Additionally, it is of tremendous pedagogical value to work out this example, since dealing with the highly non-trivial mode structure of SCET will expose many of the issues that arise when attempting to separate scales.  We will close by making connections between the toy SCET theory and the realistic version that is relevant for gauge theories with charged matter.  Then in a final concluding section, we will briefly touch on some physical applications of (Goldstone boson) EFTs and SCET.  Finally, note that a variety of \Primer~sections are provided along the way which include technical details and conventions -- while they are necessary for getting the right factors of two, we have chosen to sequester them as their concepts are likely familiar to the reader and could be skipped in a quick first pass reading of these lectures.

Before leaving the introduction, we will provide a general discussion of the scalar fields and interactions that will be used throughout what follows.  This will allow us to normalize notation and emphasize some aspects of our approach using toy model EFTs.

\subsection{The Scalar Playground}
As has already been emphasized, the goal is to focus our attention on the EFT techniques of matching and running.  Essentially all of the physics of interest\footnote{The only example that will fall outside this set of toy models (where we will need to introduce an extra scalar state $\chi$) will appear briefly in \cref{sec:RegionsMassiveSudakov}.} can be exposed in the context of simple toy theories\footnote{Our scalar examples do yield a few minor idiosyncrasies, which will be highlighted as they are encountered.} built out of a light real scalar field $\phi$ and a heavy real scalar field $\Phi$.  

We will define the ``\s\FT\s'' using a Lagrangian that provides a good description above a scale $\mu > M$.  Our various \FT~models will utilize a subset of the following terms
\begin{align}
\mathcal{L}^\textsc{Full}_{\text{Kin}} &=  \frac{1}{2} \big(\partial_\mu \phi\big)\big(\partial^\mu \phi\big)  - \frac{1}{2}\s m^2\, \phi^2 + \frac{1}{2} \big(\partial_\mu \Phi\big)\big(\partial^\mu \Phi\big) - \frac{1}{2}\s M^2\, \Phi^2 \notag\\[8pt]
\mathcal{L}^\textsc{Full}_{\text{Int}} &=-\frac{1}{3!}\s a\,\phi^3- \frac{1}{2}\s b\,\phi^2\, \Phi - \frac{1}{4}\s \kappa\,\phi^2\,\Phi^2\,- \frac{1}{3!}\s \rho\,\phi^3\,\Phi\,-\frac{1}{4!}\s \eta\,\phi^4.
\label{eq:ToyFullLagrangian}
\end{align}
Note that we are already breaking the EFT rules in the \FT~(before we have even stated them properly, see \cref{sec:PowerCountingSyms}) in that we have set many terms to zero that are allowed by the symmetries, \emph{e.g.}~$\phi$, $\Phi^4$, $\phi\,\Phi^3$, and so on.  Furthermore, as you will see below, we are only going to analyze the impact on the IR dynamics from turning on one or two interaction terms at a time.  We are not claiming at any point to have completely captured all of the physics implied by the \FT, but are simply using this toy approach to explore the scalar one-loop integral structures of interest.  Everything that will be done in what follows is self consistent, it is just not generic. 

Our goal will be to match a given \FT~onto a low energy EFT.  We do not write down a general form for the EFT Lagrangian here, since the symmetries and power counting depend on the interaction structure of the \FT, along with the chosen process we want to compute; as we will see, our relativistic EFTs look very (very) different from SCET.  

We will quickly make contact with a simple example so that the path forward is clear.  Assume that there are two interactions, $\mathcal{L}^\textsc{Full}_{\text{Int}} =- \frac{1}{4}\s\kappa\,\phi^2\,\Phi^2 - \frac{1}{3!}\s \rho\, \phi^3\,\Phi$, which have the associated Feynman rules
\begin{align}
\includegraphics[width=0.13\textwidth, valign=c]{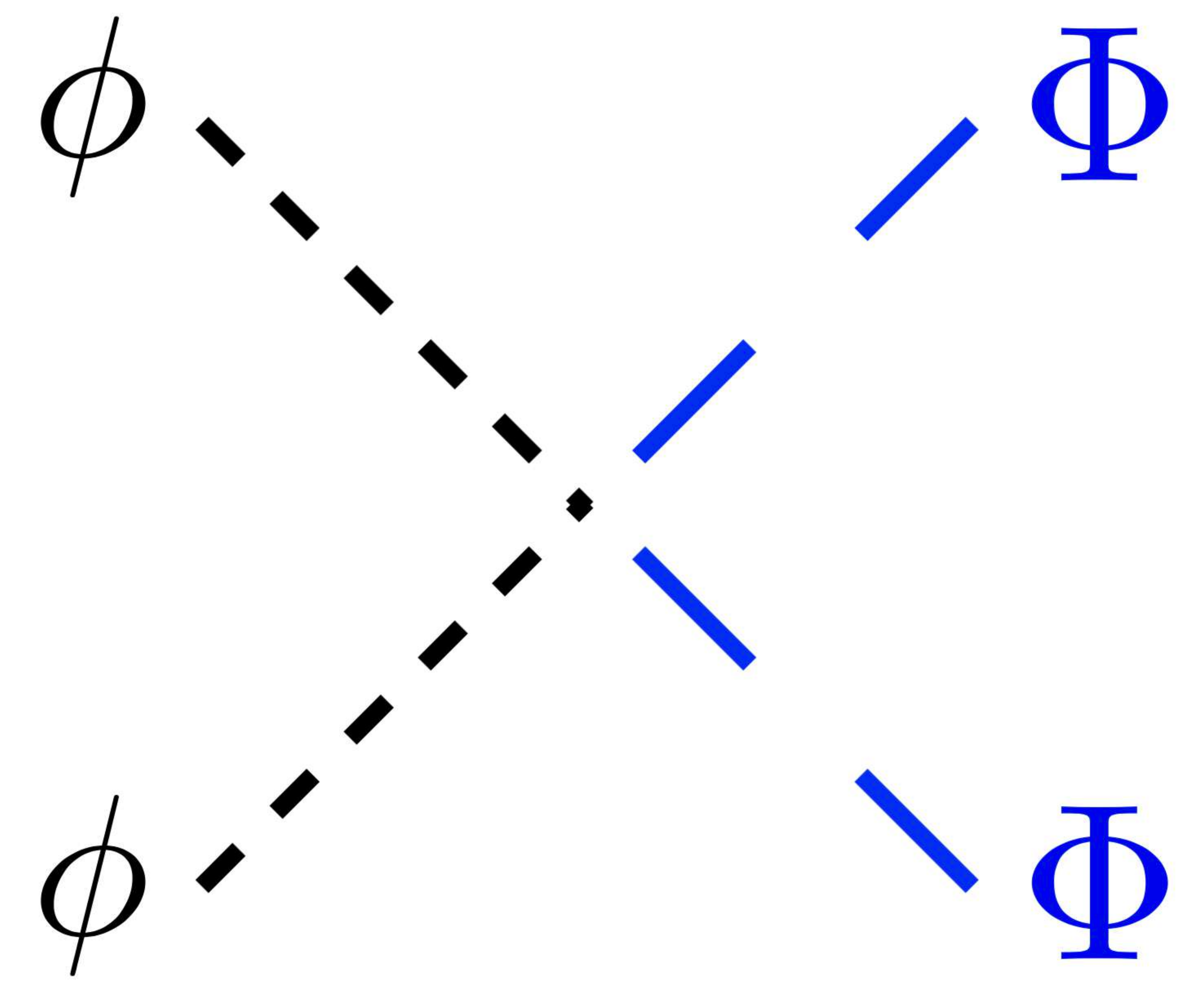} = -i\s \kappa \qquad \qquad \qquad \includegraphics[width=0.14\textwidth, valign=c]{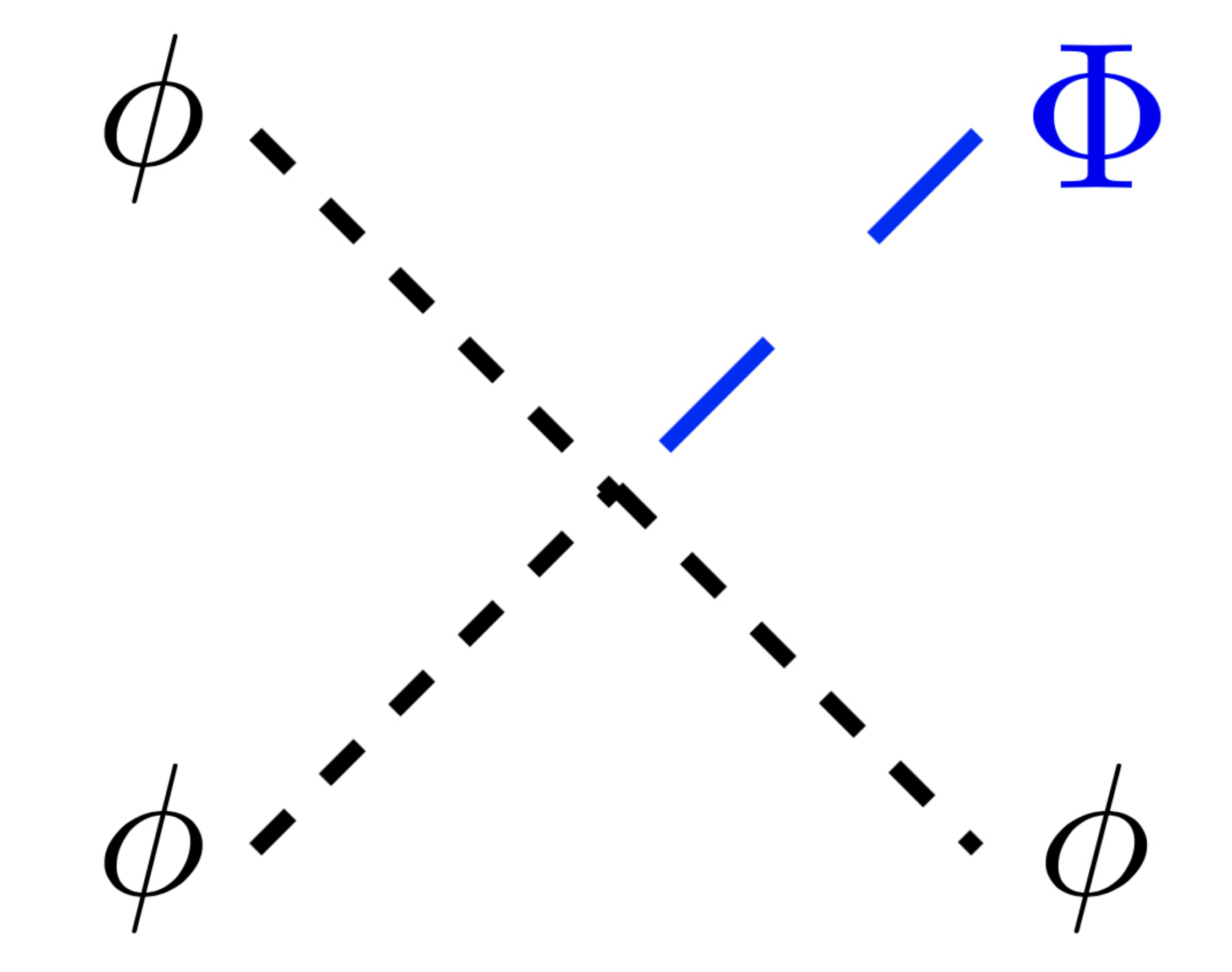} = -i\s \rho \,.
\end{align}
Then we can match this theory at a scale $\mu \sim M$ onto an EFT with only $\phi$ as a propagating degree of freedom.  We emphasize that technically this is a different $\phi$ field than the one in the \FT.  Diagrammatically, we see that $\mathcal{L}^\textsc{Full}_{\text{Int}}$ induces a tower of local contact operators in the EFT as one flows to the IR.  For example, at tree-level in the \FT~we have diagrams like
\vspace{-5pt}
\begin{align}
\hspace{5pt}\includegraphics[width=0.24\textwidth, valign=c]{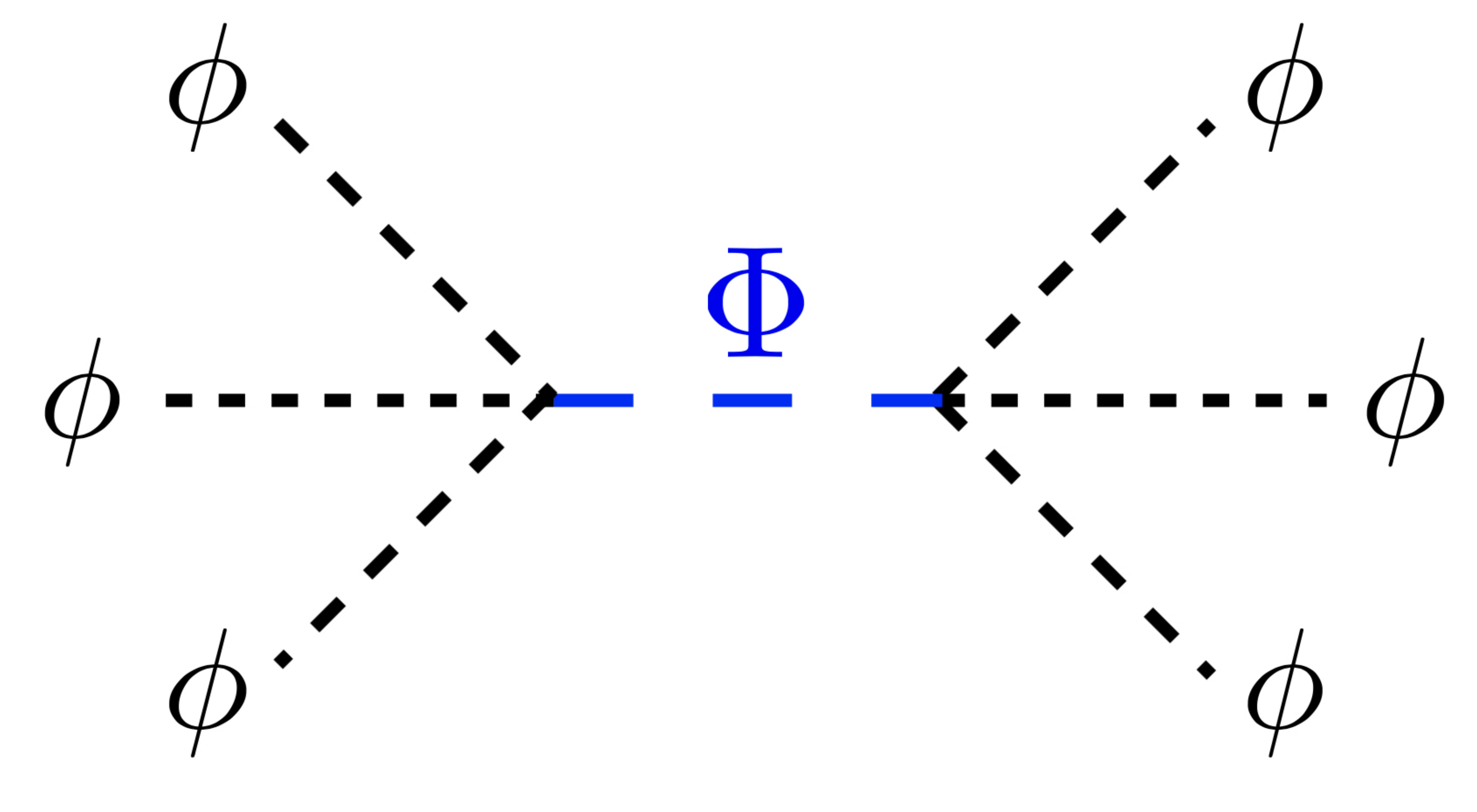} \hspace{80pt} \includegraphics[width=0.18\textwidth, valign=c]{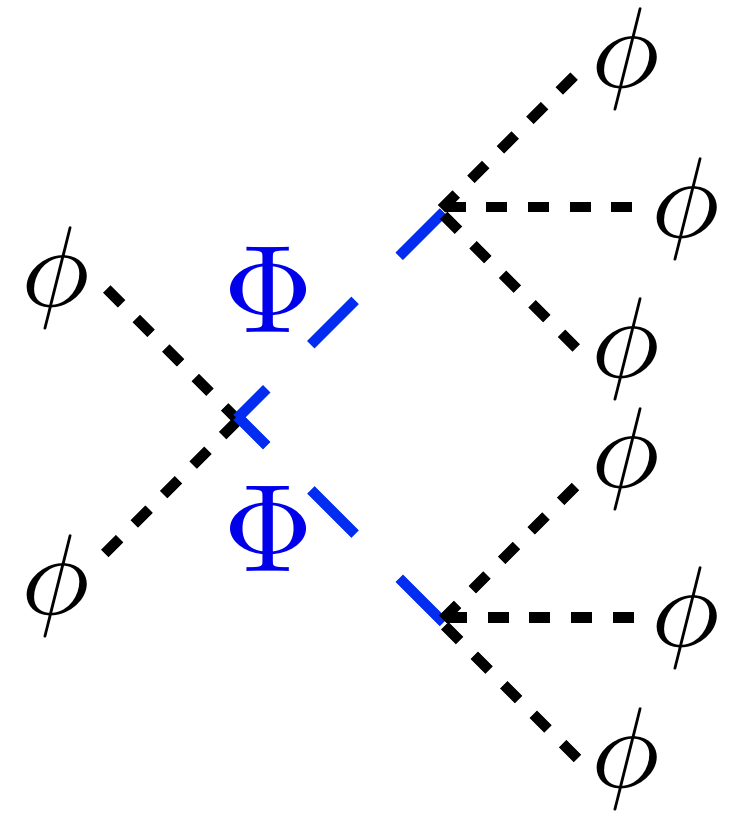} \,,\\[-30pt]
\notag
\end{align}
while at loop level, we have diagrams like
\begin{align}
\hspace{15pt}\includegraphics[width=0.22\textwidth, valign=c]{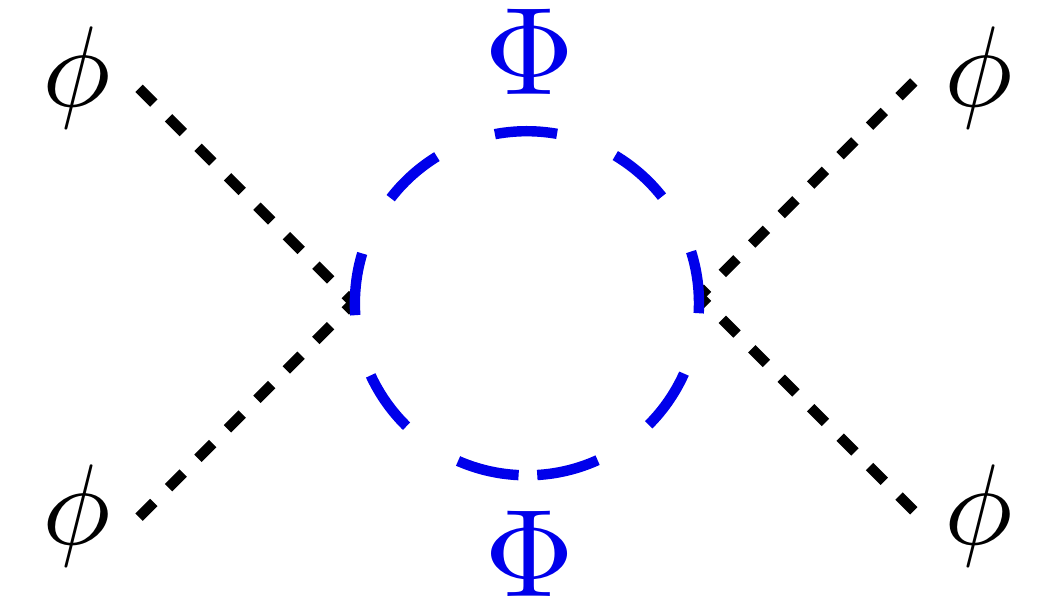} \qquad \qquad \qquad \includegraphics[width=0.22\textwidth, valign=c]{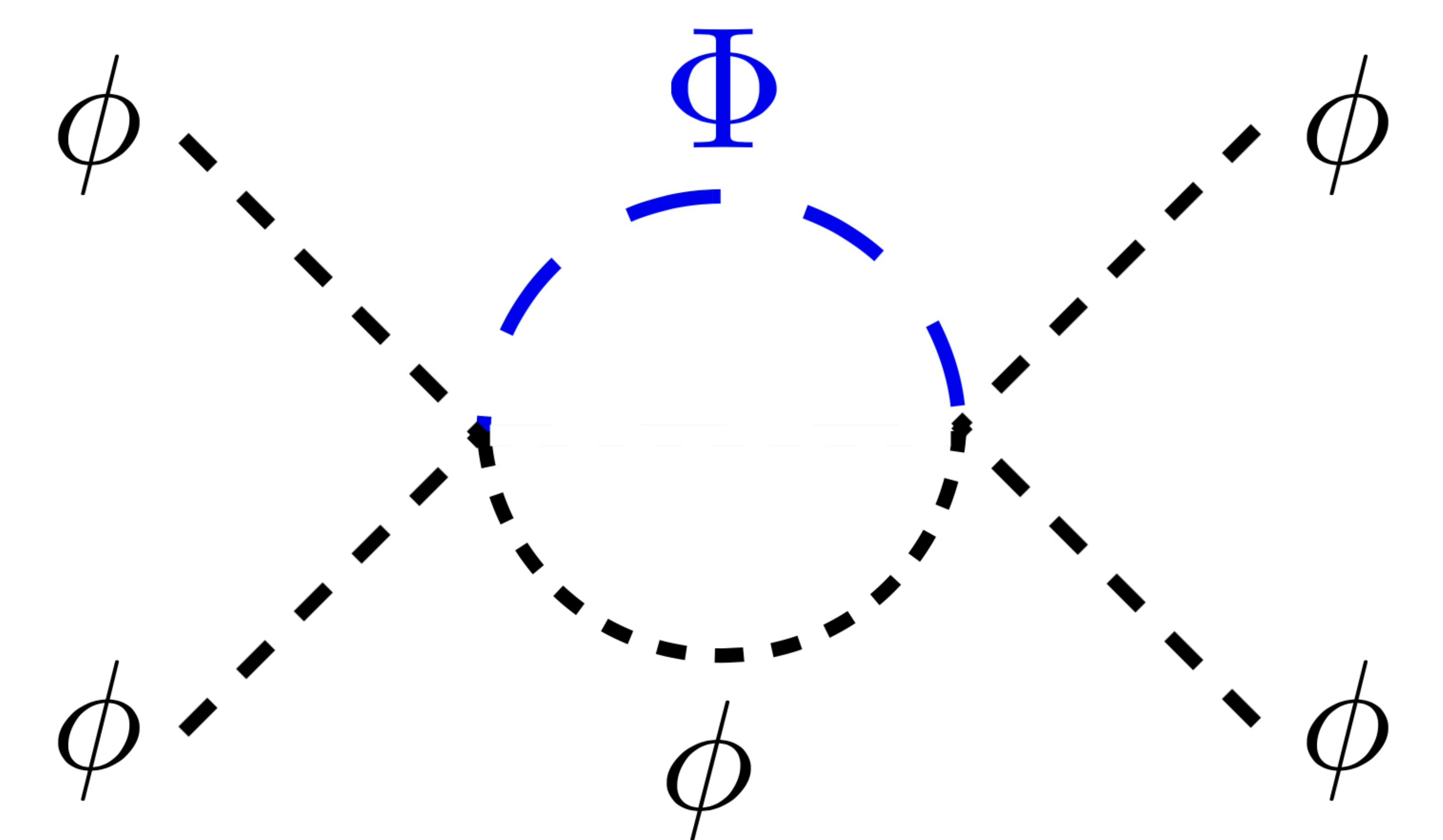} \,.
\end{align}
All of these processes can be modeled using the EFT Lagrangian
\begin{align}
\mathcal{L}^\text{EFT} =  \frac{1}{2} \big(\partial_\mu \phi\big)\big(\partial^\mu \phi\big)  - \frac{1}{2}\s m^2 \phi^2 -\frac{1}{4!}\s C_4 \,\phi^4 -\frac{1}{6!}\frac{1}{M^2}\s C_6\,\phi^6 -\frac{1}{8!}\frac{1}{M^4}\s C_8\,\phi^8 + \cdots\,,
\end{align}
where we are neglecting higher dimension operators involving derivatives. Determining how to map the \FT~parameters $\kappa$ and $\rho$ onto the EFT parameters $C_k$ is exactly the kind of matching calculations we are going to perform in detail in what follows.  Note that the non-genericity of the UV theory leads to $C_4$ being generated at loop level, while $C_6$ and $C_8$ are generated by tree diagrams.  If we had included $\eta\, \phi^4$ in the UV, then $C_4$ would clearly be generated at tree level as well.  Also, note that there is a $\mathbb{Z}_2$ symmetry in the UV where $\phi \leftrightarrow -\phi$ and $\Phi \leftrightarrow -\Phi$ simultaneously, which implies that no $\phi$-odd terms are generated in the IR (assuming no vevs are induced by loop effects).

Our hope is that this discussion has established the context of these lectures, and motivated the reader to chase us down the rabbit hole.  However, for the unconvinced, we provide two appendices with more information and resources regarding the study of EFTs. \hyperlink{sec:EFTZoo}{Appendix~A}~provides a lightning fast overview of some important EFTs that appear in nature.  These examples highlight physical systems where the technical apparatus we are going to develop can be applied.  Then in \hyperlink{sec:AnnBib}{Appendix~B},~we provide an annotated bibliography to help the reader navigate the extraordinary resources that already exist on the topic of EFTs, and in particular SCET.   For those who are not interested in this detour, we turn to the first topic of these lectures.  Our starting point is to explain how to utilize a notion of power counting to construct an EFT Lagrangian that additionally respects some set of symmetries relevant to the physical process of interest.

\newpage
\section{Power Counting and Symmetries}
\label{sec:PowerCountingSyms}
We begin by setting up the fundamental rules for building an EFT.  The goal is to write down a field theoretic description of the modes that continue to propagate as one flows to the IR.  Given these ``light'' degrees of freedom, one is then tasked to construct a kinetic term that models their fluctuations, and to write down all allowed interactions, ideally following some organizing principles.  In these notes, we will emphasize situations where we not only know the \FT, but it can be described perturbatively at the high scale where we match onto the EFT.  This allows us to utilize a matching procedure to determine the Wilson coefficients at this matching scale, and then we can use the RGE to evolve the EFT down to the low scale, thereby summing large logarithms and improving the convergence of perturbation theory.  

The two central concepts are ``power counting'' and ``symmetry.''  The idea of enforcing a symmetry should be familiar to the reader from their background in quantum field theory, and as such we will assume a working knowledge of this topic.  Power counting generalizes the idea of organizing a Lagrangian as an expansion in operator mass dimension.  With an appropriate identification of a power counting parameter $\lambda$, we perform this organization by tracking an operator's order in $\lambda$.  For example, $\lambda \sim m/M$ will be our power counting parameter for the example provided in \cref{eq:SepScalesHLlog} below.  First, we will explain the rules for constructing and organizing the EFT Lagrangian, followed by their application to a number of illustrative examples.

\subsection{The Rulebook}
\label{sec:TheRulebook}
If we are interested in probing the physics around energies $E \sim m$, it is relatively straightforward to calculate in field theories that contain only the single scale $m$.  However, our desire to model processes in the real world often forces us to solve more complicated multi-scale problems.  Therefore, it would be ideal to develop an approach for expanding away scales such that we are left with a single-scale EFT to work with.   As we will see, this hope is realizable when we are working with theories whose  scales are hierarchical.   This procedure will be organized around a power counting parameter $\lambda \sim m/M$ for $m \ll M$, such that observables are a Laurent expansion\footnote{It is a Laurent expansion since inverse powers of $\lambda$ can also appear.} in terms of $\lambda$.  

We want a systematic procedure that avoids the issues raised above when we compared the unexpanded \cref{eq:preTaylorExpand} with the improperly expanded \cref{eq:postTaylorExpand}.  We can assume that there are some other couplings $\alpha$ in the theory, such that we will ultimately perform a dual expansion in $\lambda$ and $\alpha$.  Logarithms will generically be generated by the $\alpha$ expansion.\footnote{Note that for an EFT like Heavy Quark Effective Theory, $\alpha$ is non-perturbative leading to the so-called incalculable ``brown muck.''  However, it is not only possible to show that these contributions factorize from the hard process, but one can also power count the non-perturbative corrections.  This implies that to a given power, only a finite number of non-perturbative parameters are required to predict an (in principle) infinite number of observables.}  When we encounter a term that schematically takes the form $\alpha^r\log^s \lambda$ for some integers $r$ and $s$, we will power count it as $\mathcal{O}(1)$, which tells us that if you want to improve the convergence of the perturbative expansion, you must sum it to all orders.  

The whole framework rests on the assumption that physics is local, in that poles in the $S$-matrix can only be due to light particles going on-shell.  This implies that we should be able to integrate out heavy modes and encode their influence on the low energy physics as an expansion in local interactions of the light fields suppressed by powers of $1/M$, or equivalently suppressed by additional powers of $\lambda$.  Assuming we have perturbative control of the \FT, we can ``match'' it onto an EFT as long as the IR limits for both reproduces the same physics.  The matching and running procedure is defined by~\cite{Georgi:1991ch} 
\begin{enumerate}[label=\roman*)]
\item Pick a physical process $\sigma$ that only has IR degrees of freedom as external states.
\item Compute the couplings in the EFT at a matching scale $\mu_M \sim M$ by relating the calculation of $\sigma$ in the \FT~and the EFT
\begin{align}
\delta \sigma_\text{EFT} = \sigma_\textsc{Full} - \sigma_\text{EFT}\,.
\label{eq:matchingSchematic}
\end{align}
and extracting a relationship between the couplings to set $\delta \sigma_\text{EFT}$ to zero.
\item Evolve the EFT Wilson coefficients down to a low scale $\mu_L \sim m$ using the RGEs derived within the EFT.
\item Compute the RG improved observable using the low energy EFT parameters. 
\end{enumerate}
The choice to match at a scale $\mu_M \sim M$ is made to minimize the logs that appear in the matching procedure.  Then $\delta \sigma_\text{EFT}$ will have a well defined Taylor expansion in terms of the power counting parameter $\lambda$.  Notice that we are ignoring mass dimension, contrary to the way one traditionally orders operators.  A term in the Lagrangian should be multiplied by appropriate powers of $M$ to yield a mass dimensionless action (when using natural units).  It can be helpful to interpret this $M$ as defining the ruler by which everything else is measured.  In fact, EFT practitioners will often work in units where the dimensionful heavy scale $M = 1$, since all that matters is the power counting.

As a concrete example, we can consider the low energy limit of a toy theory with a light scalar $\phi$ constructed using \cref{eq:ToyFullLagrangian}.  We want to probe the theory at a low scale $s_{ij}/M^2 \sim \lambda^2$, where $s_{ij} = (p_i+p_j)^2$ are the Mandelstam invariants, in the parameter space where the momenta are small so that $\lambda \sim m/M \ll 1$.  In order to devise a power counting scheme that respects Lorentz invariance, we need all the components of the $\phi$ momentum $p_\mu$ to scale homogeneously
\begin{align} 
p_\mu \sim M \big(\lambda,\lambda,\lambda,\lambda\big) \sim m \big(1,1,1,1\big)\,,
\label{eq:pExamplePowerCounting}
\end{align}
which implies that $p^2 \sim M^2\, \lambda^2 \sim m^2$.  Note that if $p_\mu$ scales non-trivially with power counting, then the canonically conjugate variable $x_\mu$ must also scale such that our quantum mechanical canonical commutation relations are unsuppressed: $[x,p] = i \sim \mathcal{O}(1)$.  This means that the power counting for $x$ scales as the inverse of the power counting for $p$.  In our example, \cref{eq:pExamplePowerCounting} implies that $\D^4 x \sim 1/\lambda^4$.  

After a brief detour where we define many of the conventions that will be used throughout, we will construct the unique kinetic term for our light scalar $\phi$, by relying on symmetry and power counting arguments alone.  Then we will address interactions and local operators to complete our picture of the EFT Lagrangian structure.

\scenario{Conventions}
\addcontentsline{toc}{subsection}{\color{colorTech}{{Primer~\thescenario.} Conventions}}
\label{tech:Conventions}
One goal of these notes is to make the source of subtle minus signs and factors of two we will encounter as obvious as possible.  To this end, before we get our hands dirty with some actual calculations, we should establish some conventions.  If ever in doubt, assume we are following the conventional choices made in~\cite{Schwartz:2013pla}, \emph{except} that we will define dimensional regularization with $d = 4 - 2\s \epsilon$.  Obviously, $\hbar = c = 1$, and we will take the standard Fourier transform convention that the measure in momentum space is $\D^4 p/(2\s\pi)^4$.  We use the mostly minus metric $g_{\mu\nu} = \text{diag}(+1,-1,-1,-1)$, and the index $\mu = 0,1,2,3 = t,x,y,z$.  

We will be computing $S$-matrix elements using momentum space Feynman diagrams.  Specifically, a Feynman diagram should be interpreted as $i\s\mathcal{A}$, where $\mathcal{A}$ is the amplitude or ($S$-)matrix element. The $i\s\mathcal{A}$ are derived by applying the LSZ reduction procedure to the time-ordered products of fields that can be extracted from the path integral using
\begin{align}
(-i)^n \frac{1}{Z[\s 0\s]} \frac{\partial^n Z[J]}{\partial J(x_1) \cdots \partial J(x_n)} \bigg|_{J=0} = \,\vev{0 | T \phi(x_1)\cdots \phi(x_n)| 0}\,,
\end{align}
where $Z[J]$ is the generating functional 
\begin{align}
Z[J] = \int \mathcal{D} \phi \exp\left(i\s S[\phi] + i\int \D^4 x\, J(x)\, \phi(x) \right)\,,
\label{eq:pathInt}
\end{align}
and $J(x)$ are external currents.  The action is computed using $S = \int \D^4 x\, \mathcal{L}$, which implies that $\mathcal{L}$ carries mass dimension four.

The Lagrangian for a free scalar field $\phi_0$ is
\begin{align}
\mathcal{L}_\text{Kin} = \frac{1}{2}\big(\partial_\mu \phi_0\big)\big(\partial^\mu \phi_0\big)-\frac{1}{2} m^2\,\phi_0^2\,,
\end{align}
from which one can derive the propagator 
\begin{align}
\vev{0\big|\s T\s\phi_0(x) \s\phi_0(y)\big|0} = \int \frac{\D^4 p}{(2\s\pi)^4} \frac{i}{p^2-m^2 + i0}\,e^{i\s p\s\cdot (y-x)}\,,
\label{eq:twoPtFn}
\end{align}
where the $i0$ factor is shorthand for taking the limit to zero from above -- this choice yields a time-ordered causal pole prescription.\footnote{We use the notation $i0$ instead of $i\epsilon$ since we will reserve $\epsilon$ for the regulator in dim reg with $d = 4-2\s\epsilon$.}
Then clearly the momentum space scalar Feynman propagators are 
\begin{align}
\includegraphics[width=0.13\textwidth, valign=c]{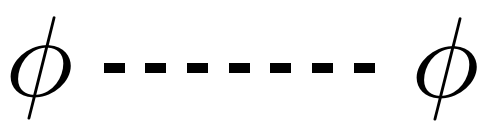}\,\, =\,\, \frac{i}{p^2 - m^2 +i0} \,.
\label{eq:Propagator}
\end{align}
We will often drop $i0$ unless we are evaluating a loop integral using contours.

Speaking of contour integration, it is worth stating our prescription for integrating a function $f(z)$ along a contour in the complex plane that does not cross any branch cuts:
\begin{align}
\oint_\gamma \D z f(z) = 2\s\pi \s i \sum_k \text{Res}(f,z_k)\,,
\end{align}
where the closed contour $\gamma$ is taken to be counterclockwise, $z_k$ are the poles contained within $\gamma$, and the residue of $f$ at $z_k$ is
\begin{align}
\text{Res}\big(f,z_k\big) = \lim_{z\rightarrow z_k} \Big[ \big(z-z_k\big)\,f(z)\Big]\,,
\end{align}
when the $z_k$ are simple poles.

Finally, vertices get a factor of $i$ from expanding the exponent of the path integral.  For example, if the interaction Lagrangian contains a term $\mathcal{L}_\text{Int} = -\kappa\,\phi^4/4!$, or equivalently $V = \kappa \,\phi^4/4!$, then
\begin{align}
\includegraphics[width=0.15\textwidth, valign=c]{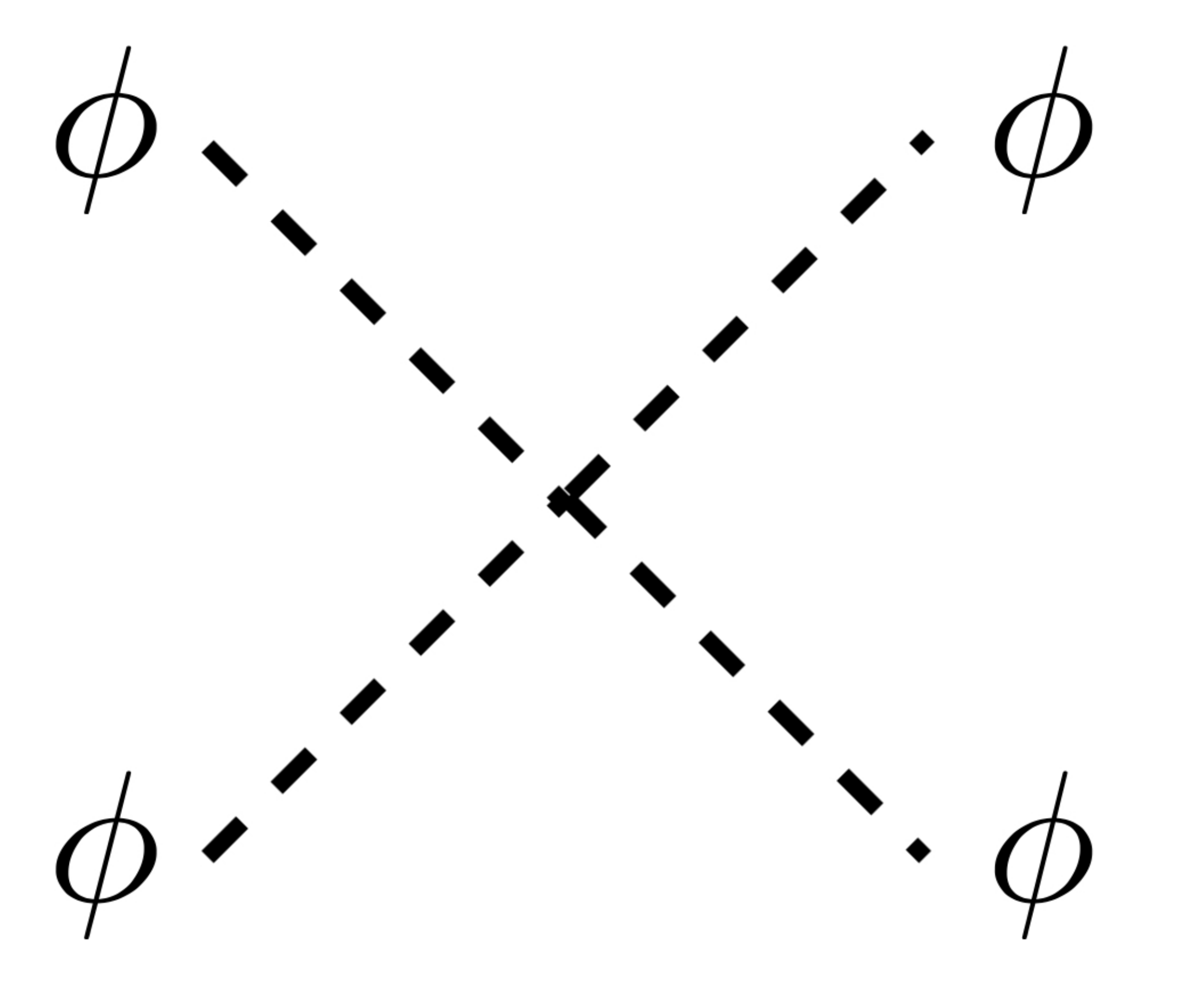} \,= \,-i\s \kappa
\end{align}
is the Feynman rule.

\subsubsection*{Gauge Theory}
We will not encounter gauge theory until \cref{sec:RealSCET}.  But since we are defining conventions, we might as well include these here.  The covariant derivative is given by
\begin{align}
D_\mu \phi_i = \partial_\mu \phi_i - i\s g\,A^a_\mu\,\big(T^a\big)_{ij}\,\phi_j\,,
\end{align}
where the generators satisfy the commutation relation
\begin{align}
\Big[T^a,T^b\Big] = i \s f^{abc}\,T^c\,.
\end{align}
The Yang-Mills Lagrangian is 
\begin{align}
\mathcal{L}_\text{YM} = -\frac{1}{4} \sum_a F_{\mu\nu}^a\,F^{a\mu\nu} =  - \frac{1}{4} \sum_a \big(\partial_\mu\,A^a_\nu - \partial_\nu\,A^a_\mu + g\,f^{abc}\,A^b_\mu\,A^c_\nu\big)^2 \,.
\end{align}
We will find it useful below to express the field strength tensor in terms of covariant derivatives:
\begin{align}
F_{\mu\nu} = \frac{i}{g}\Big[D_\mu,D_\nu\Big] \,,
\label{eq:FmunuFromDD}
\end{align}
where for the non-Abelian case, this should be interpreted as a matrix expression with $F_{\mu\nu} = F_{\mu\nu}^a\,T^a$ and $A_\mu = A_\mu^a\,T^a$.
Under a gauge transformation, 
\begin{align}
A_\mu(x) \quad \longrightarrow \quad &U(x)\,A_\mu(x)\,U(x)^\dag - \frac{i}{g} \,\big(\partial_\mu\s U(x)\big)\, U(x)^\dag\,.
\label{eq:YMgaugeTrans}
\end{align}
where 
\begin{align} 
U(x) = \exp\big(i\s \alpha^a(x)\,T^a\big)\,,
\label{eq:UxGaugeT}
\end{align}
and $\alpha^a(x)$ is the gauge transformation parameter.  In the infinitesimal limit, and expressing this in terms of the adjoint gauge field with explicit indices, this reduces to
\begin{align}
A_\mu^a(x) \quad \longrightarrow \quad &A_\mu^a(x) + \frac{1}{g} \,\partial_\mu\,\alpha^a(x) - f^{abc}\,\alpha^b(x)\,A_\mu^c(x)\,.
\end{align}
Then for charged matter
\begin{align}
\phi_i \quad \longrightarrow \quad U_{ij}\,\phi_j\,.
\end{align}

\subsubsection*{Field Redefinitions}
One immediate consequence of the path integral formulation of quantum field theory is that physical predictions are independent of field redefinitions~\cite{Chisholm:1961tha, Kamefuchi:1961sb}
\begin{align}
\phi \rightarrow \phi + \phi\,f\left(\frac{\phi}{\Lambda}\right)\,,
\end{align}
where $\Lambda$ is a dimensionful parameter.  This redefinition has no impact on observables since \cref{eq:pathInt} integrates over all configurations of $\phi(x)$, and the measure $\mathcal{D}\phi$ compensates any change of variables due to the Jacobian.  There is one additional effect, which is that this change of variables can introduce new couplings to sources, but this does not typically lead to any complications, see~\emph{e.g.}~\cite{Arzt:1993gz, Criado:2018sdb} for a discussion.

After performing a field redefinition, it is typically convenient to then redefine the fields such that their kinetic terms are diagonalized and canonically normalized.  This is done with an invertible matrix  
\begin{align}
\phi_i  \quad \longrightarrow \quad K_{ii'} \,\phi_{i'}\,,
\end{align}
where the $K_{ii'}$ matrix is derived by requiring that
\begin{align}
\hspace{-10pt}\mathcal{L} \supset \frac{1}{2}\, Z_{ij} \big(\partial_\mu \phi_i\big)\big(\partial^\mu \phi_j\big) \quad \longrightarrow \quad \mathcal{L} \supset\frac{1}{2}\,Z_{ij} K_{ii'}\,K_{jj'}\,\big(\partial_\mu \phi_{i'}\big)\big(\partial^\mu \phi_{j'}\big) = \frac{1}{2} \big(\partial_\mu \phi_i\big)\big(\partial^\mu \phi_i\big)\,,\hspace{-4pt}
\end{align}
where $Z_{ij}$ encodes the initial non-diagonal kinetic term.  

Additionally, it is typically useful to diagonalize the mass matrix, so that the propagators are fully diagonal.  Fortunately, if one is given a set of diagonal and normalized kinetic terms, there is still the freedom to rotate the fields with a unitary transformation (or an orthogonal transformation for the special case when $\phi$ is real)
\begin{align}
\mathcal{L} \supset - \frac{1}{2}\, m^2_{ij}\, \phi_i\,\phi_j \quad \longrightarrow \quad \mathcal{L} \supset-\frac{1}{2} \,U^{T}_{ii'} \,U_{jj'} \, m^2_{i'j'} \,\phi_{i'}\,\phi_{j'} = -\frac{1}{2}\, m_i^2\, \phi_i^2\,,
\end{align}
where $m_i^2$ is the diagonalized mass matrix.  This $U$ transformation does not induce any kinetic mixing terms since the unitary matrix passes through the diagonal kinetic terms.  Now that we have these terms in canonical form, one can still perform further field redefinitions that will only change the structure of interacting Lagrangian.  

\subsubsection*{Equations of Motion}
The principle of least action holds in quantum field theory as an operator statement:
\begin{align}
\vev{0\,\bigg|\frac{\partial \mathcal{L}}{\partial \phi} - \partial_\mu \frac{\partial \mathcal{L}}{\partial(\partial^\mu \phi)}\bigg|\,0} = 0\,.
\end{align}
It is straightforward to prove this by noting that the contribution to the path integral from a variation of the action $\delta S$ induced by a variation in $\delta\phi$ must vanish since $\phi \rightarrow \phi + \delta \phi$ is simply a field redefinition.  This implies that one can always apply the equations of motion to reduce the Lagrangian to a simpler or more useful form.  A demonstration of how this works practically at loop level can be found in~\cite{Arzt:1993gz}. 

\subsubsection*{Integration by Parts}
As always in quantum mechanics, integration by parts is extremely useful:
\begin{align}
f(\phi_i) \big[\partial_\mu\s g(\phi_i)\big] = -  \big[\partial_\mu\s f(\phi_i)\big] g(\phi_i)\,,
\end{align}
where the fields are assumed to vanish at $\pm \infty$.\footnote{Under the assumption that the field configuration does not carry any non-trivial topological structure.}

\vspace{5pt}\mybox{
\begin{itemize}
\item {\bf Exercise:}  We conclude this section with an amusing exercise that demonstrates these principles in action.  Starting with a free scalar theory, we make the field redefinition $\phi \rightarrow \phi + \phi^2/\Lambda$, where $\Lambda$ is some unspecified dimensionful parameter:
\begin{align}
\mathcal{L} =\,&\frac{1}{2}  \big(\partial_\mu \phi\big)\big(\partial^\mu \phi\big) - \frac{1}{2}\s m^2\,\phi^2+\frac{2}{\Lambda}\,\phi\,\big(\partial_\mu \phi\big)\big(\partial^\mu \phi\big) - \frac{m^2}{\Lambda}\, \phi^3 \notag\\[7pt]
&+ \frac{2}{\Lambda^2}\, \phi^2\, \big(\partial_\mu \phi\big)\big(\partial^\mu \phi\big) - \frac{m^2}{2\,\Lambda^2}\, \phi^4\,.
\end{align}
Now it naively appears that the amplitude for the process $\phi\,\phi \rightarrow \phi\,\phi$ is non-zero.  However, we know that it must be zero since we have simply performed a field redefinition of a free scalar.  Show that the amplitude for $\phi\,\phi \rightarrow \phi\,\phi$ is zero at tree-level by explicit calculation.
\end{itemize}}

\subsection{Constructing a Kinetic Term}
In this section, we will take an overly pedantic walk through the steps for constructing the kinetic term in the Lagrangian for our EFT scalar field $\phi(x)$.  First, it must be quadratic in the fields since the time ordered insertion of a field operator $\phi(x)$ followed by $\phi(y)$ yields the propagations of a particle in spacetime from $y$ to $x$.  One approach to deriving a notion of propagation is to start with a time foliation.  This user specified choice is tied to the definition of the Hamiltonian $H$, since it is this operator that generates time translations in quantum mechanics.  For a Minkowski frame, the obvious choice is the zero component of a spacetime vector.\footnote{When we work with light-cone coordinates below to construct the kinetic term for a collinear fermion, the choice of a time coordinate will be less obvious, see~\cref{sec:CollinearFermions}.}  Therefore, a kinetic term must involve at least one derivative with respect to time, $\partial_0\s \phi(x)$, and since we obviously want to build it covariantly, we must lift this to the full partial derivative $\partial_\mu\s \phi(x)$.

Next, we impose the symmetries of our theory.  Since $\phi(x)$ is a Lorentz scalar and $\mathcal{L}$ is also a Lorentz scalar,\footnote{Technically $\mathcal{L}$ is only required to be a Lorentz scalar up to the parity and time discrete spacetime transformations.} we must contract $\partial_\mu \phi(x)$ with another Lorentz vector if it is going to contribute to the Lagrangian.  The only available vector is $\partial_\mu$.  Putting all these pieces together yields the following candidate Lagrangian for the kinetic term
\begin{align}
\mathcal{L} = Z_1\,\partial_\mu \s \partial^\mu\s \phi^2 + Z_2\,\phi\, \partial_\mu\s \partial^\mu\s \phi + Z_3\, \big(\partial_\mu\s \phi\big) \big(\partial^\mu\s \phi\big)\,,
\end{align}
where $Z_1$, $Z_2$, and $Z_3$ are constants.

Take the first term:
\begin{align}
\partial_\mu\s \partial^\mu\s \phi^2 = \partial_\mu \big(2\, \phi \,\partial^\mu\s \phi\big) = 2\Big(\phi \,\partial_\mu\s \partial^\mu\s \phi + \big(\partial_\mu\s \phi\big) \big(\partial^\mu\s \phi\big)\Big)\,,
\end{align}
implying that this term can be absorbed by simply redefining $Z_2$ and $Z_3$.  Next we can apply integration by parts to the second term, which yields
\begin{align}
\int \text{d}^4 x\, \phi\, \partial_\mu\s \partial^\mu\s \phi = - \int \text{d}^4 x\, \big(\partial_\mu\s \phi\big)  \big(\partial^\mu\s \phi\big)\,,
\end{align}
such that we can absorb $Z_2$ into a redefinition of $Z_3$.  The result is the unique kinetic term  
\begin{align}
\mathcal{L}_\text{Kin} =  Z_\phi\, \frac{1}{2} \big(\partial_\mu\s \phi\big) \big(\partial^\mu\s \phi\big)\,,
\end{align}
where $Z_\phi$ is the wave-function renormalization factor for the kinetic term, the sign choice is fixed to yield positive kinetic energy, and the $1/2$ yields the canonical normalization of the propagator if we assume that $Z_\phi = 1 + \text{corrections}$, as it does in perturbation theory.  Note that $Z_\phi$ does not carry any mass dimension, since the EFT kinetic term operator is mass dimension four -- of course this is by definition, since the requirement that the kinetic term carry mass dimension four is what we use to determine the mass dimension for $\phi$.

\subsection{Power Counting for Fields}
\label{sec:PowerCountingForFields}
Now we can turn to the power counting for our field $\phi$.  We can interpret the propagator as summing an infinite number of insertions of the kinetic operator.  Therefore, we must be able to insert a factor of this operator at no power counting cost.   Assuming the power counting for the $\phi$ momentum defined in \cref{eq:pExamplePowerCounting}, 
\begin{align}
p_\mu \sim M \big(\lambda,\lambda,\lambda,\lambda\big) \sim m \big(1,1,1,1\big)\,,
\end{align}
we find that that the kinetic term scales as $\lambda^2\,\lambda_\phi^2$, where each derivative scaling like $\lambda$ is inherited from the scaling of the $\phi$ momentum, and $\lambda_\phi$ is the power counting parameter for $\phi$.  Recall from above that our power counting choice implies that $\D^4 x \sim 1/\lambda^4$, such that $\mathcal{L}_\text{Kin} \sim \lambda^4$ so that the action for the kinetic term scales as $\mathcal{O}(1)$.  We have derived 
\begin{align}
\lambda_\phi = \lambda\,.
\label{eq:PowercountPhi}
\end{align} 
We can additionally infer that marginal or leading power terms in our EFT will power count as $\lambda^4$.    The same kind of argument will be revisited in SCET below, where we will see that not all propagating EFT fields will have the same power counting.  However, in our relativistic EFT, we see that power counting is identical to organizing by mass dimension.

Note that there is another quadratic term consistent with the symmetries, namely the mass term
\begin{align}
\mathcal{L} \supset \frac{1}{2}\,M^2\, C_{\phi^2}\s \phi^2\,.
\label{eq:EFTmass}
\end{align}
Note that the coefficient of this operator must carry mass dimension two.  But the only parameter around with mass dimension is $M$, which is why it appears explicitly in \cref{eq:EFTmass}.  However, this is the scale we have integrated out to generate the EFT.  But if $C_{\phi^2} \sim \mathcal{O}(1)$, our ``light'' field $\phi$ has a mass of order the cutoff of the EFT, and the whole setup breaks down!  This argument is the conceptual source of the hierarchy problem -- we will see this same conundrum appear dynamically in \cref{sec:HierarchyProb} below.  

Obviously, we need $\phi$ to have a small mass.  One consistent (albeit fine-tuned) choice, is to take $C_{\phi^2} \sim \lambda^2 \sim m^2/M^2$.  Then it becomes sensible to sum the mass into the propagator, yielding the massive propagator given in \cref{eq:Propagator}.  It is worth emphasizing that nothing in this setup requires $\phi$ to have any mass at all.  Once we tune the mass to zero at $\mu_M \sim M$, it will stay zero within the EFT as emphasized in \cref{sec:HierarchyProb} below.\footnote{For a massless theory, it could be useful to define power counting in terms of $s_{ij}/M^2$, depending on the process we are interested in modeling.}

\subsection{Interactions and Local Operators}
\label{sec:ConstructOps}
For the sake of simplicity (and to emphasize the role of symmetry), we will assume that the \FT~has a $\mathbb{Z}_2$ symmetry that sends $\phi \rightarrow - \phi$ and $\Phi \rightarrow -\Phi$ simultaneously.  In other words, we are only allowing $\kappa$ and $\eta$ to be non-zero, using the notation defined in \cref{eq:ToyFullLagrangian} for $\mathcal{L}_\text{Int}^\textsc{Full}$.  Since the physics we have integrated out at the scale $\mu_M \sim M$ does not violate the $\mathbb{Z}_2$ symmetry, the EFT should manifest a $\mathbb{Z}_2$ as well, so we will only consider terms with even powers of $\phi$.

The leading interaction is
\begin{align}
\mathcal{L}^\text{EFT}_\text{Int} \supset C_{\phi^4}\s \phi^4\, \sim \lambda^4\,,
\end{align}
so it is a marginal operator in our EFT.  We can do an arbitrary number of insertions of this operator at no power counting cost.  Marginal operators are usually considered part of the interacting EFT Lagrangian, as opposed to being classified as a local operator.

When we go to higher power, we begin to encounter irrelevant operators.  Due to the $\mathbb{Z}_2$ symmetry of our EFT, they must scale as $\sim\lambda^6$ or higher.  At next order, there are two possible terms
\begin{align}
\mathcal{L} \supset \frac{1}{M^2} \s C_{\phi^6}\s \phi^6 + \frac{1}{M^2}\s C_{\partial^2 \phi^4}\, \phi^2\,\partial^2 \phi^2 \,,
\end{align}
where $M$ is the dimensionful high scale, and since $\phi \sim \lambda$ and $\partial\sim \lambda$, both terms scale as $\lambda^{6}$.  These are usually referred to as the ``local operators'' or ``contact operators,'' in that they encode non-trivial interactions among many fields simultaneously evaluated at a single spacetime  point.  Their role in the EFT is primarily to model the residual influence of the heavy scale on the light system, that is mediated by off-shell physics near the scale $M$.  In this context, the division into a ``propagating EFT'' (that includes the propagating modes and their super-renormalizable and marginal interactions) and the ``local interactions'' is a reframing of the more familiar ideas of ``renormalizable'' versus ``non-renormalizable'' theories.  While this distinction might seem overly pedantic, for a more complicated EFT like SCET, the UV and IR theories have dissimilarities, and furthermore the operator building blocks for the local interaction Lagrangian take a still different form.  For now, we can be satisfied that power counting provides a consistent way to categorize our operators into super-renormalizable, marginal, and irrelevant -- keeping with the spirit of EFTs as we are developing them, we will often refer to the classification of operators as instead being super-leading power, leading power, and sub-leading power respectively.

\subsubsection*{The Operator Basis and the Hilbert Series}
As should now be clear, there can be tremendous freedom when choosing a basis of operators to work with, especially in situations with a large number of fields, \emph{e.g.}~the Standard Model.  In particular, the redundancies implied by integration by parts and the equations of motion complicate the characterization of a basis.  Ideally, one would carefully pick the basis that would make a given calculation as straightforward as possible.  Furthermore, the RGE will tend to mix operators.  If one needs to evolve the theory between scales, then a complete basis is required, see \emph{e.g.}~\cite{Jenkins:2013zja, Jenkins:2013wua, Alonso:2013hga, Alonso:2014zka} in the context of the Standard Model EFT up to dimension 6.  

Just to emphasize the point concretely, note that the equations of motion in the Standard Model can be used to relate currents of fermions to derivatives of gauge/Higgs bosons as 
\begin{align}
J_{G\s \mu}^a &= g_s \sum_\text{quarks} \bar{f}\,\gamma_\mu\,f \quad \,\,\longleftrightarrow \quad D^\nu G^a_{\mu\nu} \notag\\
J^a_{W\s \mu} &= g \sum_\text{left} \bar{f}\,\gamma_\mu \frac{\sigma^a}{2}\,f \quad \longleftrightarrow \quad D^\nu W_{\mu\nu}^a - \frac{i}{2}\, g\,H^\dag \sigma^a\, \overleftrightarrow{D}_\mu\, H\notag\\
J_{B\s\mu} &= \sum_\text{matter} Y_f\, \bar{f}\gamma_\mu \,f  \quad\,\s \longleftrightarrow \quad \partial^\nu B_{\mu\nu} - \frac{i}{2}\,g'\,H^\dag\, \overleftrightarrow{D}_\mu\, H\,,
\end{align}
for SU(3), SU(2), and U(1) currents respectively, and where all variable are taken as the canonical Standard Model definitions, see \emph{e.g.}~\cite{Wells:2015uba} for details and applications to the Standard Model EFT.  Clearly, this implies that  care must be taken when performing calculations that depend on higher dimension operators in the Standard Model.

Given this complexity, an algorithm that would count the number of independent operators for a theory of interest was an unsolved problem until recently.  While the details are beyond the scope of this review, it is worth briefly mentioning the ideas that underlie this counting~\cite{XL}.  There is a technique which is colloquially known as the Hilbert series approach that can be used to count the independent elements in the operator basis.  This approach was first applied to higher dimension operators in the Standard Model to construct flavor and CP invariants~\cite{Jenkins:2009dy, Hanany:2010vu}, which do not involve derivatives.  Later, it was first shown how to accommodate derivatives with this method as applied to a toy scalar EFT that lives in one-dimensional spacetime~\cite{Henning:2015daa}.\footnote{The choice to work in one-dimension implies that the expansion in derivatives truncates at $\partial^2 \phi = 0$, dramatically simplifying the problem.}  Now the task of classifying higher dimension operators in the Standard Model that include derivatives has been achieved~\cite{Lehman:2015via, Lehman:2015coa, Henning:2015alf}.  There are in fact two underlying approaches for dealing with operators involving momentum~\cite{Henning:2015daa, Henning:2017fpj}.  The first is a momentum space approach where polynomials of the momenta form an algebraic structure known as a ``commutative ring.''  In this language, the equations of motion and integration by parts redundancies are an ``ideal'' of the ring -- the set of polynomials with these redundancies removed is then the ``quotient ring.''  There is a complementary position space picture for dealing with the redundancies that appear when operators include derivatives.  The first step is to identify each Standard Model field as a representation of the four-dimensional conformal group.  Then accounting for the equations of motion redundancies is equivalent to ``shortening'' these multiplets.  One then constructs all reducible tensors with a given mass dimension by taking products of the shortened representations.  Next, one decomposes these reducible tensors into irreducible representations.\footnote{For recent progress automating this step, see \emph{e.g}~\cite{Criado:2019ugp}.}  The last step is to extract the primary operator for each of these irreducible tensors using the Molien-Weyl formula, which accounts for the integration by parts redundancies.  The result is that there are $2, 84, 30, 993, 560, 15456, \dots$ higher dimension operators for the Standard Model EFT assuming a single generation of fermions, where this list corresponds to the number of independent operators ordered in terms the operator mass dimension, starting at 5, see~\cite{Henning:2015alf} for details.  Converting the output of the Hilbert series into explicit operators requires a final step, see~\emph{e.g}~\cite{Hays:2018zze, Gripaios:2018zrz}. 

\subsubsection*{Accidental Symmetries and EFTs}
One of the miraculous properties of the Standard Model is that baryon and lepton number are accidentally conserved.  Specifically, this means that if one takes the Standard Model matter fields and writes down the most general set of operators that are allowed by gauge invariance up to dimension four, global U(1) symmetries for lepton and baryon number emerge.  This idea of an accidental symmetry is ubiquitous in field theory.  In the Standard Model, there is a unique operator at dimension 5~\cite{Weinberg:1979sa}:\footnote{This operator accounts for the ``2'' referred to in the previous section~\cite{Henning:2015alf}, where the doubling is due to their convention of taking the operator and its hermitian conjugate as independent.}
\begin{align}
\mathcal{L} \supset \frac{1}{\Lambda} \Big(H\,\bar{L}\Big)^2 + \text{h.c.}\,,
\end{align}
where $L = (\nu, e)$ is the lepton doublet, $H$ is the Higgs doublet, we are assuming only one generation for simplicity, and $\Lambda$ is a dimensionful scale suppressing the operator.  This operator violates lepton number, since the gauge invariant contraction is proportional to $L^2$ as opposed to $L^\dag\,L$.  When the Higgs is expanded about its electroweak symmetry breaking vacuum, this operator leads to a Majorana mass for the neutrinos, $m_\nu \sim v^2/\Lambda$.  It has a simple interpretation as being generated by integrating out a heavy right handed neutrino using techniques that will be discussed below in \cref{sec:TreeMatching}, \emph{i.e.}, the seesaw mechanism.  Going to dimension 6, it becomes clear that baryon number can also be violated by new physics at a UV scale, which for example can lead to proton decay.  One compelling explanation for $\Lambda$ is to associate it with the scale of grand unification~\cite{Georgi:1974sy}.  We see that the accidental symmetries of the Standard Model, augmented by EFT reasoning, gives a compelling explanation for both the smallness of the neutrino masses and the long lifetime of the proton.

More recently, accidental symmetries have been used to solve the hierarchy problem (see~\cref{sec:HierarchyProb} below) in the context of the Twin Higgs mechanism~\cite{Chacko:2005pe}.  These models rely on a global SU(4) symmetry in the Higgs sector (which requires the addition of new ``twin'' Higgs fields).  However, the matter sector of the Lagrangian only respects a $\mathbb{Z}_2$ exchange symmetry, and in particular does not require the new matter states to be charged under the Standard Model gauge groups -- for this reason, the study of models and signatures associated with this clever application of accidental symmetries is often referred to as ``neutral naturalness.''  The magic of this mechanism is that the SU(4) is maintained by the one-loop corrections to the Higgs potential.  In particular, a light pseudo-Goldstone Higgs state dynamically emerges whose mass parameter is protected by the global symmetry breaking pattern $\text{SU}(4)\rightarrow \text{SU}(3)$, \emph{i.e.}, it does not receive a large quadratic contribution to its mass parameter.  However, this accidental symmetry is violated by higher loop effects.  While the Twin Higgs approach and its extensions~\cite{Burdman:2006tz, Craig:2014aea, Craig:2014roa, Craig:2015pha, Cohen:2018mgv} only postpone the hierarchy problem by a loop factor, they intriguingly motivate novel phenomenological observables at the LHC, in cosmology, and for dark matter detection, see~\emph{e.g.}~\cite{Craig:2013xia, Burdman:2014zta, Craig:2015pha, Curtin:2015fna, Freytsis:2016dgf, Craig:2016lyx, Cheng:2016uqk, Chacko:2018vss, Kilic:2018sew}.  

Now that we have laid the foundation for how to construct an EFT, we will turn to our main line of inquiry.  We will develop the technology to match a \FT~onto an EFT at both tree and loop level, and then run the couplings within the EFT in order to maintain precision control of perturbation theory.

\newpage
\section{Matching and Running}
\label{sec:MatchingRunning}
In this section, we explain the matching and running approach to scale separation.  Matching was defined schematically in \cref{eq:matchingSchematic}.  The procedure is formalized more carefully in this section.  First we work out examples  at tree level (see \cref{sec:TreeMatching}) and then at one loop (see \cref{sec:LoopMatching} through \cref{eq:SepScalesHLlog}).  Matching has two main purposes:  it serves to connect the coefficients of the EFT to a more fundamental UV description, and it eliminates any non-analytic dependence on the IR parameters that the \FT~Feynman diagram expansion might manifest.  Once matching has been performed, we will show how to derive a set of RGEs that can be integrated to run couplings from the high scale $\mu_H$ to a low scale $\mu_L$.  The perturbative expansion in terms of the low-scale EFT parameters displays improved convergence, since logarithms are absorbed into the running of these couplings.

We will encounter a number of technical and conceptual issues along the way.  After our tree-level example, we will detour into a series of two \Primers.  The first provides a review of dimensional regularization, the modified minimal subtraction scheme, and then explains the vanishing of dimensionally regulated scaleless integrals.  This is followed by a \Primer~on RG evolution, where we show how to derive RGEs.  Then we move back into the lectures, where we provide a simple computation of an anomalous dimension of an operator and use it to run the relevant coupling.  We then turn to a number of interesting subtleties that appear when matching at loop level.  First, we will explore the EFT approach to decoupling the contributions to the anomalous dimensions from heavy particles, when using a massless regulator such as dim reg.  This is followed by a calculation to highlight the implications of integrating out a heavy state in a model with a light particle whose mass is not protected by a symmetry -- the hierarchy problem.  We will additionally see how EFT reasoning clarifies some confusing aspects of how this problem manifests when using dim reg.  Finally, we will present the most important calculation of this section, where we separate scales for a ``heavy-light'' logarithm.  Up until this point, the need for the RG will have been an obvious consequence of the apparent dependence on an unphysical renormalization scale.  In \cref{eq:SepScalesHLlog}, we will explore a toy model from which emerges a logarithm that is a function of only physical scales.  This provides the opportunity to explore how matching can be used to introduce scale dependence, paving the way to derive a set of RGs.  We will end this section with an introduction to the method of regions using our heavy-light log as an example application.  Although this technique is overkill for the simple case we are presenting here, it will be critical to our understanding of and ability to calculate within SCET.

\subsection{Tree-level Matching}
\label{sec:TreeMatching}
Our goal in this section is to learn how to integrate out a heavy particle at tree level,  to express the \FT~dynamics as an EFT expansion in terms of the light particle interactions.  We will match these two theories at a scale $\mu_M \sim M$.  For simplicity, the only interaction we will turn on in the \FT~is\footnote{\textbf{Disclaimer:}  For the last time, we will note that this (and all of our examples) will involve a non-generic choice of the couplings in the UV \FT. In particular, if we run the couplings in the \FT~to a scale much higher than $\mu_M$ additional operators would be induced.  This effect will play no role in the physics of interest here.}
\begin{align}
\mathcal{L}_\text{Int}^\textsc{Full} = - \frac{1}{2}\, b \,\phi^2\, \Phi\,.
\label{eq:TreeMatchLInt}
\end{align}
The first step is to pick a process.\footnote{While this is not strictly necessary when integrating out a heavy particle relativistically, our goal is to present the more familiar example in a way that parallels the SCET approach discussed below.}  We will match the two theories by equating $\phi\, \phi \rightarrow \phi\, \phi$ at the scale $\mu_M$.  Note that while we will not be careful to distinguish them, the \FT~field $\phi_\textsc{Full}$ is not the same as the EFT field $\phi_\text{EFT}$.  In the case of matching across a relativistic threshold at tree-level, this distinction is not critical, but we will see it manifest non-trivially below in SCET, where the EFT degrees of freedom take a different form than those in the \FT.  However, note that if we were investigating the detailed loop structure of this example, these differences would manifest in a variety of ways, \emph{e.g.}~the wave-function renormalization for $\phi_\textsc{Full}$ would be different from that of $\phi_\text{EFT}$.  

Next, we need to write down an EFT Lagrangian that can capture the physics at scales $\mu_L \ll M$.  Clearly our EFT cannot include $\Phi$ as a dynamical degree of freedom, since by construction the EFT is valid when there is not enough energy available to create this heavy state.\footnote{There are techniques for treating the fluctuations of a heavy field $\Phi$ at low energies.  This is known as heavy quark effective theory, and it is mentioned in \hyperlink{sec:EFTZoo}{Appendix A} below.}  However, when working with loop diagrams one is still supposed to integrate over all momenta, so one might naively expect that $\Phi$ must still be included in loop calculations.  To see why this is not the case, we can appeal to unitarity.  Specifically, the optical theorem implies that if there are no external $\Phi$ particles in Feynman diagrams, then any loop involving $\Phi$ cannot be put on-shell.  This in turn tells us that all dependence on the mass of $\Phi$ must be analytic, \emph{i.e.}, there can be no logs that depend on $M$ generated within the EFT, and all $M$ dependence that contributes to $\phi\, \phi \rightarrow \phi\, \phi$ can only result from the matching procedure that determines the local operator structure of the theory at low energies.  In other words, it is completely sensible to use an EFT to describe the physics at low energies to arbitrary order in perturbation theory.

Since our \FT~respects a $\mathbb{Z}_2$ symmetry that acts on the light field, we should enforce $\phi \rightarrow - \phi$ within the EFT as well.\footnote{Of course, we do not need to enforce this.  We could instead include all possible higher dimension operators and then matching would set the Wilson coefficients of operators that violate this symmetry to zero.}  This restriction (plus Lorentz invariance) implies that our EFT Lagrangian takes the form 
\begin{align}
\mathcal{L}^\text{EFT} &= \frac{1}{2}\big(\partial_\mu \phi\big)\big(\partial^\mu \phi\big)-\frac{1}{2} m^2\,\phi^2 - \frac{1}{4!}\,C_{(4,0)} \,\phi^4 \notag\\[5pt]
&\hspace{12pt}- \frac{1}{6!} \,\frac{1}{M^2}\,C_{(6,0)}\,\phi^6 - \frac{1}{4}\,\frac{1}{M^2}\,C_{(4,2)} \,\phi^2\big(\partial^2\s\phi^2\big) + \cdots\,,
\label{eq:EFTLagTreeMatch}
\end{align}
where $C_{(i,j)}$ are the Wilson coefficients specified by the number of fields $i$ and the number of derivatives $j$, and we have included factors of $M$ to make all the Wilson coefficients dimensionless.\footnote{We caution the reader that one can get confused about dimensional analysis when restoring factors of $\hbar$ due to the fact that the choice of dimensionful suppression scale is often defined to include masses and couplings, such that the scale fundamentally does not have the same units as mass.}  Next, we define our power counting in the $m\ll M$ limit as we did above in \cref{eq:pExamplePowerCounting}:
\begin{align}
p_\phi \sim M \big(\lambda,\lambda,\lambda,\lambda\big)\,.
\end{align}
Following the same logic that led to \cref{eq:PowercountPhi}, we find that $\phi \sim \lambda$ and $\partial \sim \lambda$.  This allows us to power count our operators, and we see that $\phi^4 \sim \lambda^4$, $\phi^6 \sim \lambda^6$, and $\partial^2\s \phi^4 \sim \lambda^6$.  Our operator expansion organized by mass dimension is ordered in power counting.

Now we are ready to compute $\phi\, \phi \rightarrow \phi\, \phi$ at low energies in both the \FT~and the EFT.  We begin with the \FT~calculation
\begin{align}
\includegraphics[width=0.23\textwidth, valign=c]{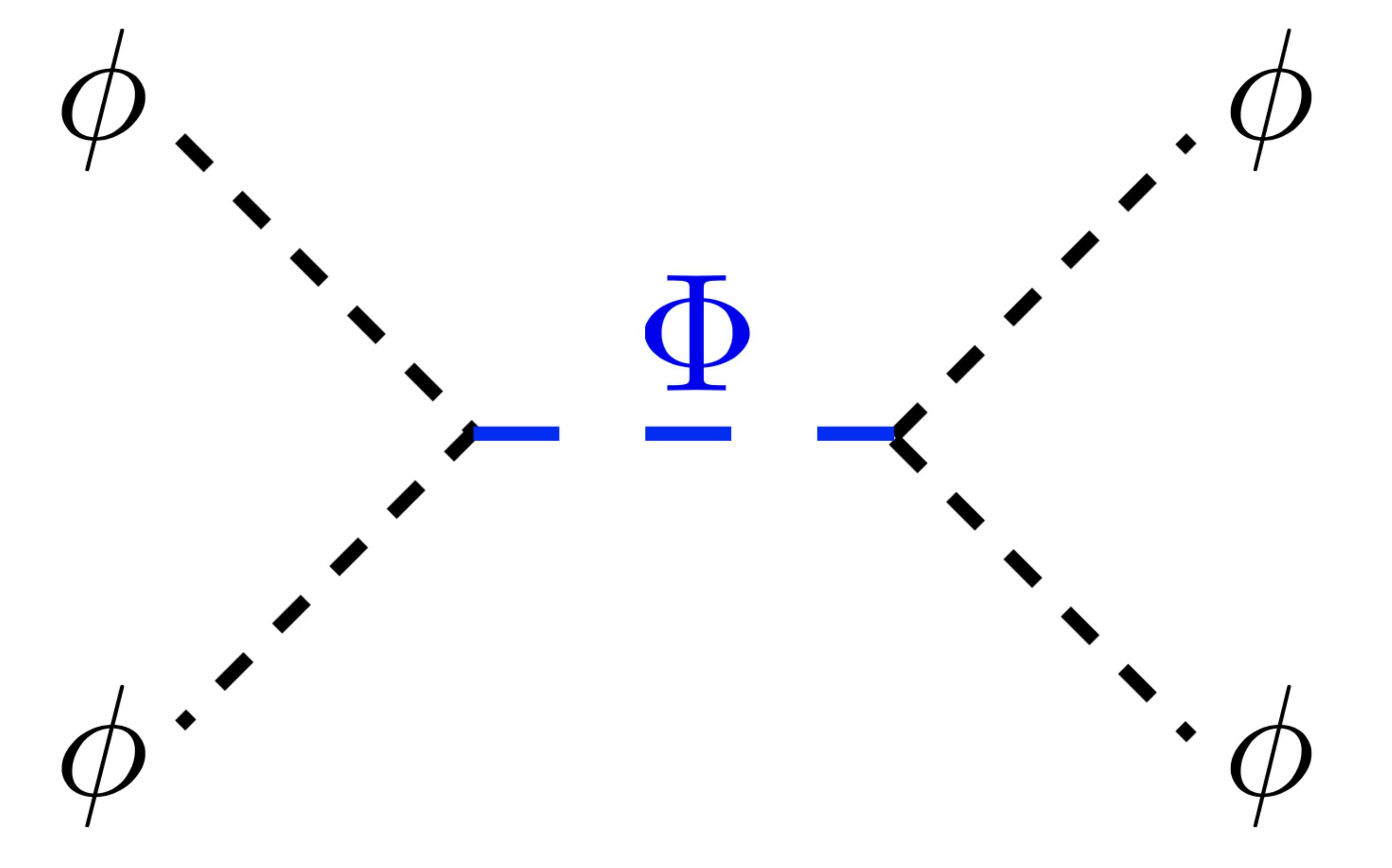}  \quad\begin{array}{l}+\,t\,\text{-channel} \\+\,u\,\text{-channel}\end{array} 
&= i\s\mathcal{A}^\textsc{Full} =  -i\s b^2 \sum_{s,t,u} \frac{1}{p_\Phi^2 - M^2} \notag \\[-7pt]
&= -i\s b^2 \left(\frac{1}{s - M^2}+\frac{1}{t - M^2}+\frac{1}{u - M^2}\right)\notag\\[8pt]
&= -i\s b^2\left(-\frac{1}{M^2}\right) \left(3+\frac{4\s m^2}{M^2} +\frac{s^2+t^2+u^2}{M^4} + \cdots \right) \notag \\[8pt]
&\sim \frac{b^2}{M^2}\,\big(1+ \lambda^2 + \lambda^4 + \cdots\big)\,,
\label{eq:PropExp}
\end{align}
where in the second line we have power expanded the amplitude assuming our external fields have small momentum $s,t,u\sim m^2 \sim \lambda^2$, and we have used the fact that $s+t+u = 4\s m^2$.  Recall that $b$ carries mass dimension and the appearance of $M$ ensures that amplitude has the same dimensions as the four point amplitude in the EFT.  This diagram produces a Taylor expansion in $\lambda$ as expected.

Next, we compute $\phi\,\phi \rightarrow \phi\,\phi$ in the EFT.  At tree-level, this is given by the $\phi^4$ interaction in the EFT Lagrangian,
\begin{align}
 \includegraphics[width=0.16\textwidth, valign=c]{Figures/LLLL4Pt.pdf} = i \s \mathcal{A}^\text{EFT} = -i\s C_{(4,0)}\,.
\end{align}
Finally, we are ready to match.  At tree level, we simply equate the two amplitudes
\begin{align}
i\s\mathcal{A}^\textsc{Full} = i\s\mathcal{A}^\text{EFT}\,,
\label{eq:matchTree}
\end{align}
order by order in $\lambda$, where we have chosen the same kinematics within both the descriptions (this is trivial here, but is important for more complicated examples).  Since we are working with tree-level amplitudes, there are no logarithms to worry about and this matching is completely straightforward:
\begin{align}
i\s\mathcal{A}^\textsc{Full} = \includegraphics[width=0.16\textwidth, valign=c]{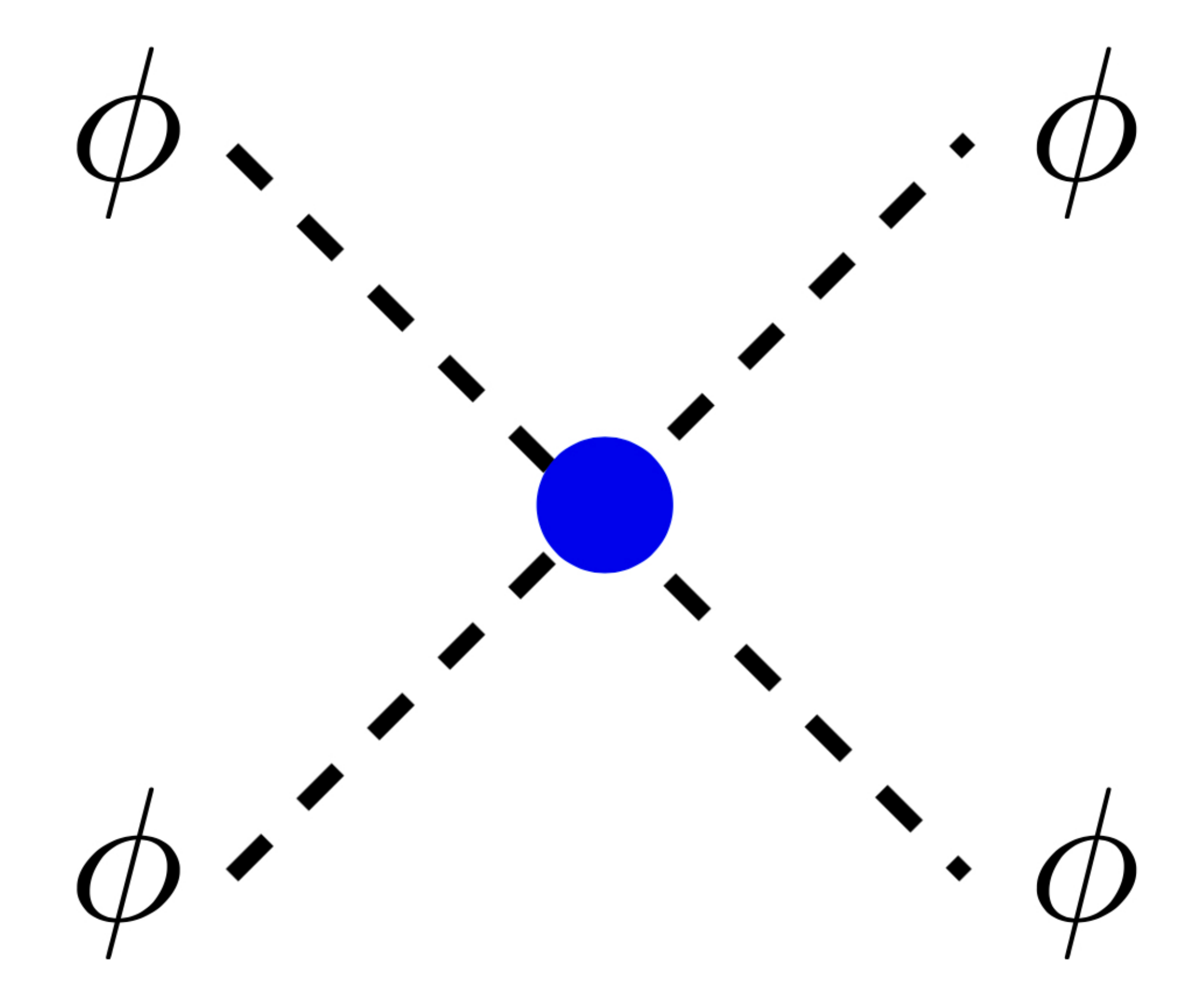} = 3\s i\,\frac{b^2}{M^2}  = -i\s C_{(4,0)} = \includegraphics[width=0.16\textwidth, valign=c]{Figures/LLLL4Pt.pdf} = i\s\mathcal{A}^\text{EFT}\,,
\label{eq:DeriveC4}
\end{align}
where we have truncated to leading order in $\lambda$.  Here the blue dot represents the fact that we have shrunk a heavy propagator to a point.  Technically, we should keep in mind that we match this onto the EFT at the scale $\mu_M = M$, implying that our matching procedure yields
\begin{align}
C_{(4,0)}\big(\mu_M\big) = -3\left(\frac{b\big(\mu_M\big)}{M\big(\mu_M\big)}\right)^2\,.
\end{align}
However, since we are working at tree-level, there is no scale dependence to keep track of -- we will have to keep careful track of these scales when we match at loop level in \cref{sec:LoopMatching} below.  Additionally, for a more complicated matching calculation one must address the choice of EFT operator basis, as discussed in \cref{sec:ConstructOps} above.

\vspace{5pt}\mybox{
\begin{itemize}
\item {\bf Exercise:}  The higher power terms computed in \cref{eq:PropExp} include a contribution to $C_{(4,0)}$ and to $C_{(4,2)}$.  You should first convince yourself that we have captured the complete basis of operators with two derivatives and four fields, and then determine both the $C_{(4,0)}$ and $C_{(4,2)}$ Wilson coefficients at $\mathcal{O}(\lambda^2)$ power.
\end{itemize}}

Finally, there is one more interesting subtlety to point out.  When matching, we should only include the contributions that are one-$\phi$-particle irreducible.  For example, one might be tempted to compute a matching coefficient $C_{(6,0)}$ using a diagram in the \FT~with six external $\phi$ states.  However, this process is already contained within the EFT through diagrams
\begin{align}
\includegraphics[width=0.2\textwidth, valign=c]{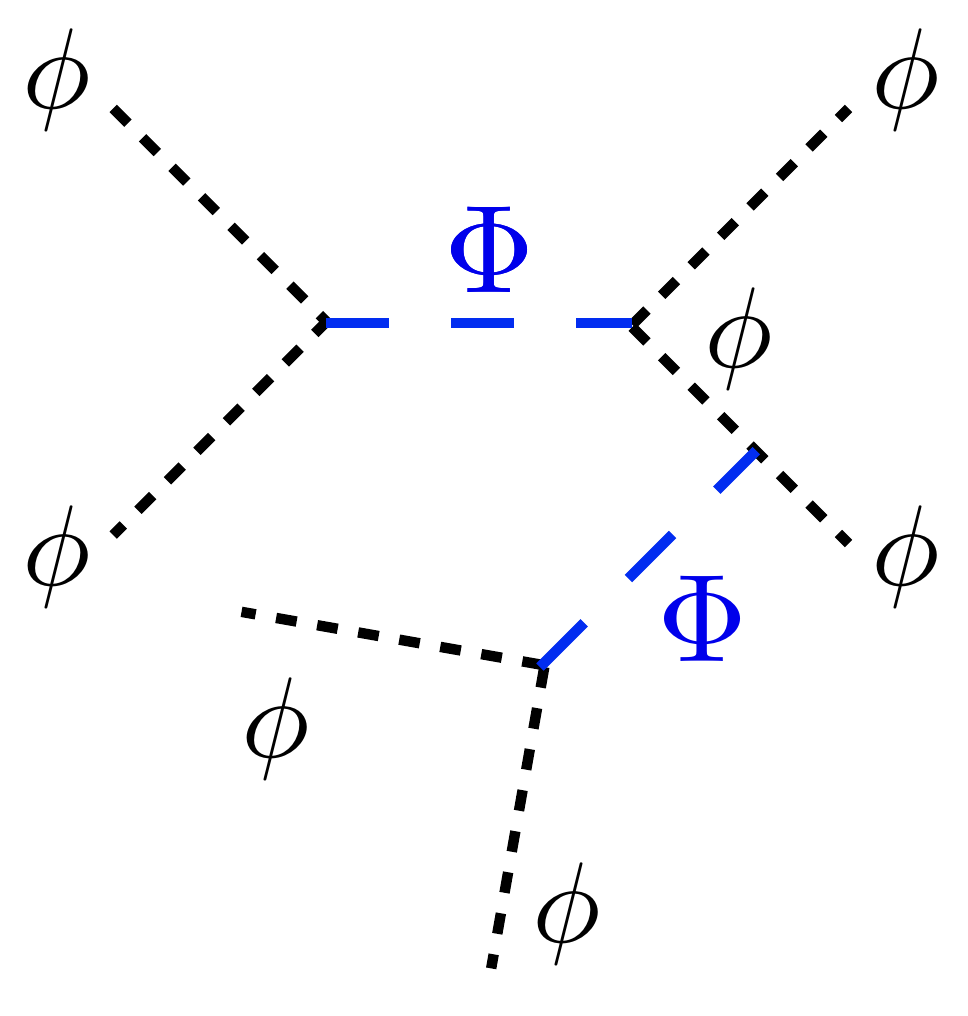} \quad = \quad \includegraphics[width=0.2\textwidth, valign=c]{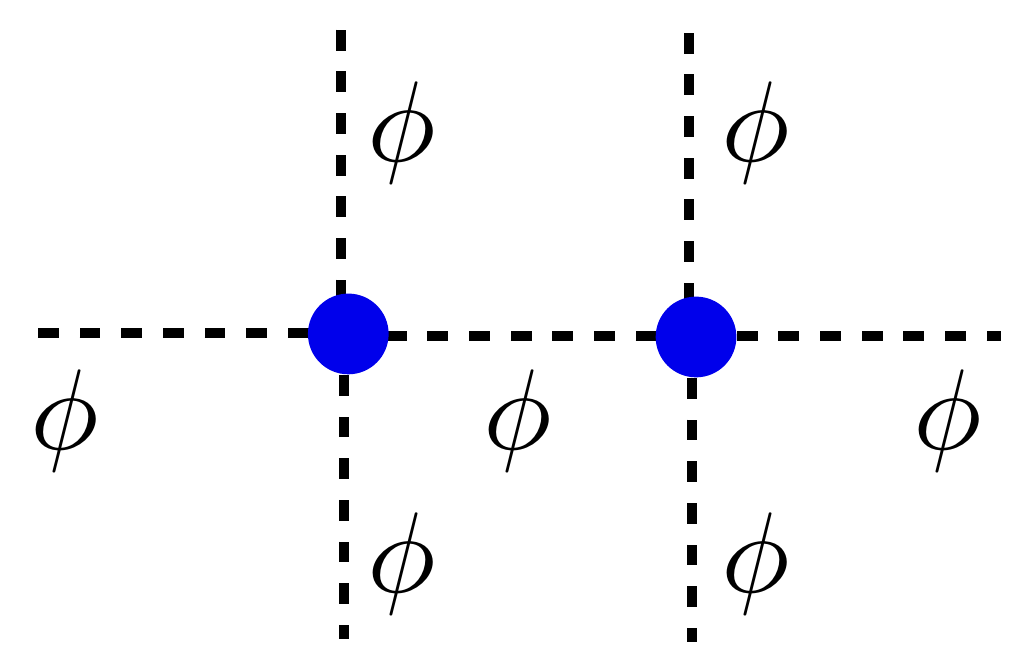}\,,  
\end{align}
where one must permute the labels to account for all the channels on both sides of the equals sign to reproduce detailed agreement.  Physically, the EFT captures the effects of shrinking $\Phi$ propagators to a point and Taylor expands around that limit, so that all $\phi$ lines must be left propagating at the matching step.  Therefore, $C_{(n,0)}= 0$ for all $n >4$ in this EFT, which is another non-generic aspect of this example.  Next, we will demonstrate a technique for computing the matching coefficients that relies on the equations of motion for $\Phi$, making the concept of ``integrating out'' the heavy particle manifest.

\subsubsection*{Matching and the Equations of Motion}
In this section, we connect the diagrammatic approach that was just explained with a convenient method that relies on the equations of motion for $\Phi$.  Starting with the \FT~Lagrangian, whose interacting part is given in \cref{eq:TreeMatchLInt}, we can derive the equations of motion for $\Phi$ following the standard Euler-Lagrange procedure:
\begin{align}
\big(\Box - M^2\big)\Phi = \frac{b}{2}\, \phi^2 \qquad \Longrightarrow \qquad \Phi = \frac{b}{2}\, \frac{1}{ \Box - M^2}\, \phi^2\,,
\label{eq:PhiSol}
\end{align}
where dividing by the propagator is interpreted as formally solving for $\Phi$; since we are interested in the EFT where $\Phi$ is far off-shell, we will not encounter any subtleties (such as contact terms) when interpreting this expression.  Plugging~\cref{eq:PhiSol} back into the Lagrangian yields
\begin{align}
\mathcal{L}^\textsc{Full}_\text{Int} = -\frac{\,\,b^2}{8}\, \phi^2\, \frac{1}{\Box - M^2}\, \phi^2\,.
\end{align}
This is still equivalent to the \FT~Lagrangian, since we have only used the equations of motion to rewrite it.  Although this is an obscure way to write the theory if we are interested in physics that depends on external $\Phi$ fields, it is useful for taking the low energy limit.  Using our expansion parameter $\lambda$, we have $\Box/M^2 \sim \lambda^2$ since the (inverse) derivatives act on $\phi$.\footnote{Inverse derivatives might seem disturbing at first, but they can be understood by Fourier transforming, and working at finite $p^2$.  Note that an inverse derivative is often a sign that a theory is non-local.  A non-local theory would emerge if we attempted to integrate out a light mode that should have been kept as a propagating degree of freedom to low energies.}  Then the EFT can be derived by simply Taylor expanding in $\lambda$
\begin{align}
\mathcal{L}^\text{EFT}_\text{Int} =-\frac{\,\,b^2}{8} \,\phi^2\left(-\frac{1}{M^2}\right)\s \phi^2 -\frac{\,\,b^2}{8} \, \phi^2 \left(-\frac{\Box}{M^4}\right)\s \phi^2 + \cdots\,\,\,.
\label{eq:EOMEFTResult}
\end{align}
This gives another way to see that the EFT is simply a Taylor expansion in $\lambda$ of the \FT.  Just to make sure all the factors work out correctly, we can derive a Feynman rule for the $\phi^4$ interaction from this Lagrangian, which has a coefficient   $-3\s b^2/M^2$.  This agrees with the derivation in \cref{eq:DeriveC4} that utilized diagrammatics.

The equation of motion approach is extremely useful at tree level, but becomes much less straightforward at loop level.  However, this technique has a beautiful path integral interpretation.  Recently, it has been shown how to perform matching and running at one-loop order using the path integral directly~\cite{Gaillard:1985uh, Cheyette:1987qz, Henning:2016lyp}, with applications to the higher dimensional operators for the Standard Model~\cite{Henning:2014wua, Drozd:2015rsp, Zhang:2016pja, Ellis:2017jns, Jiang:2018pbd}.  However, we will stick to diagrammatics as we move on to loop-level effects.  Before we are ready to calculate any loops, we will need the technology discussed in the following two \Primers.

\scenario{Dimensional Regularization}
\addcontentsline{toc}{subsection}{\color{colorTech}{Primer~\thescenario.} Dimensional Regularization}
\label{sec:dimreg}
Dimensional regularization (dim reg)~\cite{Bollini:1972ui, tHooft:1972tcz} is otherworldly.  We are taught to worship it in our quantum field theory courses primarily because it respects gauge invariance, but also since it can be applied with relative ease.  Furthermore, the way in which dim reg regulates integrals respects EFT power counting.  Related to this fact is the remarkable property that scaleless integrals vanish when using dim reg.  So much of our EFT formalism beyond tree level is made simpler by this wonderful fact, and we will see the connection to maintaining manifest power counting concretely as we work through some examples.  This feature of dim reg also underlies the difference between Wilsonian and continuum RG approaches as discussed in \cref{sec:RGE} below.

The idea for dim reg is to start with an integral that diverges in $d=4$ dimensions, and regulate it by deviating away from four dimensions by a small amount $\epsilon$.  We will use the convention $d = 4-2\s \epsilon$ throughout.\footnote{A word of caution here: many books including Peskin \& Schroeder~\cite{Peskin:1995ev} and Schwartz~\cite{Schwartz:2013pla} use $d= 4-\epsilon$, such that they always get the combination $1/\epsilon + \log\mu/\mathcal{M}$.}  As we will see below, this implies that our divergences will always have companion logarithms in the combination $1/\epsilon + \log\mu^2/\mathcal{M}^2$, where $\mu$ is the renormalization scale and $\mathcal{M}$ is in general a dimensionful combination of parameters.

When we change the dimension of spacetime, we also alter the mass dimension of our fields.  To derive this effect, we start with the fact that the action is dimensionless for any choice of spacetime dimension.  Then the kinetic term for a scalar field is simply
\begin{align}
S = \int \D^4 x\, (\partial_\mu\phi)(\partial^\mu \phi) \qquad \longrightarrow \qquad S=\int \D^{4-2\s\epsilon} x\, (\partial_\mu\phi)(\partial^\mu \phi)\,.
\end{align}
Using $\big[S\big] = 0$ and $\big[\D^{4-2\s\epsilon} x\big] = -(4-2\s\epsilon)$, where we are using the standard notation that $[...]$ returns the mass dimension of the object inside the brackets, we find
\begin{align}
\Big[(\partial_\mu\phi)(\partial^\mu \phi)\Big] = 4-2\s\epsilon\,.
\end{align}
Note that the mass dimension of derivatives does not depend on spacetime dimension since $\big[-i\s \partial_\mu\big] = \big[p_\mu\big] = 1$ for any $d$.  So the mass dimension of our field is
\begin{align}
\big[\phi\big] = 1- \epsilon\,.
\end{align}
We can then apply this to interactions:
\begin{align}
\mathcal{L} \supset \frac{1}{4!}\,C_4\,\phi^4\hspace{20pt} \quad &\longrightarrow \quad \mathcal{L}\supset \frac{1}{4!}\,C_4 \,\mu^{2\s\epsilon}\,\phi^4\notag\\[5pt]
\mathcal{L} \supset \frac{1}{6!}\,\frac{1}{M^2}\,C_6\,\phi^6 \quad &\longrightarrow \quad \mathcal{L}\supset \frac{1}{6!}\, \frac{1}{M^2}\, C_6 \,\mu^{4\s\epsilon}\,\phi^6\,,
\label{eq:muDependOperators}
\end{align}
where the dimensionful parameter $\mu$ appears so that the couplings remain dimensionless when $d = 4-2\s\epsilon$.  Note that the small $\epsilon$ expansion of $\mu^{2\s n\s\epsilon} \rightarrow n\,\log \mu^2$ for an integer $n$.

We will simply quote the following general result, since it is derived in any number of introductory field theory books, \emph{e.g.}~\cite{Peskin:1995ev, Schwartz:2013pla}:
\begin{align}
\int \frac{\text{d}^d\ell}{(2\s\pi)^d} \frac{1}{(\ell^2-\mathcal{M}^2)^b} = \frac{i}{(4\s\pi)^{d/2}}\frac{(-1)^{b}\,\Gamma(b-d/2)}{\Gamma(b)}\,\big(\mathcal{M}^2\big)^{d/2-b}\,,
\label{eq:dimRegEvalGeneral}
\end{align}
where $\mathcal{M}$ is some combination of variables that carries mass dimension.  Since we will only perform scalar integrals with trivial numerators, this will be the only form that appears below.  This evaluation utilized a Wick rotation~\cite{Wick:1954eu} such that a one-loop integral over one (two) propagator picks up a factor of $ -i$ ($i$).  These annoying factors of $i$ and signs will be important for the matching calculations that follow.  For the reader who wishes to take their understanding of dim reg to the next level, the Collins book on renormalization~\cite{Collins:1984xc} is an incredible resource that contains many formal details.

Since we will use them extensively, it is worth presenting the small $\epsilon$ expansion for two specific cases:
\begin{align}
\hspace{-10pt}\mu^{2\epsilon}\int \frac{\text{d}^d\ell}{(2\s\pi)^d} \frac{1}{\ell^2-\mathcal{M}^2} &= \frac{-i}{(4\s\pi)^{2-\epsilon}} \frac{\Gamma(\epsilon-1)}{\Gamma(1)}\left(\frac{\muT^2}{4\s\pi\,e^{-\gamma_E}}\right)^\epsilon\left(\frac{1}{\mathcal{M}^2}\right)^{\epsilon-1}\notag\\[7pt] 
&= \frac{i}{16\s\pi^2}\mathcal{M}^2 \left(\frac{1}{\epsilon}+\log\frac{\muT^2}{\mathcal{M}^2}+1\right) + \mathcal{O}(\epsilon) \qquad\!\! \big[\text{one propagator}\big]\,,
\end{align}
and
\begin{align}
\hspace{-10pt}\mu^{2\epsilon} \int \frac{\text{d}^d\ell}{(2\s\pi)^d} \frac{1}{(\ell^2-\mathcal{M}^2)^2} &= \frac{i}{(4\s\pi)^{2-\epsilon}} \frac{\Gamma(\epsilon)}{\Gamma(2)}\left(\frac{\muT^2}{4\s\pi\,e^{-\gamma_E}}\right)^\epsilon\left(\frac{1}{\mathcal{M}^2}\right)^{\epsilon}\notag\\[7pt] 
&= \frac{i}{16\s\pi^2} \left(\frac{1}{\epsilon}+\log\frac{\muT^2}{\mathcal{M}^2}\right) + \mathcal{O}(\epsilon) \qquad\hspace{26pt} \big[\text{two propagators}\big]\,,
\label{eq:dimRegEvalTwoProps}
\end{align}
where $\mathcal{M}$ has mass dimension one, and we have introduced the notation 
\begin{align}
\muT^2 = 4\s\pi\,e^{-\gamma_E}\,\mu^2\,,
\label{eq:DefMuMSBar}
\end{align} 
where $\gamma_E$ is the Euler-Mascheroni constant.  Then the minimal subtraction scheme \big($\s\overline{\text{MS}}\s\big)$ is implementing by first writing integrated results in terms of the $\overline{\text{MS}}$ scale $\muT$, and then defining the $n^\text{th}$ loop counterterms so that they cancel terms proportional to $1/\epsilon^n$ for $n>0$, see more discussion below in \cref{sec:RGE}.  Note that if you have used Feynman parameters (discussed next) to manipulate your integral into the general form in \cref{eq:dimRegEvalGeneral}, you must integrate over these parameters before expanding in $\epsilon$, assuming your goal is to correctly capture all of the logarithmic dependence.

\subsubsection*{Combining Denominators}
We will use the Feynman parameter trick for combining propagators:
\begin{align}
\frac{1}{A\,B} = \int_0^1 \text{d}x\, \frac{1}{\big[x\,A + (1-x)\,B\big]^2}\,,
\label{eq:FeynParam}
\end{align}
which generalizes in the case of $n$ propagators to
\begin{align}
\frac{1}{A_1\,A_2\,\cdots\,A_n} = \int_0^1 \D x_1\,\cdots \, \D x_n\, \delta\Big(\sum x_i-1\Big) \,\frac{(n-1)!}{\big[x_1\,A_1+\cdots + x_n\,A_n\big]^n}\,.
\label{eq:NormalFeynmanParamsGeneral}
\end{align}

We will also use a less well known variation of this trick that is useful when one encounters linear denominator factors.  Starting with the simple case where $A$ is a standard quadratic propagator and $b$ is linear,
\begin{align}
\frac{1}{A\,b} = \int_0^\infty \D y\, \frac{1}{(A + b\,y)^2}\,,
\end{align}
where the $y$ integral runs from zero to infinity, and $y$ can carry mass dimension as needed for consistency.  This generalizes to the case of $n$ linear propagators as
\begin{align}
\frac{1}{A\,b_1\,b_2\,\cdots\,b_n} = \int_0^\infty \D y_1\,\cdots\,\int_0^\infty \D y_n\, \frac{n!}{\big[A+b_1\,y_1 + \cdots + b_n\,y_n \big]^{n+1}}\,,
\label{eq:CombineLinProps}
\end{align}
and for higher powers of both propagators
\begin{align}
\frac{1}{A^n\,b^{m}} = \frac{\Gamma(m+n)}{\Gamma(m)\,\Gamma(n)} \int_0^\infty \D y \,\frac{y^{m-1}}{(A+y\,b)^{n+m}}\,,
\end{align}
where again $A$ is a quadratic Feynman propagator and $b$ is linear, and $n$ and $m$ are positive integers.  We will use these formulas to compute loop corrections in SCET.

\subsubsection*{Scaleless Integrals Vanish}
This subsection will devote some effort to convincing you that scaleless integrals vanish in dim reg.  The phrase ``scaleless integral'' refers to a loop integral whose integrand has no dependence on a physical scale that carries mass dimension.  Conceptually, one reason that scaleless integrals must vanish is that dim reg generates logarithms of the RG dimensionful scale $\mu$, and, since the argument of a log must be dimensionless, there must be some other scale around to produce a consistent non-zero result.  The absence of such a scale implies that the integral must return zero.

Take a scaleless integral that is naively both UV and IR divergent, and simply evaluate it by enforcing a UV and an IR cutoff, $\Lambda_\text{UV}$ and $\Lambda_\text{IR}$ respectively:
\begin{align}
\mathcal{I} = \int \frac{\D^4 \ell}{(2\s\pi)^4} \frac{1}{\ell^4}  = \frac{i}{8\s\pi^2}\int_{\Lambda_\text{IR}}^{\Lambda_\text{UV}} \!\D \ell\, \frac{1}{\ell} = \frac{i}{16\s\pi^2} \log\frac{\Lambda_\text{UV}^2}{\Lambda_\text{IR}^2} \,,
\label{eq:ScalelessExCutoff}
\end{align}
where in the second step we have Wick rotated. Note that when using a cutoff regulator, the integral is non-vanishing.  However, the logarithm only depends on unphysical $\Lambda_\text{IR}$ and $\Lambda_\text{UV}$ regulator scales, so care must be taken when interpreting this result, since it is clearly scheme dependent.

Next, we can see that this integral vanishes when using dim reg.  We can rewrite \cref{eq:ScalelessExCutoff} so that it is regulated by dim reg (see \emph{e.g.}~\cite{Petrov:2016azi})
\begin{align}
\mathcal{I} = \mu^{2\epsilon}\int \frac{\D^d \ell}{(2\s\pi)^d} \frac{1}{\ell^4} \,.
\label{eq:ScalelessExDimReg}
\end{align}
By introducing a mass scale $m$, we can rewrite the integrand as 
\begin{align}
 \frac{1}{\ell^4} = \frac{\ell^2}{\ell^4(\ell^2-m^2)} - \frac{m^2}{\ell^4 (\ell^2-m^2)} \,,
\end{align}
which allows us to break the integral into UV and IR divergent terms\footnote{Note that for both of these integrals we are using $d = 4-2\s \epsilon$ with $\epsilon > 0$.  This is a reasonable choice for $\mathcal{I}_\text{UV}$ since it converges for $d <4$.  However, $\mathcal{I}_\text{IR}$ only converges if $d > 4$, and thus it should be regulated using $d = 4 + 2\s \epsilon_\text{IR}$.  Then by analytically continuing $\epsilon_\text{IR} \rightarrow - \epsilon_\text{IR}$, we derive the result in \cref{eq:UVIRIntSeparated}. } 
\begin{align}
\mathcal{I}_\text{UV} &= \mu_\text{UV}^{2\epsilon}\int \frac{\D^d \ell}{(2\s\pi)^d} \frac{1}{\ell^2(\ell^2-m^2)}  = \frac{i}{16\s\pi^2}\left(\frac{1}{\epsilon_\text{UV}} +\log\frac{\muT^2_\text{UV}}{m^2}+1\right) + \mathcal{O}\big(\epsilon_\text{UV}\big)\notag\\[10pt]
\mathcal{I}_\text{IR} &=  \mu_\text{IR}^{2\epsilon}\int \frac{\D^d \ell}{(2\s\pi)^d} \frac{m^2}{\ell^4 (\ell^2-m^2)} = \frac{i}{16\s\pi^2}\left(\frac{1}{\epsilon_\text{IR}} +\log\frac{\muT^2_\text{IR}}{m^2}+1\right) + \mathcal{O}\big(\epsilon_\text{IR}\big)\,,
\label{eq:UVIRIntSeparated}
\end{align}
where $\mathcal{I} = \mathcal{I}_\text{UV} - \mathcal{I}_\text{IR}$, we have used the three denominator version of \cref{eq:NormalFeynmanParamsGeneral} to simplify the second integral, and we then applied \cref{eq:dimRegEvalGeneral} to evaluate the loop integral, and integrated over the Feynman parameters.  Since we want to interpret this as two contributions to the same integral, we should take $\muT^2_\text{UV} = \muT^2_\text{IR}$ 
\begin{align}
\mathcal{I} = \frac{i}{16\s\pi^2}\left(\frac{1}{\epsilon_\text{UV}}-\frac{1}{\epsilon_\text{IR}}\right) = 0\,,
\label{eq:UVIRint}
\end{align}
where the last equality is true when we take $\epsilon_\text{UV} = \epsilon_\text{IR}$.  This makes precise the notion that scaleless integrals vanish in dim reg.  

\vspace{5pt}\mybox{
\begin{itemize}
\item {\bf Exercise:} Derive \cref{eq:UVIRIntSeparated}.
\end{itemize}}

The feature that scaleless integrals vanish is very generic, and includes situations where the integrand is also a function of single components of the loop momentum.  For example, integrals that depend on $\ell\cdot v$, with $v^\mu = (1,0,0,0)$, show up in heavy particle EFTs, or factors like $\ell \cdot n$, with $n^\mu = (1,0,0,1)$, appear in SCET.

\vspace{5pt}\mybox{
\begin{itemize}
\item {\bf Exercise:} Show that 
\begin{align}
\int \frac{\D^d \ell}{(2\s\pi)^d}\frac{1}{\ell^2\,(v\cdot\ell)^2} = 0\,,
\end{align}
where $v = (1,0,0,0)$.
\end{itemize}}

Note that this also sharpens the notion of introducing an IR regulator as a way to isolate the UV divergent part of an integral.  If we wanted to extract the dim reg UV divergence of our integral in \cref{eq:ScalelessExDimReg}, we can regulate the IR by hand through the introduction of a small mass parameter $m$.  One might choose to regulate the integral as $\mathcal{I}_\text{UV}$ given in \cref{eq:UVIRIntSeparated}, or one could make an equally reasonable choice
\begin{align}
\mathcal{I} = \mu^{2\epsilon}\int \frac{\D^d \ell}{(2\s\pi)^d} \frac{1}{(\ell^2-m^2)^2} = \frac{i}{16\s\pi^2}\left(\frac{1}{\epsilon_\text{UV}} +\log\frac{\muT^2_\text{UV}}{m^2}\right) + \mathcal{O}\big(\epsilon_\text{UV}\big)\,,
\end{align}
which is just the integral given in \cref{eq:dimRegEvalTwoProps}.  Both approaches yield the same coefficient of $1/\epsilon_\text{UV}$, the avatar of the UV divergence, which could then be used to derive an anomalous dimension.  However, note that the finite terms are different.  This implies that if one is interested in the physics associated with the finite terms (as we will be in what follows), then care is required to ensure that the IR has been consistently regulated in the same way across all diagrams, for both the \FT~and EFT descriptions.

\scenario{Renormalization Group Evolution}
\addcontentsline{toc}{subsection}{\color{colorTech}{Primer~\thescenario.} Renormalization Group Evolution}
\label{sec:RGE}
Quantum field theory predicts the behavior of observables once a set of reference measurements have been fixed.  This allows one to derive couplings that are finite to a given order in perturbation theory, and can in turn be used to make finite predictions.  Logarithmic dependence can be absorbed into running couplings, whose evolution is governed by a set of RGEs.

From a more practical perspective, renormalization and RG evolution are an artifact of one of the inherent inefficiencies of the Feynman diagram expansion.  As opposed to being able to calculate finite matrix elements directly, Feynman loop integrals are often divergent, and counterterms must be included to derive a  physical result.  This procedure leaves behind logarithms, which can in principle become large enough to disturb the perturbative expansion.  Fortunately, RG techniques allow us to derive a set of RGEs that can be integrated to sum these large logarithms and improve the convergence of perturbation theory.  Specifically, we will see that the RG improved coupling $C$ as a function of a scale $\mu$ can schematically take the form (\emph{e.g.}~see \cref{eq:C4runningSol} below)
\begin{align}
C(\mu) = \frac{C(\Lambda)}{1 -  \frac{\gamma_C\,C(\Lambda)}{16\s\pi^2} \log \frac{\mu^2}{\Lambda^2}}\,,
\label{eq:RGSolToy}
\end{align}
where $\Lambda$ is some reference scale and $\gamma_C$ is the anomalous dimension (up to normalization).  If our coupling $C$ is perturbative, and we take $\mu \sim \Lambda$, then the logs are small and so we could choose to expand \cref{eq:RGSolToy} and truncate to finite order depending on the precision required 
\begin{align}
C(\mu)_\text{Expanded} = C(\Lambda) \times \left[1 + \frac{\gamma_C\,C(\Lambda)}{16\s\pi^2} \log \frac{\mu^2}{\Lambda^2} + \left(\frac{\gamma_C\, C(\Lambda)}{16\s\pi^2} \log \frac{\mu^2}{\Lambda^2} \right)^2 + \cdots \right] \simeq C(\mu)\,.
\end{align}
However, when $\mu \gg \Lambda$ or vice versa, the scale dependence of the coupling can no longer be neglected, and the use of the RG improved coupling in \cref{eq:RGSolToy} dramatically improves our control of the perturbative expansion by accounting for the largest logarithms that appear to all-orders.  

This discussion puts phrases like ``spoil the convergence of perturbation theory'' into context.  This choice of language should not be interpreted as literally implying that perturbation theory no longer holds.  It is worth noting that a scale does exist where the couplings truly become non-perturbative as a result of the presence of large logarithms.  This leads to an effect known as ``dimensional transmutation,'' and will be discussed around \cref{eq:LandauPole} below.\footnote{This statement is true of UV logs, where a breakdown of perturbation theory is tied to the running coupling becoming larger than $4\s\pi$.  However, when working with IR logs, the NLO cross sections can become negative due to a large contribution schematically of the form $1-\alpha/(16\s\pi^2) \log^2(M^2/m^2)$ while $\alpha$ is still perturbative.  This is an obvious breakdown of the perturbative expansion.  In such a case, one can tame this issue by RG improving perturbation theory, which provides a sensible controlled prediction.}   However, even when probing the theory far from the Landau pole, one is typically interested in making the most precise predictions possible, and so it is clear that making a scale choice to minimize the size of logarithms is ideal.  

There is additionally a non-trivial interplay between the RG improved coupling and the fixed order corrections that will appear in the improved perturbative expansion, as will be emphasized many times in the examples that follow.  We will see one important consequence of these fixed order effects when we study cases where large logarithms arise that are not a function of the RG scale $\muT$.  We will learn to interpret these large logarithms as the manifestation of working with the ``wrong'' effective description, and to apply the techniques of matching and running to sum them.  Furthermore, we will emphasize the appealing conceptual picture of flowing from a UV \FT~in terms of one set of degrees of freedom to an IR description that can appear very different.  The goal of this \Primer~is to remind us how to derive a set of RGEs for a given theory.

When reading about renormalization, one might encounter the statement, which is usually credited to  Georgi~\cite{Georgi:1994qn}, that there are two types of RGs.  The ``Wilsonian RG''  comes with an intuitive physical picture of integrating out successive momentum shells.  This was put on even more rigorous formal footing in a beautiful paper by Polchinski~\cite{Polchinski:1983gv}.  The idea of integrating out degrees of freedom to arrive at a coarse grained description also plays a role in condensed matter physics and the theory of phase transitions, see \emph{e.g.}~the classic Wilson and Kogut review article~\cite{Wilson:1973jj}.  And as has been emphasized here, the use of the RG to connect different EFTs is tied to the principle of locality in quantum field theory.

Unfortunately, the Wilsonian RG point of view introduces some calculational complexity.  One obvious issue is that integrating out a momentum shell imposes a hard cutoff which breaks gauge invariance.  Hence, when using this approach one must carefully restore gauge invariance through the appropriate choice of renormalization prescription order by order in perturbation theory.  We will avoid these complications by utilizing what Georgi calls the ``continuum RG'' picture, which is essentially an interpretation of the RG as implemented when using dim reg.  At its core, this approach relies on the vanishing of scaleless integrals to allow the user to integrate over all momenta.  So even when working within an EFT that is defined with respect to a cutoff $\Lambda_\text{UV}$, one can still integrate the loop momenta over the infinite domain.  The claim is that the regions of momentum that lie outside the regime of validity for the EFT only yield scaleless contributions to the total integral, so we might as well integrate over them, simplifying our calculations dramatically.  Specifically, we can naively apply the dim reg formulas discussed in the previous \Primer, even when we are working with an EFT that is only valid up to some finite cutoff scale.  We will put this vague discussion on much stronger mathematical footing when we introduce the ``method of regions''~\cite{Beneke:1997zp} for expanding loop integrals, first in \cref{sec:RegionsHeavyLight} and then in \cref{sec:RegionsMasslessSudakov} and \cref{sec:RegionsMassiveSudakov} below.  For now, the rest of this \Primer~is devoted to technical aspects of the continuum RG implemented in the $\overline{\text{MS}}$ scheme.

\subsubsection*{Deriving Renormalization Group Equations}

A bare Wilson coefficient $C^0$ is related to a renormalized one $C^r$ through the use of a counterterm $Z$:\footnote{\textbf{Disclaimer:} Technically this expression is missing the wave function renormalization factors associated with the fields that appear in the operator.  For all the examples considered in these lectures, the one-loop wave function renormalization vanishes, and so we will not include them for simplicity.  We caution the reader that these factors appear at one-loop order in most realistic cases, and therefore must be tracked.}
\begin{align}
C^0 = Z\,\mu^{n  \epsilon}\,C^r\,,
\label{eq:DefCounterTerms}
\end{align}
where we have included a factor of $\mu^{n \epsilon}$ to ensure that the coupling  remains dimensionless as in \cref{eq:muDependOperators}, and $n$ is an integer that depends on the mass dimension of the operator being renormalized, see \cref{eq:muDependOperators}.
For a perturbative model, 
\begin{align}
Z = 1 + \mathcal{O}\big(C^r,\alpha^r \big)\,,
\end{align}
where $C^r$ is the renormalized Wilson coefficient, and $\alpha^r$ are additional renormalized couplings within the theory, \emph{e.g.}~a gauge coupling.\footnote{\textbf{Disclaimer:}  There are many conventional choices one can make for how to define the counterterms, which in principle modify the general form of the RGE defined below in \cref{eq:RGEgeneral}.  For example, \cite{Peskin:1995ev} defines the counterterm for $\lambda\,\phi^4$ interactions as $-i\s \delta_\lambda$ (see Fig.~10.3 and the one-loop result in Eq.~(10.24)), while~\cite{Schwartz:2013pla} defines the same counterterm vertex as $-i\s \lambda\,\delta_\lambda$ (see the inline result for $\delta_\lambda$ directly below Eq.~(23.94)).  Yet another alternative formulation writes $Z$ as a matrix, see \emph{e.g.}~Eq.~(12.108) of~\cite{Peskin:1995ev}.  This matrix formulation has the advantage that one can interpret the off-diagonal $\gamma_{ij}$ elements as capturing all the physics that results from operator mixing, when the dependence on the Wilson coefficients is linear, but this will not always be the case for our examples.  Here, we choose to follow the conventions in~\cite{Schwartz:2013pla}.  This unfortunately obscures a simple operator mixing interpretation of $\gamma_{ij}$.  As we will see explicitly in the examples that follow, both $\gamma_{ii}$ and $\gamma_{ij}$ will have contributions from operator mixing.}  Counterterms are scheme dependent.  We will use $\overline{\text{MS}}$, so our prescription is that counterterms are derived by enforcing that they cancel all the terms which diverge as $\epsilon \rightarrow 0$ then $Z-1$ is the infinite correction generated by perturbation theory.  After using \cref{eq:DefMuMSBar} to redefine $\mu \rightarrow \muT$, the $\overline{\text{MS}}$ RG scale is identified with $\muT$.

The premise of the RGE (also known as the Callan-Symanzik equation~\cite{Callan:1970yg, Symanzik:1970rt, Symanzik:1971vw}) is that the bare Lagrangian cannot depend on the unphysical RG parameter $\muT$
\begin{align}
0 = \muT \frac{\text{d}}{\text{d} \muT} C^{0} = \muT \frac{\text{d}}{\text{d} \muT}\big(\s Z\,\mu^{n  \epsilon}\,C^r\big) \,.
\label{eq:invarianceOfC}
\end{align}

From here forward, we will specialize to an example for concreteness.  We will encounter some subtle factors of two, which is a reminder that it is typically good practice to derive RGEs from scratch by starting with \cref{eq:invarianceOfC} for a given theory of interest.\footnote{If only working to one-loop order, it is often simpler to just compute the log derivative of the renormalized coupling directly to extract the RGE, and then integrate the result to exponentiate the relevant logs.}  We want to derive the RG evolution of the coupling $C_4$ in a scalar non-renormalizable EFT with interactions\footnote{For simplicity, we will not include the operators with derivative dependence, and are therefore using a simplified notation $C_4 = C_{(4,0)}$.} 
\begin{align}
\mathcal{L}_\text{EFT}^\text{Int} \supset \frac{1}{4!}\,C_4\,\phi^4 + \frac{1}{6!}\,\frac{C_6}{M^2}\,\phi^6\,,
\end{align}
where the Wilson coefficients $C_4$ and $C_6$ are dimensionless.  Then we can evaluate \cref{eq:invarianceOfC} order by order in perturbation theory.  At tree-level
\begin{align}
0 &= \muT \frac{\D}{\D\muT} C_4^0 = \muT \frac{\D}{\D\muT} \big(\s Z_4\,\mu^{2\epsilon}\,C_4^r\big)\notag\\[7pt]
&=\muT\left(\cancelto{\, 0}{\frac{1}{Z_4}\frac{\D Z_4}{\D\mu}}+ \frac{1}{C_4^r}\frac{\D C^r_4}{\D\mu}+\frac{1}{\mu^{2\epsilon}}\,2\s\epsilon\,\mu^{2\epsilon-1}\right)\,\cancelto{\,1}{Z_{}^{}}_4\,C_4^r\,\muT^{2\epsilon} \notag\\[10pt]
&\Longrightarrow \quad \frac{\D C^r_4}{\D\log\muT^2} = - \epsilon\,C_4^r \qquad\big[\text{LO}\big]\,,
\label{eq:DDLogMuC4}
\end{align}
where in the last step we rewrote the derivative to be with respect to $\log \muT^2$ and absorbed a factor of 2 in the process, and we have used the fact that $Z_4 = 1 + \cdots$, and $\epsilon\,\mu^{2\epsilon} \rightarrow \epsilon + \mathcal{O}(\epsilon^2)$ as $\epsilon \rightarrow 0$.   This classical result is simply the statement that the dimension of the operator changes as a function of the spacetime dimension\footnote{Due to our choice to define the anomalous dimension with respect to $\D/\D \log\muT^2$, this factor is actually one half of the change in the dimension, see the $2$ in the exponent of \cref{eq:solveRGEToy}.  We choose to work self-consistency with this definition and ignore this distinction in what follows.}
\begin{align}
\gamma_4^\text{classical} = \frac{1}{C_4^r}\frac{\text{d}C_4^r}{\text{d}\log \muT^2} = -\epsilon\,,
\label{eq:GammaC4Classical}
\end{align}
where $\gamma_4$ is the anomalous dimension for the operator $C_4$.  We can perform the same manipulations for the term $Z_6\,C_6\,\phi^6/M^2 \,\rightarrow Z_6\,\mu^{4\epsilon}\,C_6\,\phi^6/M^2$, which yields
\begin{align}
\gamma_6^\text{classical} = \frac{1}{C_6^r}\frac{\text{d}C_6^r}{\text{d}\log \muT^2} = -2\s \epsilon\,.
\label{eq:GammaC6Classical}
\end{align}

The next step is to evaluate \cref{eq:invarianceOfC} to the next order in perturbation theory.  This will capture the leading quantum corrections.  The example we are working out here is instructive because it requires keeping track of loop induced operator mixing.  Physically this effect occurs when loops generated by insertions of one operator contribute to matrix elements of a different operator.  We have two diagrams which renormalize $C_4$ at one loop: 
\begin{align}
 \includegraphics[width=0.22\textwidth, valign=c]{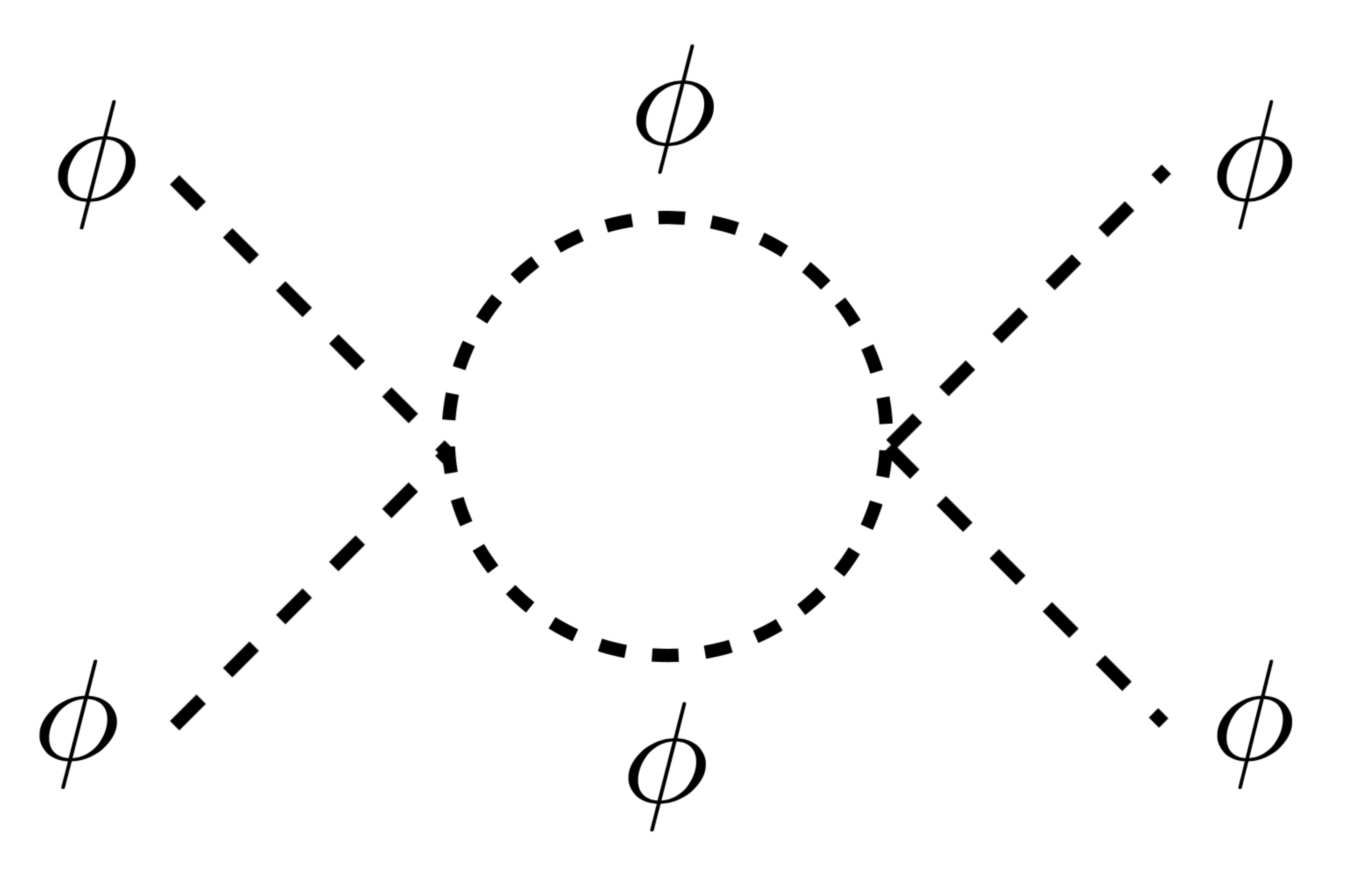} \qquad \qquad \qquad  \includegraphics[width=0.2\textwidth, valign=c]{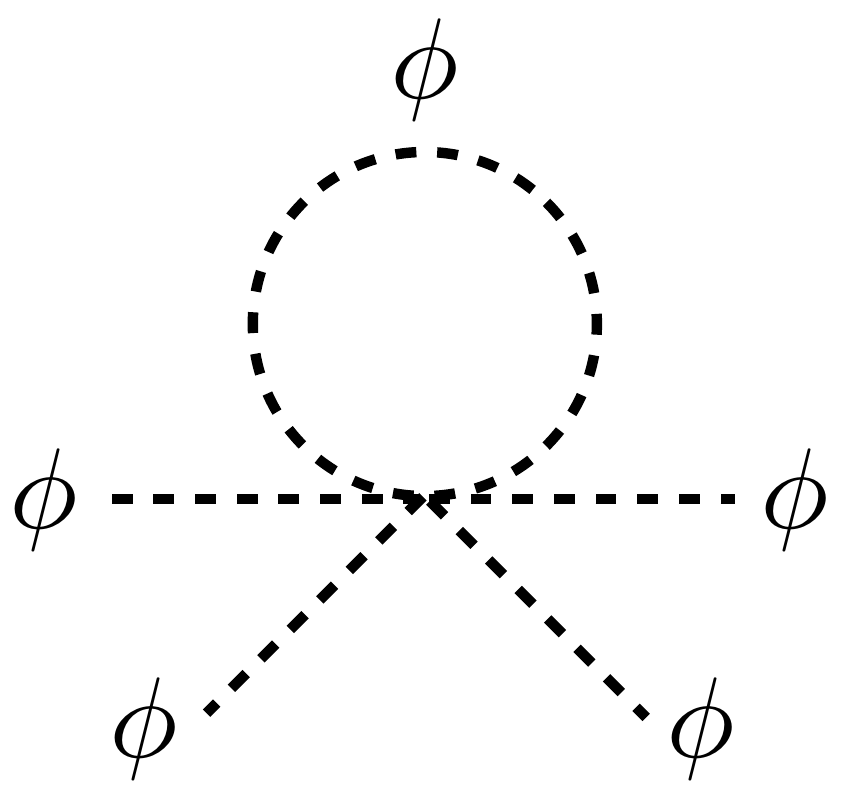}\,\,.
 \label{eq:FigsAnnDimLoops}
\end{align}
This implies that the counterterm depends on both Wilson coefficients $Z_4 = Z_4\big(C_4^r, C_6^r\big)$, and therefore \cref{eq:DDLogMuC4} generalizes to
\begin{align}
0 &=  \frac{\D}{\D\log\muT^2} \,C_4^0 =  \frac{\D}{\D\log\muT^2} \Big(Z_{4}\big(C_4^r,C_6^r\big)\,\mu^{2\epsilon}\,C_4^r\Big)\notag\\[10pt]
 &= \frac{1}{2} \left(\frac{\partial Z_{4}}{\partial C^r_4}\frac{\muT}{Z_{4}}\frac{\D C^r_4}{\D\muT}+\frac{\partial Z_4}{\partial C^r_6}\frac{\muT}{Z_{4}}\frac{\D C^r_6}{\D\muT}+ \frac{\muT}{C_4^r}\frac{\D C^r_4}{\D \muT}+2\s\epsilon\right) Z_{4}\,\mu^{2\epsilon}\,C_4^r\,.
 \label{eq:DeriveRGEC4}
 \end{align}
For a perturbative expansion $Z_{4} = 1 + \mathcal{O}(C_4,C_6)$, which implies we can expand and truncate $1/Z_4 = 1 + \cdots\,$.  We then plug in \cref{eq:GammaC4Classical} and \cref{eq:GammaC6Classical}, and solve for the log derivative of $C_4^r$, which yields
\begin{align}
\frac{\D C^r_4}{\D\log\muT^2}  = \left(\epsilon\,\big(C_4^r\big)^2\,\frac{\partial Z_{4}}{\partial C^r_4}-\epsilon\,C_4^r +2\s \epsilon\,C_4^r\,C_6^r\frac{\partial Z_{4}}{\partial C^r_6}\right) \qquad\big[\text{NLO}\big]\,.
\label{eq:RGEOneLoopC4C6}
\end{align}
Finally, we identify the anomalous dimensions including quantum effects and operator mixing as
\begin{align}
\gamma_{44} =\lim_{\epsilon\rightarrow 0}\left( \epsilon\,C_4^r\,\frac{\partial Z_{4}}{\partial C^r_4}-\epsilon\right) \qquad\qquad \qquad \gamma_{46} = \lim_{\epsilon\rightarrow 0}\left(2\s\epsilon\,C_4^r\,\frac{\partial Z_{4}}{\partial C^r_6}\right) \,,
\label{annDimPractical}
\end{align}
such that the RGE can be written in a general form
\begin{align}
\frac{\text{d}}{\text{d}\log \muT^2} C^r_n = \gamma_{nm}\,C^r_m\,,
\label{eq:RGEgeneral}
\end{align}
where now we include two indices on the anomalous dimensions to account for mixing effects.  One implication of these expressions is that $C_4^r$ will be generated by RG running as long as $C_6 \neq 0$, even if $C_4^r\big(\muT_M\big) = 0$.  This physics will play a role when we work out our heavy-light example in \cref{eq:SepScalesHLlog} below.

\subsubsection*{Implications of the Renormalization Group}

This general form of the RG makes clear why scale separation is a non-trivial problem.  The RGE sums logs that are a function of the RG scale $\muT$.  But if one encounters a large logarithm that only depends on physical scales, then it is not obvious how to apply this formalism.  The key observation emphasized over and over in these lectures is that the matching procedure will introduce a spurious dependence on $\muT$, which then allows us to derive RGEs of the form~\cref{eq:RGEgeneral}.

Before moving on, we will briefly discuss the connection between the object $\gamma$ and the phrase ``anomalous dimension.''  Imagine a case where we are interested in renormalizing a Wilson coefficient $C$ that receives corrections from some interactions implying a non-zero anomalous dimension $\gamma = \text{const}$.\footnote{Note that the anomalous dimension computed from the diagrams in \cref{eq:FigsAnnDimLoops} take a slightly different form (even with the simplifying approximation that $C_6 = 0$), since $\gamma_{44} \sim C_4 \neq \text{const}$.  The resulting solution to the RGE does not make the direct connection to the change in dimension that we are highlighting here.  The explicit calculation for $\phi^4$ theory is given in \cref{eq:gamma44} below.  For unity of notation, we will always phrase our RGE calculations in terms of anomalous dimensions, especially since this is the terminology often used by EFT practitioners.}  This could occur for example if one had a local operator whose insertion generated some external charged lines, and then the leading contribution to $\gamma$ was due to exchanges of gauge bosons, \emph{e.g.} our SCET examples below.  In this approximation, we can solve the RGE
\begin{align}
\int \frac{1}{C^r}\, \D C^r &= \int \gamma\,\text{d}\log \muT^2 \notag \\
\Longrightarrow\qquad C^r\big(\muT_H\big) &= C^r\big(\muT_L\big)\,\exp \left(\gamma \log \frac{\muT_H^2}{\muT_L^2}\right) = \left(\frac{\muT_H}{\muT_L}\right)^{2\s\gamma}\,C^r\big(\muT_L\big)\,.
\label{eq:solveRGEToy}
\end{align} 
This shows that the role of the anomalous dimension is to account for the small change in the mass dimension of the Wilson coefficient as it is evolved from the high scale to the low scale.  Also, note that the middle expression in the bottom line of \cref{eq:solveRGEToy} makes explicit what is meant by the phrase ``exponentiating logarithms,'' since we literally see the appearance of $\exp\big(\gamma\log \muT_H^2/\muT_L^2\big)$ in the solution to the RGE.

The running Wilson coefficients are a critical component of an improved perturbation theory.  Since this sums large logarithms that could otherwise cause issues with convergence.  Now we have a dual expansion, schematically as a function of $\frac{\alpha}{4\s \pi}\, \log \muT^2_H/\muT^2_L$ and $\alpha$ separately, where $\alpha$ is a proxy for some coupling constant $\alpha = g^2/(4\s \pi)$.  RG improvement becomes necessary when the logs becomes sufficiently large that $\alpha \sim \frac{\alpha^2}{4\s \pi}\, \log \muT^2_H/\muT^2_L$.  

The anomalous dimension $\gamma$ can be determined perturbatively, and the order in the log expansion that is being summed is set by the order to which the anomalous dimension has been computed -- we call this the next$^n$ leading log expansion \big(N$^{n}$LL\big).  This can be compared with the more familiar expansion in terms of the coupling $\alpha$, which we refer to as the next$^m$ leading order \big(N$^{m}$LO\big) expansion, where the careful implementation of the matching procedure ensures no double counting will occur.  Note this implies that LL begins at one-loop order, in contrast with LO which starts at tree-level.  This dual expansion can be represented schematically as
\vspace{5pt}
\begin{align}
\label{eq:NLONLLUV}
\includegraphics[width=0.55\textwidth, valign=c]{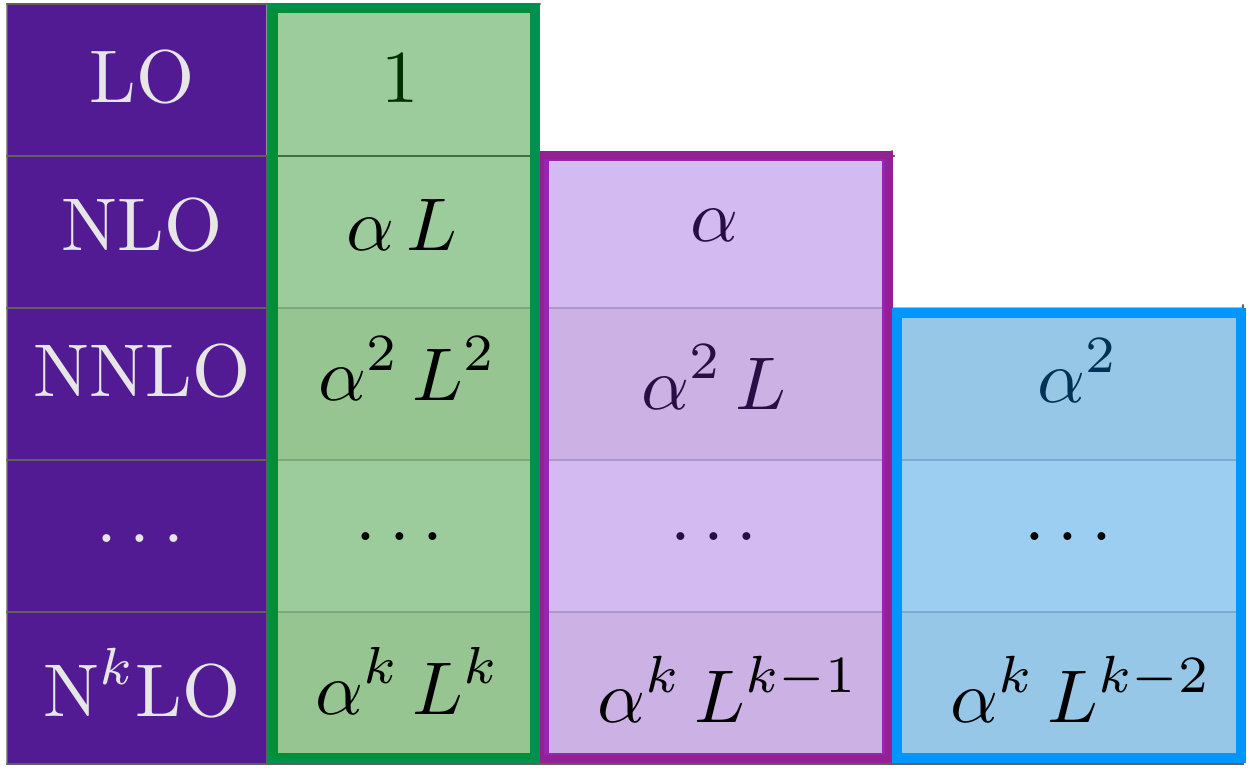} \\[-15pt] \notag
\end{align}
where $\alpha$ tracks the expansion in the coupling, while $L$ tracks the logarithms.  The rows denote the kinds of terms that appear as one expands in $\alpha$, while the shaded columns show the series that are summed at each logarithmic order, starting with leading log (LL) in green, subleading log (NLL) in magenta, and so on.  We emphasize that the complete expressions are not fully captured by resummed couplings, but also require the inclusion of fixed order terms as we will see explicitly in many examples that follow. 

While this dual expansion is straightforward to keep track of for the UV logs discussed here, we will see that the interplay of N$^{n}$LL and N$^{m}$LO becomes more complicated when summing large IR logarithms, see \cref{sec:CuspAnnDim} below.  We emphasize the remarkable fact that each order in the log expansion organizes itself this way, and we encourage the reader to be suitably impressed by this consequence of the RG approach.\footnote{The simplicity encountered here can be understood as a consequence of having a linear set of RGEs.  In principle, more complicated RGEs can appear, which would make the interpretation of the resummed couplings more complicated.}

Next, we will apply this technology to a series of examples which will allow us to explore how to sum logarithms.  The first example is an EFT that includes only a $\phi^4$ coupling. This result will not only be of use as a simple pedagogical application of RG techniques, but it will additionally be recycled when we encounter our first case of separating scales in \cref{eq:SepScalesHLlog}.  

\subsection{Summing Logs}
\label{sec:ToyModelLLResum}
Now that we have the technology for deriving RGEs, we use these techniques to sum the leading logarithm that appears in a single particle EFT with the interaction
\begin{align}
\mathcal{L}_\text{Int}^\text{EFT} = -\frac{1}{4!}\,C_4\,\phi^4\,.
\end{align}
We will compute the process $\phi\,\phi \rightarrow \phi\,\phi$ at threshold.  The one-loop diagrams will yield logarithms, and we will derive the RGE that sums them by running from $\muT^2_H$ to $\muT^2_L$:
\begin{align}
\includegraphics[width=0.3\textwidth, valign=c]{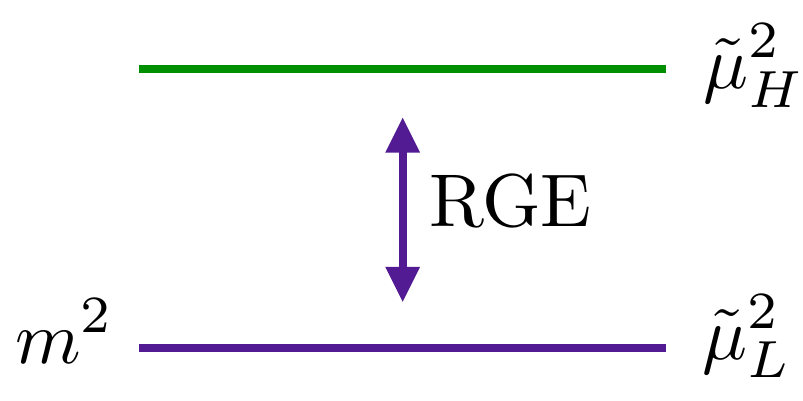} 
\end{align}
Note that our choice to run from a high scale $\muT^2_H \gg \muT_L^2\sim m^2$ will cause artificial large log problems, since this is a single scale EFT and we are at threshold which is set by the same scale.  However, we will work out the details as a pedagogical example, since many of the technical features will appear in the more complicated models that follow. 

Our tree-level Feynman rule is 
\begin{align}
 \includegraphics[width=0.15\textwidth, valign=c]{Figures/LLLL4Pt.pdf} = -i\s C_4^r\big(\muT_H\big)\,,
\end{align}
and this is also the tree-level matrix element for our process of interest.  

Next, we want to compute the one-loop diagrams to derive the counterterm from which we will extract the anomalous dimension $\gamma_{44}$.  At one loop, the $t$- and $u$-channel diagrams take the form
\begin{align}
\includegraphics[width=0.23\textwidth, valign=c]{Figures/LLLL_LLLoop.pdf} &= \frac{1}{2}\,\big(\mu_H^{2\epsilon}\,C^r_4\big)^2  \int \frac{\text{d}^d\ell}{(2\s\pi)^d} \frac{1}{(\ell^2-m^2)^2} \notag\\
&= \frac{i}{32\s\pi^2}\,\big(C^r_4\big)^2 \left(\frac{1}{\epsilon}+\log\frac{\muT_H^2}{m^2}\right)\,,
\label{eq:phi4LoopDiag}
\end{align}
where the $1/2$ is a symmetry factor.  Then additionally including the $s$-channel, we find 
\begin{align}
i\s\mathcal{A} = 
\frac{3\s i}{32\s\pi^2}\,\big(C^r_4\big)^2\left( \frac{1}{\epsilon} +\log\frac{\muT_H^2}{m^2} + \frac{2}{3}\right)\,.
\label{eq:Aphi4oneloop}
\end{align}  
\vspace{-10pt}\mybox{\begin{itemize}
\item \textbf{Exercise:} Note that our choice to study the process $\phi\,\phi\rightarrow \phi\,\phi$ at threshold implies  $\qquad p = \big(m,\vec{0}\,\big)$ for all the external states.  Hence, there is finite momentum flowing through the $s$-channel loop, such that this diagram has a different $m^2$ dependence in the denominator than \cref{eq:phi4LoopDiag}.  This is an IR effect and can have no implications for the UV divergence structure.  It does yield the non-log factor that mysteriously appears in the previous expression, which you are encouraged to check as an exercise.
\end{itemize}}
Then we add a counterterm diagram to these three loop diagrams,
\begin{align}
3\left(i\s \frac{1}{32\s\pi^2}\,\mu_H^{2\epsilon}\,\big(C^r_4\big)^2 \frac{1}{\epsilon} +\cdots \right) - i\s \mu_H^{2\epsilon}\,C^r_4\big(Z_4-1\big) = 0\,,
\end{align}
where setting this combination equal to zero is equivalent to taking the $\overline{\text{MS}}$ scheme, and we find
\begin{align}
Z_{4} &= 1 +\frac{3}{32\s\pi^2}\,C^r_4\,\frac{1}{\epsilon} \,.
\label{eq:Z44}
\end{align}
Adding these four contributions together, sending $\mu \rightarrow \muT$, and taking $\epsilon \rightarrow 0$ yields the renormalized amplitude: 
\begin{align}
i\s\mathcal{A}^\text{EFT}\Big(\phi\,\phi \rightarrow \phi\,\phi\Big) = -i\s C_4^r\left[1 - \frac{3}{32\s\pi^2}\,C_4^r\left( \log\frac{\muT_H^2}{m^2}-\frac{2}{3}\right)\right] \qquad \text{\big[NLO\big]}\,,
\label{eq:A4ptRenorm}
\end{align}
If $\muT_H^2 \gg m^2$, this log becomes large and our perturbation expansion begins to break down.  Therefore, we would like to improve perturbation theory through the use of summation so that we can maintain a well behaved expansion across a wider range of scales.   

One resolution is to evaluate the theory at a lower scale (alternatively, we could define the coupling using a different choice of process away from threshold).  However, the nature of the multi-scale problems we are working towards will force us to deal with exactly this situation, where a simple scale choice in the \FT~will not be sufficient to keep the truncated fixed order perturbative expansion under control.  Therefore, we will spend the rest of this section deriving an RGE, which will allow us to evolve our well behaved low scale result to derive an improved high scale prediction.  This will also expose the interplay between LL and NLO, which will be another reoccurring theme in the examples that follow.

In order to derive the RGE, we plug \cref{eq:Z44} into our general expression \cref{annDimPractical} to yield
\begin{align}
 \gamma_{44} &= \frac{3}{32\s\pi^2}\,C^r_4\,.
\label{eq:gamma44}
\end{align}
Now we can use this anomalous dimension and \cref{eq:RGEgeneral} to determine the RGE that runs the $C^r_4$ coupling at LL order:
\begin{align}
\frac{\D C^r_4}{\D \log \muT^2} &=  \frac{3}{32\s\pi^2}\,\big(C^r_4\big)^2\,.
\end{align}
We integrate from our high scale $\muT_H$ to the IR scale $\muT_L$:
\begin{align}
  \int_{C^r_4(\muT_L)}^{C^r_4(\muT_H)} \frac{\D C^r_4}{\big(C^r_4\big)^2} &= \frac{3}{32\s\pi^2} \int_{\muT_L^2}^{\muT_H^2}\D \log \muT^2 \notag\\[9pt]
  \frac{1}{C^r_4\big(\muT_L\big)}-\frac{1}{C^r_4\big(\muT_H\big)} &= \frac{3}{32\s\pi^2} \log \frac{\muT_H^2}{\muT_L^2}\notag\\[9pt]
  C^r_4\big(\muT_L\big) &=  \frac{C^r_4\big(\muT_H\big)}{1+C^r_4\big(\muT_H\big)\,\frac{3}{32\s\pi^2} \log \frac{\muT_H^2}{\muT_L^2}}\,.
\label{eq:C4runningSol}
\end{align}
This provides us with a running coupling.  When we use this in our EFT calculations, we call this the ``RG improved'' perturbation theory.  For example, we can compute our process at the low scale to LL + LO accuracy using the high scale coupling
\begin{align}
i\s\mathcal{A}^\text{EFT} =  -i\s C^r_4\big(\muT_L\big)= \frac{-i\s C^r_4\big(\muT_H\big)}{1+C^r_4\big(\muT_H\big)\,\frac{3}{32\s\pi^2} \log \frac{\muT_H^2}{\muT_L^2}}\qquad \text{\big[LL\big]}\,.
\end{align}
Note that for the special choice of scale $\muT_L = m$, we can expand to subleading order in $C_4\big(\muT_H\big)$, and recover all but the finite piece at NLO, which is exactly what we would expect from \cref{eq:NLONLLUV}.

We can extend this to include NLO corrections at the low scale by repeating the calculation that led to \cref{eq:A4ptRenorm} but evaluated at $\muT_L$.  This yields (again using the RG to express our low scale result in terms of the high scale Wilson coefficient):
\begin{align}
\hspace{-10pt}i\s\mathcal{A}^{\text{EFT}} &=\frac{-i\s C^r_4\big(\muT_H\big)}{1+C^r_4\big(\muT_H\big)\,\frac{3}{32\s\pi^2} \log \frac{\muT_H^2}{\muT_L^2}}\notag\\[5pt]
&\hspace{15pt}\times\left[1 - \frac{3}{32\s\pi^2}\,\frac{C^r_4\big(\muT_H\big)}{1+C^r_4\big(\muT_H\big)\,\frac{3}{32\s\pi^2} \log \frac{\muT_H^2}{\muT_L^2}} \left(\log\frac{\muT_L^2}{m^2}+\frac{2}{3}\right)\right]\qquad \text{\big[LL + NLO\big]}\,.
\label{eq:A4ptResummedNotExpanded}
\end{align}
This expression is well behaved, and in particular the factor of $\log \muT_L^2/m^2$ is under control as long as $\mu_L \sim m$.  Next, we see that expanding this LL + NLO result pretending that $\frac{C^r_4}{16\s\pi^2}\,\log\frac{\muT_H^2}{\muT_L^2}$ is small yields\footnote{This is the first time we have encountered a result that is labeled with an ``Expanded'' subscript.  This signifies that we have first RG improved, and then expanded our RGE solution to leading order.  We are going to do this with essentially every example in what follows, so it is worth a bit of extra time to make sure you understand the motivation for why we are presenting the result this way.} 
\begin{align}
i\s\mathcal{A}^{\text{EFT}}_\text{Expanded} &= -i\s C^r_4\big(\muT_H\big)\left\{ \left(1- C^r_4\big(\muT_H\big)\,\frac{3}{32\s\pi^2} \log \frac{\muT_H^2}{\muT_L^2}\right)\right.\notag\\[8pt]
&\hspace{20pt}\times \left. \left[1 - \frac{3}{32\s\pi^2}\,C^r_4\big(\muT_H\big) \left(\log\frac{\muT_L^2}{m^2}+\frac{2}{3}\right)\right] + \mathcal{O}\left(\big(C^r_4\big)^2 \log^2 \frac{\muT_H^2}{\muT_L^2} \right)\right\} \notag\\[10pt]
&= -i\s C^r_4\big(\muT_H \big) \left\{1-C^r_4\big(\muT_H\big) \left(\log\frac{\muT_H^2}{m^2}+\frac{2}{3}\right)+ \mathcal{O}\left(\big(C_4^r\big)^2 \log^2 \frac{\muT_H^2}{\muT_L^2} \right) \right\}\notag\\[10pt]
\hspace{-100pt}&\hspace{280pt} \text{\big[LL + NLO\big]}\,,
\label{eq:A4ptResummed}
\end{align}
where the higher order terms are those that are summed by the RG at LL order.  This expanded form makes clear that the RG evolution reproduces the high scale evaluation given in \cref{eq:A4ptRenorm} to NLO order.  Furthermore, we now see how the RG solves the large log problem we artificially introduced in \cref{eq:A4ptRenorm}.  Specifically, the summed version of this expression given in \cref{eq:A4ptResummedNotExpanded} is well behaved for any choice of low scale.  If our scales were not particularly separated, then the expansion performed in \cref{eq:A4ptResummed} would be a good approximation, and there would not have been a problem to solve in the first place.  On the other hand, if $\muT_H \gg \muT_L$, then our desire for a convergent perturbation theory forces us to RG improve.  

Although we summed, note that working at LL + NLO order is only a good approximation as long as the next-to-leading log is small: $\frac{(C_4^r)^2}{16\s\pi^2} \log \frac{\muT_H^2}{m^2} \ll 1$.  If not, then we would need to include higher order corrections, and the same story would result from the interplay of NLL and NNLO logs, where summing to NLL order requires a two-loop anomalous dimension, and NNLO is the fixed order contribution at two loops.  This interplay persists so that if we were able to work to all orders in our N$^n$LL and N$^m$LO expansion, we would find that all the scale dependence would be eliminated.  Recall that this was the requirement we used to derive the RGEs in the first place, so the story is self consistent.  However, in practice we always work at finite order, so it is extremely useful to have an RG improved perturbation theory which is well behaved across a huge range of scales.  

Another lesson of \cref{eq:A4ptResummedNotExpanded} is that it provides us with the ability to estimate a ``theory error'' by varying the unphysical high and low scales that appear in the logs.  For example, instead of choosing $\muT_L = m$, which eliminates the NLO log completely, one can vary this scale choice -- typically for concreteness the variation is taken to be range from $\mu/2 \rightarrow 2\,\mu$.  This probes the higher logs which are captured by the RGEs, but are not explicitly included since the finite corrections are truncated to a given order.  Although this scale variation does not have to be the only source of theory error, it does provide a concrete test of the extent to which higher orders must be computed to achieve the desired accuracy.

There is one final interesting point to make before moving to our next example.  It turns out, even our RG improved theory can dynamically generate a non-perturbative breakdown of perturbation theory.  To see how this occurs, we can solve \cref{eq:C4runningSol} for the scale $\muT_H$ at fixed $\muT_L$,  
\begin{align}
\muT_H^2 = \muT_L^2\,\exp\left(\frac{1- \frac{C^r_4(\muT_L)}{C^r_4(\muT_H)}}{ \frac{3}{32\,\pi^2}\,C^r_4 (\muT_L)}\right)\,.
\end{align}
We can then take the obviously non-perturbative limit $C^r_4(\Lambda) \rightarrow \infty$, where this implicitly defines the dimensionful scale $\Lambda$.  This yields a finite result:
\begin{align}
\Lambda^2 = \muT_L^2\,\exp\left(\frac{1}{ \frac{3}{32\,\pi^2}\,C^r_4(\muT_L)}\right)\,.
\label{eq:LandauPole}
\end{align}
This scale $\Lambda$ is known as a Landau pole.  It tells us that the theory is no longer perturbative at a scale exponentially higher than the reference scale $\muT_L$.\footnote{A typical choice would be to set $\muT_L = m$.}  So while the RG improvements have afforded us an exponentially large region of validity for our EFT, we are not in a position to extrapolate our theory to arbitrarily high energies.  

We emphasize that this dimensionful scale $\Lambda$ was generated by studying the behavior of a dimensionless parameter $C^r_4$.  This is a non-perturbative effect known as dimensional transmutation.  The same underlying mechanism is responsible for the generation of the QCD scale (albeit this instead happens as you run to low energies due to the famous sign of the QCD $\beta$-function), and dimensional transmutation could also yield an effect known as dynamically generated supersymmetry breaking~\cite{Witten:1981nf} (assuming of course that nature is supersymmetric at some fundamental scale).

In the next section, we will calculate our first loop level matching correction, and will investigate the role of the finite terms for maintaining decoupling of heavy scales as one evolves a theory past a heavy mass threshold.

\subsection{One-loop Matching and Heavy Particle Decoupling}
\label{sec:LoopMatching}
As we have emphasized many times, heavy physics will decouple as we flow to low energies.   However, this is fact can be obscured depending on the approach one takes to extracting observables.  The detailed demonstration of decoupling for the RG evolution of gauge couplings is known as the Appelquist-Carazzone theorem~\cite{Appelquist:1974tg}.  It is straightforward to see this decoupling effect when regulating divergences with a hard cutoff or some other dimensionful regulator, see \emph{e.g.}~\cite{Manohar:1995xr, Petrov:2016azi, Manohar:2018aog} where this physics is worked out in detail.  For a massless regulator like dim reg makes, this is a more subtle issue, see \emph{e.g.}~\cite{Kazama:1979xc, Ovrut:1979pk, Weinberg:1980wa, Hagiwara:1980jm} for some early work on exploring decoupling for massless regulators.    

In this section, we will first write down a \FT~that has a large log problem at NLO.  We will furthermore see that decoupling is naively violated by our $\overline{\text{MS}}$ formulation of the RG.  The resolution to both of these issues will be achieved through the application of matching and running.  Then in the section following this one, we will demonstrate that applying the same strategy to scalar mass terms leads to a different class of decoupling violation, \emph{i.e.}, the so-called hierarchy problem.

For concreteness, we will study the process $\phi\,\phi \rightarrow \phi\,\phi$ at threshold.  We begin by defining our \FT~interaction Lagrangian
\begin{align}
\mathcal{L}^\textsc{Full}_{\text{Int}} &= - \frac{1}{4}\s \kappa\,\phi^2\,\Phi^2-\frac{1}{4!}\s \eta\,\phi^4\,,
\end{align}
where $\phi$ has mass $m$, $\Phi$ has mass $M$, and we will be interested in the limit $m\ll M$.   We will calculate the one-loop corrections including LL summation for this process, and use matching and running to make decoupling manifest. 

First we will calculate the renormalization factor for $\eta$.\footnote{\textbf{Disclaimer:} We will no longer be careful to delineate the difference between bare and renormalized quantities, so you will notice the absence of ``$\s r\s$'' superscripts.}   This has two contributions, one proportional to $\eta^2$ and one proportional to $\kappa^2$.  The first set of diagrams is identical to the ones computed in the previous section, so we can pull the answer from \cref{eq:gamma44}:
\begin{align}
3\times \includegraphics[width=0.23\textwidth, valign=c]{Figures/LLLL_LLLoop.pdf} = \frac{3\s i}{32\s\pi^2}\,\mu^{2\epsilon}\,\eta^2 \left( \frac{1}{\epsilon}  +\log\frac{\muT^2}{m^2} + \frac{2}{3} \right)\,.
\label{eq:phi4phiLoop}
\end{align}
The second class of diagrams is the same with $m \rightarrow M$ (up to subtleties with tracking the threshold momentum flow through $s$-channel diagram, which only modifies the result by power suppressed terms $\sim m^2/M^2 \sim \lambda^2$):
\begin{align}
3\times \includegraphics[width=0.23\textwidth, valign=c]{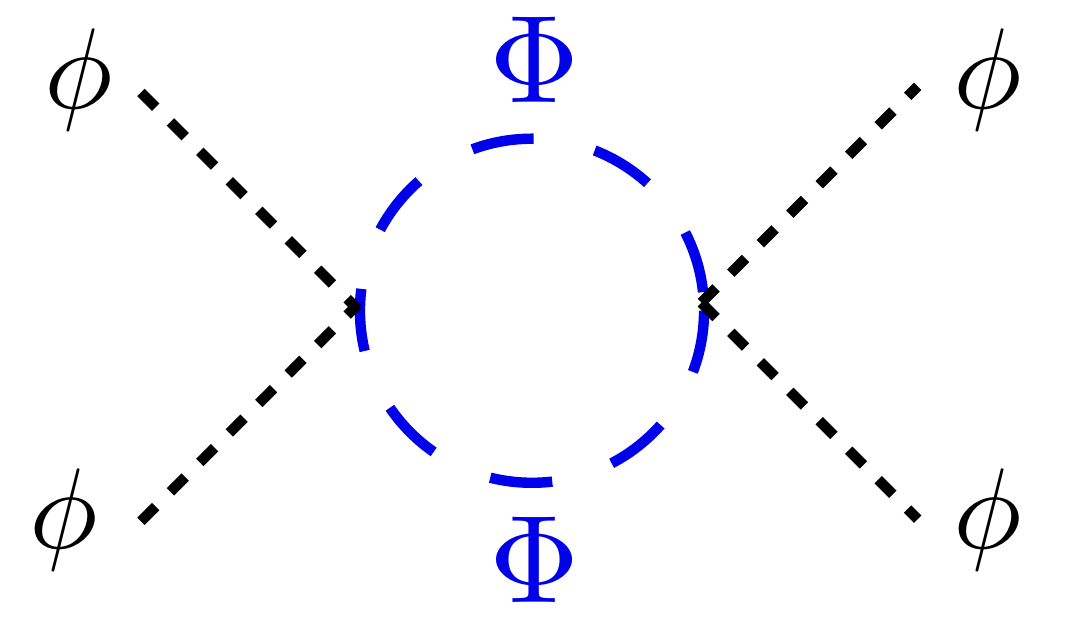} = \frac{3\s i}{32\s\pi^2}\,\mu^{2\epsilon}\,\kappa^2 \left( \frac{1}{\epsilon}  +\log\frac{\muT^2}{M^2} \right) + \mathcal{O}\big(\lambda^2\big)\,.
\end{align}
This implies\footnote{The strange looking factor of $1/\eta$ appearing in the $Z_\eta$ expression is due to our convention for counter terms defined in~\cref{eq:DefCounterTerms} above.}
\begin{align}
Z_\eta(\eta,\kappa) = 1 + \frac{3}{32\s\pi^2} \,\eta\,\frac{1}{\epsilon} + \frac{3}{32\s\pi^2} \,\frac{\kappa^2}{\eta}\,\frac{1}{\epsilon}\,.
\label{eq:Zeta}
\end{align}

We also need $Z_\kappa$, which is calculated by extracting the divergent part of the two one-loop diagrams that contribute to $\phi\,\Phi \rightarrow \phi\,\Phi$.  The first is   
\begin{align}
2\times \includegraphics[width=0.23\textwidth, valign=c]{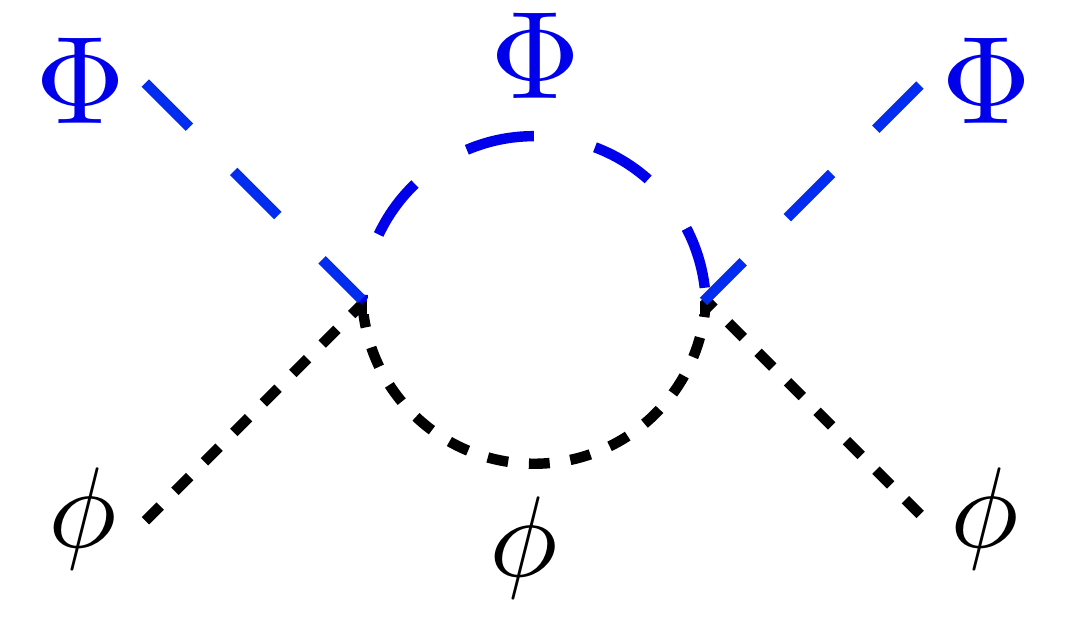}&= 2\,\big(\mu^{2\epsilon}\,\kappa\big)^2 \int \frac{\D^d \ell}{(2\s \pi)^d} \frac{1}{\ell^2-M^2}\frac{1}{\ell^2-m^2} \notag\\[8pt]
&= \frac{i}{8\s\pi^2}\,\mu^{2\epsilon}\,\kappa^2\,\frac{1}{\epsilon} + \cdots\,,
\label{eq:Phi2phi2Oneloop}
\end{align}
where the overall factor of 2 accounts for the contributions from the $s$- and $u$-channel, and we are neglecting the flow of threshold momentum through the propagator since we are only interested in the divergent terms, see the discussion around \cref{eq:Aphi4oneloop}.   There is a second diagram where only light particles appear in the loop
\begin{align}
\includegraphics[width=0.23\textwidth, valign=c]{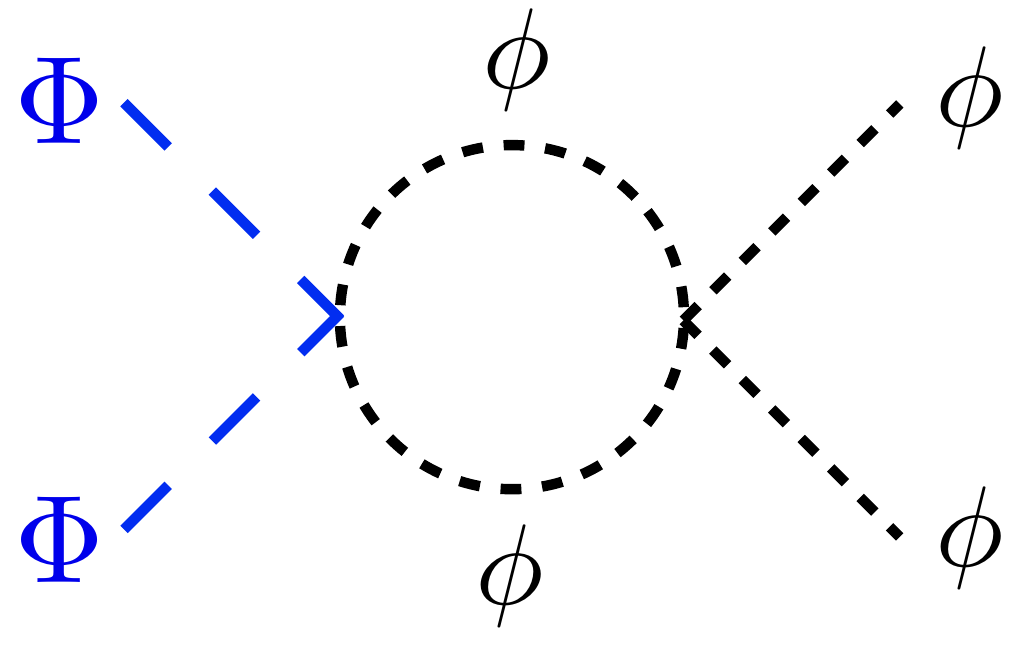}\hspace{-5pt} = \frac{1}{2}\,\big(\mu^{2\epsilon}\big)^2\kappa\,\eta \int \frac{\D^d \ell}{(2\s \pi)^d} \frac{1}{(\ell^2-m^2)^2} = \frac{i}{16\s\pi^2}\,\mu^{2\epsilon}\,\kappa\,\eta\,\frac{1}{\epsilon} + \cdots\,,
\label{eq:Phi2phi2Oneloop2}
\end{align}
where there is only a $t$-channel diagram and the $1/2$ is a symmetry factor.  Then the $Z_\kappa$ renormalization factor is
\begin{align}
Z_\kappa(\kappa, \eta) = 1+  \frac{1}{8\s\pi^2} \,\kappa\, \frac{1}{\epsilon}+\frac{1}{32\s\pi^2} \,\eta\, \frac{1}{\epsilon} \,.
\label{eq:Zkappa}
\end{align}

Now that we have $Z_\kappa$ and $Z_\eta$, we use them to renormalize our couplings.  Putting all of this together yields the NLO \FT~amplitude for $\phi\,\phi \rightarrow \phi\,\phi$: 
\begin{align}
i\s\mathcal{A}^\textsc{Full} = -i\s\eta +i\s \frac{3}{32\s\pi^2}\, \eta^2\, \left( \log\frac{\muT^2}{m^2}+\frac{2}{3}\right) +i\s \frac{3}{32\s\pi^2} \,\kappa^2\, \log\frac{\muT^2}{M^2}+ \mathcal{O}\big(\lambda^2\big) \qquad  \text{\big[NLO\big]}\,.
\label{eq:phi4FullLargeLogProb}
\end{align}
This makes the large log problem clear, since there is no choice of the scale $\muT$ that minimizes both log dependent terms when $m \ll M$.

We encountered a similar issue when we took $\muT_H \gg m$ in single scale theory discussed above, see~\cref{eq:A4ptRenorm}.  There, we were able to address this problem by working with RG improved couplings: we showed that the LL + NLO single scale result was free of any large log issues, see~\cref{eq:A4ptResummedNotExpanded}.  The reason this worked above is that we artificially introduced a large log problem through our poor choice of the unphysical RG scale.  We will now show that this is not the resolution to the problem encountered here due to the presence of multiple physical scales in the theory.  

The first step is to see that RG improvement does not resolve our decoupling confusion.  We need to compute the \FT~RGEs, which depend on the anomalous dimensions for both $\kappa$ and $\eta$.  Following the same steps as in \cref{sec:RGE} above, we derive the analog of \cref{annDimPractical} in order to extract the anomalous dimensions from the counterterms:
\begin{align}
\gamma_{\eta\eta} &=\lim_{\epsilon\rightarrow 0}\left( \epsilon\,\eta\,\frac{\partial Z_{\eta}}{\partial \eta}-\epsilon\right) \qquad\quad \qquad \gamma_{\eta\kappa} =\lim_{\epsilon\rightarrow 0}\left(  \epsilon\,\eta\,\frac{\partial Z_{\eta}}{\partial \kappa}\right)\notag\\[9pt]
 \gamma_{\kappa\eta} &=\lim_{\epsilon\rightarrow 0}\left(  \epsilon\,\kappa\,\frac{\partial Z_{\kappa}}{\partial \eta}\right) \hspace{78pt} \gamma_{\kappa\kappa} =\lim_{\epsilon\rightarrow 0}\left(  \epsilon\,\kappa\,\frac{\partial Z_{\kappa}}{\partial \kappa}-\epsilon\right) \,,
\label{annDimPracticalMix}
\end{align}
where we have used the classical anomalous dimensions for $\eta$ and $\kappa$  given in \cref{eq:DDLogMuC4}.  In this way of formulating the anomalous dimensions, both $\gamma_{\eta\eta}$ and $\gamma_{\eta\kappa}$ will contribute operator mixing terms to the RGE for $\eta$, while $\gamma_{\kappa \eta}$ will yield operator mixing for $\kappa$:
\begin{align}
\gamma_{\eta\eta} &= \frac{3}{32\s\pi^2} \,\eta -  \frac{3}{32\s\pi^2} \,\frac{\kappa^2}{\eta} \qquad\qquad\qquad \gamma_{\eta \kappa} =  \frac{3}{16\s\pi^2} \,\kappa\notag\\[8pt]
 \gamma_{\kappa\eta} &= \frac{1}{32\s\pi^2} \,\kappa  \hspace{127pt} \gamma_{\kappa\kappa} = \frac{1}{8\s\pi^2} \,\kappa \,.
\label{eq:AnnDimEtaKappa}
\end{align}
Plugging these into the general equation~\cref{eq:RGEgeneral} yields the RGEs for $\eta$ and $\kappa$:\footnote{Note that this set of RGEs are technically incorrect as written since they do not include the $\Phi^4$ coupling, and its influence on the running of $\kappa$.  Since we are assuming this coupling is zero at our UV scale, and we will only write down an expanded leading log solution, all the resulting expressions we derive are technically correct.  However, we caution that this coupling must be included if one is interested in analyzing the full RG structure.} 
\begin{align}
\frac{\D \eta}{\D \log \muT^2} &=  \frac{3}{32\s\pi^2}\,\eta^2 + \frac{3}{32\s\pi^2}\,\kappa^2\notag\\[5pt]
\frac{\D \kappa}{\D \log \muT^2} &=  \frac{1}{8\s\pi^2}\,\kappa^2 +  \frac{1}{32\s\pi^2}\,\kappa\,\eta\,,
\label{eq:phi4DecouplingRGEs}
\end{align}
whose solution sums the leading logs to all orders.  Note that this set of coupled equations manifestly violates decoupling, since it has no dependence on the scale $M$.  This implies that RG improving the \FT~will not solve the large log problem we have encountered here.  We can see this explicitly by putting together the summed LL + NLO \FT~amplitude for $\phi\,\phi \rightarrow \phi\,\phi$ is
\begin{align}
i\s\mathcal{A}^\textsc{Full} &= -i\s\eta\big(\muT\big) +i\s \frac{3}{32\s\pi^2}\, \big(\eta\big(\muT\big)\big)^2\, \left(\log\frac{\muT^2}{m^2}+\frac{2}{3}\right) \notag\\[8pt]
&\hspace{54pt}+i\s \frac{3}{32\s\pi^2} \,\big(\kappa\big(\muT\big)\big)^2\, \log\frac{\muT^2}{M^2}\hspace{50pt}\text{\big[LL+NLO\big]} \,,
\label{eq:phi4FullLargeLogProbLL}
\end{align}
where this is considered LL order because we are keeping track of the $\muT$ dependence of the couplings that are being run using the solution to \cref{eq:phi4DecouplingRGEs}.  We will also find it instructive to expand the RGE for $\eta(\muT)$ to leading order in the couplings.  The only terms that contributes to \cref{eq:phi4FullLargeLogProbLL} at leading order, evolved from a scale $\muT_H \rightarrow \muT_L$ are 
\begin{align}
\eta\big(\muT_L\big)_\text{Expanded} &= \eta\big(\muT_H \big) - \frac{3}{32\s\pi^2} \big(\eta^2 + \kappa^2\big) \log\frac{\muT_H^2}{\muT_L^2} \qquad \text{\big[LL + NLO\big]}\,.
\label{eq:etaRunningLL}
\end{align}
This expression can be derived by simply solving the $\eta$ equation in the approximation that the $\eta^2/(16\s\pi^2)$ and $\kappa^2/(16\s\pi^2)$ terms are constant, which is why we have written them here without a scale dependent argument.  This is equivalent to first solving the $\kappa$ equation, and then plugging that solution into the $\eta$ equation, followed by expanding to only keep the leading dependence on the couplings.
\vspace{0pt}\mybox{\begin{itemize}
\item \textbf{Exercise:} Show that both suggested ways of deriving \cref{eq:etaRunningLL} are equivalent.
\vspace{-5pt}
\end{itemize}}

Obviously, \cref{eq:etaRunningLL} does not depend on either $m$ or $M$, emphasizing that the RG can not cure our perturbation theory ills.  It is worth pausing to fully appreciate the nature of the problem we are trying to solve.  The claim was that RG techniques should sum large logarithms and systematically maintain the convergence of perturbation theory.  Yet, we see that no choice of $\muT$ can simultaneously make all the logs small in our LL + NLO result.  Something additional is forced upon us.  Specifically, when we encounter situations with a large separation of physical scales, we must augment our RG improved perturbation theory by matching onto an EFT that models the dynamics of the light modes in isolation.  This will allow us to run coupling within the EFT to low energies characteristic of our observable, and tame our large log problem.

\subsubsection*{Matching onto the EFT and Heavy Particle Decoupling}

In order to match, the first step is to pick a process -- unsurprisingly, we will continue to use $\phi\,\phi \rightarrow \phi\,\phi$ at threshold. The only propagating degree of freedom in the EFT is $\phi$, and the EFT will inherit a $\phi \rightarrow - \phi$ symmetry, so the interactions follow the same pattern as \cref{eq:EFTLagTreeMatch}.  For our purposes here, we will only need
\begin{align}
\mathcal{L}^\text{EFT}_\text{Int} =  - \frac{C_{4}}{4!} \,\phi^4\,.
\end{align}
To match at one-loop order, we must include counterterms and track scale dependence.  This implies that consistent matching requires the choice of an RG scale $\muT_M$.  Noting that dealing with loop effects leads to scheme dependence, we must take care to regulate integrals in the \FT~and EFT in a self-consistent way, \emph{i.e.}, we must use the same UV and IR regulators for both theories.  To this end, the generalization of the tree-level matching procedure given in~\cref{eq:matchTree} takes the form
\begin{align}
\mathcal{A}^\text{Match} = \Big[\mathcal{A}^\textsc{Full} + \mathcal{A}^\textsc{Full}_\text{c.t.}\Big] - \Big[\mathcal{A}^\text{EFT} + \mathcal{A}^\text{EFT}_\text{c.t.}\Big]\,.
\label{eq:MatchinRenorm}
\end{align}
For our chosen process, this is expressed diagrammatically as
\begin{align}
i\s\mathcal{A}^\text{Match} &\equiv \includegraphics[width=0.16\textwidth, valign=c]{Figures/LLLL_4PtEFT.pdf} \notag \\[10pt]
&= \left(\includegraphics[width=0.16\textwidth, valign=c]{Figures/LLLL4Pt.pdf} + \includegraphics[width=0.23\textwidth, valign=c]{Figures/LLLL_LLLoop.pdf} + \includegraphics[width=0.23\textwidth, valign=c]{Figures/LLLL_HHLoop.pdf} \,\,+ \,\, \text{c.t.}^\textsc{Full}\right) \notag\\[10pt]
&\hspace{30pt}-\left(  \includegraphics[width=0.16\textwidth, valign=c]{Figures/LLLL4Pt.pdf} + \includegraphics[width=0.23\textwidth, valign=c]{Figures/LLLL_LLLoop.pdf} \,\,+ \,\, \text{c.t.}^\text{EFT}\right)    \,,
\label{eq:iAMatchPhi4}
\end{align}
where $\text{c.t.}^\textsc{Full}$ and $\text{c.t.}^\text{EFT}$ denote the \FT~and EFT counterterm contributions respectively.  At leading order, the RG improved matching relation is\footnote{We have augmented our brackets we use to keep track of N$^m$LO and N$^n$LL order to now include a subscript $M$, which denotes the order to which the matching contributions have been computed.  We will use this notation below to distinguish when we have only included NLO matching corrections as opposed to the full set of NLO corrections.}
\begin{align}
C_4\big(\muT_M\big) = \eta\big(\muT_M\big) \qquad\qquad\quad \big[\text{LO$_M$}\big]\,,
\label{eq:C4MatchLO}
\end{align} 
where this equality is evaluated at the matching scale $\muT_M$, and practically serves as a boundary condition when we solve the RGEs to run the EFT couplings.  Then the one-loop matching corrections come from the diagram in \cref{eq:iAMatchPhi4} with the heavy particles in the loop (including all three channels):\footnote{Here the superscript ``Match'' denotes that this is a matching correction, which should be added to \cref{eq:C4MatchLO} to derive NLO boundary condition for the RGE, see~\cref{eq:C4muTM}.}  
\begin{align}
C_4^\text{Match}\big(\muT_M\big) = - \frac{3}{32\s\pi^2}\,\big(\kappa\big(\muT_M\big)\big)^2 \log\frac{\muT_M^2}{M^2} + \mathcal{O}\big(\lambda^2\big) \qquad\qquad\quad \text{\big[NLO$_M$\big]}\,.
\label{eq:C4Match}
\end{align}
Note that although the loop expansion in the \FT~depends on terms of the form $\log \muT^2/m^2$, \emph{e.g.}~see~\cref{eq:phi4FullLargeLogProbLL}, the matching coefficient does not.  This observation is critical to the success of matching at loop level.  If $C_4^\text{Match}$ did depend on $\log \muT^2/m^2$, then the $m^2 \rightarrow 0$ limit taken within the EFT would be non-analytic at the high scale, implying that separating scales would not be possible.  Said another way, checking for a meaningful $m^2 \rightarrow 0$ limit at the high scale is a useful test that the EFT models the detailed IR structure of the \FT.  The EFT diagrams in \cref{eq:iAMatchPhi4} are present in the matching calculation precisely to enforce this requirement.

The result is that we have now have a one-loop fixed order modification to the boundary condition for our EFT coupling:
\begin{align}
C_4\big(\muT_M\big) = \eta\big(\muT_M\big) - \frac{3}{32\s\pi^2}\,\big(\kappa\big(\muT_M\big)\big)^2 \log\frac{\muT_M^2}{M^2} \qquad \qquad \big[\text{NLO$_M$}\big]\,,
\label{eq:C4muTM}
\end{align}
and as long as we choose $\muT_M \sim M$, we avoid any issues with large logarithms for this boundary condition.  

Now we need the RGEs to run $C_4$ down to low scales within the EFT.  Note that the RGE analysis is identical to that provided in \cref{sec:ToyModelLLResum} above, so we can just take the results from there.  For completeness, we repeat the LL EFT RGE here
\begin{align}
\frac{\D C_4}{\D \log \muT^2} &=  \frac{3}{32\s\pi^2}\,\big(C_4\big)^2\,.
\end{align}
Then we can use this RGE to evolve our coupling from $\muT_M \rightarrow \muT_L$, with \cref{eq:C4muTM} as a boundary condition.   Including fixed order NLO corrections, our low scale EFT amplitude is given by 
\begin{align}
i\s\mathcal{A}^\text{EFT} =& -i\s C_4\big(\muT_L\big) +i\s \frac{3}{32\s\pi^2} \big(C_4\big(\muT_L \big)\big)^2 \left(\log\frac{\muT_L^2}{m^2} + \frac{2}{3}\right)  \qquad\quad \big[\text{LL + NLO}\big]\,.
\label{eq:A4ptFixedOrderEFT}
\end{align}
No terms with $\log M^2/\muT_L^2$ appear, and so this amplitude does not manifest any large log problems.  This is obvious when taking the canonical choice $\muT_L \sim m$.  Note that in practice, to derive the most precise answer possible at a given order, it is ideal to choose a scale that minimizes all the logarithms that appear.

At this point we have everything we need to understand the full picture.  Imagine our \FT~couplings are defined at a scale $\muT^2_H$, and we want to run them past our matching threshold $\muT_M^2$ down to a scale $\muT^2_L$, where our EFT description should be used.  Schematically, 
\begin{align}
\includegraphics[width=0.25\textwidth, valign=c]{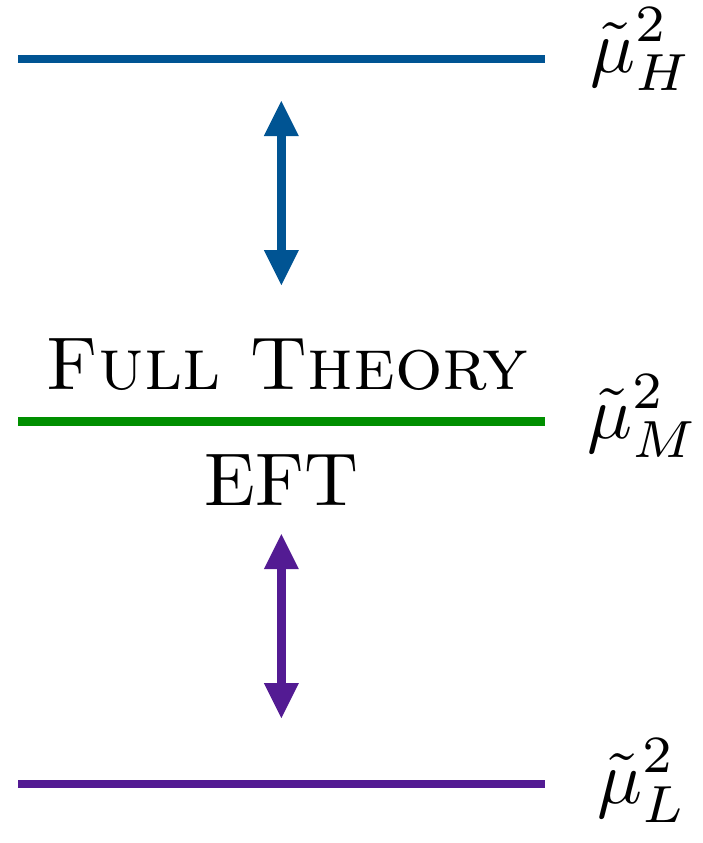} 
\label{eq:RGEScalesDecoup}
\end{align}
Then we can see exactly what has been accomplished by expanding our RGE solutions to leading order to see that we reproduce the one-loop fixed order results.  Of course, by doing this expansion we are re-introducing the problem we set out to solve, and emphasize that in practice one should obviously use the full RG solutions for the summed couplings such that the full tower of LL terms are included.  However, since this calculation provides an insightful closure test, it is worth working out in detail.  

Combining \cref{eq:etaRunningLL} and \cref{eq:C4muTM}, yields a boundary condition for the EFT RGE at the matching scale:
\begin{align}
C_4\big(\muT_M\big)_\text{Expanded} &= \eta\big(\muT_H \big) - \frac{3}{32\s\pi^2} \big(\eta^2 + \kappa^2\big) \log\frac{\muT^2_H}{\muT_M^2}  - \frac{3}{32\s\pi^2}\,\big(\kappa\big(\muT_M\big)\big)^2 \log\frac{\muT_M^2}{M^2}\,, \notag \\[6pt]
&=\eta\big(\muT_H \big) - \frac{3}{32\s\pi^2}\left(\eta^2\, \log\frac{\muT^2_H}{\muT_M^2}+ \kappa^2\, \log\frac{\muT^2_H}{M^2} \right)\qquad  \text{\big[LL + NLO$_M$\big]}\,,
\end{align}
where we have used the expanded solution to the RGE to evolve this coupling from $\muT_H$ down to the matching scale explicitly.  Then to capture the running of $C_4$ down to the EFT scale $\muT_L$, we can equate this expression to \cref{eq:C4runningSol} with the identification $\muT_H \rightarrow \muT_M$ and $\muT_L \rightarrow \muT_L$.  Expanding and solving for $C_4\big(\muT_L\big)$ gives
\begin{align}
C_4\big(\muT_L\big)_\text{Expanded} &= \eta\big(\muT_H \big) - \frac{3}{32\s\pi^2}\left( \eta^2\, \log\frac{\muT^2_H}{\muT_M^2} + C_4^2\, \log\frac{\muT_M^2}{\muT_L^2} + \kappa^2\, \log\frac{\muT^2_H}{M^2}\right)\quad\text{\big[LL + NLO$_M$\big]}\,,
 \label{eq:C4muEFinal}
\end{align}
where this now includes the term generated from running $C_4$ within the EFT (and then expanding).  

The final step is to compute the amplitude including the one-loop fixed order correction at the low scale, \emph{i.e}, plugging \cref{eq:C4muEFinal} into \cref{eq:A4ptFixedOrderEFT}: 
 \begin{align}
i\s\mathcal{A}^\text{EFT}_\text{Expanded} &=  -i\s\eta\big(\muT_H \big) + \frac{3\s i}{32\s\pi^2}\left[ \eta^2\, \log\frac{\muT^2_H}{\muT_M^2} + C_4^2\, \left(\log\frac{\muT_M^2}{m^2} + \frac{2}{3}\right)\right] \notag\\[8pt]
&\hspace{60.5pt}+ \frac{3\s i}{32\s\pi^2}\, \kappa^2\, \log\frac{\muT^2_H}{M^2} \hspace{130pt} \text{\big[LL+NLO\big]}\,,
 \label{eq:A4ptDecouplingExampleFinal}
 \end{align}
where the only operational difference between this expression and \cref{eq:C4muEFinal} (besides that this is now technically an amplitude for $\phi\,\phi \rightarrow \phi\,\phi$) is the $m$ dependence inside the logarithm and the finite factor, that are both due to the low scale fixed order corrections.

This expression is identical to our result in~\cref{eq:phi4FullLargeLogProb}, when we take $\muT_H \rightarrow \muT$ and note that $C_4 = \eta$ to the level of approximation captured when expanding.  However, the summed result yields an improved perturbation theory.  To see what was accomplished by matching and running, we remind the reader where each of the terms in \cref{eq:A4ptDecouplingExampleFinal} came from.  This expression provides the leading order evolution of our system from $\muT_H$ down to $\muT_L$.  We have consistently accounted for the running of the \FT~parameters from $\muT_H \rightarrow \muT_M$, where the scale $M$ should decouple.  Matching leaves us with an EFT description, which can be used to run the $C_4$ coupling from $\muT_M \rightarrow \muT_L$.  Critically, the EFT does not receive any dynamical contributions from  loops involving heavy particles.  This is exactly the decoupling behavior that we hoped to find by consistently flowing below the scale $M$.  Furthermore, by matching and running we have consistently eliminated any large logs that could spoil our perturbative expansion to LL + NLO order.  Obviously, this procedure is systematically improvable, and one can in principle include corrections to however high of order the reader has the strength and persistence to compute.  Finally, we again emphasize that by evaluating the summed result for various choices of the scales $\muT_H$, $\muT_\text{M}$ and $\muT_L$ one can estimate theoretical ``error bars'' that result from the truncation of perturbation theory to finite order.

Now we have seen the connection between decoupling and matching a \FT~to an EFT.  In the next section, we will explore a non-decoupling effect that appears in matching calculation for scalar masses.  We will then discuss when small scalar masses might naively be tuned, followed by a discussion of the implications for physics beyond the Standard Model.

\subsection{Quadratic Divergences and the Hierarchy Problem}
\label{sec:HierarchyProb}
At this point, we hope you are fully convinced of the need to match and run when computing with the $\overline{\text{MS}}$ scheme in the presence of a large separation of scales.  In this section, we are going to work out a famous consequence of matching and running, known as the hierarchy problem, see~\emph{e.g.}~\cite{Luty:2005sn, Skiba:2010xn} for similar treatments in the context of EFTs.  Specifically, what we will now show is that when we couple a light scalar (whose mass is not protected by a symmetry) to a heavy particle, there will be a matching correction for the light mass that is proportional to the heavy mass.  This is a necessary condition for testing if one's theory exhibits a hierarchy problem.  

There is no in principle obstruction to carefully choosing the parameters at the matching scale to tune away this large contribution in order to realize a low energy theory containing a parametrically light scalar.  However, as we will work out in detail below, if such a tuning is required \emph{and} there is a notion of a calculable UV completion, then one expects that nature should invoke some new mechanism that resolves this seeming issue with the naturalness of the underlying physical parameters.  This is particularly relevant for the Standard Model since we have now observed a light scalar, the Higgs boson~\cite{Aad:2012tfa, Chatrchyan:2012xdj}.  The Higgs mass parameter suffers the necessary condition for a hierarchy problem derived in what follows.  Therefore, it is compelling to search for extensions of the Standard Model where this tuning is ameliorated, ideally with associated experimental signatures.  Since our focus here is on EFT techniques, we will show how this necessary condition arises in detail.  Then (since I am frankly not able to help myself) the concrete calculation will be followed with some musings on the interpretation.

Our focus is on loop corrections to the mass of $\phi$.  The \FT~is the same as in the previous section:
\begin{align}
\mathcal{L}^\textsc{Full} &=\frac{1}{2} \big(\partial_\mu \phi\big)\big(\partial^\mu \phi\big)  - \frac{1}{2}\s m_\text{F}^2\, \phi^2 + \frac{1}{2} \big(\partial_\mu \Phi\big)\big(\partial^\mu \Phi\big) - \frac{1}{2}\s M^2\, \Phi^2 - \frac{1}{4}\s \kappa\,\phi^2\,\Phi^2-\frac{1}{4!}\s \eta\,\phi^4\,.
\label{eq:LFullHierarchyProblem}
\end{align}
where we are writing the mass terms explicitly since they are the focus of this section.  The only terms we need to include in the EFT Lagrangian are the mass and quartic for $\phi$:
\begin{align}
\mathcal{L}^\text{EFT} = \frac{1}{2} \big(\partial_\mu \phi\big)\big(\partial^\mu \phi\big)  - \frac{1}{2}\s m_\text{E}^2\, \phi^2 - \frac{C_{4}}{4!} \,\phi^4\,,
\end{align}
where we are being careful to distinguish the $\FT$~mass parameter $m^2_\text{F}$ from the EFT mass parameter $m_\text{E}^2$.  The next subsection shows that if there is no coupling between the light and heavy state, \emph{i.e.},~$\kappa = 0$, then there is no large correction to the light scalar mass parameter.  This is followed by a calculation with $\kappa \neq 0$, where the necessary condition for fine-tuning is derived.  The notation for the RG scales will follow the previous section, see~\cref{eq:RGEScalesDecoup}.

\subsubsection*{No Heavy Scale Means No Hierarchy Problem}

First, we match the \FT~(with $\eta \neq 0$ and $\kappa = 0$)~onto the EFT at tree-level: for this first calculation, the light scalar does not couple to the heavy state.  When matching for a mass, we use the convenient prescription of ensuring that the pole of the $\phi$ propagator is the same in both theories:
\begin{align}
-i\s\big(m_\text{F}\big)^2 = -i\s\big(m_\text{E}\big)^2 \qquad \qquad\big[\text{LO}_M\big]\,.
\label{eq:m2EFTMatchTree}
\end{align}

Next, we can work out the consequences of the $\phi$ self coupling.\footnote{This is the simplest theory that exhibits a ``quadratic divergence,'' which is straightforward to derive by simply regulating the following integral with a hard momentum cutoff regulator.}  The one-loop correction for the $\phi$ mass in the \FT~is
\begin{align}
\includegraphics[width=0.17\textwidth, valign=c]{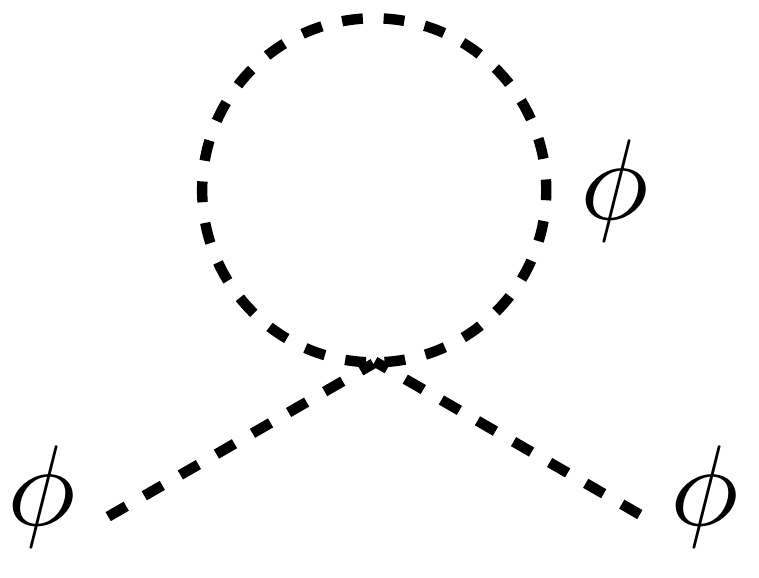} &= \frac{1}{2}\,\eta \,\mu_M^{2\epsilon}\int\frac{\text{d}^d \ell}{(2\s\pi)^d}\frac{1}{\ell^2-m_\text{F}^2} \notag\\
&= \frac{i\s \eta}{32\s \pi^2}\,m_\text{F}^2\,\left[\frac{1}{\epsilon}+\log\frac{\muT_M^2}{m_\text{F}^2} +1 + \mathcal{O}(\epsilon)\right]\,,
\label{eq:quadDivIntFdimreg}
\end{align}
where the $1/2$ is a symmetry factor and we have chosen to evaluate this loop at the matching scale $\mu_M$.  Note that this contribution is proportional to $m_\text{F}^2$, as it had to be since this is the only scale appearing in the loop integral -- this is how quadratic divergences manifest in dim reg.\footnote{A way to see the relation to quadratic divergences is to notice that there is a pole at $d = 2$, corresponding to a logarithmic divergence in two dimensions.  There is an EFT driven approach for nuclear physics that uses this fact to invent a scheme called ``power divergence subtraction''~\cite{Kaplan:1998tg}, which enables on to include  quadratically divergent contributions to the RGEs.} Note that this integral is insensitive to arbitrary physics at short distances (for example gravity becoming strong near the Planck scale), because the UV region of the dim reg integral is scaleless and therefore vanishes.  This fact can lead to misinterpretations of the hierarchy problem, which we hope this section will help to clarify.

Then we define a counterterm for the mass 
\begin{align}
m_{\text{F},\s 0}^2 = Z_{m^2_\text{F}} \, m_\text{F}^2\,,
\end{align}
which implies
\begin{align}
Z_{m^2_\text{F}} = 1 + \frac{\eta}{32\s \pi^2}\,\frac{1}{\epsilon}\,,
\end{align}
where $\eta$ has been renormalized using the results of the previous section.

Next, we calculate the mass correction at the matching scale within the EFT:
\begin{align}
\includegraphics[width=0.17\textwidth, valign=c]{Figures/LL_LLoopFull.pdf} =   \frac{i\s C_4}{32\s \pi^2}\,m_\text{E}^2\,\left[\frac{1}{\epsilon}+\log\frac{\muT_M^2}{m_\text{E}^2} +1 + \mathcal{O}(\epsilon)\right]\,.
\end{align}
As above, matching the quartic at tree level gives $\eta = C_4$, see~\cref{eq:C4MatchLO}.  Then we can perform a one-loop matching calculation for the mass of $\phi$ at the matching scale $\muT_M$:
\begin{align}
-i\s m^2_\text{Match} &= -i\s m_{\text{F}}^2+\frac{i\s \eta}{32\s \pi^2}\,m_{\text{F}}^2\,\left[\log\frac{\muT_M^2}{m_{\text{F}}^2} +1\right]\notag \\[7pt]
&\hspace{13pt}- \left(-i\s m_{\text{E}}^2+\frac{i\s C_4}{32\s \pi^2}\,m_{\text{E}}^2\,\left[\log\frac{\muT_M^2}{m_{\text{E}}^2} +1\right]\right) = 0 \qquad \big[\text{NLO}_M\text{ with } \kappa = 0\big]\,.
\label{eq:matchmSqEta}
\end{align}
Unsurprisingly, we see that the correction generated in the \FT~is compensated by the analogous correction in the EFT, implying that the EFT mass does not receive a large correction at the matching scale, and so there is no issue with large logarithms or fine-tuning.  This is to be expected, since we are working with a single scale EFT, which should be well behaved at all scales (at least once it has been RG improved).  Said another way, although this the diagram does introduce a one-loop correction at low scales $\muT_L$, 
\begin{align}
\hspace{-10pt}\includegraphics[width=0.12\textwidth, valign=c]{Figures/ScalarProp.pdf} \,\,+ \includegraphics[width=0.17\textwidth, valign=c]{Figures/LL_LLoopFull.pdf} & =   -i\s m_\text{E}^2\big(\muT_L\big)+\frac{i\s C_4\big(\muT_L\big)}{32\s \pi^2}\,m_\text{E}^2\big(\muT_L\big)\left[\log\frac{\muT_L^2}{m_\text{E}^2\big(\muT_L\big)} +1\right]\notag\\
&\hspace{200pt}\big[\text{LL+NLO}\big]\,,
\label{eq:mSqCorrectionLowScale}
\end{align}
one will not encounter any issues with perturbation theory, as long as the RG improved mass $m_\text{E}^2\big(\muT_L\big)$ is used.

\subsubsection*{Matching Across a Heavy Threshold and The Hierarchy Problem}

Now we turn on $\kappa$ in order to explore the impact of a non-trivial coupling between our light state $\phi$ and a heavy particle $\Phi$.  The one-loop correction to the $\phi$ mass in the \FT~at the matching scale $\muT_M$ receives another contribution:
\begin{align}
\includegraphics[width=0.17\textwidth, valign=c]{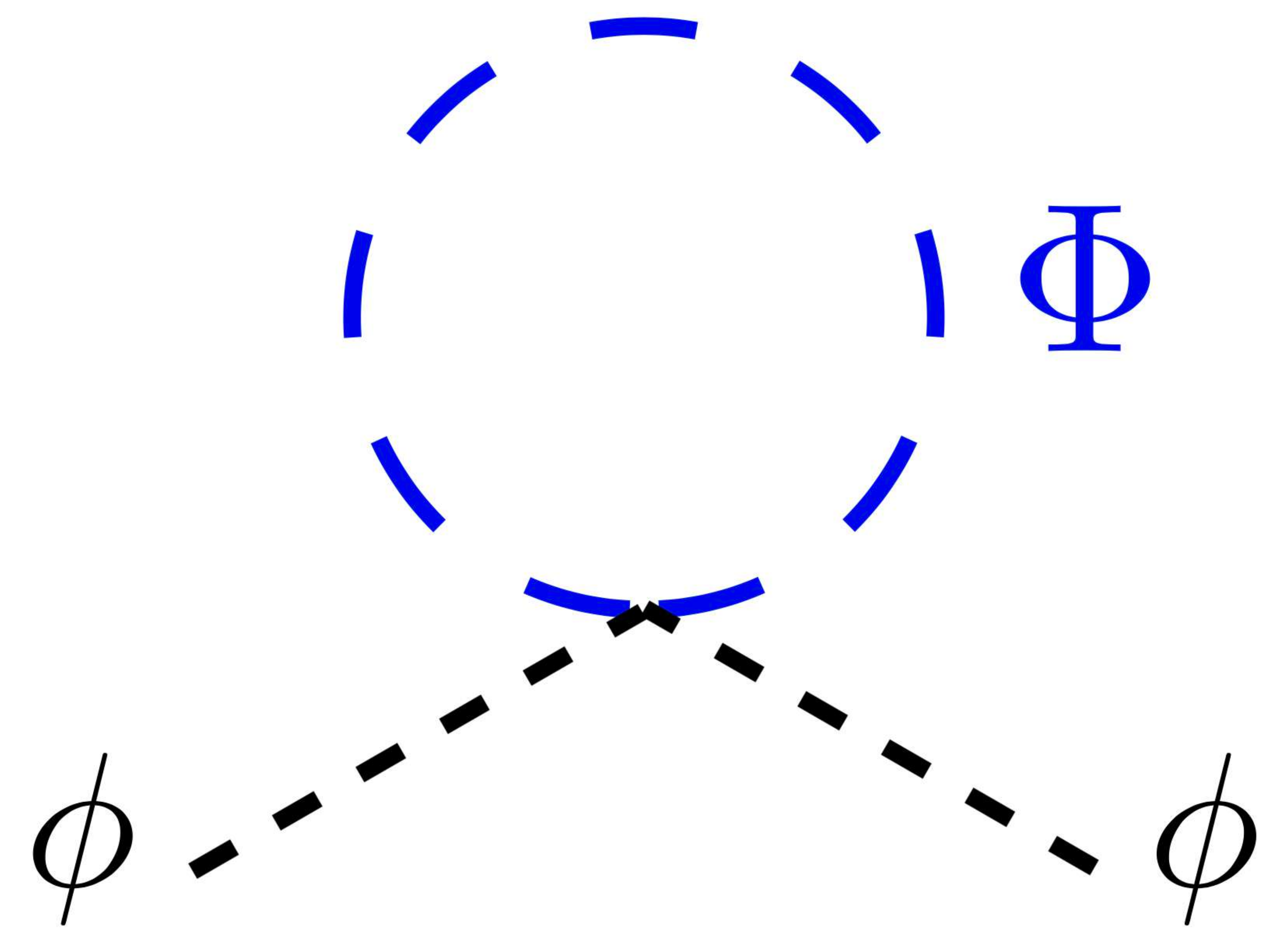} = \frac{i\s \kappa}{32\s \pi^2}\,M^2\,\left[\frac{1}{\epsilon} + \log\frac{\muT_M^2}{M^2} +1 + \mathcal{O}(\epsilon) \right]\,.
\label{eq:phiMassMloop}
\end{align}
In this theory, the mass counterterm is
\begin{align}
Z_{m^2_\text{F}} = 1  + \frac{\eta}{32\s \pi^2}\,\frac{1}{\epsilon}+ \frac{\kappa}{32\s \pi^2}\,\frac{M^2}{m_{\text{F}}^2}\frac{1}{\epsilon}\,,
\label{eq:Zm}
\end{align}
where now all the parameters have been renormalized.\footnote{The counterterms $Z_\eta$ and $Z_\kappa$ can be found in \cref{eq:Zeta} and \cref{eq:Zkappa} respectively, and we did not derive the counterterm factor $Z_M$ since it does not play any role in our analysis.}  Matching between the \FT~and the EFT~implies:
\begin{align}
-i\s m^2_\text{Match}\big(\muT_M\big) &= \left(-i \s m_{\text{F}}^2 + \frac{i\s\kappa}{32\s \pi^2}\,M^2\,\left[\log\frac{\muT_M^2}{M^2} +1\right]\right) - \big(-i \,m_{\text{E}}^2\big) \notag \\[7pt]
&=\frac{i\s \kappa}{32\s \pi^2}\,M^2\,\left[\log\frac{\muT_M^2}{M^2}+1\right]\qquad\qquad \text{\big[NLO$_M$\big]}\,,
\label{eq:m2EFTMatch}
\end{align}
where we have not included the contributions from the light loop at an intermediate step since we have already shown this contribution cancels between the \FT~and the EFT in \cref{eq:matchmSqEta}.  As expected, this matching correction does not manifest any non-analytic dependence on $m_\text{F}^2$, such as $\log m_\text{F}^2/\muT_M^2$.

Although the necessary condition for the hierarchy problem is already apparent, we postpone our interpretation of this result and instead will push our analysis forward to achieve LL + NLO accuracy by computing the RG evolution of the mass.  We need to derive an RGE for the mass parameter, starting with
\begin{align}
0 &= \frac{\D}{\D \log\muT^2} \,m_{\text{F},0}^2 = \frac{\D}{\D \log\muT^2} \Big( Z_{m_\text{F}^2}\,m_\text{F}^2\Big) \notag\\[7pt]
&= m_\text{F}^2\,Z_{m_\text{F}^2} \left(\frac{1}{m_\text{F}^2} \frac{\muT}{2} \frac{\D}{\D\muT}\, m_\text{F}^2 + \frac{1}{Z_{m_\text{F}^2}}\frac{\muT}{2}  \frac{\D}{\D\mu}\, Z_{m_\text{F}^2} \right) \,,
\end{align}
where we have used the fact that the dimension of the mass term does not change when we shift to $d=4-2\s\epsilon$ dimensions.  The $\muT$ dependence of the mass is inherited from the $\eta$ and $\kappa$ couplings, since $\D m_\text{F}^2/\D\log\muT^2 = 0$ at tree-level.  Then noting that the leading order relationship \cref{eq:GammaC4Classical} applies, we can write
\begin{align}
\frac{1}{Z_{m_\text{F}^2}}\frac{\muT}{2}\frac{\D}{\D\muT} Z_{m_\text{F}^2} = \frac{1}{Z_{m_\text{F}^2}}\frac{\partial Z_{m_\text{F}^2}}{\partial \eta}\frac{\muT}{2} \frac{\D \eta}{\D \muT} + \frac{1}{Z_{m_\text{F}^2}}\frac{\partial Z_{m_\text{F}^2}}{\partial \kappa}\frac{\muT}{2} \frac{\D \kappa}{\D \muT} = - \epsilon\s \eta\,\frac{\partial Z_{m_\text{F}^2}}{\partial \eta_r}-\epsilon \s\kappa\,\frac{\partial Z_{m_\text{F}^2}}{\partial \kappa}\,,
\end{align}
where in the last step we have used \cref{eq:GammaC4Classical} which also applies here, and we have expanded $1/Z_{m_\text{F}^2} = 1 + \cdots\,\,$.  Finally, we plug this into the previous equation to arrive at our RGE for the mass:
\begin{align}
\frac{\D}{\D\log\muT^2}\, m_\text{F}^2  = \gamma_{m_\text{F}^2}\, m_\text{F}^2 \qquad \text{with} \qquad \gamma_{m_\text{F}^2} = \lim_{\epsilon\rightarrow 0} \left( \epsilon\s \eta\frac{\partial Z_{m_\text{F}^2}}{\partial \eta}+\epsilon \s\kappa \frac{\partial Z_{m_\text{F}^2}}{\partial \kappa}\right)\,,
\end{align}
which we can apply to \cref{eq:Zm} to find
\begin{align}
\gamma_{m_\text{F}^2} = \frac{\eta}{32\s\pi^2}  + \frac{\kappa}{32\s\pi^2} \frac{M^2}{m_{\text{F}}^2}\,.
\label{eq:gammam2}
\end{align}
The resulting RGE sums logarithmic corrections to the mass of $\phi$ in the \FT~to LL order.  Note that it contains a contribution proportional to $M^2$, which is one of the sources of the hierarchy problem.

In order to expose the physics, we solve the LL RGE and expand keeping only the leading contribution:
\begin{align}
\hspace{-8pt}m^2_\text{F}\big(\muT_M\big)_\text{Expanded} & = m_{\text{F}}^2\big(\muT_H\big) - \frac{\eta}{32\s\pi^2}\,m_{\text{F}}^2\,\log\frac{\muT_H^2}{\muT_M^2} - \frac{\kappa}{32\s\pi^2}\, M^2\log\frac{\muT_H^2}{\muT_M^2} \,\,\quad \text{\big[LL + NLO$_M$\big]}\,,
\end{align}
where we have dropped the scale dependent arguments of the coefficients of the log terms since these corrections are higher order.  A similar calculation in the EFT yields\footnote{The EFT RGE is identical to the one derived for the \FT~calculation with $\eta = C_4$ and $\kappa = 0$.}
\begin{align}
m_{\text{E}}^2\big(\muT_M\big)_\text{Expanded} = m_{\text{E}}^2\big(\muT_L\big) - \frac{C_4}{32\s\pi^2}\,m_{\text{E}}^2\log\frac{\muT_L^2}{\muT_M^2}  \qquad \text{\big[LL\big]}\,.
\end{align}
The last required piece is the boundary condition for the EFT mass at the matching scale using \cref{eq:m2EFTMatchTree} and \cref{eq:m2EFTMatch}:
\begin{align}
m_{\text{E}}^2\big(\muT_M\big) = m^2_\text{F}\big(\muT_M\big) -\frac{\kappa}{32\s\pi^2}\,M^2\,\left[\log\frac{\muT_M^2}{M^2}+1\right] \qquad \big[\text{NLO}_M\big]\,.
\label{eq:massBC}
\end{align}
Now in exact analogy with \cref{eq:C4muEFinal}, we can run our mass from the scale $\muT_H$ in the \FT~(noting that $\muT_H$ is a proxy for the UV scale where the fundamental parameters are defined) down to a scale $\muT_L$ passing through the mass threshold $M$:
\begin{align}
m^2_{\text{E}}\big(\muT_L\big)_\text{Expanded} 
 =\,\s &m_{\text{F}}^2\big(\muT_H\big)  -\frac{\kappa}{32\s \pi^2}\,M^2 -\frac{C_4}{32\s \pi^2}\,m_\text{E}^2 \notag\\[5pt]
 &\,- \frac{\eta}{32\s\pi^2}\,m_{\text{F}}^2\,\log\frac{\muT_H^2}{\muT_M^2}-\frac{C_4}{32\s\pi^2}\,m_{\text{E}}^2\log\frac{\muT_M^2}{m_\text{E}^2} \notag\\[5pt]
 &\, - \frac{\kappa}{32\s\pi^2}\, M^2\log\frac{\muT_H^2}{M^2} \qquad\qquad\qquad\big[\text{LL + NLO}\big]\,, 
\end{align}
where we have used \cref{eq:mSqCorrectionLowScale} to include the fixed order correction within the EFT in analogy with \cref{eq:A4ptDecouplingExampleFinal} above.  This expression gives the (expanded) RGE solution + fixed order corrections, and is therefore under perturbative control when the full RG solution is utilized.

This demonstrates that the heavy state does not contribute once you evolve below the scale $M$, as it must have been due to the requirement of decoupling.  However, there is a very important non-decoupling difference here as compared to the quartic interaction example worked out previously.  Specifically, there are both log enhanced and finite matching corrections that are proportional to $M^2$.  This implies that the UV scale $M$ contributes to the light mass -- the one-loop corrections to this small parameter are not parametrically under control.  We conclude that, as opposed to our example in \cref{sec:ToyModelLLResum} above, the RG evolution does not provide any compensating effects to cancel the fine-tuning required when matching.  This is a necessary condition for a theory to manifest a hierarchy problem.

To summarize, we can start with the boundary condition for our mass parameter in~\cref{eq:massBC}, which was derived by matching the \FT~to the EFT, and we can run it to a low scale where we compute the NLO corrections within the EFT to yield
\begin{align}
m^2_{\text{F}}\big(M\big) -\frac{\kappa}{32\s \pi^2}\,M^2= m_{\text{E}}^2\big(m\big) - \frac{C_4}{32\s\pi^2}m_{\text{E}}^2\left(1+\log\frac{M^2}{m_\text{E}^2}\right)\,,
\label{eq:theHierarchyProblem}
\end{align}
where for simplicity we have taken the scale choices $\muT_H = \muT_M = M$, and $\muT_L = m$, which can be interpreted as assuming that the UV parameters are defined at $\mu_H = M$.  This implies that if we wish to maintain $m \ll M$, we must carefully choose the \FT~mass parameter at the matching scale $m_\text{F}^2\big(\muT_M\big)$.  This concludes the technical demonstration of this issue, thereby laying the groundwork for the philosophizing that appears next.

\subsubsection*{A Brief Discourse on the Hierarchy Problem}
For the reader who has not yet had enough of the hierarchy problem, I will now provide my personal perspective on the interpretation of the necessary condition derived in~\cref{eq:theHierarchyProblem}.  First, I will clarify some technical issues associated with the way the hierarchy problem appears when using dim reg.  Then, I will interpret \cref{eq:theHierarchyProblem} as motivating the need for new physics.  One key point that will be emphasized throughout is that promoting the fine-tuning to a physical effect requires a calculable UV framework, where the mass parameters in the IR can be interpreted as predictions of an underlying theory.  This is intrinsically a UV issue, and as such there is no model independent interpretation.

The hierarchy problem for the Higgs boson in the Standard Model\footnote{The implications for naturalness due to the presence of a fundamental scalar Higgs boson in the Standard Model were first realized in~\cite{Susskind:1978ms}.} is often presented in a not particularly technically accurate way.\footnote{I have certainly been guilty of this doing this myself more than once.}  In particular, it is easy to argue schematically that a top-loop would imply a correction to the Higgs mass 
\begin{align}
m_H^2 \sim -i\s y_t^2 \int \frac{\D^4 \ell}{(2\s\pi)^4} \frac{1}{\ell^2-m_t^2} = -\frac{y_t^2}{8\s\pi^2} \int_0^{\Lambda_\text{UV}} \D \ell\, \frac{\ell^3}{\ell^2+m_t^2} = -\frac{y_t^2}{16\s\pi^2} \,\Lambda_\text{UV}^2 + \cdots\,,
\label{eq:HiggsMassCorrection}
\end{align}
where $y_t$ is the top Yukawa coupling, and $\Lambda_\text{UV}$ is some hard UV cutoff that regulates this UV divergent integral.  This back-of-the-envelope calculation gives the appearance that the Higgs mass receives a large quadratic correction from some unknown UV scale.  However, note that if this integral were evaluated using dim reg by applying \cref{eq:quadDivIntFdimreg}, one would find a result that is proportional to the only scale that appears in the integral, namely $m_t$.  It might then be tempting to conclude that the hierarchy problem is a hoax.  However, there is no inconsistency here, but instead the interpretation of \cref{eq:HiggsMassCorrection} requires treating $\Lambda_\text{UV}$ with care.  Specifically, one should be very cautious when claiming that a cutoff dependent result has a physical interpretation.  As I have emphasized in the technical part of this section, there is no fine-tuning issue if the only scale in the problem is $m$.  This is why I coupled $\phi$ to a heavy state $\Phi$ in order to expose the necessary condition for the hierarchy problem -- without the physical mass scale $M$, there is no problem.  Note that one of the remarkable aspects of the Standard Model is that gauge invariance plus chiral symmetry forbids mass parameters for the fermions -- the Higgs mass parameter is the only dimensionful scale around.  This implies that if there are no additional UV scales beyond the Standard Model, there are no fine-tuning issues associated with the Higgs mass.  This is made plain in the careful treatment above, see \cref{eq:mSqCorrectionLowScale}.

However, there are many outstanding issues with the Standard Model that very likely introduce new scales\footnote{The most obvious ones are dark matter, baryogenesis, unification, strong CP, \dots\,\,.} that are likely to couple to the Higgs at some loop order (due to gravitational interactions if nothing else).  In addition, we lack a detailed picture for what occurs at the physical scale $M_\text{Pl}$, where gravity becomes strong.  This has convinced many of us that 
\begin{enumerate}[label=\roman*)]
\item The Standard Model should UV complete into a calculable\footnote{Here I mean in-principle calculable, since one compelling possibility is that the new physics is strongly coupled.} framework such that low energy parameters are predictions in terms of fundamental parameters.
\item In the absence of a protection mechanism, new high scales would feed into the Higgs mass parameter.\footnote{The argument is so robust that it has even been extended to show that the Higgs would receive a contribution to its mass parameter in a situation where the only new scale is associated with a non-perturbative deviation in the anomalous dimensions of the theory~\cite{Tavares:2013dga}.} 
\end{enumerate}
This motivates me to take the fine-tuning problem seriously as an argument for new physics.  

Before I discuss specific approaches, I want to distinguish the two ``kinds'' of naturalness that are often discussed.  The first is ``aesthetic naturalness'' (or ``Dirac naturalness''~\cite{Dirac:1938mt}, which could arguably be credited to Gell-Mann~\cite{Gell-Mann:1956iqa}).  It is the statement that given a theory with a fundamental scale $\Lambda$, all dimensionful quantities should be proportional to $\Lambda$ with some order one coefficient:
\begin{align}
\mathcal{L} \supset \sum_{\{O\}}\,c_O \times \Lambda^{4-[\s O\s ]}\times O\,,
\end{align}
where $O$ is some operator with mass dimension $\big[O\big]$, and $c_O \sim \mathcal{O}(1)$ is a Wilson coefficient, and the sum is over the set of operators of relevance ${\{O\}}$.  This can be compared with a more nuanced (or legalistic) definition of natural choices for Wilson coefficients, usually referred to as ``technical naturalness'' (or 't Hooft naturalness~\cite{tHooft:1979rat}).  For theories in this class, a small parameter is considered natural as along as an additional symmetry is restored in the limit that it is taken to zero:
\begin{align}
\mathcal{L} \supset  \sum_{\{O\}}\,c_O \times \Lambda^{4-[\s O\s ]}\times O\,\,\, +\,\,\,  \sum_{\{\tilde{O}\}}\,s\times c_{\tilde{O}} \times \Lambda^{4-[\s \tilde{O} \s ]}\times \tilde{O} \,,
\label{eq:TechNaturalness}
\end{align}
where now a special set of operators $\{\tilde{O}\}$ have been separated off, with Wilson coefficients $c_{\tilde{O}}$.  The additional factor $s$ is a ``spurion'' in that it tracks the breaking of whatever symmetry is restored in the limit that $s \rightarrow 0$.  This implies that these coefficients can in principle take values $s\,c_{\tilde{O}} \ll 1$ while maintaining technical naturalness.  

There have been many recent complaints that we should abandon the notion that fine-tuning in the context of the Standard Model is a serious issue.  This can largely be traced to the (as of yet) non-observation of new physics at the LHC and other experiments, which is driven by the feeling that aesthetic naturalness should have been realized by nature.  However, many technically natural ideas remain viable, and it is my point of view that they are to be taken seriously.

\vspace{-15pt}
\subsubsection*{A Supersymmetric Detour\footnote{A working knowledge of the relevant supersymmetry tools (superpotentials and their renormalization properties, and gauge mediated supersymmetry breaking in particular) is required to follow these arguments. For introductions, see \emph{e.g.}~\cite{Martin:1997ns, Terning:2006bq}.}}

At this point, a toy model would illuminate the ideas I are trying to convey.  I will work in the context of weakly coupled supersymmetry, since this is a simple field theoretic framework that makes scalar masses calculable.  The purpose of the choices made here is to mimic \cref{eq:LFullHierarchyProblem}, to see how the hierarchy problem emerges.  This is followed by a more sophisticated example that manifests masses that are calculable in terms of gauge couplings.

Take a four-dimensional $\mathcal{N}=1$ supersymmetric model with a light superfield $\bm{\phi}$ and a heavy superfield $\bm{\Phi}$.  The K\"ahler potential is canonical, and the superpotential is\footnote{I will resist the temptation to call the superpotential coupling $\sqrt{\kappa}$, even though this would more closely resemble the notation above.}
\begin{align}
W = \frac{1}{2}\,m\,\bm{\phi}^2 + \frac{1}{2}\,M\,\bm{\Phi}^2 + \frac{1}{2}\,\kappa\,\bm{\phi}^2\,\bm{\Phi}\,.
\end{align}
Famously, the superpotential does not receive any quantum corrections.  Therefore, the limit $m \ll M$ is radiatively stable, and this theory does not have any hierarchy problem.  This is true even though it contains the same scalar interactions (and more) as our example above with the Lagrangian given in \cref{eq:LFullHierarchyProblem}:
\begin{align}
\mathcal{L}_\text{Int} = \big|m\,\phi + \kappa\,\phi\,\Phi\big|^2 + \big|M \, \Phi + \kappa\,\phi^2\big|\,.
\end{align}
In particular, if one were to match onto an EFT with only the $\bm{\phi}$ superfield at a scale near $M$, one would find that the extra scalar diagrams that result from terms like $M\,\kappa\,\Phi^\dag\,\phi^2 + \text{h.c.}$ along with the loops involving the fermionic superpartners would all conspire to result in no matching corrections for the light mass.  

In order to discover a toy model with a calculable hierarchy problem, one must softly break supersymmetry.  In particular, adding the following additional terms to the Lagrangian at the matching scale
\begin{align}
\mathcal{L}_\text{Soft} = -M\,\kappa\,\Phi^\dag\,\phi^2  -m\,\kappa\,\phi^\dag\,\phi\,\Phi + \text{h.c.}\,,
\label{eq:Lsoft}
\end{align}
would yield the exact same diagrammatic structure in the scalar sector as before (of course there are still the fermion diagrams which contribute).  Then this version of the model would realize a large correction to the mass of $\phi$ proportional to $\kappa^2\,(M^2+m^2)/(16\s \pi^2)$.  If one needed to maintain the small mass for $\phi$, the UV parameters would need to be chosen with care.\footnote{Note that for the sake of simplicity, I have taken the soft breaking parameters to be exactly the same as their supersymmetric counterparts, even though there is no mechanism to maintain their equality.  In fact, RG evolution makes this an unstable choice.  However, it is straightforward to see that the argument presented here does not require that this choice be precisely satisfied.}

\vspace{5pt}\mybox{\begin{itemize}
\item \textbf{Exercise:}  Work out the matching calculation for the mass of $\phi$ in the theory where supersymmetry is maintained.  Then do the same when the soft-breaking Lagrangian \cref{eq:Lsoft} is included to verify my conclusions.
\end{itemize}}

At the risk of overdoing it, I will introduce a more sophisticated toy model to draw attention to one flaw in the example I just presented.  In particular, since I added soft supersymmetry breaking by hand, one might argue that this does not satisfy the requirement that the parameters are fully calculable in the UV.  To that end, I will rely on a scheme for deriving calculable soft breaking masses known as gauge mediation, see~\cite{Giudice:1998bp} for an introduction.  

The model has three superfields $\bm{\phi}$, $\bm{\Phi}_1$, and $\bm{\Phi}_2$, who couple to each other with the superpotential
\begin{align}
W = \kappa \,\bm{\phi}\,\bm{\Phi}_1\,\bm{\Phi}_2\,,
\end{align}
which gives cross quartic couplings among all three scalar fields.  Next, assume there are two gauged U$(1)$ symmetries, labeled by $A$ and $B$, under which the fields are charged as
\begin{align}
\renewcommand{\arraystretch}{1.4}
\setlength{\arrayrulewidth}{.3mm}
\setlength{\arraycolsep}{1 em}
\begin{array}{c|c|c}
\text{field} & Q_A & Q_B\\
\hline
\bm{\phi}  & +1 & 0 \\
\bm{\Phi}_1  & -1 & +1 \\
\bm{\Phi}_2  & 0 & -1  
\end{array}
\end{align}
Then imagine that these fields receive gauge mediated masses, which are generated at two-loops.  Parametrically, this implies $m^2 \sim \big[(g^2_{A,B}/16\s\pi^2)\,F/M_{A,B}\big]^2$, where $F$ breaks supersymmetry, which is coupled to a set of ``messenger'' fields with mass $M_{A,B}$ for the superfield carrying U$(1)_{A,B}$ charges respectively. This in turn assumes that the $F$-term derives from a fully calculable source, \emph{e.g.} dynamical supersymmetry breaking~\cite{Witten:1981nf}.  Then at leading order, this model could realize a scalar mass spectrum $m^2_\phi \ll m^2_{\Phi_{1,2}}$ if there is a hierarchy in the gauge couplings $g_A \ll g_B$, assuming $M_A \sim M_B$.  This choice is technically natural, since taking a gauge coupling to zero enhances the global symmetry structure of the model.  However, if one then calculated with this theory to the next order in perturbation theory, RG evolving to scales below $m^2_{\Phi_{1,2}}$ would yield a non-trivial threshold correction to $m^2_\phi$ in exact analogy with above.  Then in order to maintain the lightness of the $\phi$ field, toy model nature would have to fine-tune the two gauge couplings $g_A$ and $g_B$.  This is a concrete realization of the hierarchy problem in terms of UV parameters, and serves to highlight how bizarre it would be if the smallness of the weak scale were ultimately traced back to a fundamental physics conspiracy of this ilk.

\vspace{5pt}
\begin{center}
\textsc{End Detour}
\end{center}
\newpage

These kinds of arguments have motivated many of us to seek natural theories for the Higgs mass.  One way to classify these models is through their reliance on either a field-theoretic or a cosmological mechanism.   The field-theoretic approaches introduce a new symmetry\footnote{One might also classify approaches that solve the problem by lowering the cutoff of the theory to the TeV scale, although these models do not rely on a new symmetry, see \emph{e.g.}~\cite{Dvali:2007hz}, as field-theretic.} that can be used to forbid the Higgs mass parameter.  They then reintroduce this parameter in a controlled way by softly breaking the symmetry, \emph{i.e.}, with a spurion as in \cref{eq:TechNaturalness}.  The two classic approaches incorporate supersymmetry (following the same logic underlying the two examples I just presented), where the Higgs mass parameter inherits the chiral symmetry protection mechanism of its Higgsino superpartners, and/or a shift-symmetry, where the Higgs is assumed to be a pseudo-Nambu-Goldstone boson (a composite state), and therefore has a small protected mass parameter.  For a detailed discussion, see the classic review of the supersymmetric Standard Model~\cite{Martin:1997ns}, and for some reviews of composite Higgs and other non-supersymmetry approaches, see~\cite{Contino:2010rs, Panico:2015jxa, Csaki:2018muy}.

The cosmological solutions instead solve the problem outside of field theory.  Starting with a Higgs that suffers a fundamental fine-tuning, these approaches propose that some dynamics in the early universe selects a small Higgs mass parameter.  The classic example is anthropic selection~\cite{Barrow:1988yia} (especially since this is essentially the only plausible explanation for the small cosmological constant\footnote{Here I are going to display uncanny hubris by relegating a brief mention of the cosmological constant fine-tuning problem to a footnote.  But if I am are honest, this choice is due to the fact that I are not nearly clever enough to have anything useful to say about this monumental failure of naturalness driven arguments.  To this end, I defer to Weinberg for wisdom~\cite{Weinberg:1988cp}.}~\cite{Weinberg:1987dv}), but recent years have seen the emergence of many new ideas, \emph{e.g.}~the relaxion~\cite{Graham:2015cka} which couples the Higgs mass parameter to the potential of a new axion-like field (which is now understood to have observable consequences, see \emph{e.g.}~\cite{Flacke:2016szy}), $N$naturalness~\cite{Arkani-Hamed:2016rle} with its reliance on many new degrees of freedom and a novel approach to reheating, and a more recent proposal that correlates the vacuum energy during inflation with the value of the Higgs mass parameter~\cite{Geller:2018xvz} and also~\cite{Cheung:2018xnu}.  It is even possible that some kind of break down of decoupling, \emph{a.k.a.}~UV/IR mixing, could solve the problem as well, although I am not aware of any plausible example model on the market that incorporates this notion into the Standard Model.

I want to acknowledge that it is fair to be skeptical of these arguments as a motivation for new physics since there is no known physical observable directly associated with an unnatural Higgs mass parameter.  Said another way, discovering a hierarchy problem requires the context of a UV model.  As such, one can maintain a perfectly consistent point of view that the perceived fine-tuning is simply an artifact of the way we do calculations and should be ignored since it has no implications for experiment.  However, I would push back by pointing out that there are many historical examples where interpreting a quadratic divergence as the harbinger of new physics has borne fruit: the power law divergent classical self energy of the electron is screened by the presence of the positron at short distances~\cite{Weisskopf:1996bu}; the quadratically divergent pion mass splitting is cutoff by the $\rho$-meson; the quadratically divergent mass difference between the long and short neutral Kaon states were used to predict the charm quark mass~\cite{Gaillard:1974hs} -- these are all summarized in \emph{e.g.}~\cite{Giudice:2008bi}.  Additionally, from a theory space point of view, every attempt to build a model where the Higgs mass is protected by a symmetry (thereby making it a calculable parameter) has yielded a fine tuning of the Higgs mass parameter when the \FT~parameter space with $M \gg m$ is explored.  Finally, we note that there are cases where one has control of analogs of the mass parameters in an experiment: when tuning an external magnetic field in a condensed matter system (where quantum field theory provides a good description near a phase transition) the presence of fine-tuning often manifests, see \emph{e.g.} the discussion in~\cite{Polchinski:1992ed}.  

To summarize, we have seen an ad-hoc Higgs mechanism appear before in the form of superconductors, and this model turned out to have its origins in a microscopic theory such that its parameters are in-principle calculable.  The least radical option is that the same is true for the Higgs sector of the Standard Model.  Furthermore, claiming that the weak scale is simply an incalculable quantity that is set by a renormalization condition would have tremendous implications for the reductionist philosophy that has driven progress in fundamental physics since its inception.  So while at the time these notes were written there is still no experimental experimental evidence for a resolution of the hierarchy problem, I am of the strong opinion that the motivations still stand.  Addressing electroweak naturalness in new ways (especially if they yield novel connections to experiment) is a worthwhile pursuit.     

The next section gets back to technical aspects of EFTs,\footnote{For the reader who is interested in more modern discussions of the hierarchy problem, I recommend these recent lectures on BSM physics~\cite{MMLecture}, this essay on ``post-modern naturalness''~\cite{Giudice:2017pzm}, this essay on ``naturalness under stress''~\cite{Dine:2015xga}, and this essay on the interpretation of fine-tuned theories~\cite{Wells:2018yyb}.} and presents our first example of separating scales inside a log that has no explicit $\muT$ dependence.

\subsection{Separation of Scales for a Heavy-light Log}
\label{eq:SepScalesHLlog}
We expect that many of the manipulations and techniques utilized thus far would be somewhat familiar to the reader.  In fact, the purpose of providing so much detail for these calculations was in large part to set a strong foundation for this section.  Here we work with our first non-trivial example of a logarithm that is not an explicit function of the RG scale $\muT$.  Specifically, we will construct a toy scalar model whose perturbative structure will generate a factor of $\log m^2/M^2$.  We will then show how to sum this log by matching and running.  This is relevant to the parameter space where power counting will be effective, \emph{i.e.}, $\lambda \sim m/M \ll 1$.  Along the way, we will encounter new features associated with power counting that will be emphasized.\footnote{Specifically, our example actually yields a power suppressed log, $\lambda^2\,\log\lambda^2$.  Clearly, the power suppression keeps it from actually becoming large enough to cause a problem for perturbation theory.  Nonetheless, this simple example will show how the procedure works, and will then generalize to more complicated cases, \emph{e.g.}~SCET as discussed below.}

The \FT~contains $\phi$ and $\Phi$ as propagating degrees of freedom, and we only need one interaction
\begin{align}
\mathcal{L}^\textsc{Full}_{\text{Int}} &=- \frac{1}{3!}\s \rho\,\phi^3\,\Phi\,.
\label{Eq:LIntHL}
\end{align}
Schematically, our approach will involve the following scales: 
\begin{align}
\includegraphics[width=0.30\textwidth, valign=c]{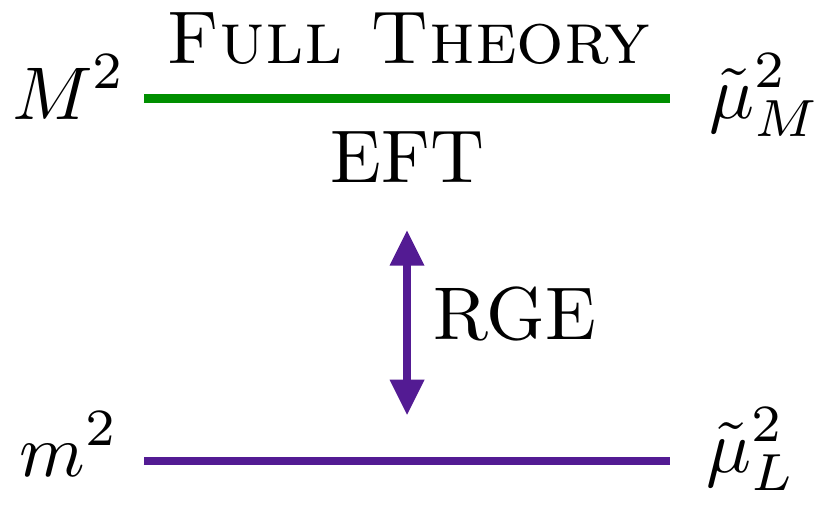}
\end{align}
where in this figure we are correlating the RG scales with their natural size.  We want to match our \FT~at a scale $\muT_M$~onto the EFT for the light modes alone.  Therefore, we must pick a process, and -- surprise! -- we will use  $\phi\,\phi \rightarrow \phi\,\phi$ at threshold.  There is a contribution from the \FT~diagrams, 
\begin{align}
\begin{array}{c}t\,\text{-channel}\\+\\u\,\text{-channel}\end{array} : \quad\includegraphics[width=0.25\textwidth, valign=c]{Figures/LLLL_HLLoop.pdf} = 2\,\big(\mu_M^{2\epsilon}\,\rho\big)^2 \int \frac{\text{d}^d \ell}{(2\s\pi)^d} \frac{1}{(\ell^2-m^2)(\ell^2-M^2)}\,,
\label{eq:HLInt_others}
\end{align}
We will call this a ``heavy-light'' loop for reasons that we hope are self-evident.  There is also an $s$-channel contribution which we will evaluate below, see~\cref{eq:HLInt_schannel}.  For now we will focus on these diagrams because they do not have external momentum flowing through the loop.  Using the Feynman parameter trick from \cref{eq:FeynParam}, this integral becomes 
\begin{align}
\!\!2\times\!\!\includegraphics[width=0.25\textwidth, valign=c]{Figures/LLLL_HLLoop.pdf} 
&= 2\,\big(\mu_M^{2\epsilon}\,\rho\big)^2 \int \frac{\text{d}^d \ell}{(2\s\pi)^d} \int_0^1 \text{d}x \,\frac{1}{\big[\ell^2 - \big(x\,m^2 +(1-x)M^2\big)\big]^2} \notag\\
&= \frac{2\s i\,\mu_M^{2\epsilon}\,\rho^2}{16\s\pi^2}\int_0^1 \text{d}x \,\left[\frac{1}{\epsilon}+\log\left(\frac{\muT_M^2}{x\,m^2+(1-x)\,M^2}\right)\right] \notag\\[10pt]
&= \frac{2\s i \,\mu_M^{2\epsilon}\,\rho^2}{16\s\pi^2} \left[\frac{1}{\epsilon} - \frac{m^2}{M^2-m^2}\log\frac{\muT_M^2}{m^2}+ \frac{M^2}{M^2-m^2}\log\frac{\muT_M^2}{M^2}\right]\notag\\[10pt]
&= \frac{2\s i\,\mu_M^{2\epsilon}\,\rho^2}{16\s\pi^2} \left[\frac{1}{\epsilon} + \log \frac{\muT_M^2}{M^2} + \frac{m^2}{M^2}\log\frac{m^2}{M^2} +1 + \mathcal{O}\left(\frac{m^4}{M^4}\right)\right]\,,
\label{eq:IntHeavyLight}
\end{align}
where in the last line, we used\footnote{Note that it is easy to derive this expression from the original one by taking $\muT_M = M$.  However, since we are interested in tracking the $\muT$ dependence, we will maintain it as a free parameter.} 
\begin{align}
\vspace{-10pt} &\left( - \frac{m^2}{M^2-m^2}\log\frac{\muT_M^2}{m^2}+ \frac{M^2}{M^2-m^2}\log\frac{\muT_M^2}{M^2}\right)+ \left(\frac{m^2}{M^2 - m^2}\log\frac{\muT_M^2}{M^2} - \frac{m^2}{M^2 - m^2}\log\frac{\muT_M^2}{M^2} +1 \right) \notag \\[10pt]
&\hspace{30pt}= \log \frac{\muT_M^2}{M^2} + \frac{m^2}{M^2-m^2}\log\frac{m^2}{M^2}= \log \frac{\muT_M^2}{M^2} + \frac{m^2}{M^2}\log\frac{m^2}{M^2} + 1 + \mathcal{O}\left(\frac{m^4}{M^4}\right)\,.
\end{align}
Before expanding in small $\lambda \sim m/M$, this result is $m \leftrightarrow M$ symmetric as it must be.  However, since we are interested in the limit $\lambda \ll 1$, the expanded result is relevant.

Intriguingly, this generates a term with the structure $\lambda^2 \log\lambda^2$.  So we have produced a log that is only a function of physical scales as promised, but this log is power suppressed.  This power suppression can be understood from a physical perspective.  Going back to the original expression in~\cref{eq:IntHeavyLight}, we see that this integral is IR finite in the limit that $m\rightarrow 0$.  Therefore, if $m$ appears as the argument of a logarithm, it must do so in a way that is also finite as $m\rightarrow 0$, explaining the power suppression. 

Next, we turn to the $s$-channel diagram.  Since we are evaluating our process at threshold, $p_1 = p_2 = \big(m,\vec{0}\,\big)$.  Above, we neglected the potential for this momentum flowing through the diagram to change our results since we were focused on the UV divergent terms.  Here our focus is on a logarithm whose argument depends on $m$, so we should treat this carefully:\footnote{which will lead us to discover that being careful will in fact not change the result for the term of interest...}
\begin{align}
s\text{-channel:}\quad &\includegraphics[width=0.25\textwidth, valign=c]{Figures/LLLL_HLLoop.pdf}  = \big(\mu_M^{2\epsilon}\,\rho\big)^2 \int \frac{\text{d}^d \ell}{(2\,\pi)^d} \frac{1}{((\ell+p_1+p_2)^2-m^2)(\ell^2-M^2)}\notag\\[10pt]
&\hspace{40pt} = \frac{i\s\mu_M^{2\epsilon}\,\rho^2}{16\,\pi^2} \left[\frac{1}{\epsilon}+\log \frac{\muT_M^2}{M^2} +1+\frac{m^2}{M^2}\left( \log\frac{m^2}{M^2} +2\right)+\mathcal{O}\big(\lambda^4\big)\right] \,.
\label{eq:HLInt_schannel}
\end{align}
So we see that the $s$-channel diagram does yield a different result from \cref{eq:IntHeavyLight}, but it contributes the same $1/\epsilon$ pole and log terms as the other two channels, yielding a total contribution of three $\lambda^2 \log\lambda^2$ when we sum channels.

\vspace{5pt}\mybox{
\begin{itemize}
\item {\bf Exercise:} Derive \cref{eq:HLInt_schannel}.
\end{itemize}}

All together, we find 
\begin{align}
3\times \includegraphics[width=0.25\textwidth, valign=c]{Figures/LLLL_HLLoop.pdf} &=   \frac{3\s i\,\mu_M^{2\epsilon}\,\rho^2}{16\s\pi^2} \left[\frac{1}{\epsilon}+\log \frac{\muT_M^2}{M^2} + 1 \right. \notag\\
 & \hspace{75pt}\left. +\, \frac{m^2}{M^2}\left(\log\frac{m^2}{M^2}  + \frac{2}{3}\right) + \mathcal{O}\big(\lambda^4\big)\right]\,.
\end{align}
There is another class of diagram that contributes:   
\begin{align}
4\times\includegraphics[width=0.25\textwidth, valign=c]{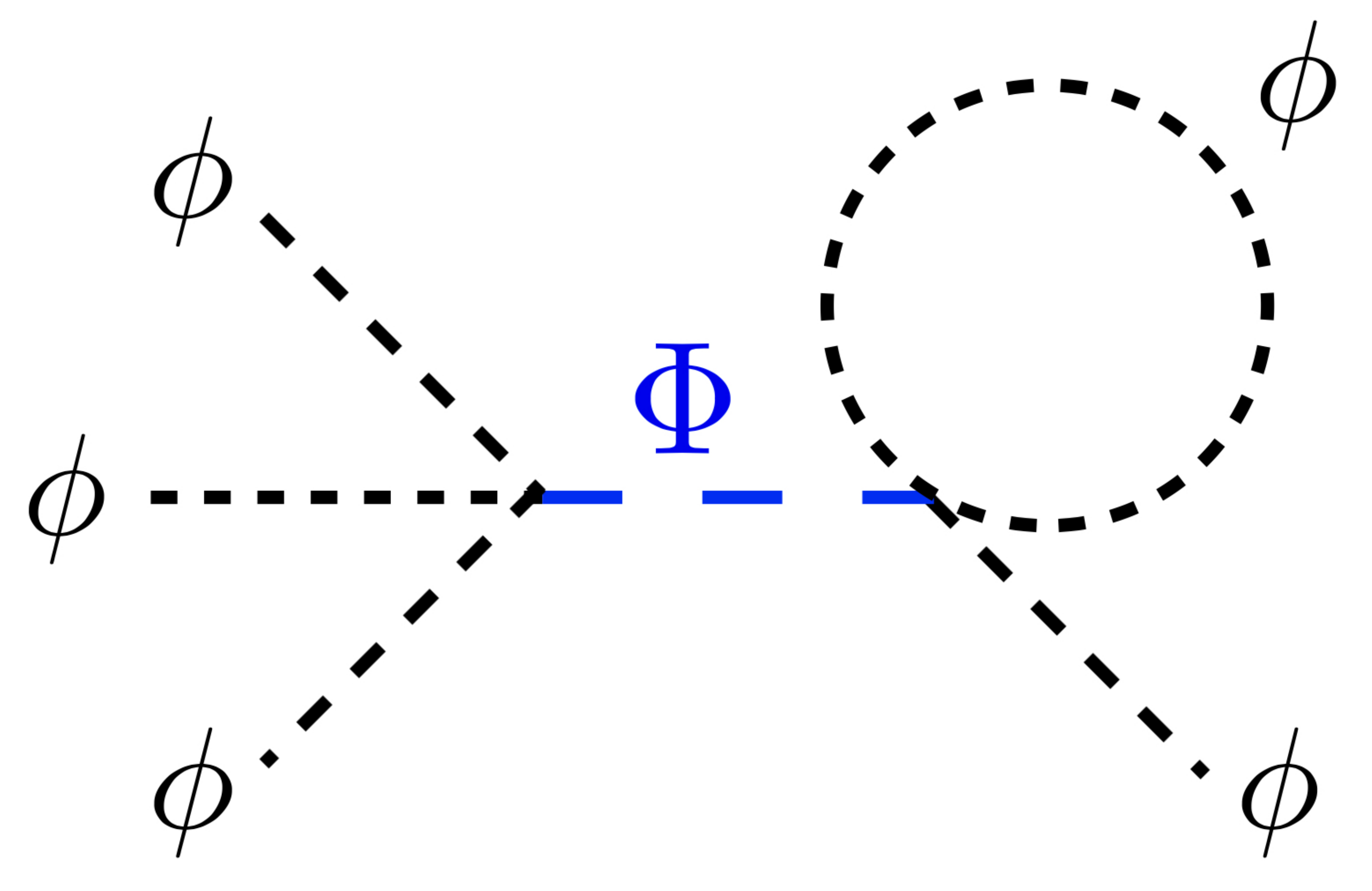} &= \big(\mu_M^{2\epsilon}\,\rho\big)^2\, \frac{4}{2} \,\frac{1}{p_\Phi^2- M^2} \int \frac{\D^d\ell}{(2\s\pi)^d} \frac{1}{\ell^2 - m^2}\notag\\
&= -\frac{i\s \mu_M^{2\epsilon}\,\rho^2}{8\s\pi^2} \frac{m^2}{M^2} \left(\frac{1}{\epsilon} - \log\frac{m^2}{\muT^2_M}+1\right) + \mathcal{O}\big(\lambda^4\big)\,,
\label{eq:assholeDiagrams}
\end{align}
where the $1/2$ is a symmetry factor, and the overall factor of 4 counts that there are 4 independent diagrams, one for each choice of momentum assignment.  Note that in moving from the first to the second line, we have expanded $p_\Phi^2 \sim \lambda^2$ for the off-shell heavy scalar propagator that appears before the integral.  These higher power corrections play no role here.

These diagrams in \cref{eq:assholeDiagrams} are perhaps unfamiliar because they are not naively one-particle irreducible.  However, since we are performing a matching calculation, we should only apply our one-particle irreducible criterion to the light degrees of freedom.  The justification for this assertion comes from realizing that by working in the $\lambda \ll 1$ limit, any $\Phi$ propagator is always far off-shell, and so it is necessary to include diagrams of this form.  Note that the power suppression is clear at the diagrammatic level due to the presence of the heavy off-shell propagator.\footnote{There is yet another way to interpret the physics associated with this diagram.  If one were working consistently to one-loop order in the \FT, then taking our interaction in \cref{Eq:LIntHL} and closing a $\phi$ loop would generate a mass mixing between $\phi$ and $\Phi$.  Then the proper procedure would be to diagonalize the \FT~scalar masses, which would yield the same result when expanded in $\lambda$.  Additionally, note that at higher loop order the fields would mix due to wavefunction renormalization.}

After renormalization using the $\overline{\text{MS}}$ scheme, the NLO \FT~matrix element for $\phi\,\phi \rightarrow \phi\, \phi$ at threshold is 
\begin{align}
i\s\mathcal{A}^\textsc{Full} = \frac{i\s \rho^2}{16\,\pi^2} \left[3\s\log \frac{\muT_M^2}{M^2}+ 3 + \frac{m^2}{M^2}\left(3\s\log\frac{m^2}{M^2} + 2 \s\log \frac{m^2}{\muT_M^2} \right) + \mathcal{O}\big(\lambda^4\big)\right]\qquad \text{\big[NLO\big]}\,.
\label{eq:PhiScatFullTheoryRenorm}
\end{align}
The first log can be summed by introducing and subsequently running $\eta$, following exactly as in \cref{sec:ToyModelLLResum} above.  Our focus here is on separating the scales appearing in the second and third logs, and showing how the $m$ dependent log is regenerated at a low scale using EFT techniques. Again we emphasize that this power suppressed log will not ever cause an issue for perturbation theory, although it could be important if one needs the prediction to a particularly high accuracy.  However, this model provides an excellent pedagogical case study for applying the matching and running procedure.  The power suppression will also impact our interpretation of the RG, as we will show in what follows.

\subsubsection*{Matching onto the EFT and the Separation of Scales}
Now we are ready to match onto the EFT up to one-loop order.  We will need the EFT interactions
\begin{align}
\mathcal{L}^\text{EFT}_\text{Int} \supset \frac{1}{4!}C_4\,\phi^4 + \frac{1}{6!} \frac{C_6}{M^2}\,\phi^6\,.  
\end{align}
Our first step is to match the \FT~to the EFT at tree-level.  Note that although $C_4 = 0$ at tree-level due to our UV choice, it will receive a one-loop matching correction from \cref{eq:PhiScatFullTheoryRenorm}.  There is a non-zero tree-level contribution to $C_6$:
\begin{align}
\frac{1}{2}\begin{pmatrix}6 \\3\end{pmatrix} &\times \includegraphics[width=0.28\textwidth, valign=c]{Figures/LLLLLL_IntOutH.pdf} = -i\s 10\,\rho^2\left(-\frac{1}{M^2}\right) + \cdots \notag\\[5pt]
&= \includegraphics[width=0.22\textwidth, valign=c]{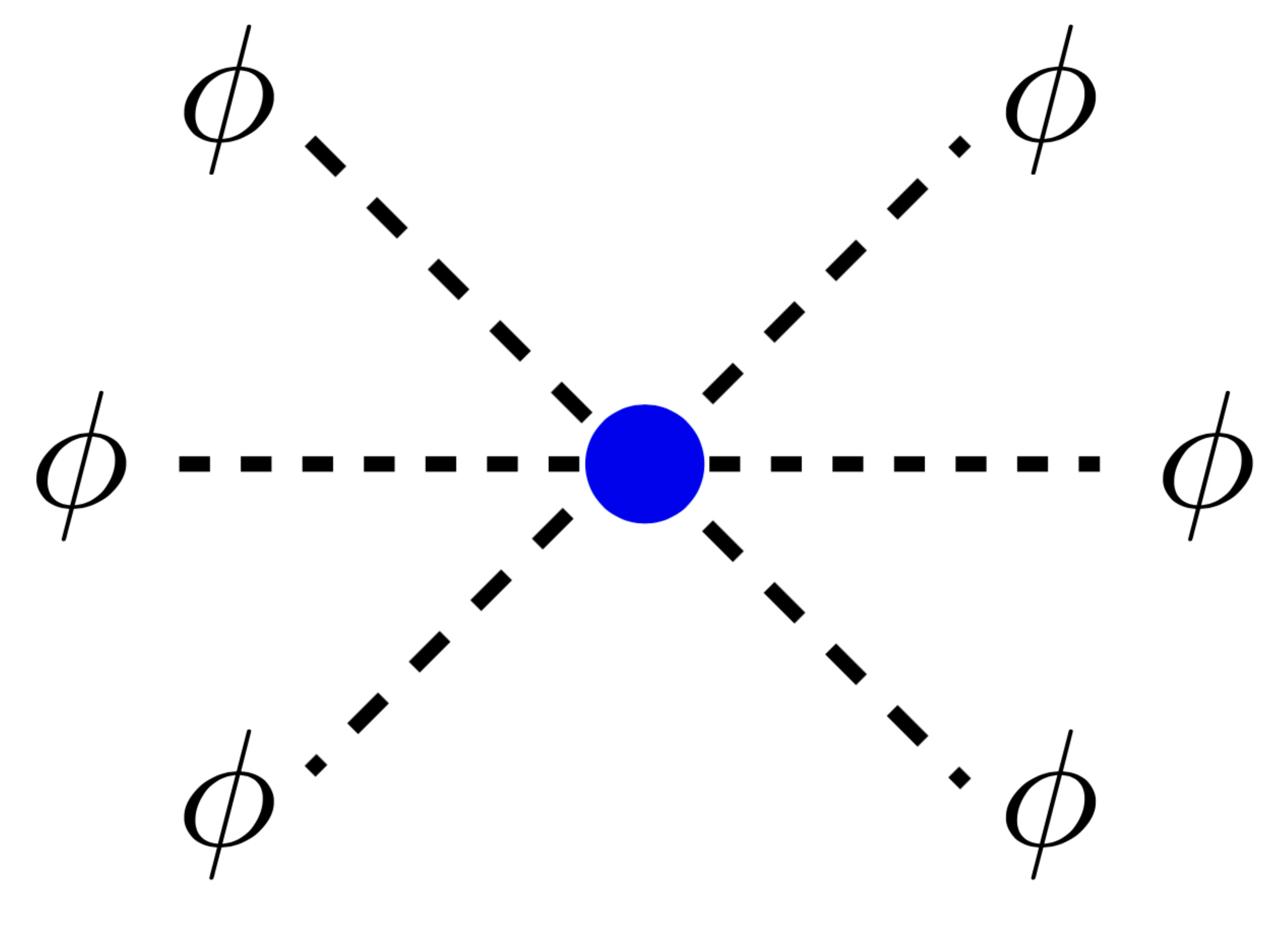} + \cdots =  -i\s\frac{C_6}{M^2}\,,
\end{align}
where the factor of (6 choose 3) accounts for all possible momentum routings.   So tree-level matching yields 
\begin{align}
C_4 = 0 \qquad\qquad C_6 = -10\s\rho^2 \qquad\qquad \big[\text{LO}_M\big]\,.
\label{eq:C6TreeMatch}
\end{align}

\vspace{5pt}\mybox{
\begin{itemize}
\item {\bf Exercise:} Derive \cref{eq:C6TreeMatch} by using the equations of motion to integrate out $\Phi$ in the \FT, following the same steps that led to \cref{eq:EOMEFTResult}.
\end{itemize}}

Now we are ready to match the one-loop power expanded \FT~result \cref{eq:PhiScatFullTheoryRenorm} onto the EFT.  We need the EFT to generate the same non-analytic structure as a function of the IR parameter $m^2$ as in the \FT.  This implies that the matching correction will be analytic in $m^2$. It turns out there is only one class of EFT diagrams to compute
\begin{align}
\includegraphics[width=0.23\textwidth, valign=c]{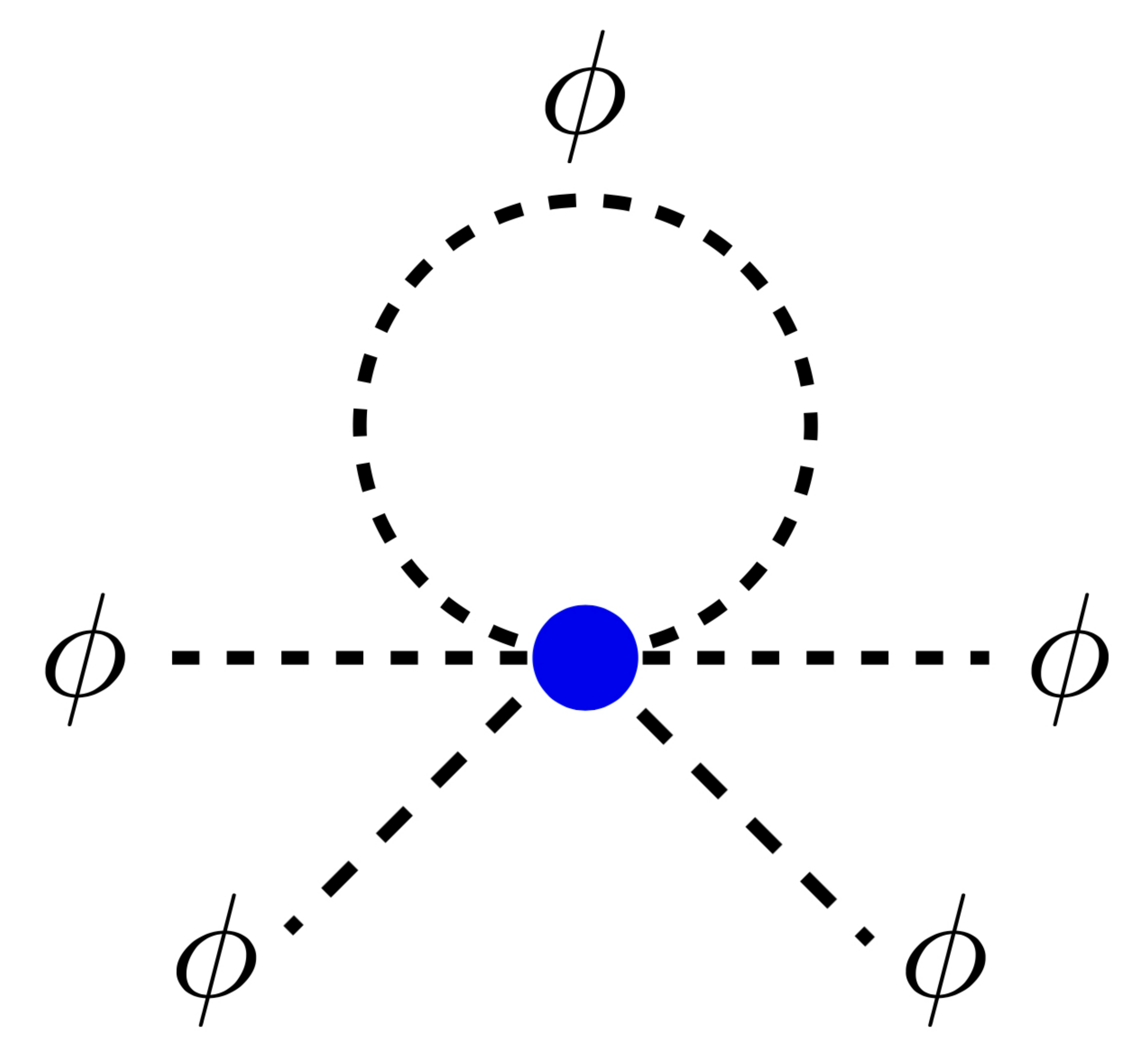} &= \frac{1}{2} \frac{C_6}{M^2}\, \mu_M^{4\epsilon}\int\frac{\text{d}^d \ell}{(2\,\pi)^d}\frac{1}{\ell^2-m^2} \notag\\
&=\frac{ i\s \mu_M^{2\epsilon}\,C_6}{32\s \pi^2}\frac{m^2}{M^2}\,\left[\frac{1}{\epsilon}+\log\frac{\muT_M^2}{m^2} +1 + \mathcal{O}(\epsilon)\right]\,,
\label{eq:phiFourQuadDivLoopEFT}
\end{align}
where the $1/2$ is a symmetry factor, and the RG scale is $\muT_M^2$ because we are matching at the high scale where \cref{eq:C6TreeMatch} is appropriate.  This can be interpreted as an operator mixing effect, since $C_6$ generates a contribution to $C_4$ at loop level, and as such will lead to a non-homogeneous RGE, see \cref{eq:RGEHLmodel} below.  Note that this diagram will only contribute to the RGE at $\mathcal{O}(\lambda^2)$, \emph{i.e.}, it is power suppressed.

Since it is higher power, we will not need to renormalize $C_6$ when working at one-loop.\footnote{If we wanted to run the $C_6$ coupling, we would need to include operators that power count like $\lambda^4$.  In other words, we would need to include a $\phi^8$ local operator in our Lagrangian, which would then both contribute to $C_4$ (at two loops) and to $C_6$ (at one loop), implying that we would have to renormalize $C_6$.}  We renormalize the $C_4$ coupling in the low energy EFT by setting the $Z_4$ counterterm to cancel the $1/\epsilon$ term as usual.  This gives us a counterterm 
\begin{align}
Z_4 = 1+ \frac{1}{C_4}\frac{C_6}{32\s\pi^2}\frac{m^2}{M^2}\frac{1}{\epsilon}\,.
\label{eq:Z4CT}
\end{align}
Now we can apply \cref{eq:MatchinRenorm} to match the \FT~and EFT at one-loop:
\begin{align}
\hspace{-10pt} C_4^\text{Match}\big(\muT_M\big) = \,&- \frac{\rho^2}{16\s\pi^2} \left[3\,\log \frac{\muT_M^2}{M^2}+ 1 + \frac{m^2}{M^2}\left(3\log\frac{m^2}{M^2} + 2 \,\log \frac{m^2}{\muT_M^2} \right)\right]\notag\\[6pt]
& \hspace{45pt}+ \left[-\frac{5i\s\,\rho^2}{16\s \pi^2}\frac{m^2}{M^2}\,\left(\log\frac{\muT_M^2}{m^2} +1 \right)\right]\notag\\[10pt]
= &\,\frac{1}{10} \frac{C_6}{16\s\pi^2} \left[3\,\log \frac{\muT_M^2}{M^2}+3 +\frac{m^2}{M^2}\left( 3\,\log\frac{\muT_M^2}{M^2}+5\right) \right]  \quad \,\,\big[\text{NLO}_M\big] \,.
\label{eq:C4MatchHLlog}
\end{align}
Recalling that we are assuming that the \FT~tree-level $\phi^4$ coupling is zero at the matching scale, see \cref{eq:C6TreeMatch}, we have now derived the boundary condition for the RGEs, $C_4\big(\muT_M\big)=C_4^\text{Match}\big(\muT_M\big)$.  Again, we emphasize that this matching coefficient is fully analytic in $m$ as it must be, see the discussion below \cref{eq:C4Match}.  Note that from the EFT point of view, the $\log \muT_M^2/M^2$ term is not absorbed by RG evolution, and instead simply contributes as a threshold correction, exactly as in our previous examples.    

Before we finish the calculation, it is worth pausing to appreciate that a non-trivial \emph{separation of scales} has now been performed.  Inspecting the power suppressed term in \cref{eq:C4MatchHLlog}, we see that our $\log m^2/M^2$ has now been separated into a $\log \muT_M^2/M^2$.  The other part, $\log m^2/\muT_L^2$ will be generated by EFT loops.   This is what we need to apply RG techniques within the EFT to sum our log!  The rest of this section will do just that, and in particular will show how the LL + NLO result expanded to leading log order will recombine back into \cref{eq:PhiScatFullTheoryRenorm}.  We reiterate that what matching has done is to take the $\log m^2/M^2$ apart, resulting in a contribution at the matching scale that depends explicitly on the associated scale $\muT_M$.  This provides a boundary condition for our RG evolution within the single scale EFT down to a low scale $\muT_L$, where we then can compute the one-loop EFT fixed order corrections to derive a summed result that no longer suffers from any (in principle) large log problems.

Moving forward in the calculation, our next step is to derive an RGE that can be used to evolve the matched $C_4\big(\muT_M\big)$ to a low scale $\muT_L$.  We will need the anomalous dimension for the $\phi^4$ operator in our EFT.  Fortunately, we have already computed the running of $C_4$ due to its self interaction, see \cref{eq:C4runningSol} above.  All we need to do is extract the running due to $C_6$ from \cref{eq:phiFourQuadDivLoopEFT}.  This is straightforward to accomplish through the application of the formulas for the anomalous dimensions we derived above in \cref{annDimPractical}:
\begin{align}
\gamma_{44} = \frac{3}{32\s\pi^2}C_4 -\frac{1}{32\s\pi^2}\frac{C_6}{C_4}\frac{m^2}{M^2} \qquad\qquad\qquad \gamma_{46} =  \frac{1}{16\s\pi^2}\,\frac{m^2}{M^2}\,.
\end{align}
Then we use the general form of the RGE in \cref{eq:RGEgeneral} along with $\gamma_{44}$ as computed in \cref{eq:gamma44} to evolve our Wilson coefficient from the high scale $\muT_M$ to the low scale $\muT_L$: 
\begin{align}
\frac{\D}{\D\log \muT^2} C_4 = \frac{3}{32\s\pi^2}\,C_4^2 + \frac{1}{32\s\pi^2}\frac{m^2}{M^2}\,C_6\,.
\label{eq:RGEHLmodel}
\end{align}
Note that we must be careful to track that each insertion of the $\phi^6$ operator comes with a power suppression.  Since we have truncated our expansion to only capture terms of $\mathcal{O}\big(\lambda^2\big)$, we can only consistently include one insertion of this operator when integrating the RGE.  Note that this is not an issue for $\phi^4$ since this operator power counts as $\mathcal{O}(1)$, implying that we can do an infinite number of $\phi^4$ insertions at no cost (which is how we interpret what integrating the RGE is doing).  Therefore, to solve the RGE in a way that is consistent with power counting, we take our full $C_4$ solution,  given in~\cref{eq:C4runningSol} above, and perturb it with a single $\phi^6$ insertion.  If we wanted to go to $\mathcal{O}\big(\lambda^4\big)$, we would need to include the effects of the $\phi^8$ operator.  This is how renormalization proceeds for a ``non-renormalizable'' theory.

We can track power counting order $n$ with a superscript $C_4^{(n)}$, so that \cref{eq:C4runningSol} solves for $C_4^{(0)}$.  To derive the first sub-leading power contribution due to an insertion of $\phi^6$, we expand $C_4 = C_4^{(0)} + C_4^{(2)}$ and plug this into the RGE
\begin{align}
\frac{\D}{\D\log \muT^2} \Big(C_4^{(0)} + C_4^{(2)} \Big) = \frac{3}{32\s\pi^2}\,\Big(C_4^{(0)}\Big)^2 + \frac{1}{32\s\pi^2}\frac{m^2}{M^2}\,C_6\,.
\end{align}
Note that the when we expand $C_4^2$ consistently, the $C_4^{(0)} C_4^{(2)}$ cross term can be neglected since it is both coupling and power suppressed.\footnote{If one were interested in extending this calculation to higher loop order (but same order in power counting), it would no longer be consistent to ignore the cross term between the leading and power suppressed pieces of $C_4$.  One could still solve the exact RGE with $C_4/C_6$ mixing and then truncate in $\lambda$, but another strategy would be to factorize the RGE into separate differential equations with uniform power counting.
}  Then we can use the fact that $C_4^{(0)}$ solves the RGE with $C_6=0$ to derive the leading power correction to the RGE 
\begin{align}
\frac{\D}{\D\log \muT^2}\, C_4^{(2)} =  \frac{1}{32\s\pi^2}\frac{m^2}{M^2}\,C_6\,,
\end{align}
which is trivial to solve. The solution for the running coupling $C_4\big(\muT_L\big)$ to $\mathcal{O}(\lambda^2)$ is\footnote{Now the $\text{LL}_{\lambda^2}$ in the brackets is there to track that we are including the sub-leading power correction to the LL RGE.}  
\begin{align}
C_4\big(\muT_L\big) &=  \frac{C_4^{(0)}\big(\muT_M\big)}{1-C_4^{(0)}\big(\muT_M\big)\,\frac{3}{32\s\pi^2} \log \frac{\muT_M^2}{\muT_L^2}}\notag\\[7pt]
&\hspace{20pt} +\left(C_4^{(2)}(\muT_M)+ \frac{1}{32\s\pi^2}\frac{m^2}{M^2}\,C_6 \,\log\frac{\muT_M^2}{\muT_L^2}\right) \qquad \text{\big[LL}_{\lambda^2}\big]\,,
\label{eq:RGEsolEFT}
\end{align}
where $C_4^{(2)}\big(\muT_M\big)$ is the power suppressed $C_6$ dependent contribution from matching given in \cref{eq:C4MatchHLlog}, while the non-power suppressed terms from the matching are absorbed into $C_4^{(0)}\big(\muT_M\big)$.

The last step is to compute the perturbative corrections within the EFT at the low scale.  Specifically, these come from a set of diagrams proportional to $\big(C_4\big)^2$, see \cref{eq:A4ptRenorm}, and a set proportional to $C_6$, see \cref{eq:phiFourQuadDivLoopEFT}, but now evaluated at the scale $\muT_L$:  
\begin{align}
 i\,\mathcal{A}^\text{EFT}  = -i\s C_4\big(\muT_L\big)\left[1 - \frac{3}{32\s\pi^2}\,C_4\big(\muT_L\big) \log\frac{\muT_L^2}{m^2}\right] + \frac{i\s C_6}{32\s \pi^2}\frac{m^2}{M^2}\,\left[\log\frac{\muT_L^2}{m^2} +1\right]  \qquad \big[\text{NLO}\big]\,.
\end{align}

So now we have all the necessary pieces required to derive an LL$_{\lambda^2}$ + NLO summed amplitude, although we will not do so explicitly.  Instead, it will prove to be illuminating to expand our RGE solution and put everything together to show that we recover the \FT~result \cref{eq:PhiScatFullTheoryRenorm} when expanding the RG solution to leading log order (recall that $C_6\big(\muT_M\big) = -10\,\rho^2$):
\begin{align}
i\,\mathcal{A}^\textsc{Full} &= i\s\frac{-10\,\rho^2}{16\s\pi^2}\frac{1}{10}\frac{m^2}{M^2} \left[3\,\log\frac{m^2}{M^2}  +2 \,\log \frac{m^2}{\muT_M^2} \right] +\cdots \qquad \big[\text{NLO}\big]\,;\notag\\[14pt]
i\,\mathcal{A}_\text{Expanded}^\text{EFT} &= i\s\frac{1}{10}\frac{1}{16\s\pi^2}\,C_6\,\frac{m^2}{M^2}\left(3\,\log\frac{\muT_M^2}{M^2} + 2\,\log\frac{\muT^2_M}{\muT^2_M}\right) \notag\\[5pt]
&\hspace{13pt} - i\s\frac{1}{32\s\pi^2}\frac{m^2}{M^2}\, C_6 \,\log\frac{\muT_M^2}{\muT_L^2} \notag\\[5pt]
&\hspace{13pt}  - i\s\frac{1}{32\s\pi^2}\frac{m^2}{M^2}\,C_6\,\log\frac{\muT_L^2}{m^2} +\cdots\notag\\[5pt]
& = i\s\frac{C_6}{32\s\pi^2}\frac{1}{10}\frac{m^2}{M^2} \left[3\,\log\frac{m^2}{M^2}  +2 \,\log \frac{m^2}{\muT_M^2} \right]+\cdots \qquad \big[\text{LL}_{\lambda^2}\text{ + NLO}\big]\,,
\end{align}
where the \FT~result on the first line only reproduces the power suppressed log dependent terms from~\cref{eq:PhiScatFullTheoryRenorm}. The first line of the EFT result is the matching contribution, the second line is the RG contribution expanded to keep only the leading logs, and the final line is the EFT low scale loop correction.  The final equality clearly shows agreement with the \FT~result, demonstrating that our procedure for separating scales works as anticipated.
\vspace{5pt}\mybox{\begin{itemize}
\item \textbf{Exercise:} Track the finite terms to show that the complete \FT~result in \cref{eq:PhiScatFullTheoryRenorm} can be built by putting together the matching + running + EFT loop contributions.
\end{itemize}}

We see that this exercise exposes all the promised features of matching and running.  All integrals were expanded such that they only were functions of a single scale.  Once the scales were separated, the application of the RGE was completely straightforward, thereby summing all dangerous logarithms and connecting the high scale description to the low scale.

\vspace{5pt}\mybox{
\begin{itemize}
\item {\bf Exercise:}  Famously, the Higgs quartic receives a finite log correction within the Minimal Supersymmetric Standard Model (MSSM) whose argument is the ratio of the stop mass to the top mass~\cite{Haber:1990aw}.  This is a very important effect since the Higgs-boson mass is bounded by the $Z$-boson mass at tree level in the MSSM, and so the stop-top loop corrections are critical to achieve a 125 GeV physical Higgs mass.  An excellent application of matching and running is to apply the techniques discussed here to separate scales inside that log and sum it at one-loop order.
\end{itemize}}

We close this section with one final analysis.  We will define the ``method of regions'' and apply it to our heavy-light integral in \cref{eq:HLInt_others}.  This will lay the groundwork for our investigation of more complicated integrals that yield Sudakov logarithms, as discussed in \cref{sec:RegionsMasslessSudakov} and \cref{sec:RegionsMassiveSudakov} below.

\subsection{Method of Regions for a Heavy-light Integral}
\label{sec:RegionsHeavyLight}
In this section, we introduce the so-called method of regions~\cite{Beneke:1997zp},\footnote{Some useful general references are~\cite{Smirnov:2002pj} and Chapter 9 of~\cite{Smirnov:2012gma}.  For a demonstration of problems that can arise at higher loop order, see~\cite{Smirnov:1999bza}, and for some work towards a proof of the method, see~\cite{Jantzen:2011nz, Semenova:2018cwy}. In particular,~\cite{Jantzen:2011nz} demonstrates how to use this approach in the presence of zero-bins (which are discussed in \cref{sec:ZeroBin} below).} and apply it to~\cref{eq:HLInt_others}, the heavy-light integral from the previous section.  The purpose of this method is to decompose a complicated multi-scale integral into many simpler single scale integrals by utilizing an expansion in a power counting parameter $\lambda$.  The result is that one dimensionally regulated integral over the full domain is expressed as many integrals whose limits of integration are localized about the region where they have non-trivial support.  Then relying on the fact that scaleless integrals vanish in dim reg, one can extend each of these expanded integrals to the full domain of integration, and then apply dim reg to regulate them individually.  Initially, this will be used as a method for computing integrals in the limit that $\lambda \ll 1$, but then in \cref{sec:ToySCET} we will see that we can interpret each non-vanishing on-shell region\footnote{A way to see if a mode can go ``on-shell'' is to check if it can satisfy $p^2 = 0$ when $\lambda\neq 0$.  This condition holds for the hard, collinear, and soft modes of $\phi$.  Regions that do not have an on-shell interpretation will yield effects such as ``potential'' or ``Glauber'' exchanges, see \cref{eq:AllTheModes} for the scalings.  These effects will only be briefly mentioned as they are beyond the scope of these lectures.} as contributing a propagating degree of freedom to the EFT.

We begin with our heavy-light integral
\begin{align}
\mathcal{I} = \mu^{2\epsilon} \int\frac{\D^d \ell}{(2\s\pi)^d}\,\frac{1}{\ell^2-m^2}\,\frac{1}{\ell^2-M^2}\,.
\end{align}
Define a power counting parameter $\lambda \sim m/M$.  Next, assume a particular $\lambda$ scaling for $\ell$, and Taylor expand the denominator factors assuming this $\ell$ scaling before integrating.  Since we want to reproduce our $\lambda^2 \log \lambda^2$ heavy-light log, we will need to carry these expansions to $\lambda^2$ subleading order.

The most obvious region has hard scaling
\begin{align}
\ell_h^\mu \sim M\,\big(1,1,1,1\big)\,,
\end{align}
so that our integral can be expanded as
\begin{align}
\frac{\text{d}^4\ell}{\big(\ell^2-M^2\big)\big(\ell^2-m^2\big)}\quad \longrightarrow \quad \frac{\text{d}^4\ell_h}{\big(\ell_h^2-M^2\big)\ell_h^2} +  \frac{m^2\,\text{d}^4\ell_h}{\big(\ell_h^2-M^2\big)\ell_h^4}\,,
\end{align}
which go as $\sim \mathcal{O}\big(1\big)$ and $\sim\lambda^2$ respectively.  We also can expand our integrand assuming a soft scaling
\begin{align}
\ell_s^\mu \sim M\,\Big(\lambda,\lambda,\lambda,\lambda\Big) \,,
\end{align}
yielding
\begin{align}
\frac{\text{d}^4\ell}{\big(\ell^2-M^2\big)\big(\ell^2-m^2\big)} \quad \longrightarrow \quad \frac{\text{d}^4\ell_s}{-M^2\big(\ell_s^2-m^2\big)}\,,
\end{align}
which scales as $\sim \lambda^2$.

Next we integrate each of these expanded integrands:
\begin{align}
\mathcal{I}_h^{(0)} &= \mu^{2\epsilon} \int\frac{\D^d \ell_h}{(2\s\pi)^d}\frac{1}{\big(\ell_h^2-M^2\big)\ell_h^2} = \frac{i}{16\s\pi^2}\left(\frac{1}{\epsilon} +\log\frac{\muT^2}{M^2}+1\right)\notag \\[10pt]
\mathcal{I}_h^{(2)} &=\mu^{2\epsilon} \int\frac{\D^d \ell_h}{(2\s\pi)^d}\frac{m^2}{\big(\ell_h^2-M^2\big)\ell_h^4} = \frac{i}{16\s\pi^2}\frac{m^2}{M^2}\left(\frac{1}{\epsilon} +\log\frac{\muT^2}{M^2}+1\right)\notag \\[10pt]
\mathcal{I}_s^{(2)} &=\mu^{2\epsilon} \int\frac{\D^d \ell_s}{(2\s\pi)^d}\frac{1}{-M^2\big(\ell_s^2-m^2\big)} = -\frac{i}{16\s\pi^2}\frac{m^2}{M^2} \left(\frac{1}{\epsilon}+\log\frac{\muT^2}{m^2}+1\right)\,,
\label{eq:HLregionsInts}
\end{align}
where the superscripts $n$ on $\mathcal{I}^{(n)}$ correspond to the order in power counting that this integral will contribute, and we used the same evaluation as was needed for \cref{eq:UVIRint} above.  Then simply summing the results in \cref{eq:HLregionsInts} yields
\begin{align}
\mathcal{I} = \mathcal{I}_h^{(0)} + \mathcal{I}_h^{(2)} + \mathcal{I}_s^{(2)} = \frac{i}{16\s\pi^2}\left[\left(\frac{1}{\epsilon} +\log\frac{\muT^2}{M^2}+1 \right) +\frac{m^2}{M^2}\left(\log\frac{\muT^2}{M^2} - \log\frac{\muT^2}{m^2}\right)\right]\,.
\end{align}
We see that the method of regions approach reproduces our direct calculation of the full integral up to $\mathcal{O}\big(\lambda^2\big)$, see~\cref{eq:IntHeavyLight}.

\vspace{5pt}\mybox{\begin{itemize}
\item \textbf{Exercise:} Convince yourself that any other scalings for $\ell$ will yield  a scaleless integral.  Since our EFT is Lorentz invariant, we can restrict ourself to homogeneous scalings of the loop momentum $\ell_h^\mu \sim M\,\big(\lambda^a,\lambda^a,\lambda^a,\lambda^a\big)$, for some choice of $a$.
\end{itemize}}

\vspace{5pt}\mybox{\begin{itemize}
\item \textbf{Exercise:} Perform the same type regions expansion to evaluate the $s$-channel diagram given in \cref{eq:HLInt_schannel} to $\lambda^2$ order.
\end{itemize}}

As was mentioned above, identifying the regions has a corresponding interpretation from the EFT point of view.  The hard region contributes to the local Wilson coefficient via matching, while the soft region corresponds to dynamics in the EFT, modeled by the scalar $\phi$ and its interactions.  This procedure is unnecessarily sophisticated for this simple heavy-light example (which is why we are doing it at the end), but it will be critical for our understanding of SCET in what follows.

The next section reviews the physics of soft and collinear divergences, including the identification of large IR logarithms that can require summation.  This will prime us for our toy version of SCET that follows in \cref{sec:ToySCET}.

\newpage
\section{Soft and Collinear Divergences}
\label{sec:SoftCollinearDiv}
Now that we have developed a strong foundation for applying matching and running to separate scales, we can turn our attention to some more involved examples.  The goal for the rest of these lectures is that the reader will learn to sum (double) logarithms associated with soft and collinear divergences using EFT techniques.  To that end, this section is devoted to setting up the relevant background, before introducing SCET in \cref{sec:ToySCET}.  Before we get into the physics, we explain light-cone coordinates in the next \Primer.  This is the natural setting for working with collinear divergences.  A \Primer~on the systematics of IR divergences in field theory then follows.  The main results for this section are provided in \cref{sec:RegionsMasslessSudakov} and \cref{sec:RegionsMassiveSudakov}, where we will use the method of regions to evaluate two archetypical integrals that yield the so-called Sudakov double logs~\cite{Sudakov:1954sw} we are interested in summing.  

\scenario{The Light Cone}
\addcontentsline{toc}{subsection}{\color{colorTech}{Primer~\thescenario.} The Light Cone}
\label{sec:LightCone}
We will be interested in taking the collinear limit of momenta, so it is useful to work with respect to a light-like reference direction.  To this end, we introduce so-called light-cone coordinates, where our coordinate system is decomposed along a ``collinear'' direction $n^\mu$, an ``anti-collinear'' direction $\bar{n}^\mu$, and the two ``perp'' directions specified with a $\perp$ symbol, whose defining feature is that they are perpendicular to $n$ and $\bar{n}$.  Calculating in these coordinates is also known as field theory in the infinite momentum frame or null plane field theory.\footnote{Essentially all of the light-cone formalism that follows can be found in~\cite{Kogut:1969xa}, where they show how to perform QED calculations when expressed in the infinite momentum frame.  They additionally show that the elements of the subgroup of Poincar\'e which leave $\bar{n} \cdot x$ invariant are the familiar non-relativistic Galilean symmetries.}

When we need to be explicit, we will always make the standard choice to point $n^\mu$ along the $+\hat z$ light-cone direction: 
\begin{align}
n^\mu = \big(1,0,0,1\big)\,,
\label{eq:n}
\end{align}  
such that $n^2=0$.  Then the anti-collinear direction is defined by the condition that it is light-like, $\bar{n}^2 = 0$, and that it points opposite to $n$, such that $n \cdot \bar{n} = 2$ (the 2 is a normalization convention, and will lead to many subtle factors of two that are easy to miss).\footnote{A choice of conventions that avoids many of these factors of 2 is to instead define $n^\mu = (1/\sqrt{2},0,0,1/\sqrt{2})$ and similar for $\bar{n}^\mu$ so that $n\cdot \bar{n} = 1$.  Here, we have chosen to follow the conventions that are typically used in the SCET literature.}  When we need to be explicit, we will always take the convenient choice 
\begin{align}
\bar{n}^\mu =\left(1,0,0,-1\right)\,.
\label{eq:nBar}
\end{align}
Then the other two directions are defined by having a vanishing dot product with $n$ and $\bar{n}$.  In our explicit frame, this implies
\begin{align}
p^\mu_\perp = \big(0,p^2,p^3,0\big)\,.
\end{align}
Then the metric tensor can be written as
\begin{align}
g^{\mu\nu} = \frac{n^\mu \s \bar{n}^\nu}{2} + \frac{\bar{n}^\mu \s n^\nu}{2} + g_\perp^{\mu\nu}\,,
\label{eq:gmunuLC}
\end{align}
where $g_\perp^{\mu\nu}$ is defined implicitly by this equation.
Lorentz four vectors are then\footnote{It is common to introduce additional notation $p_+ = n\cdot p$ and $p_- = \bar{n}\cdot p$.  We chose to leave the $n$ and $\bar{n}$ dependence explicit for pedagogical reasons, at the expense of bulkier expressions.}
\begin{align}
p^\mu =  \frac{\bar{n}^\mu}{2}  \, n \cdot p + \frac{n^\mu}{2}\,\bar{n}\cdot p + p_{\perp}^{\mu}, 
\label{eq:lightconemomentum}
\end{align}
such that 
\begin{align}
p^2 = (n\cdot p) (\bar{n}\cdot p) + p_\perp^2\,,
\label{eq:pSqLC}
\end{align} 
where $p_\perp^2 \equiv p_\perp \cdot p_\perp = p_\perp^\mu \, p_{\perp\s\mu} = -p_1^2 - p_2^2$.  Note the opportunity for minus sign mistakes here; the ``\,$\cdot$\," denotes a four-vector dot-product.  Another confusing fact is that $n\cdot p$ is the component of the momentum that points in the $\bar{n}^\mu$ direction and vice versa.

It will also be useful to take light-cone projections of the derivatives.  For concreteness, we can work in the explicit frame specified by \cref{eq:n} and \cref{eq:nBar}, so that
\begin{align}
n\cdot p = p^0 - p^3\qquad \qquad \qquad \bar{n}\cdot p = p^0 + p^3 \,.
\label{eq:cDotxExplicit}
\end{align}
Then 
\begin{align}
n^\mu \frac{\partial}{\partial p^\mu} &= \frac{\partial}{\partial p^0} + \frac{\partial}{\partial p^3}= \frac{\partial n\cdot p}{\partial p^0} \frac{\partial}{\partial n\cdot p} + \frac{\partial \bar{n}\cdot p}{\partial p^0} \frac{\partial}{\partial \bar{n}\cdot p} + \frac{\partial n\cdot p}{\partial p^3} \frac{\partial}{\partial n\cdot p} + \frac{\partial \bar{n}\cdot p}{\partial p^3} \frac{\partial}{\partial \bar{n}\cdot p} \notag \\[7pt]
&= \frac{\partial}{\partial n\cdot p} + \frac{\partial}{\partial \bar{n}\cdot p}  - \frac{\partial}{\partial n\cdot p} + \frac{\partial}{\partial \bar{n}\cdot p} = 2\, \frac{\partial}{\partial \bar{n}\cdot p}\,.
\label{eq:DecomposeDer}
\end{align}
This is another subtle factor of two to watch out for.

\vspace{5pt}\mybox{\begin{itemize}
\item \textbf{Exercise:} Show that \cref{eq:DecomposeDer} is self consistent by writing 
\begin{align}
\frac{\partial}{\partial p^\mu} = n_\mu\, \frac{\partial}{\partial n\cdot p} + \bar{n}_\mu\,\frac{\partial}{\partial \bar{n}\cdot p} + \frac{\partial}{\partial p_\perp^\mu}\,,
\end{align}
and evaluating this for $\partial/\partial p^0$.  Note that this will require inverting \cref{eq:cDotxExplicit} and using the chain rule to evaluate terms like $\partial p^0/\partial n\cdot p$.
\end{itemize}}

Next, we introduce light-cone notation for the components of a Lorentz vector:
\begin{align}
p^\mu = \big(n\cdot p, \bar{n} \cdot p , p_\perp\big)\,,
\label{eq:pLightConeNotation}
\end{align}
which will be utilized extensively in what follows.  For example, the frame specified by \cref{eq:n} and \cref{eq:nBar} is written as
\begin{align}
n^\mu = (0,2,0) \qquad \qquad \qquad \bar{n}^\mu = (2,0,0)\,,
\end{align}
which is another factor of two to track.   

We will see that SCET is not explicitly Lorentz invariant.  One way this manifests is that momentum can power count non-trivially in different directions.  We will account for this fact by letting $p^\mu \sim \big(\lambda^a, \lambda^b, \lambda^c \big)$ for some $a$, $b$, and $c$ below -- this shorthand assumes that $\perp$ is treated homogeneously when expanding around the light cone.  Again, we emphasize that this is not a three-momentum, but should instead be interpreted using \cref{eq:pLightConeNotation}.

\subsubsection*{Reparametrization Invariance}
Working in light cone coordinates obscures Lorentz invariance, since we have explicitly chosen a frame.  Obviously, the underlying physics is still Lorenz invariant -- a theory expressed in light-cone coordinates must know about Lorentz invariance.  Operationally, one must work with operators that respect the transformations that are induced by the broken Lorentz generators, order by order in the power counting expansion.  This is known as reparameterization invariance (RPI).  When working with SCET, there is a non-trivial interplay between Lorentz invariance and power counting, since we will be truncating our mode expansion around the light cone.  Note that if all orders in $\lambda$ were included, SCET must be equivalent to the \FT, where full Lorentz symmetry would be restored.  Therefore, the EFT must track the broken Lorentz generators order-by-order in power counting, so that RPI serves as an additional consistency condition for SCET. When working from the top down, RPI will automatically be preserved.   Alternatively, when working from the bottom up, one should constrain the operator structure to respect RPI.  Due to its simple nature, RPI will play a minimal role in our toy scalar SCET theory studied in the next section.  Therefore, the rest of this \Primer~is most relevant to~\cref{sec:RealSCET}, where we discuss SCET theories involving gauge bosons and fermions.

Practically, RPI can be characterized by noting that any choice of $n^\mu$ and $\bar{n}^\mu$ which satisfy the conditions $n \cdot \bar{n} = 2$ and $n^2 = 0=\bar{n}^2$ must yield the same physics. These conditions are invariant under three different kinds of reparameterization:\\
\vspace{-15pt}
\begin{center}
\renewcommand{\arraystretch}{1.2}
\setlength{\arrayrulewidth}{.3mm}
\setlength{\tabcolsep}{1.2em}
\begin{tabular}{c|c|c}
RPI-I & RPI-II & RPI-III \\
\hline
$n_\mu \rightarrow n_\mu + \Delta_\mu^{\perp}$ & $n_\mu \rightarrow n_\mu$ & $n_\mu \rightarrow e^\alpha\, n_\mu$ \\
$\bar{n}_\mu \rightarrow \bar{n}_\mu$ & $\bar{n}_\mu \rightarrow \bar{n}_\mu + \epsilon_\mu^\perp$ & $\bar{n}_\mu \rightarrow e^{-\alpha}\, \bar{n}_\mu$
\end{tabular} 
\end{center}
where $\Delta_\mu^{\perp}$, $\epsilon_\mu^\perp$, and $\alpha$ are the transformation parameters for RPI-I, RPI-II, and RPI-III respectively, and $\bar{n}\cdot \epsilon^\perp = n \cdot \epsilon^\perp = \bar{n} \cdot \Delta^\perp = n \cdot \Delta^\perp = 0$.  

To make the connection with Lorentz invariance clear, we can express the RPI generators in terms of the full Lorentz group structure.\footnote{For a formulation of RPI in terms of a spinor helicity representation of the light cone, see~\cite{Cohen:2018qvn}.}  These transformations were first worked out for SCET in~\cite{Manohar:2002fd}.  A novel approach for dealing with EFT operator structures in the presence of reference vectors was developed in~\cite{Heinonen:2012km}.

We begin with the Poincar\'e algebra:
\begin{align}
\label{eq:poincare1}
\Big[P^\mu, P^\nu \Big] &= 0 \notag\\[7pt] 
\Big[M^{\mu\nu},P^\rho \Big] &= i\s g^{\mu\rho}\s P^\nu-i\s g^{\nu\rho}\s P^\mu  \notag\\[7pt]
\Big[M^{\mu \nu}, M^{\kappa \rho}\Big] &= - i\s g^{\mu \kappa}\s M^{\nu \rho} - i\s g^{\nu \rho}\s M^{\mu \kappa} + i\s g^{\mu \rho}\s M^{\nu \kappa} + i\s g^{\nu \kappa}\s M^{\mu \rho}\, ,
\end{align}
where $P_\mu = i\s  \partial_\mu$ is the generator of translations, and $M^{\mu \nu}$ is the usual anti-symmetric matrix of Lorentz generators which has an interpretation as generating the rotations $M^{ij} = - \epsilon_{ijk}\, J^k$ and boosts $M^{0i} = K^i$:
\begin{align}
\Big[J_i , J_j \Big] = i \s\epsilon_{ijk}\s J_k\qquad\qquad \Big[ J_i, K_j \Big] = i \s \epsilon_{ijk} \s K_k\qquad \qquad \Big[ K_i, K_j\Big] = - i\s \epsilon_{ijk} \s J_k\,, 
\end{align}
where $\epsilon_{ijk}$ is the anti-symmetric Levi-Civita tensor, \emph{e.g.}~$\epsilon_{123} =1$, $\epsilon_{213} =-1$, $\epsilon_{122} = 0$.

To expose the broken generators, we project $M^{\mu\nu}$ onto $n$ and $\bar{n}$ as defined by \cref{eq:n} and \cref{eq:nBar}: 
\begin{align}
& R_{\rm I}^{\s\nu_\perp} = \bar{n}_\mu\s M^{\mu \nu_\perp} = M^{0 \nu_\perp} + M^{3 \nu_\perp} = \left\{\begin{array}{l} K^1 - J^2\\K^2+J^1 \end{array}\right. \notag\\[7pt]
& R_{\rm II}^{\s\nu_\perp} =  n_\mu\s M^{\mu \nu_\perp} = M^{0 \nu_\perp} - M^{3 \nu_\perp}= \left\{\begin{array}{l} K^1 + J^2\\K^2-J^1 \end{array}\right. \notag \\[7pt]
& R_{\rm III} = n_\mu\s \bar{n}_\nu\, M^{\mu \nu} = 2\, M^{03} = 2\, K^3\,,
\label{eq:LCgenbreaking}
\end{align}
where $\nu_\perp = 1,2$, yielding the five generators of the RPI transformations.  Note that the $J_3$ generator does not appear.  This could have been anticipated since rotations about the $z$-axis leave our $n_\mu$ and $\bar{n}_\mu$ vectors unchanged.  When we discuss collinear fermions in \cref{sec:CollinearFermions} below, we will see how these definitions can be used to infer the fermion RPI transformations from knowledge of the full Lorentz transformations.

Using the explicit matrix forms of the $J_i$ and $K_i$ generators for a four-vector basis, it is straightforward to check that the identification made in \cref{eq:LCgenbreaking} corresponds to the RPI transformations for $n_\mu$ and $\bar{n}_\mu$.  Then we can check the commutators
\begin{align}
\begin{array}{ll}
  \Big[ R^{\s\mu_\perp}_{\rm I}, R^{\nu_\perp}_{\rm I} \Big] = 0 &\qquad\qquad\qquad   \Big[ R^{\s\mu_\perp}_{\rm II}, R^{\nu_\perp}_{\rm II} \Big] = 0 \\[12pt]
 \Big[ R_{\rm I}^1, R_{\rm II}^1 \Big] =  i\s R_{\rm III} &\qquad\qquad\qquad  \Big[ R_{\rm I}^2, R_{\rm II}^2 \Big] = i\s R_{\rm III} \\[12pt]
  \Big[ R_{\rm I}^1, R_{\rm II}^2 \Big] =  -2\s i \s J_3 &\qquad\qquad\qquad  \Big[ R_{\rm I}^2, R_{\rm II}^2 \Big] = 2\s i\s J_3 \\[12pt]
  \Big[ R_{\rm I}^1, R_{\rm III} \Big] = -2\s i\s R_{\rm I}^1 &\qquad\qquad\qquad  \Big[ R_{\rm I}^2, R_{\rm III} \Big] = -2\s i\s R_{\rm I}^2 \\[12pt]
    \Big[ R_{\rm II}^1, R_{\rm III} \Big] = 2\s i\s R_{\rm II}^1 &\qquad\qquad\qquad  \Big[ R_{\rm II}^2, R_{\rm III} \Big] = 2\s i\s R_{\rm II}^2 \,\,\,.
\end{array}
  \label{eq:RPIComm}
\end{align}
This explains the sense in which the RPI transformations encode the full structure of the Lorentz group, since $J_3$ appears when taking commutators of particular component of RPI-I and RPI-II.  Again, we emphasize that RPI is enforced as a constraint on our EFT Lagrangian.  Furthermore, performing an RPI transformation mixes orders in our power counting parameter, which is a sign that restoring full Lorentz invariance requires including terms to all orders in $\lambda$.  If one were to sum up operators to all orders in power counting, the full theory (with the accompanying Lorentz invariance) would emerge.

\vspace{5pt}\mybox{\begin{itemize}
\item \textbf{Exercise:} Check the identifications made in \cref{eq:LCgenbreaking} reproduce the RPI transformations, and reproduce the commutation relations in \cref{eq:RPIComm}.
\end{itemize}}

In the next section, we will discuss IR divergences generally.  By taking the soft and collinear limits of a concrete tree and loop process, we will develop some physical intuition for how these divergences emerge.

\scenario{IR Logarithms}
\addcontentsline{toc}{subsection}{\color{colorTech}{Primer~\thescenario.} IR Logarithms}
\label{sec:IRLogs}
This \Primer~is devoted to exploring the systematic properties of soft and collinear IR divergences.  Our starting point is the statement that the calculation of a physical process cannot be IR divergent order-by-order in perturbation theory (when computing with true asymptotic states).  This must be true since there are no IR analogs of the counterterms used to eliminate UV divergences.   However, amplitudes can individually manifest IR divergences, which are the harbinger of potentially large physical IR logarithms that can emerge.  Measuring a fully inclusive final state will not lead to any IR logs, while restrictions of the final state kinematics or particle type can expose the Sudakov double log.  Exploring these concepts is the goal of this \Primer.

For concreteness, we will focus on a simple example process.  Our \FT~consists of a heavy scalar $\Phi$ and a light scalar $\phi$ with an interaction Lagrangian
\begin{align}
\mathcal{L}^\textsc{Full}_{\text{Int}} &=-\frac{1}{3!}\s a\,\phi^3- \frac{1}{2}\s b\,\phi^2\, \Phi\,.
\end{align}
Our process of interest will be $\Phi \rightarrow \phi\,\phi$, \emph{i.e.}, the injection of large energy into a system of nearly massless particles.  At LO, there is only one diagram
\begin{align}
\includegraphics[width=0.16\textwidth, valign=c]{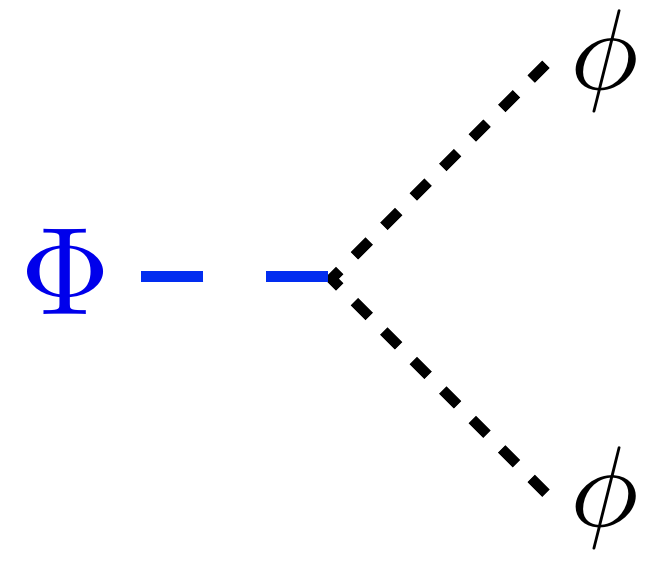} \,,
\end{align}
which is trivially finite.  At NLO, this process gets contributions from three additional diagrams
\begin{align}
\includegraphics[width=0.16\textwidth, valign=c]{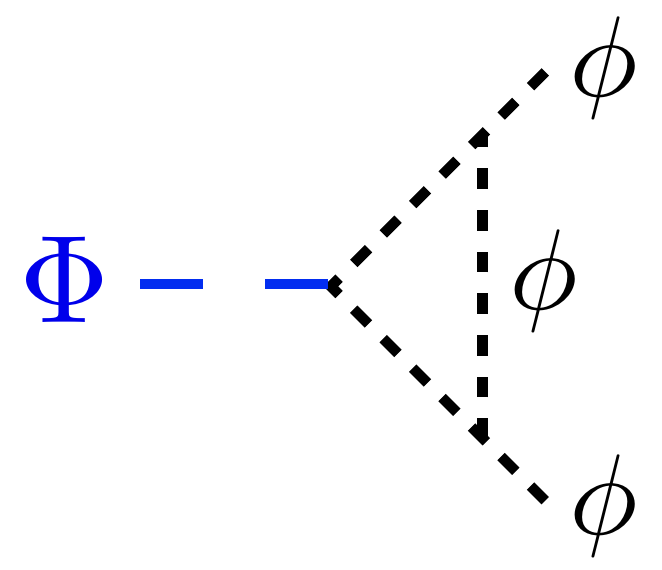} \qquad\qquad
\includegraphics[width=0.16\textwidth, valign=c]{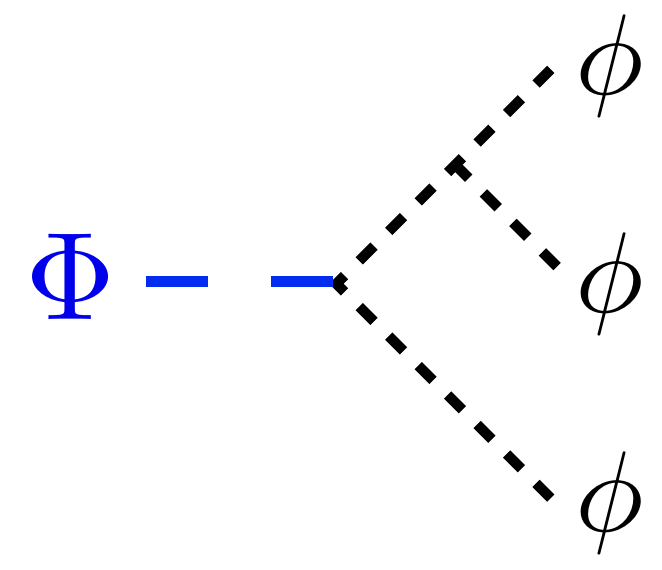} \qquad\qquad \includegraphics[width=0.16\textwidth, valign=c]{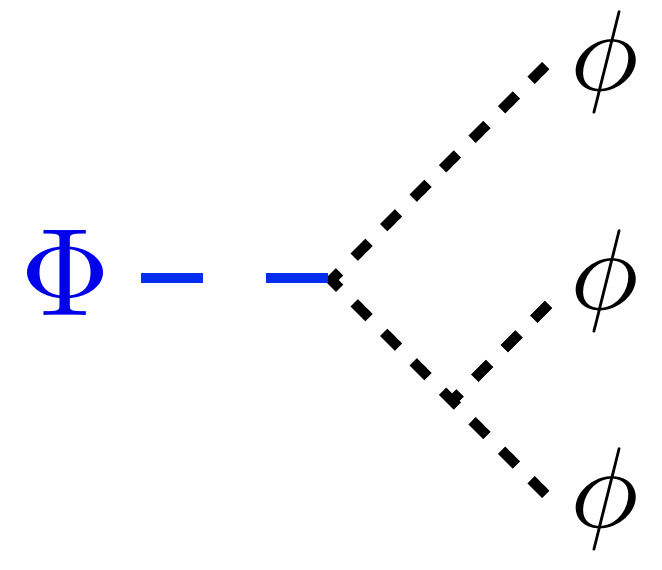}\,\,\,,
\end{align}
where the first diagram has a soft and collinear divergence from the loop integration (the final state kinematics is fully specified by momentum conservation), and the second and third diagrams generate IR divergences when integrated over phase space.  The statement of IR finiteness holds at the level of the observable.  In our case, this is the total width for $\Phi$, which is given schematically at NLO by
\begin{align}
\Gamma_\Phi = \mathop{\mathlarger{\mathlarger{\int}}} \left(\left|\includegraphics[width=0.1\textwidth, valign=c]{Figures/IRDivTree.pdf}  \right|^2 + 2\,\textbf{Re}\!\left[\includegraphics[width=0.1\textwidth, valign=c]{Figures/IRDivTree.pdf} \times \includegraphics[width=0.1\textwidth, valign=c]{Figures/IRDivLoop.pdf}^{\mathop{\mathlarger{\mathlarger{\mathlarger{\dag}}}}}\,\right]\right) + \mathop{\mathlarger{\mathlarger{\int}}} \left|\includegraphics[width=0.1\textwidth, valign=c]{Figures/IRDivTreeTop.pdf}  + \includegraphics[width=0.1\textwidth, valign=c]{Figures/IRDivTreeBottom.pdf}  \right|^2\,,
\label{eq:GammaPhiNLO}
\end{align}
where the integrals are over the final state phase space, such that the IR divergences emerging from the loop integration cancel against the phase space singularities associated with the extra emissions.

This interplay among the IR divergences from different diagrams follows from unitarity.  While a derivation to all orders is beyond the scope of these lectures, see \emph{e.g.}~\cite{Collins:1989gx, Collins:2011zzd}, it is straightforward to understand the IR structure of our one-loop calculation of $\Gamma_\Phi$ by studying the ``unitarity cuts'' of the two-loop diagram 
\begin{align}
\includegraphics[width=0.22\textwidth, valign=c]{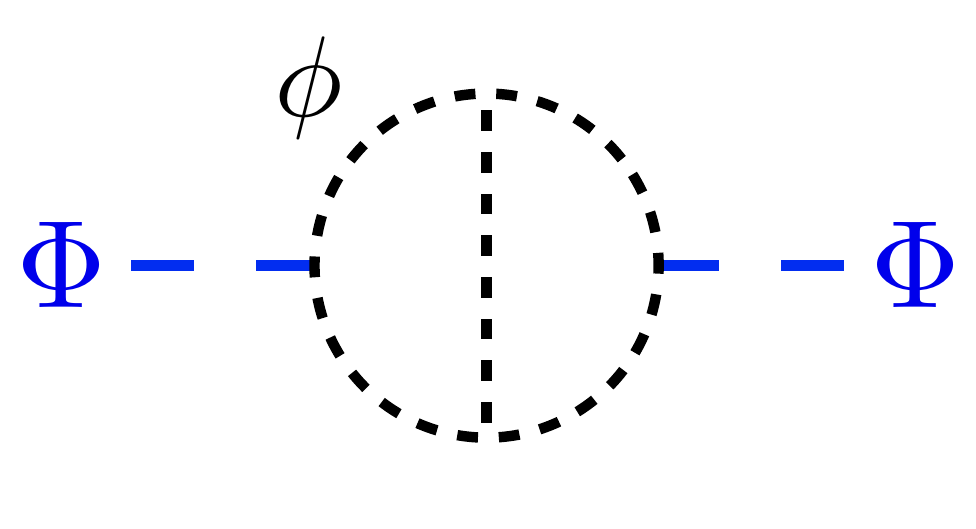} = \frac{i}{2}\s a^2\,b^2\!\int \frac{\D^4 \ell}{(2\s\pi)^4} \frac{\D^4 k}{(2\s\pi)^4} \,\frac{1}{(p_\Phi+\ell)^2}\,\frac{1}{(\ell-k)^2}\,\frac{1}{(p_\Phi+k)^2}\,\frac{1}{\ell^2}\,\frac{1}{k^2}\,,
\label{eq:PhiPhi2Loop}
\end{align}
where $\Phi$ is on-shell, and the 1/2 is a symmetry factor.  First, we note that there is no UV divergence, since taking $|\ell| \rightarrow \infty$ while holding $k$ finite yields 
\begin{align}
\includegraphics[width=0.22\textwidth, valign=c]{Figures/DaveExample_noCut.pdf} \sim  \frac{\D^4 k}{(2\s\pi)^4} \, \frac{1}{(p_\Phi+k)^2}\,\frac{1}{k^2}\int \frac{\D^4 \ell}{(2\s\pi)^4}  \,\frac{1}{\ell^6} \quad \longrightarrow \quad \text{UV finite}\,,
\end{align}
and the same holds for $|\ell|$ small and $|k| \rightarrow \infty$.  Next, we check the limit where $|\ell|$ and $|k|$ are simultaneously large, which yields\footnote{Note that the potentially confusing limit with $\ell = k$ naively leads to a log divergence, which does not contribute when preforming the full integral.}
\begin{align}
\includegraphics[width=0.22\textwidth, valign=c]{Figures/DaveExample_noCut.pdf} \sim  \int \frac{\D^4 \ell}{(2\s\pi)^4} \frac{\D^4 k}{(2\s\pi)^4} \,\frac{1}{(\ell-k)^2}\,\frac{1}{\ell^4}\,\frac{1}{k^4}
\quad \longrightarrow \quad \text{UV finite}\,,
\end{align}
Next, we can check the IR structure by taking the soft limit for $|\ell| \rightarrow 0$ with $|k|$ fixed
\begin{align}
\includegraphics[width=0.22\textwidth, valign=c]{Figures/DaveExample_noCut.pdf}\sim \frac{\D^4 k}{(2\s\pi)^4} \,\frac{1}{k^4}\,\frac{1}{(p_\Phi+k)^2} \,\frac{1}{M^2} \int \frac{\D^4 \ell}{(2\s\pi)^4}  \,\frac{1}{\ell^2} \quad \longrightarrow \quad \text{IR finite}\,,
\end{align}
and the same holds for $|\ell|$ fixed and $|k| \rightarrow 0$.  Finally, we check the soft limit where $|\ell|$ and $|k|$ are taken to zero simultaneously
\begin{align}
\includegraphics[width=0.22\textwidth, valign=c]{Figures/DaveExample_noCut.pdf}\sim \frac{1}{M^4} \int \frac{\D^4 \ell}{(2\s\pi)^4} \frac{\D^4 k}{(2\s\pi)^4} \,\frac{1}{(\ell-k)^2}\,\frac{1}{\ell^2}\,\frac{1}{k^2} \quad \longrightarrow \quad \text{IR finite}\,.
\end{align}
Note that there are no collinear divergences, since there are no external light-like lines in the initial or final state that one could take a collinear limit with respect to.  We conclude that the integral in \cref{eq:PhiPhi2Loop} is finite.
\vspace{5pt}\mybox{\begin{itemize}
\item \textbf{Exercise:} Integrate \cref{eq:PhiPhi2Loop} explicitly to demonstrate that it is finite.  Note that since there are no divergences to regulate, one can just combine denominators and integrate without encountering any subtleties.
\end{itemize}}

Next, we can make the connection with \cref{eq:GammaPhiNLO} explicit by taking the non-zero unitarity cuts of our two-loop diagram.\footnote{The optical theorem relates the imaginary part of a loop diagram to a set of integrated amplitudes squared.  This can be implemented diagrammatically through the use of the symbol ``\,\includegraphics[width=0.015\textwidth, valign=c]{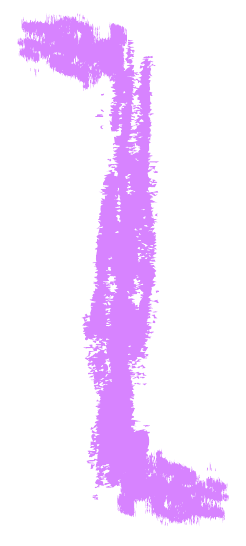}\s ,'' which denotes a unitarity cut: replace each propagator that crosses with $1/(p^2-m^2 + i 0) \,\rightarrow\, -2\s i\s\pi\,\delta\big(p^2-m^2\big)\,\theta\big(p^0\big)$, see \emph{e.g}~Sec.~24.1.2 of~\cite{Schwartz:2013pla}.  Then simply integrate over the on-shell phase space to derive the optical theorem.}  Explicitly,
\begin{align}
2 \,\text{Im}\Bigg[\includegraphics[width=0.22\textwidth, valign=c]{Figures/DaveExample_noCut.pdf}\Bigg] &= 
\includegraphics[width=0.22\textwidth, valign=c]{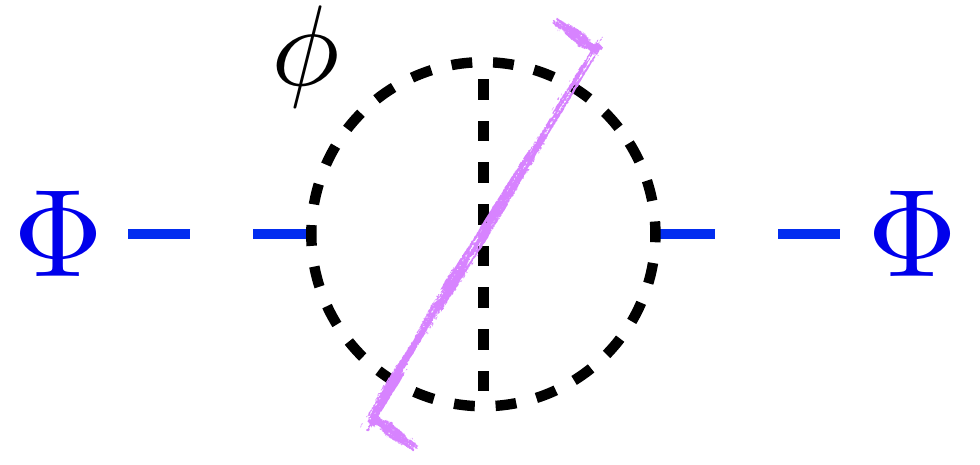} + \includegraphics[width=0.22\textwidth, valign=c]{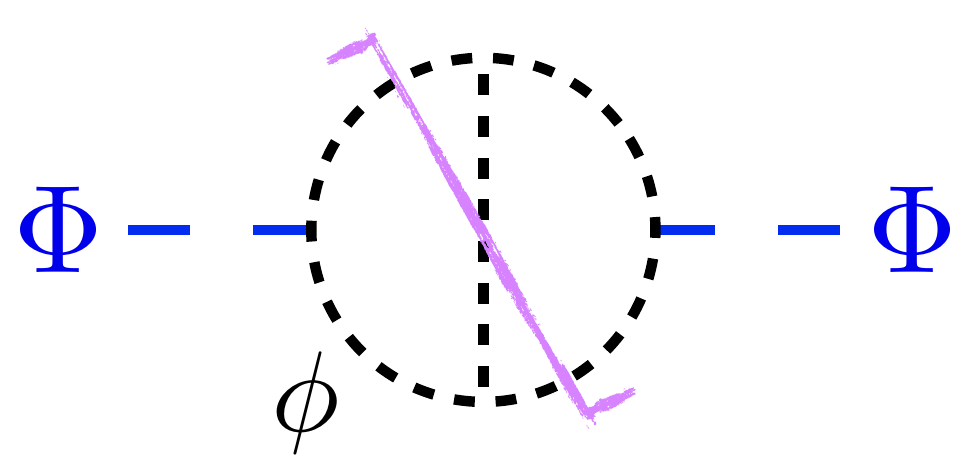} \notag \\[5pt]
&
\hspace{1pt}+\includegraphics[width=0.22\textwidth, valign=c]{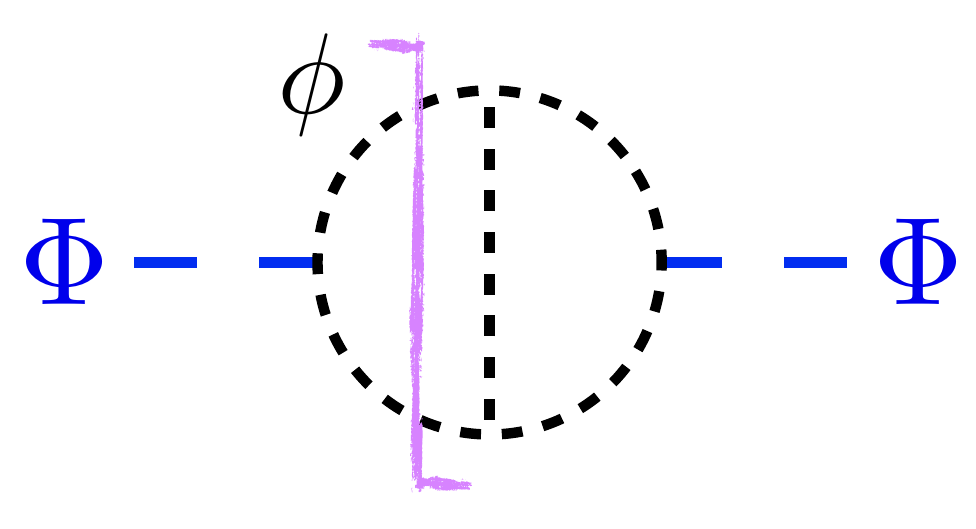} + \includegraphics[width=0.22\textwidth, valign=c]{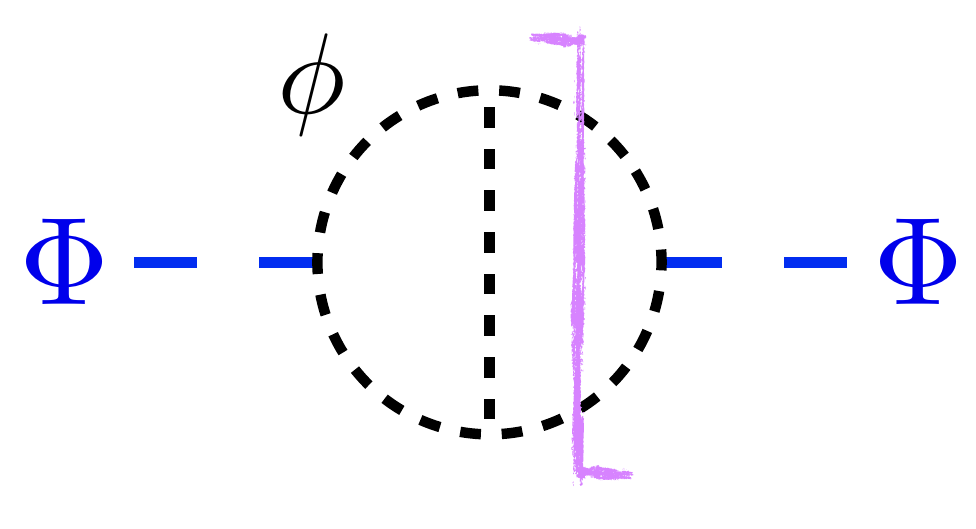} \,,
\label{eq:DaveExample_RCut}
\end{align}
which must be finite since the full two-loop diagram is finite.  Up to normalization (specifically, a factor of $M$), these diagrams are equivalent to the NLO calculation represented schematically in~\cref{eq:GammaPhiNLO}.  Although the individual pieces are IR divergent, they sum to an IR finite result.\footnote{If our goal was to perform a complete calculation, we would also need diagrams of the form \includegraphics[width=0.09\textwidth, valign=c]{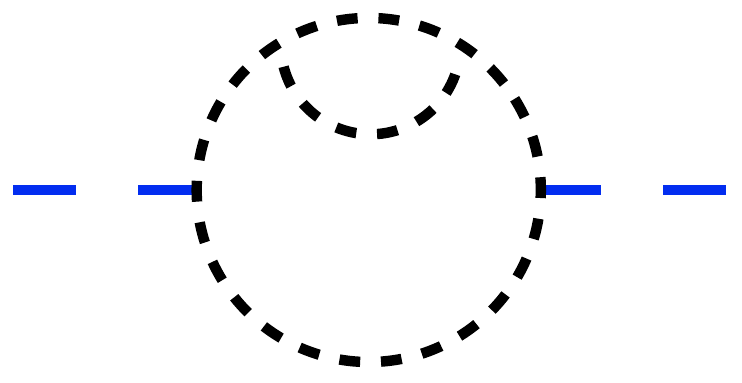}\,, which is both UV and IR divergent.  The UV divergence yields corrections to the scalar propagator.  The IR divergence is more subtle.  For example, one choice of cuts gives what looks like an additional contribution to the extra emission diagram, but this is spurious and can be understood in terms of subtleties with defining LSZ reduction in a theory with massless charged external states.  The resolution in this case is to define ``dressed states'' and the IR divergences are removed at the LSZ step, as discussed in \emph{e.g.}~\cite{Weinberg:1995mt}~Ch.~13.  One can avoid all of these subtleties by regulating the IR with a mass at intermediate steps.}

With this insight into the IR divergence structure of our diagrammatic expansion, it is straightforward to intuit what physical situations will yield functions of potentially large IR logarithms.  So far, we have been discussing the fully inclusive situation, where one should integrate over the full three-body final state phase space.  Then the IR logarithms that are associated with the divergences we have been discussing will cancel completely between the virtual and real emission contributions.  However, when performing a measurement, one is free to make kinematic restrictions on the final state.  For example, if one were to perform a set of cuts on the decay products of $\Phi$ such that any event with more than two particles was discarded, the logarithmic contribution from the real emission process would be eliminated (of course one must still include the full set of diagrams when calculating in order to yield an IR finite observable).  We conclude that our prediction could depend on a physical Sudakov log whose argument is a function of the cut parameter.  This is exactly the physical picture one should keep in mind as we study the summation of the Sudakov double logarithms.  But first, we will explore the origin of these loop and tree divergences to help clarify their physical origin.

\subsubsection*{Soft and Collinear Divergences at Tree Level}
First we will analyze the real emission diagram\footnote{It is worth noting that one should be cautious when analyzing individual diagrams since spurious divergences can cancel when considering full amplitudes.  For example, one can move divergences among different diagrams with gauge choices, see \emph{e.g.}~\cite{Feige:2013zla}.} for both the soft and collinear limits.  The Feynman diagram yields
\begin{align}
i\s \mathcal{A}_\text{Tree}^\text{NLO} = \includegraphics[width=0.18\textwidth, valign=c]{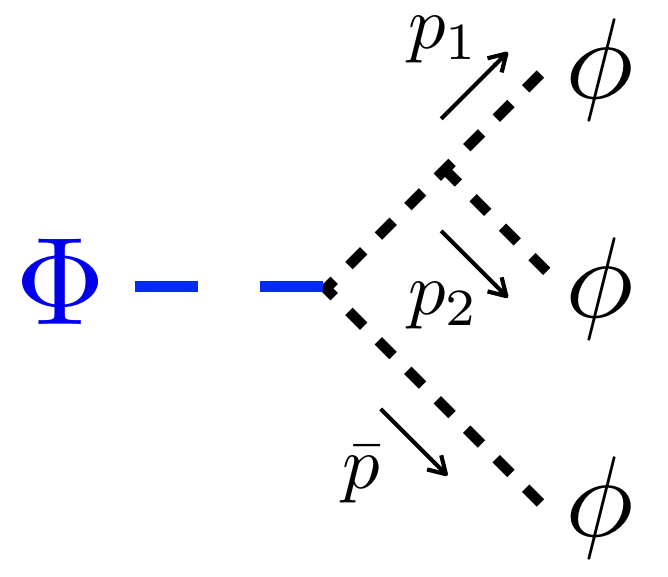} = -i\s a\,b \frac{1}{(p_1+p_2)^2} = -i\s a\,b \frac{1}{2\,p_1\cdot p_2} =  -i\s a\,b \frac{1}{M\,x\,n\cdot p_2}\,,
\label{eq:IRDivTree}
\end{align}
where we are using a notation that distinguishes $p$ collinear momenta from $\bar{p}$ anti-collinear momenta (noting that either can then be taken to be soft as this limit is independent of direction).  In the last step, we pointed $p_1$ parallel to $n$, specifically $p_1^\mu = (M/2)\,x\,n^\mu$, where $x$ is the momentum fraction carried by $p_1$.  Then taking $p_2$ soft is the limit $x\rightarrow 1$, while taking  collinear means taking $p_2 \sim (1-x)$ for $x\neq 0$ and $x \neq 1$.   Then we can assume a scaling for $p_2$ that allows us to take either the collinear or the ultrasoft limit:\footnote{Since this scaling is usually referred to as ultrasoft, we will use the same terminology here.}
\begin{align}
\text{ultrasoft:} &\quad p_2 \sim \frac{M}{2}\,\big(\lambda^2,\lambda^2,\lambda^2\big)\notag\\[7pt]
 \text{collinear:} &\quad p_2 \sim (1-x)\,\frac{M}{2}\,\big(\lambda^2,1,\lambda\big)\,,
\label{eq:DefUSandC}
\end{align}
where we interpret the ultrasoft scaling as homogeneously sending the momentum to zero, while the collinear scaling is chosen so that it approaches the $n$ direction as $\lambda \rightarrow 0$, such that its virtuality is $p^2 \sim \lambda^2$ for $\lambda \neq 0$, $x \neq0$, and $x\neq 1$.   The point is to approach the singularities so that the virtuality of the mode is always parametrically smaller than the power counting parameter, while also ensuring that the massless mode can go on-shell. This ensures that one can write down a propagating mode that captures the relevant singular behavior.  A similar story holds for the soft limit, in that its homogenous scaling allows it to go on-shell.

These scalings set the foundation for what follows.  Since the collinear scaling might appear strange at first, it is worth unpacking the physics.  Begin with $p$ pointing exactly in the $n$ direction: $p \sim (0,1,0)$.  We wish to take $p$ just slightly off-shell using our power counting parameter.  If we deform it first in the perp direction, we would have $p \sim (0,1,\lambda)$.  Then using the form for $p^2$ in light-cone coordinates, \cref{eq:pSqLC}, we should scale $\bar{n}\cdot p \sim \lambda^2$ such that $p \sim \big(\lambda^2,1,\lambda\big)$ can go on-shell.  Another way to derive this scaling is to take a particle with virtuality $\lambda^2$ in its rest-frame so that $p \sim (\lambda,\lambda,\lambda)$, and then to boost it in the $-\hat{z}$ direction by an amount $\beta \sim 1/\lambda$, which shifts $\bar{n}\cdot p \rightarrow \beta\, \bar{n}\cdot p$ and $n\cdot p \rightarrow n\cdot p/\beta$ and leaves the $\perp$ component unchanged.  This again yields the collinear scaling in \cref{eq:DefUSandC}.

Going back to the amplitude in \cref{eq:IRDivTree}, we can take the two limits to find
\begin{align}
i\s\mathcal{A}\,\Big|_{p_2 \,\rightarrow\, \text{collinear}} \sim\,\, \frac{1}{\lambda^{2}} \qquad \qquad\qquad i\s\mathcal{A}\,\Big|_{p_2\, \rightarrow\, \text{ultrasoft}}  \sim\,\, \frac{1}{\lambda^{2}}\,.
\end{align}
We see that there are divergences in either of these limits as we take the virtuality of the second scalar to zero, \emph{i.e.}, $\lambda^2 \rightarrow 0$. 

\subsubsection*{Soft and Collinear Divergences at Loop Level}
Now that we have understood the tree-level divergence structure due to the additional emissions that are generated at NLO, we turn to an analysis of the divergence structure of the one-loop diagram
\begin{align}
i\s\mathcal{A}_\text{Loop}^\text{NLO} = \includegraphics[width=0.18\textwidth, valign=c]{Figures/IRDivLoop.pdf} =  a^2\,b
\int \frac{\D^4\ell}{(2\,\pi)^4} {\frac{1}{\ell^2}} {\frac{1}{ (\ell+p)^2}}{\frac{1}{ (\ell+\bar{p})^2} } = a^2\,b\,\mathcal{I} \,,
\label{eq:int}
\end{align}
where again we are using a notation that distinguishes $p$ collinear momenta from $\bar{p}$ anti-collinear momenta, and we are assuming that $p$ flows out of the diagram while $\bar{p}$ flows in to simplify the signs appearing in the loop integral.  

First we note that this diagram is UV finite, since in the limit that $\ell \rightarrow \infty$, it reduces to $\int \D \ell\,\ell^3/\big(\ell^2\, \ell^2\, \ell^2\big)$ which converges in the UV.  However, as discussed above, we expect this diagram to exhibit both ultrasoft and collinear IR divergences, as defined by \cref{eq:DefUSandC} in the limit that $\lambda \rightarrow 0$.  

We will refer to the diagram in \cref{eq:int} as the massless Sudakov integral, and while we will solve it below in \cref{sec:RegionsMasslessSudakov} in a different kinematic configuration to leading power with $p$ and $\bar{p}$ taken slightly off-shell to regulate the IR divergences. Our goal here is to explore the IR divergence structure when $p^2 = \bar{p}^2 = 0$:
\begin{align}
p^\mu = \frac{M}{2}\,n^\mu = \left(0,M,\vec{0}\right)\qquad \qquad\qquad \bar{p}^\mu = -\frac{M}{2}\,\bar{n}^\mu = \left(-M,0,\vec{0}\right)\,.
\end{align}

It is straightforward to take the ultrasoft limit for the loop momentum:
\begin{align}
\lim_{\lambda\rightarrow 0} \D^4 \ell\, {\frac{1}{\ell^2}} {\frac{1}{ (\ell+p)^2}}{\frac{1}{ (\ell+\bar{p})^2} }\,\, \sim\,\,  \big(\lambda^6\, \D \lambda^2\big) {\frac{1}{\lambda^4}} {\frac{1}{ \lambda^2\, \bar{n} \cdot p}}{\frac{1}{\lambda^2\,n\cdot\bar{p}} }\,\, \sim\,\, \frac{\D \lambda^2}{\lambda^2}\,,
\label{eq:SoftIRDiv}
\end{align}
where in the second step we have Wick rotated, rescaled $|\ell| = \lambda^2 \,M$, and dropped the angular dependence and any overall factors.  This demonstrates the presence of the expected log divergence in the soft limit.

Next, we can take the collinear limit.  This will prove to be more involved.  The reason is that collinear divergences are inherently Lorentzian, since there is no notion of collinear null directions in Euclidean space.  Furthermore, there is the subtlety that a zero energy collinear particle is also soft, and as such one must be careful to ensure that one has actually isolated the collinear limit of the integral.  Another concern is that apparent divergences could actually integrate to zero, which will be true for some ranges of the integration parameters as we will show below.   This implies that we can not simply Wick rotate and take a limit as we did in the soft case, but must instead carefully manipulate the Minkowski integral to explore the collinear singularity.\footnote{This was first systematically understood by reframing the question of what divergences emerge in the massless limit into the task of finding ``pinch-singularity surfaces'' in~\cite{Sterman:1978bi, Libby:1978bx} using the Landau criterion following~\cite{Coleman:1965xm}, see~\emph{e.g.}~\cite{Sterman:1994ce,Collins:2011zzd} for introductory discussions.}

We begin by aligning $\ell$ with $p$: 
\begin{align}
\ell = \Big(n\cdot \ell, \,-x\,M,\, \ell_\perp\Big)\,,
\end{align}  
where $x$ tracks the fraction of $\ell$ that points in the $p$ direction, and the sign on $x$ is chosen for later convenience.  Note that we are looking for the behavior when $x \neq 0$ to be sure that this is a collinear divergence.  We will revisit this issue of overlapping soft and collinear divergences in \cref{sec:ZeroBin} below.  Then our three denominator factors are
\begin{align}
\ell^2 &= -M\,(n\cdot \ell)\,  x + \ell_\perp^2\notag \\[5pt]
\big(\ell + p\big)^2 &= \ell^2 + M\,(n\cdot \ell) = M (n\cdot \ell)(1-x)+ \ell_\perp^2\notag \\[5pt]
\big(\ell + \bar{p}\big)^2 &= \ell^2 + M^2\,x \simeq M^2\,x\,.
\label{eq:denFactors}
\end{align}
In the last line, the approximation we are taking holds in the collinear limit, where $\ell^2 \rightarrow 0$ for fixed $x$.  Our integration measure is then given by: 
\begin{align}
\D^4 \ell = \D\big(n\cdot \ell\big)\, \D\big(\bar{n}\cdot \ell\big)\, \D^2 \ell_\perp= -M\,\D x\,  \D\big(n\cdot \ell\big)\,\D\phi\,  \ell_\perp \,\D \ell_\perp\,,
\end{align}
where $\ell_\perp = |\ell_\perp|$ in the last expression and $\D \phi$ is the angular part of $\D^2\ell_\perp$.  Then the integral reduces to 
\begin{align}
\mathcal{I} &\simeq  \frac{1}{(2\s\pi)^4}\,\frac{1}{M}\int_0^{2\s\pi} \!\! \D\phi \int_{-\infty}^{\infty}\!\D x\,\frac{1}{x} \int_0^\infty \!\ell_\perp\, \D \ell_\perp \int_{-\infty}^\infty \D(n\cdot \ell)\notag\\[5pt]
&\hspace{100pt}\times \frac{1}{M\,(n\cdot \ell)\, x  - \ell_\perp^2- i0}\,\frac{1}{M\,n\cdot \ell\,(1-x)+ \ell_\perp^2+i0}\,,
\label{eq:CollinearIRDiv}
\end{align}
where we have restored the $+i0$ factors in the denominators so that we can do a contour integral over $n\cdot \ell$.  Also, note that the critical flipped sign on $i0$ in the first denominator is due to multiplying through by a minus sign.  

We are now situated to analyze the analytic structure of the $\D(n\cdot \ell)$ integral, which will be used to evaluate it by contour integration.  There are poles in $n\cdot \ell$ at
\begin{align}
(n\cdot \ell)_\text{pole} = \frac{\ell_\perp^2+i0}{M\,x} \qquad\quad \qquad \qquad\quad
(n\cdot \ell)_\text{pole} = \frac{-\ell_\perp^2 - i 0}{M\,(1-x)}\,.
\end{align}
Notice that integrand vanishes for $n\cdot \ell \rightarrow \pm \infty$, implying that the contour at infinity does not contribute.  Then we find qualitatively different behavior in the complex plane as one varies $x$:
\begin{align}
\includegraphics[width=0.3\textwidth, valign=c]{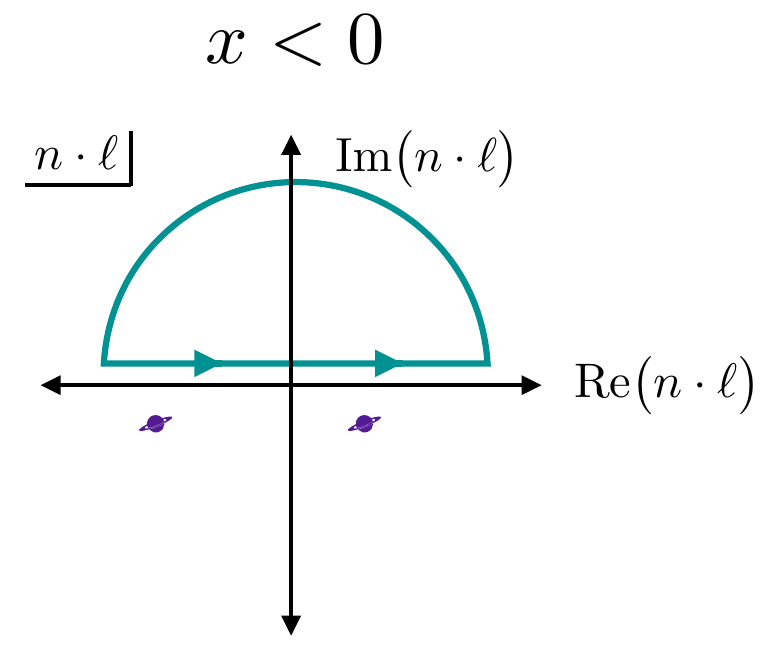} \qquad \includegraphics[width=0.3\textwidth, valign=c]{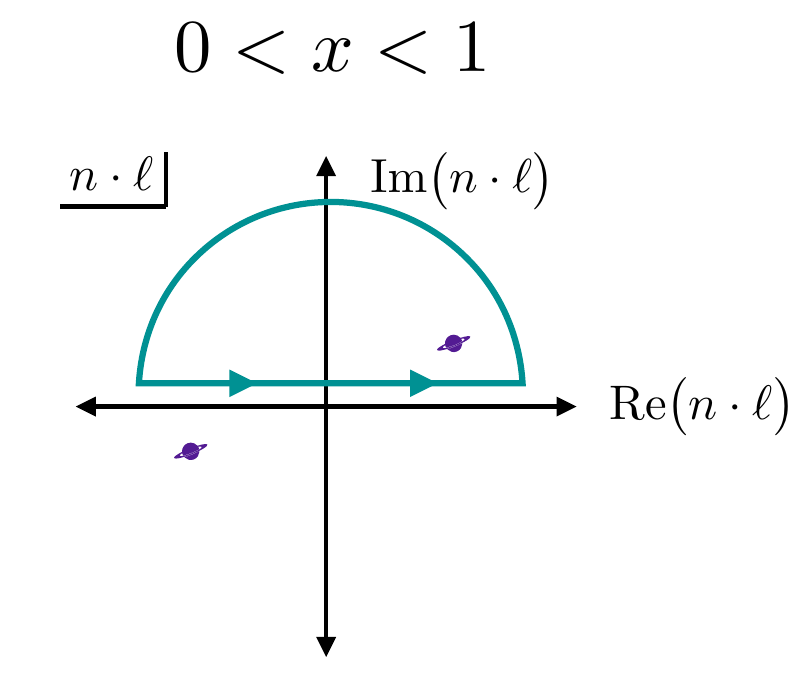} \qquad \includegraphics[width=0.3\textwidth, valign=c]{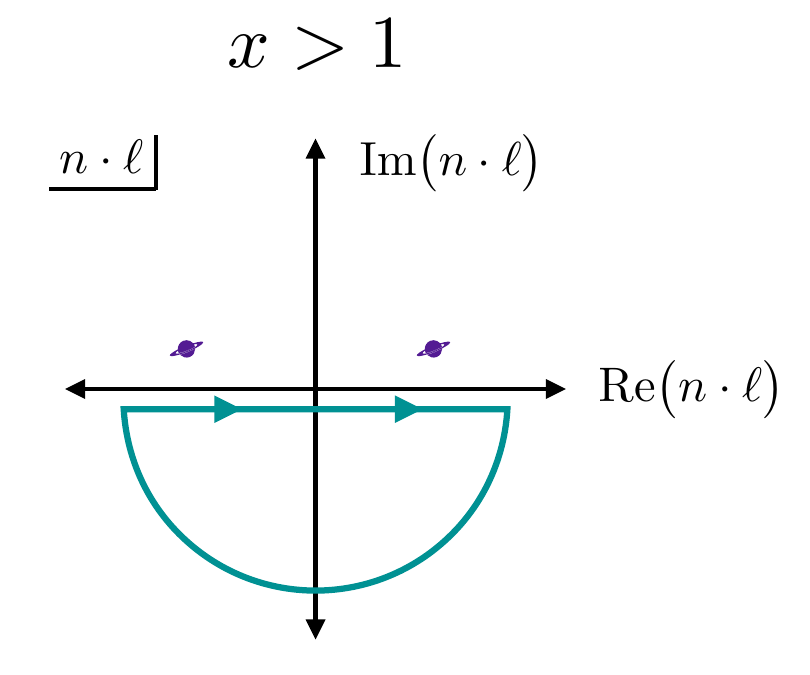} \notag
\end{align}
Taking $x < 0$, both poles are below the real axis, and we can close the contour above to return $\mathcal{I} =0$.     When $x >1$, both poles are above the real axis, and we can close the contour below to again yield $\mathcal{I} =0$.  However, in the range $0<x<1$ the poles are on opposite sides of the real axis; either contour will imply $\mathcal{I} \neq 0$.  For concreteness, we choose to evaluate the integral using the contour above the real axis, letting $z = M\,(n\cdot \ell)\,x$
\begin{align}
\oint_\text{upper} \frac{1}{x}\,\D z \,\frac{1}{z - \ell_\perp^2- i0}\,\frac{1}{z\, (1-x)/x + \ell_\perp^2+i0} = 2 \s\pi\s i\, \frac{1}{\ell_\perp^2} \quad\qquad \big[0<x<1\big]\,,
\label{eq:ColDivExContour}
\end{align}
where we have sent $i0 \rightarrow 0$ at the end.\footnote{For an analysis of the analytic structure of Feynman diagrams, and especially for the intuitive classical picture of how to interpret the physical region of the integral for the charged pion electromagnetic form factor, see~Ch.~18.6 and 18.7 of~\cite{Bjorken:1965zz}.}   

We are finally primed to see the collinear divergence.  Taking $x \neq 0$ and integrating \cref{eq:ColDivExContour} over the $\ell_\perp$ direction, we find
\begin{align}
\mathcal{I} \sim \int_0^\infty \D\ell_\perp \frac{\ell_\perp}{\ell_\perp^2}\,,
\end{align}
which is a scaleless integral that has both a UV and an IR divergence -- the lower bound of this integral yields the collinear divergence we have been working to expose.  As mentioned before, we assumed $x \neq0$ to ensure that we are not simply accidentally just rediscovering the soft divergence.  

Now we can put the pieces together (where the limits of integration on $x$ reflect the results of our contour analysis) to find
\begin{align}
\mathcal{I} \simeq \frac{i}{16\s\pi^2}\,\frac{1}{M^2}\, \int_{0}^{1}\!\D x\,\frac{1}{x} \int_0^\infty \! \D \ell_\perp \frac{1}{\ell_\perp}\,,
\end{align}
where we emphasize that in deriving this result we expanded the propagator in the collinear limit, see \cref{eq:denFactors}.  We see there is in fact a double IR log divergence when both $\ell_\perp \rightarrow 0$ and $x \rightarrow 0$, which we interpret as the collinear and soft divergences respectively.  This also tells us that our collinear analysis recovered the soft divergence, which is not surprising since it should be possible to take the energy of a collinear state to zero, thereby making it soft.  This double counting is a generic feature of these kinds of calculations, and will be revisited in \cref{sec:ZeroBin}, where we discuss the idea of zero-bin subtraction.  But first, we turn to the next section where we perform a regions analysis to discover the regions that contribute to the massless Sudakov integral, and then use the resulting expanded integrands to evaluate the Sudakov integral in the leading power approximation.

\subsection[Method of Regions for a Massless Sudakov Integral]{Method of Regions for a Massless Sudakov Integral\footnote{I am grateful to Michael Peskin for feedback and corrections in \cref{sec:RegionsMasslessSudakov} through \cref{sec:RegionsMassiveSudakov}.}}
\label{sec:RegionsMasslessSudakov}
In this section, we will study the massless Sudakov integral
\begin{align}
i\s\mathcal{A}_\text{Loop}^\text{NLO} = \includegraphics[width=0.18\textwidth, valign=c]{Figures/IRDivLoop.pdf} =  a^2\,b
\int \frac{\D^4\ell}{(2\,\pi)^4} {\frac{1}{\ell^2}} {\frac{1}{ (\ell+p)^2}}{\frac{1}{ (\ell+\bar{p})^2} } = a^2\,b\,\mathcal{I} \,,
\label{eq:intAgain}
\end{align}
now with the final state legs taken slightly off-shell, $p^2 \neq \bar{p}^2 \neq 0$, such that the IR divergences are explicitly regulated.  We will also work a different kinematic regime than in the previous discussion.  Here we take $p^2 < 0$ and $\bar{p}^2 < 0$, which can be thought of as a toy model for deep inelastic scattering, \emph{i.e.}, the $\Phi$ line is in the $t$-channel.  For context, one can think of this as a toy model for jets, where $p^2$ is measuring the mass of the final state jet.\footnote{Note that one perhaps confusing aspect of working with this example is that the choice of kinematics which yields real amplitudes has $p^2 < 0$, see \cref{eq:SudakovKinematics}.  This can be understood by recalling that since the $\phi$ states are massless, there are always imaginary contributions from the loops since these particles can go on-shell.  If one is interested in computing the kinematics appropriate for $\Phi$ decay, then they can analytically continue the results we will derive here by being careful about the location of the poles as one takes $-p^2 \to p^2$. }  Then large logs appear because we are restricting our final state to only contain a pair of two small mass jets, which essentially cuts away the contribution from the real emission diagrams.  

The integral in \cref{eq:intAgain} is difficult to evaluate, as it depends on multiple scales $M^2$, $p^2$, and $\bar{p}^2$.  However, since we are interested in the parameter space where both external momenta are parametrically small, $p^2/M^2 \sim \lambda^2 \ll1$ and $\bar{p}^2/M^2 \sim \lambda^2 \ll1$, we can utilize the method of regions to expand the integral into a set of single scale integrals that are straightforward to compute.  We will use this procedure to explicitly derive a Sudakov double log that is only a function of physical scales.  First we will argue for the scalings that capture all of the non-zero regions, and then we will state the integrated result~\cite{Becher:2014oda}.  For transparency, we will then provide our own detailed calculation of the collinear integral -- the pedagogical evaluation of all four integrals is provided in Appendix~B of~\cite{Becher:2014oda}.

Allowing the final state momenta to be slightly away from the (anti-)collinear direction with virtuality $p^2 \sim \lambda^2$ and $\bar{p}^2 \sim \lambda^2$ implies
\begin{align}
p \sim \big(\lambda^2, 1, \lambda\big) \qquad\qquad \qquad \bar{p} \sim \big(1,\lambda^2,  \lambda\big)\,.
\label{eq:CollandAntiColl}
\end{align}
We will focus on the kinematics in a physical region where
\begin{align}
P^2 = - p^2 \sim \lambda^2 \qquad\qquad \bar{P}^2 = -\bar{p}^2 \sim \lambda^2 \qquad\qquad M^2 = -(p -\bar{p})^2 \sim 1\,.
\label{eq:SudakovKinematics}
\end{align}
In order to expand the integral using the method of regions, we will need to identify all the possible scalings of $\ell$ that yield an $\mathcal{O}(1)$ contribution to $\mathcal{I}$.  Consider the following
\begin{align}
\big[\text{hard}\big] \quad \quad h: &\quad \ell_h \,\sim \big(1,1,1\big) \notag \\[2pt]
\big[\text{collinear}\big]\, \quad \quad c: &\quad \ell_c \hspace{3pt}\sim \big(\lambda^2,1,\lambda\big) \notag\\[2pt]
\big[\text{anti-collinear}\big]\, \quad\quad \overline{c}: &\quad \ell_{\bar{c}}\hspace{3pt} \sim \big(1,\lambda^2,\lambda\big)  \notag\\[2pt]
\big[\text{ultrasoft}\big] \,\,\,\quad us: &\quad \ell_{us} \!\sim \big(\lambda^2,\lambda^2,\lambda^2\big)\,.
\label{eq:scaling}
\end{align}
Note that the measure scales as $\D^4 \ell \sim \ell^4$, such that
\begin{align}
\D^4 \ell_h \sim 1\qquad \qquad\quad \D^4 \ell_c \sim \lambda^4\qquad \qquad\quad \D^4 \ell_{us} \sim \lambda^8\,.
\end{align}
Next, we check that these scalings contribute at $\mathcal{O}(1)$ in terms of the power counting parameter $\lambda$.  Take the full integral in \cref{eq:intAgain}, and expand each denominator keeping the leading terms according to the power counting in \cref{eq:scaling}.  Explicitly in the collinear case, the denominators are expanded as
\begin{align}
\ell^2 = \ell_c^2 \sim \lambda^2 \qquad\qquad  (\ell+p)^2 = (\ell_c+p)^2 \sim \lambda^2 \qquad \qquad (\ell+\bar{p})^2 \simeq \bar{n}\cdot \ell_c\,n\cdot \bar{p} \sim 1 \,.
\end{align}
Following the same logic for all four regions yields
\begin{align}
 h: &\quad \mathcal{I}\quad  \longrightarrow\quad \mathcal{I}_h =  \int \frac{\D^4\ell_h}{(2\s\pi)^4}\, \frac{1}{\ell_h^2}\,\frac{1}{\ell_h^2 +  \bar{n}\cdot \ell_h\, n\cdot p}\,\frac{1}{\ell_h^2+ n\cdot \ell_h\, \bar{n}\cdot \bar{p}} \,\, \sim \,\, \mathcal{O}(1)\notag \\[8pt]
 c: &\quad\mathcal{I} \quad\longrightarrow\quad \mathcal{I}_c = \int \frac{\D^4\ell_c}{(2\s\pi)^4}\, \frac{1}{\ell_c^2}\frac{1}{(\ell_c+p)^2}\,\frac{1}{\bar{n}\cdot \ell_c\, n\cdot \bar{p}}  \,\, \sim \,\,  \mathcal{O}(1)\notag \\[8pt]
 \bar{c}: &\quad\mathcal{I}\quad  \longrightarrow \quad\mathcal{I}_{\bar{c}} = \int \frac{\D^4\ell_{\bar{c}}}{(2\s\pi)^4}\, \frac{1}{\ell_{\bar{c}}^2}\,\frac{1}{n\cdot \ell_{\bar{c}}\, \bar{n}\cdot p}\,\frac{1}{(\ell_{\bar{c}} +\bar{p})^2}  \,\, \sim \,\,  \mathcal{O}(1) \notag \\[8pt]
 us: &\quad\mathcal{I} \quad \longrightarrow \quad \mathcal{I}_{us} =\int \frac{\D^4\ell_{us}}{(2\s\pi)^4}\, \frac{1}{\ell_{us}^2}\,\frac{1}{ n\cdot \ell_{us}\, \bar{n}\cdot p+p^2}\, \frac{1}{ \bar{n}\cdot \ell_{us}\, n\cdot \bar{p}+\bar{p}^2} \,\, \sim \,\,  \mathcal{O}(1)\,.
 \label{eq:MasslessSudakovRegions}
\end{align}

\clearpage

\vspace{5pt}\mybox{
\begin{itemize}
\item {\bf Exercise:}  There is one additional scaling that yields an $\mathcal{O}(1)$ contribution, specifically the Glauber region where $\ell_G \sim M\big(\lambda^2,\lambda^2,\lambda\big)$.  First derive the integral that results from this region.  Show by explicit calculation that it vanishes.  Then convince yourself that any scaling of the form $\ell^\mu \sim M\big(\lambda^a,\lambda^b,\lambda^c\big)$ for $a,b,c$ that are different from \cref{eq:scaling} yields vanishing scaleless integrals.  This had to be the case as discussed in Sec.~2.2 of~\cite{Feige:2014wja}.
\end{itemize}}

These integrals are straightforward to evaluate~\cite{Becher:2014oda}:\footnote{\textbf{Disclaimer:} From here forward, we will no longer track the difference between $\mu$ and $\muT$.} 
\begin{align}
\mathcal{I}_h &= -\frac{i}{16\s\pi^2}\frac{1}{M^2}\left(\frac{1}{\epsilon^2} + \frac{1}{\epsilon} \log \frac{\mu^2}{M^2} + \frac{1}{2} \log^2 \frac{\mu^2}{M^2} - \frac{\pi^2}{6} + \mathcal{O}\big(\lambda^2\big)\right) \notag\\[7pt]
\mathcal{I}_c &= - \frac{i}{16\s\pi^2}\frac{1}{M^2}\left(-\frac{1}{\epsilon^2} - \frac{1}{\epsilon} \log \frac{\mu^2}{P^2} - \frac{1}{2} \log^2 \frac{\mu^2}{P^2} + \frac{\pi^2}{6} + \mathcal{O}\big(\lambda^2\big)\right) \notag\\[7pt]
\mathcal{I}_{\bar{c}} &= -\frac{i}{16\s\pi^2}\frac{1}{M^2}\left(-\frac{1}{\epsilon^2} - \frac{1}{\epsilon} \log \frac{\mu^2}{\bar{P}^2} - \frac{1}{2} \log^2 \frac{\mu^2}{\bar{P}^2} + \frac{\pi^2}{6} + \mathcal{O}\big(\lambda^2\big)\right) \notag\\[7pt]
\mathcal{I}_{us} &=- \frac{i}{16\s\pi^2} \frac{1}{M^2}\left(\frac{1}{\epsilon^2} + \frac{1}{\epsilon} \log \frac{\mu^2\,M^2}{P^2\,\bar{P}^2} + \frac{1}{2} \log^2 \frac{\mu^2\,M^2}{P^2\,\bar{P}^2} + \frac{\pi^2}{6} + \mathcal{O}\big(\lambda^2\big)\right) \,,
\label{eq:MasslessSudakovRegionsEval}
\end{align}
where the factor of $\gamma_E - \log 4\,\pi$ has been absorbed into the renormalization parameter $\mu^2$ as per the $\overline{\text{MS}}$ scheme.  These integrals all have $1/\epsilon^2$ (double log), and $1/\epsilon$ (single log) divergences.  The scales that appear inside each of the logs for the integrated results reflect their origin:  the hard logs are a function of $\mu^2/M^2$, the collinear logs depend on $\mu^2/P^2\sim \mu^2/\big(M^2\,\lambda^2\big)$,  the anti-collinear logs depend on $\mu^2/\bar{P}^2\sim \mu^2/\big(M^2\,\lambda^2\big)$, and the ultrasoft logs are a function of $\mu^2\,M^2/\big(P^2\,\bar{P}^2\big)\sim \mu^2/\big(M^2\,\lambda^4\big)$.  So power counting has provided us with a tool to reduce our complicated multi-scale integral into four pieces that are each functions of their own single characteristic scale.  We will see that the EFT approach will provide a systematic way to introduce multiple renormalization scales, such that no large logs will remain.\footnote{It is worth emphasizing that there is a long history of summing IR logarithms pioneered by~\cite{Parisi:1979se, Collins:1981uk}.  These calculations rely on RG scale invariance and gauge invariance to derive their RG improved predictions.} 

To achieve our final result, we can sum the integrals for each of the four regions to yield the power expanded integrated result
\begin{align}
\mathcal{I} = \mathcal{I}_h + \mathcal{I}_c + \mathcal{I}_{\bar{c}} + \mathcal{I}_{us} + \mathcal{O}\big(\lambda^2\big) = -\frac{i}{16\s\pi^2}\frac{1}{M^2}\left(\log\frac{M^2}{P^2}\log\frac{M^2}{\bar{P}^2} + \frac{\pi^2}{3} \right)+ \mathcal{O}\big(\lambda^2\big)\,.
\label{eq:IMasslessSudakovCombined}
\end{align}
As expected, no $\epsilon$ dependence remains: this original integral is UV finite, and we have regulated the IR divergences with $P^2\neq 0$ and $\bar{P}^2 \neq 0$.  This implies that no explicit factors of $\mu^2$ appear.  Moreover, all the log squared terms from the individual integrals combine through what seems like a magical conspiracy into the Sudakov double log~\cite{Sudakov:1954sw}, whose physical origin can be traced to the soft and collinear divergences discussed in \cref{sec:IRLogs} above.  While this kind of log reshuffling can typically be done in one's head when working with single logs, it is significantly less trivial to see how everything combines for double logs.

\vspace{5pt}\mybox{
\begin{itemize}
\item {\bf Exercise:}  Show that 
\begin{align}
\frac{1}{2} \left(\log^2 \frac{\mu^2}{M^2} - \log^2 \frac{\mu^2}{P^2} - \log^2 \frac{\mu^2}{\bar{P}^2} + \log^2 \frac{\mu^2\,M^2}{P^2\,\bar{P}^2}\right) = \log\frac{M^2}{P^2}\log\frac{M^2}{\bar{P}^2} \,,
\end{align}
being careful to track the (obvious) complicating fact:
\begin{align}
 \log^2 \frac{\mu^2}{M^2} =  \log^2\mu^2  + \log^2 M^2 - 2 \log \mu^2 \log M^2\,,
\end{align}
with similar expressions for the other double logs.
\end{itemize}}

The result in \cref{eq:IMasslessSudakovCombined} exhibits a double log as a function of a ratio of physical scales, that can become large in the limit that $\lambda \ll 1$.  In particular, if we want to compute the NLO matrix element squared
\begin{align}
\big|\mathcal{A}_\text{NLO}\big|^2 &= \left|\includegraphics[width=0.1\textwidth, valign=c]{Figures/IRDivTree.pdf}\right|^2 + 2\, \textbf{Re}\!\left[\includegraphics[width=0.1\textwidth, valign=c]{Figures/IRDivTree.pdf} \times \includegraphics[width=0.1\textwidth, valign=c]{Figures/IRDivLoop.pdf}^{\mathop{\mathlarger{\mathlarger{\mathlarger{\dag}}}}}\,\right]\simeq b^2 - \frac{b^2}{4\s\pi^2}\frac{a^2}{M^2}\log^2\frac{M^2}{P^2}\,,
\end{align}
where we have used that the leading order diagram gives $-i\s b$ to derive the relative sign between the tree and one-loop contributions here, we have set $P^2 = \bar{P}^2$ and have not included the finite fixed-order terms for brevity.  Note that for $M^2 \gg P^2$, the second term can overwhelm the first, and our NLO matrix element squared \emph{can become negative}!  This signals a complete breakdown of perturbation theory.  Note that as opposed to the situation with only UV logs, where the breakdown of perturbation theory was associated with the presence of a Landau pole (see \cref{eq:LandauPole}), here all couplings are perturbative.  Fortunately, these potentially catastrophic large IR logs can be summed using the EFT approach, and doing so will result in a manifestly positive LL + NLO amplitude squared.  In this sense, the RG takes a failed expansion and restores the efficacy of  perturbative theory.

Our goal in what follows will be to peel apart the log squared derived in \cref{eq:IMasslessSudakovCombined} using matching and running.  This will introduce scale dependence such that we can apply RG techniques.  Separating the non-divergent integral into divergent sub-pieces is a key aspect of the EFT trick for separating scales.  As we will see in \cref{sec:ToySCET} below, the soft, collinear, and anti-collinear regions will each be associated with an independent propagating degree of freedom.  These modes make up SCET, with a corresponding set of Feynman rules and power counting structure that reproduces the individual integrals in \cref{eq:MasslessSudakovRegions}, and whose RGEs sum the large Sudakov double log.  Before we move into developing SCET, we will first provide a detailed evaluation of the collinear integral, $\mathcal{I}_c$ in \cref{eq:MasslessSudakovRegions} above, and then will explore the physics of the massive Sudakov log.

\subsubsection*{Evaluating the Collinear Sudakov Integral}
\label{sec:evalSudakovInt}
For the sake of demystifying the results in the previous section, we will show how to evaluate the $\mathcal{I}_c$ integral in \cref{eq:MasslessSudakovRegions} explicitly, following Appendix~B.2 of~\cite{Becher:2014oda}.  The first step is to combine denominators using \cref{eq:CombineLinProps}:  
\begin{align}
\frac{1}{A\,b_1\,b_2} = \int_0^\infty \D y_1 \int_0^\infty \D y_2 \frac{2}{\big[A + b_1\,y_1 + b_2\,y_2\big]^3}\,.
\end{align} 
Then taking $A = \ell^2$, $b_1 = (\ell + p)^2$, and $b_2 = \bar{n}\cdot \ell\,n\cdot \bar{p}$, we have
\begin{align}
\mathcal{I}_c = 2\,\mu^{2\epsilon}\int_0^\infty \D y_1 \int_0^\infty \D y_2 \int \frac{\D^{d}\ell}{(2\s\pi)^{d}} \frac{1}{\Big[\big(1+y_1\big)\left(\ell^2 + \ell^\mu \cdot \frac{2\,y_1\,p_\mu + y_2\,n\cdot \bar{p}\, \bar{n}_\mu}{1+y_1} + \frac{y_1}{1+y_1}\,p^2\right)\Big]^3}\,.
\end{align}
Defining
\begin{align}
\tilde{\ell} = \ell^\mu + \frac{y_1\,p^\mu + \frac{1}{2} y_2\,n\cdot \bar{p}\, \bar{n}^\mu}{1+y_1} \,,
\end{align}
we can replace $\ell \rightarrow \tilde{\ell}$ to yield
\begin{align}
\mathcal{I}_c = 2\,\mu^{2\epsilon}\int_0^\infty \D y_1 \frac{1}{\big(1+y_1\big)^3} \int_0^\infty \D y_2 \int \frac{\D^{d}\tilde{\ell}}{(2\s\pi)^{d}} \frac{1}{\left[\tilde{\ell}^2 - \left(\frac{y_1\,p^\mu + \frac{1}{2}y_2\,n\cdot \bar{p}\, \bar{n}^\mu}{1+y_1}\right)^2 + \frac{y_1}{1+y_1}\,p^2\right]^3}\,,
\end{align}
with $d = 4-2\s\epsilon$.  Next, we integrate over $\tilde{\ell}$ using the general result for dim reg given in~\cref{eq:dimRegEvalGeneral}: 
\begin{align}
\mathcal{I}_c =&\, 2\,\mu^{2\epsilon}\int_0^\infty \D y_1 \frac{1}{\big(1+y_1\big)^3} \int_0^\infty \D y_2\, \frac{(-1)^{3}i}{(4\s\pi)^{2-\epsilon}}\frac{\Gamma(1+\epsilon)}{\Gamma(3)}\notag\\[2pt]
&\hspace{160pt} \times\left(\frac{y_1^2\,p^2 +y_1\,y_2\,\bar{n}\cdot p\,n\cdot \bar{p}}{\big(1+y_1\big)^2} - \frac{y_1}{1+y_1}\,p^2\right)^{-1-\epsilon} \notag \\[10pt]
=&\,\frac{- i\s\mu^{2\epsilon}}{(4\s\pi)^{2-\epsilon}}\,\Gamma(1+\epsilon)\,\int_0^\infty \D y_1 \frac{y_1^{-1-\epsilon}}{\big(1+y_1\big)^{1-2\epsilon}} \int_0^\infty \D y_2 \left(P^2 + y_2\,\bar{n}\cdot p\,n\cdot \bar{p}\,\right)^{-1-\epsilon} \notag \\[10pt]
=&\,\frac{-i\s\mu^{2\epsilon}}{(4\s\pi)^{2-\epsilon}}\,\frac{\Gamma(1+\epsilon)}{\epsilon}\,\frac{P^{-2\epsilon}}{\bar{n}\cdot p\,n\cdot \bar{p}}\int_0^\infty \D y_1 \frac{y_1^{-1-\epsilon}}{\big(1+y_1\big)^{1-2\epsilon}}\notag\\[10pt]
=&\,\frac{-i}{(4\s\pi)^{2-\epsilon}}\left(\frac{\mu^2}{P^2}\right)^{\epsilon}\frac{1}{\bar{n}\cdot p\,n\cdot \bar{p}} \,\frac{\Gamma(1+\epsilon)}{\epsilon} \,\frac{\Gamma(1-\epsilon)\Gamma(-\epsilon)}{\Gamma(1-2\,\epsilon)} \notag \\[10pt]
=& \,\frac{-i}{(4\s\pi)^{2-\epsilon}}\frac{1}{M^2}\left(\frac{\mu^2}{P^2}\right)^{\epsilon} \,\frac{\Gamma(1+\epsilon)\Gamma(-\epsilon)^2}{\Gamma(1-2\,\epsilon)}\,,
\end{align}
where in the last line, we used that to leading power $\bar{n}\cdot p = -n\cdot \bar{p} = M$.  When this result is expanded in $\epsilon$, it yields the result given in \cref{eq:MasslessSudakovRegionsEval}.  Notice at an intermediate step, we replaced $p^2 = - P^2$ to keep our expressions real for simplicity.

Next, we will provide a \Primer~to discuss a technique known as zero-bin subtraction, which is required to ensure that collinear integrals do not receive a contribution from the soft region.

\scenario{Overlapping Regions and Zero-Bin Subtraction}
\addcontentsline{toc}{subsection}{\color{colorTech}{Primer~\thescenario.} Overlapping Regions and Zero-Bin Subtraction}
\label{sec:ZeroBin}
Recall that when we exposed the collinear divergence of the Sudakov integral in \cref{sec:IRLogs} above, we introduced a momentum fraction parameter $x$ and required that it be non-zero to ensure that we were not simply re-discovering the soft divergence in a complicated way.  This notion of double counting the soft divergence in the limit $x\rightarrow 0$ of the collinear integral is a subtlety to be wary of when expanding integrals using the method of regions or constructing them directly within an EFT.  

The point is that an EFT is built out of modes with particular momentum scalings.  However, diagrams generically involve loops of these modes and we integrate over all possible momenta, which violates the fundamental power counting rules of our setup.  The reason this approach is not dead on arrival is that essentially all of these contributions are scaleless so that dim reg sends them to zero.  This is how dim reg does not introduce contributions to an integrated result with spurious power counting.   However, one can encounter a ``zero-bin'' if there are non-trivial contributions that scale differently than the power counting assumed when separating the integrand.  In other words, if there are additional scaleful structures in the IR of a (regulated) integral, these zero-bin contributions should be subtracted from the result to avoid double counting regions by accident.\footnote{Furthermore, if one were careful to interpret the source of this zero-bin contribution by keeping track of $1/\epsilon_\text{UV}$ versus $1/\epsilon_\text{IR}$, they would realize that it is in fact due to an IR divergence and as such does not make sense to treat it as a UV contribution to an anomalous dimension.} Managing zero-bins can also be accomplished through a clever choice of regulator that treats the regions differently, as we will see explicitly in this section.

First we revisit the calculation of the massless Sudakov form factor in the previous section to explain why we did not discuss the zero-bin then.  An approach to identifying if there are overlapping regions is to take the collinear integrand and re-expand it assuming the scalings of other modes that contribute at the desired order in power counting.  Inspecting~\cref{eq:MasslessSudakovRegions}, the $\bar{p}^2$ $\big(p^2\big)$ dependence from the collinear (anti-collinear) denominators has been expanded away; the ultrasoft integral includes a factor of $\big[(n\cdot \ell_{us}\, \bar{n}\cdot p+p^2)\, ( \bar{n}\cdot \ell_{us}\, n\cdot \bar{p}+\bar{p}^2)\big]^{-1}$, which is not recoverable from either collinear integral.  We could also re-expand assuming a soft scaling, and would find only scaleless contributions.  Therefore, we could simply evaluate the integrals to derive the correct answer.  However, we emphasize that the soft region is always there, and one must keep track of it to ensure that it only contributes to the final result once.

There is a physical way to understanding why no overlap issues arose in the massless Sudakov case.  The regions analysis we performed identified that the IR physics could be captured by collinear regions with virtuality $\sim \lambda^2$ and an ultrasoft region with virtuality $\sim \lambda^4$.  These regions are separated by virtuality, and as such they do not overlap (although we caution that one should take care when applying this type of argument generically).  By taking $p^2 \neq 0$ and $\bar{p}^2 \neq 0$ we ensured that this physical distinction was present in the integrands.  However, as we will see in the next section, the massive Sudakov integral does not have an ultrasoft contribution.  It instead requires the inclusion of a soft region, which scales as $\ell_s \sim (\lambda,\lambda,\lambda)$, such that $\ell_s^2 \sim \lambda^2$.  Furthermore, we will evaluate the massive Sudakov integral with $p^2 = \bar{p}^2 = 0$, so we will find non-trivial overlaps that must be addressed.

A nice way to capture this effect visually is to sketch the hyperboloids traced by these regions in the $n\cdot \ell$ versus $\bar{n}\cdot \ell$ plane~\cite{Manohar:2006nz}:  
\begin{align}
\includegraphics[width=0.45\textwidth, valign=c]{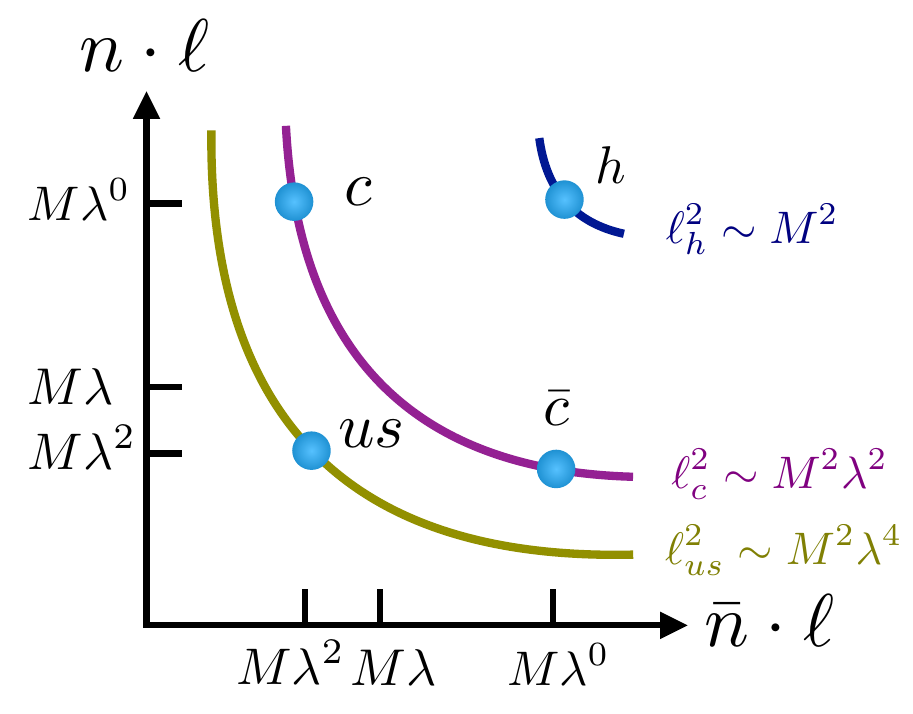} \hspace{40pt} \includegraphics[width=0.45\textwidth, valign=c]{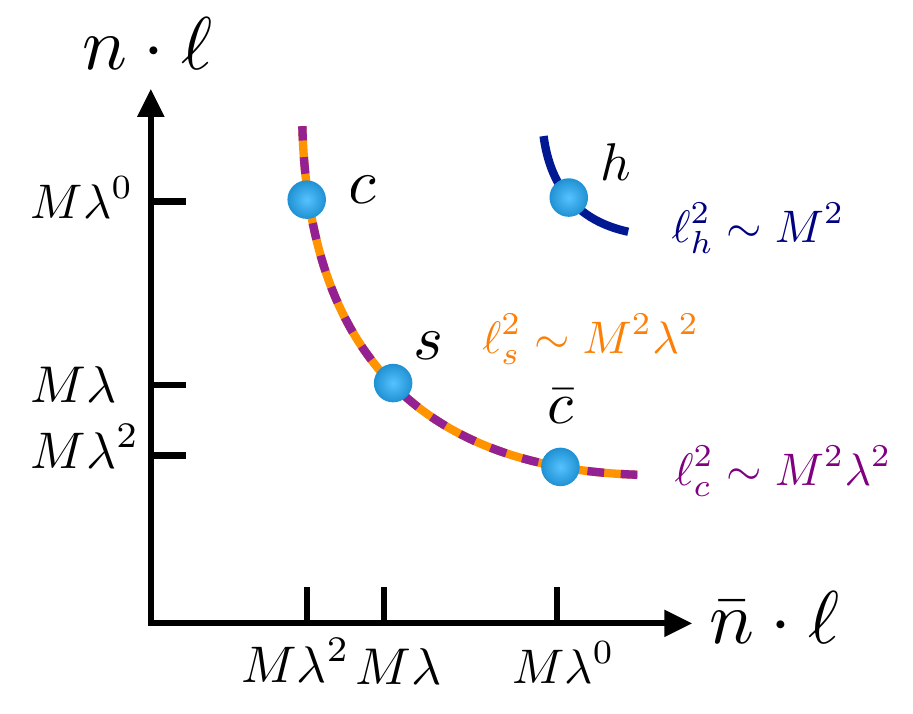}  \notag
\end{align}
On the left, we have illustrated the regions that contribute to the massless Sudakov case, the relevant EFT is known as SCET$_\text{I}$, while the right figure describes the regions for the massive Sudakov integral, which yields SCET$_\text{II}$.  The curves show the variation in rapidity along contours of constant virtuality, and the blue dots show the location of the regions that generate our logarithmic divergences.  The left diagram clearly demonstrates that virtuality does distinguish collinear from ultrasoft.  However, the right diagram illustrates that one can use a Lorentz boost to move from collinear to soft.  This implies that they overlap in virtuality.  In order to properly separate them, we will need to introduce some additional rapidity dependence into our integrals that comes in the form of a new regulator.  

We can see this effect appear mathematically, by noting that the collinear integrand derived in the next section for the massive Sudakov form factor~\cref{eq:IChiExpanded} now includes the soft integrand. Taking the soft scaling and re-expanding such that $\ell_c^2 \rightarrow 0$ so that $\ell_c \rightarrow \ell_s$:
\begin{align}
\frac{1}{\ell_c^2 +  n\cdot \ell_c\, \bar{n}\cdot p}\frac{1}{\bar{n}\cdot \ell_c \,n\cdot \bar{p}}  \quad \supset\quad \frac{1}{  n\cdot \ell_{s}\, \bar{n}\cdot p} \frac{1}{ \bar{n}\cdot \ell_{s}\, n\cdot \bar{p}}\,.
\label{eq:ZeroBinLimitIntegrand}
\end{align}

Removing this soft contribution from the collinear integral is known as zero-bin subtraction.  We define the ``zero-bin'' of the collinear integral as the limit of the denominator of the collinear integrand that reproduces the soft integrand, as in~\cref{eq:ZeroBinLimitIntegrand} above:
\begin{align}
\mathcal{I}_c \quad \longrightarrow \quad \mathcal{I}_{c,0} = \mathcal{I}_s \,,
\label{eq:ZeroBinEqSoft}
\end{align}
where $\mathcal{I}_{c,0}$ is the collinear zero-bin, which is equal to the soft integral in the simplest implementation of zero-bin subtraction.\footnote{Note that in general the soft region of the collinear integrand does not have to precisely yield the soft integral, since at there can be differences at the integrand level due to the choice of regulator, from including (and subsequently expanding) a ``measurement function'' (briefly mentioned in \cref{sec:RemainingConcepts} below), and so on.  We will encounter this explicitly in the next section where we use a regulator that treats soft and collinear sectors differently.}  Then to implement the zero-bin subtraction procedure, one simply removes this double counted piece by hand~\cite{Manohar:2006nz}
\begin{align}
\mathcal{I} &= \mathcal{I}_h + \Big(\mathcal{I}_c - \mathcal{I}_{c,0}\Big) +  \Big(\mathcal{I}_{\bar{c}} - \mathcal{I}_{\bar{c},0}\Big) +\mathcal{I}_{s} = \mathcal{I}_h + \mathcal{I}_c + \mathcal{I}_{\bar{c}} - \mathcal{I}_{s}\,,
\label{eq:ZeroBinProcedure}
\end{align}
where the last step is only appropriate when~\cref{eq:ZeroBinEqSoft} holds.  

If possible, it is convenient to chooses a regulator so that the soft limit of the collinear integrand vanishes, implying that zero-bin subtraction becomes trivial.  This is the approach taken in~\cite{Becher:2014oda} when they evaluate the massive Sudakov integral.  We will instead take a different approach in the next section, where we will use a rapidity regulator that treats collinear and soft differently and has the added benefit that it yields a rapidity RG interpretation~\cite{Chiu:2011qc, Chiu:2012ir}. 

\subsection{Method of Regions for a Massive Sudakov Integral}
\label{sec:RegionsMassiveSudakov}
Before we introduce the framework of SCET, it is worth working out the regions and subsequent leading power result for another process that exhibits a Sudakov double log.  The toy model for this example will require the presence of an additional real scalar state, $\chi$, whose mass power counts\footnote{Note that if we had instead been working in a parameter space where $m_{\chi}^2 \sim 1$, then this integral would be power suppressed, \emph{i.e.}, $\chi$ would be non-propagating at these low scales of interest.  If we had taken $m_{\chi}^2 \sim \lambda^4$ or smaller, then the mass would play no important role, \emph{i.e.}, we would just re-derive the massless Sudakov example.} as $m_{\chi}^2 \sim \lambda^2$, so that the $\chi$ particle is a propagating mode at energies far below the scale $M$.  The interacting Lagrangian is
\begin{align}
\mathcal{L}_\text{Int}^\textsc{Full} \supset \frac{1}{2} \,b\, \Phi\,\phi^2 + \frac{1}{2}\,b_\chi\, \chi\, \phi^2\,.
\end{align}
Again, we are interested in a process where $\Phi \rightarrow \phi \,\phi$ with no extra emissions and taking the same deep inelastic scattering-like kinematic configuration as in the massless Sudakov calculation, such that the relevant diagram is
\begin{align}
\includegraphics[width=0.2\textwidth, valign=c]{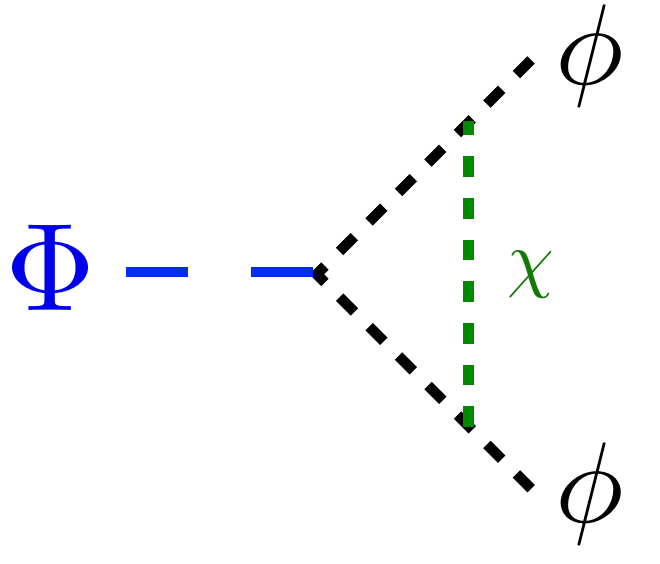} &= b^2\, b_{\chi} 
\int \frac{\D^4\ell}{(2\s\pi)^4} {\frac{1}{\ell^2-m_{\chi}^2}} {\frac{1}{ (\ell+p)^2}}{\frac{1}{ (\ell+\bar{p})^2} } = b^2\, b_{\chi} \,\mathcal{I}^{\chi} \,.
\label{eq:intMassive}
\end{align}

Next, we can perform a regions analysis of this integral, which will expose the differences with the massless Sudakov case above.  Note that we are taking the same momentum routing as above in~\cref{eq:CollandAntiColl}: the collinear momentum $p$ is outgoing while the anti-collinear momentum $\bar{p}$ is incoming.  However, here we  will take the external lines to be on-shell $p^2 = \bar{p}^2 =0$, since having $m_\chi^2 \neq 0$ already ensures that this \FT~integral is IR finite.

\vspace{5pt}\mybox{\begin{itemize}
\item \textbf{Exercise:} Following the same logic that led to \cref{eq:SoftIRDiv} and \cref{eq:CollinearIRDiv}, where we demonstrated that the massless Sudakov integral diverges in both the soft and collinear limits, show that $m_\chi^2 \neq 0$ regulates the analogous divergences for the massive Sudakov integral.  Since $\mathcal{I}^{\chi}$ is both UV and IR finite, it can be evaluated directly without the need to regulate.  Derive that
\begin{align}
\mathcal{I}^{\chi} &= -\frac{i}{16\s\pi^2}\frac{1}{M^2}\bigg[\frac{1}{2} \log^2 \frac{M^2}{m_\chi^2} + \frac{\pi^2}{6} + \text{Li}_2\bigg(1-\frac{m_\chi^2}{Q^2}\bigg)\bigg]\notag\\[5pt]
&= -\frac{i}{16\s\pi^2}\frac{1}{M^2}\bigg[\frac{1}{2} \log^2 \frac{M^2}{m_\chi^2} + \frac{\pi^2}{3} + \mathcal{O}\big(\lambda^2\big)\bigg]\,.
\label{eq:MassiveSudakovFullEval}
\end{align}
\end{itemize}}

We consider the following regions:
\begin{align}
\big[\text{hard}\big]\quad \quad h: &\quad \ell_h \sim \big(1,1,1\big) \notag \\[2pt]
\big[\text{collinear}\big]\quad \quad c: &\quad \ell_c \sim \big(\lambda^2,1,\lambda\big) \notag\\[2pt]
\big[\text{anti-collinear}\big] \quad \quad \bar{c}: &\quad \ell_{\bar{c}} \sim \big(1,\lambda^2,\lambda\big)  \notag\\[2pt]
\big[\text{soft}\big] \quad\quad s: &\quad \ell_s \sim \big(\lambda,\lambda,\lambda\big)\,.
\end{align}
As before, we can check that these regions contribute at $\mathcal{O}(1)$.  Expanding \cref{eq:intMassive}, assuming the appropriate scalings for each of the relevant loop momenta, yields four integrals:
\begin{align}
h: &\quad \mathcal{I}^\chi\quad  \longrightarrow\quad \mathcal{I}^\chi_h =  \int \frac{\D^4\ell_h}{(2\s\pi)^4} \,\frac{1}{\ell_h^2} \, \frac{1}{\ell_h^2 + n\cdot \ell_h\, \bar{n}\cdot p} \, \frac{1}{\ell_h^2+\bar{n}\cdot \ell_h\, n\cdot \bar{p}}  \,\,\sim\,\, \mathcal{O}(1)\notag \\[7pt]
c: &\quad\mathcal{I}^\chi\quad \longrightarrow\quad \mathcal{I}^\chi_c = \int \frac{\D^4\ell_c}{(2\s\pi)^4} \,\frac{1}{\ell_c^2-m_\chi^2}\, \frac{1}{\ell_c^2 + n\cdot \ell_c \,\bar{n}\cdot p} \, \frac{1}{\bar{n}\cdot \ell_c\, n\cdot \bar{p}}  \,\,\sim\,\, \mathcal{O}(1)\notag \\[7pt]
\bar{c}: &\quad\mathcal{I}^\chi\quad  \longrightarrow\quad\mathcal{I}^\chi_{\bar{c}} = \int \frac{\D^4\ell_{\bar{c}}}{(2\s\pi)^4} \, \frac{1}{\ell_{\bar{c}}^2-m_\chi^2} \, \frac{1}{n\cdot \ell_{\bar{c}} \,\bar{n}\cdot p} \, \frac{1}{\ell_{\bar{c}}^2+n\cdot \ell_{\bar{c}} \, \bar{n}\cdot \bar{p}}  \,\,\sim\,\, \mathcal{O}(1) \notag \\[7pt]
s: &\quad\mathcal{I}^\chi \quad \longrightarrow \quad \mathcal{I}^\chi_{s} =\int \frac{\D^4\ell_{s}}{(2\s\pi)^4} \, \frac{1}{\ell_{s}^2-m_\chi^2} \, \frac{1}{ n\cdot \ell_{s}\, \bar{n}\cdot p} \, \frac{1}{ \bar{n}\cdot \ell_{s}\, n\cdot \bar{p}} \,\,\sim\,\, \mathcal{O}(1)\,.
\label{eq:IChiExpanded}
\end{align}
As promised, the soft region contributes at leading power.  

Since it contributed for the massless Sudakov case, it is worthwhile to check what happens for the ultrasoft scaling $\ell_{us} \sim \big(\lambda^2, \lambda^2, \lambda^2\big)$:
\begin{align}
 us: \quad\mathcal{I}^\chi\quad \longrightarrow \quad \mathcal{I}^\chi_{us} =\int \frac{\D^4\ell_{us}}{(2\s\pi)^4} \, \frac{\,-1\,\,\,}{m_\chi^2} \, \frac{1}{  n\cdot \ell_{us}\, \bar{n}\cdot p} \, \frac{1}{ \bar{n}\cdot \ell_{us}\, n\cdot \bar{p}} \,\,\sim\,\, \mathcal{O}\big(\lambda^2\big)\,,
\end{align}
where we have used that $m_\chi^2 \sim \lambda^2$.  This demonstrates that the ultrasoft region makes a sub-leading power contribution (and in fact, this integral can be shown to be zero, see~\cite{Becher:2014oda}). 

Before we get to the solution and interpretation of these integrals, there are a few interesting features that are worth highlighting.  Naively, one might have thought that the soft and collinear limits would be universal, especially given the similarity between \cref{eq:intAgain} and \cref{eq:intMassive}.  However, we have now shown that the IR of the massive Sudakov integral has a different regions expansion when compared to the massless Sudakov integral.  This will imply that the EFT one must use to separate scales and sum \emph{depends on the observable}!   As mentioned in \cref{sec:ZeroBin}, the most common versions of SCET are known as SCET$_\text{I}$ and SCET$_\text{II}$.  However, when sufficiently complicated situations are encountered, the reader is cautioned that modes beyond the (anti-)collinear, soft, and/or ultrasoft could be required, see \cref{sec:RemainingConcepts} below which includes a list of possibilities.  This justifies why have been pedantic about specifying a process when matching throughout these lectures: when working out a new example, one cannot blindly assume the same IR description holds for any observable, and instead must be careful to ensure that the correct EFT is being utilized that captures the full richness of the IR.  

Next, we recall from~\cref{sec:ZeroBin} on zero-bin subtraction that this issue was a concern for the massive Sudakov integral.  In particular, we have a physical process for which collinear, anti-collinear, and soft all contribute, and all three of these modes have the same virtuality.  We must therefore be careful to account for overlapping regions so that we avoid double counting when evaluating the integrals.  To find the zero-bin of the collinear integral, we must re-expand it now assuming soft scaling for its momentum $\ell_{c,0} \sim (\lambda, \lambda,\lambda)$, implying $\ell^2_{c,0} \sim \lambda^2$ such that $\bar{n}\cdot \ell_{c,0} \,n\cdot p \sim \lambda$.  The collinear zero-bin integral is then
\begin{align}
\mathcal{I}^\chi_{c,0}\, = \int \frac{\D^4\ell_{c,0}}{(2\s\pi)^4}\, \frac{1}{\ell_{c,0}^2-m_\chi^2}\,\frac{1} {n\cdot \ell_{c,0} \,\bar{n}\cdot p}\,\frac{1}{\bar{n}\cdot \ell_{c,0} \,n\cdot \bar{p}}\, =\, \mathcal{I}_s^\chi\,,
\end{align}
where in the last equality we have made the identification with the soft integral from \cref{eq:IChiExpanded} above.

In order to understand the properties of this soft integral in more detail, we can integrate over $\ell_\perp$ using dim reg~\cite{Chiu:2011qc, Chiu:2012ir}
\begin{align}
\mathcal{I}_s^\chi\,\, \sim\,\, \int \big(\D n\cdot \ell_s\big) \big(\D \bar{n}\cdot \ell_s\big) \big(n\cdot \ell_s \, \bar{n}\cdot \ell_s- m_\chi^2\big)^{-2\epsilon} \,\frac{1}{n\cdot \ell_s +i0}\,\frac{1}{\bar{n}\cdot \ell_s +i0}\,.
\label{eq:PartiallyIntegratedSoft}
\end{align}
This partially integrated result exposes two features.  First, the integral only receives a non-trivial contribution for regions where $n\cdot \ell \,\bar{n}\cdot \ell \sim m_\chi^2$; otherwise the integral is scaleless and hence vanishes.  Second, there are spurious divergences that are not regulated by dim reg when either $n\cdot \ell \rightarrow 0$ or $\bar{n}\cdot \ell \rightarrow 0$.  These two divergences correspond to where the soft integral overlaps with the collinear or anti-collinear regions.   This implies that we need an additional regulator beyond dim reg, namely a rapidity regulator, to extract the physical prediction of the massive Sudakov process.  

We now turn to the explicit evaluation of the integrals in~\cref{eq:IChiExpanded}.  As we have already emphasized, we must be careful not to double count the soft region.  A related issue is that we need to regulate the associated rapidity divergences discussed below \cref{eq:PartiallyIntegratedSoft}. To this end, we will use the rapidity regulator introduced in~\cite{Chiu:2011qc, Chiu:2012ir}:
\begin{align}
\mathcal{I}^\chi = \mu^{2\s \epsilon} \int \frac{\D^d \ell}{(2\s\pi)^d} \frac{|2\, \ell_3|^{-\eta}}{\nu^{-\eta}}\frac{1}{\ell^2-m_{\chi}^2} \frac{1}{ (\ell+p)^2}\frac{1}{ (\ell+\bar{p})^2} \, ,
\end{align}
where $\ell_3 = (n\cdot \ell - \bar{n} \cdot \ell)/2$ is the $z$-component of the loop momentum, $\eta$ is the rapidity regulator parameter (analogous to $\epsilon$ in dim reg), and $\nu$ is a dimensionful parameter that must be introduced to absorb the change in mass dimension of the integrals (analogous to $\mu$ in dim reg).

Expanding this regulated form in the various regions, we find:\footnote{This same integral was analyzed in~\cite{Becher:2014oda}, where they instead used a so-called ``analytic regulator''~\cite{Smirnov:1997gx, Smirnov:1998vk, Beneke:2003pa}.  The essential idea of their approach is to pick one of the internal propagators and to deform it in all the relevant diagrams by changing its exponent slightly, \emph{e.g.}~$(\ell-p)^{-2} \rightarrow (\ell-p)^{-2(1+\alpha)}$.  This approach yields $\mathcal{I}_s = 0$, thereby avoiding the zero-bin subtraction issues in a different way.}
\begin{align}
h: &\quad \mathcal{I}^\chi \quad \longrightarrow \quad \mathcal{I}^\chi_h = \int \frac{\D^d\ell}{(2\s\pi)^d} \, \frac{1}{\ell_h^2 + n\cdot \ell_h\, \bar{n}\cdot p} \, \frac{1}{\ell_h^2+\bar{n}\cdot \ell_h\, n\cdot \bar{p}} \notag \\[7pt]
c: &\quad\mathcal{I}^\chi\quad \longrightarrow\quad \mathcal{I}^\chi_c = \int \frac{\D^d\ell}{(2\s\pi)^d} \,\frac{|\bar{n}\cdot \ell|^{-\eta}}{\nu^{-\eta}} \frac{1}{\ell^2-m_\chi^2}\,\frac{1}{\ell^2 +  n\cdot \ell \, \bar{n}\cdot p}\,\frac{1}{\bar{n}\cdot \ell\, n\cdot \bar{p}} \notag \\[7pt]
\bar{c}: &\quad\mathcal{I}^\chi \quad \longrightarrow\quad\mathcal{I}^\chi_{\bar{c}} = \int \frac{\D^d\ell}{(2\s\pi)^d} \,\frac{|n\cdot \ell|^{-\eta}}{\nu^{-\eta}} \frac{1}{\ell^2-m_\chi^2}\,\frac{1}{n\cdot \ell \,\bar{n}\cdot p}\,\frac{1}{\ell^2+ \bar{n}\cdot \ell \, n\cdot \bar{p}}   \notag \\[7pt]
s: &\quad\mathcal{I}^\chi \quad \longrightarrow \quad \mathcal{I}^\chi_{s} =\int \frac{\D^d\ell}{(2\s\pi)^d}\, \frac{|2\, \ell_3|^{-\eta}}{\nu^{-\eta}}\, \frac{1}{\ell^2-m_\chi^2}\,\frac{1}{ n\cdot \ell \,\bar{n}\cdot p}\, \frac{1}{ \bar{n}\cdot \ell \, n\cdot \bar{p}} \,,
\label{eq:MassiveSudakovWithRapadityReg}
\end{align}
where we have set $\nu = 0$ for the hard integral since this integral is fully regulated by dim reg.
The soft and collinear integrals will diverge in both the $\epsilon \rightarrow 0$ and $\eta \rightarrow 0$ limits, but it is critical that the order of limits be taken so that $\eta/\epsilon^n \rightarrow 0$ for all $n > 0$, \emph{i.e.} take the limit $\eta \rightarrow 0$ first.  This will ensure that the soft and collinear limits are taken along the same mass hyperbola as they are separated in rapidity by the presence of $\eta \neq 0$.

For the (anti-)collinear integral, the rapidity regulator effectively acts as a slight deformation of the exponent for the propagator factor $\bar{n}\cdot \ell$ ($n\cdot \ell$).  From the partially integrated form of the collinear zero-bin in \cref{eq:PartiallyIntegratedSoft}, we see that this deformation eliminates the simple pole at $\bar{n}\cdot \ell$; the modified anti-collinear integrand in the soft limit does not have a simple pole at $n \cdot \ell$.  This implies that one can choose a contour that does not enclose any poles for the soft limit of these integrands; this is how the regulator eliminates the zero-bin.  Similarly, the regulator for the soft integral has the right form to eliminate the divergences that appear in the collinear/anti-collinear limits of this integrand.

The integrals in \cref{eq:MassiveSudakovWithRapadityReg} can be evaluated through the judicious application of contour integration.  For the collinear integral, one can first evaluate the integral over $n\cdot \ell_c$ using contours; following an analysis similar to the one that led to \cref{eq:ColDivExContour} above, we find that the only integral is only non-zero for $-n\cdot p < n\cdot \ell < 0$, which corresponds to a collinear splitting with $0 < x < 1$, where $x$ is the momentum fraction.  A similar conclusion holds for the anti-collinear integral.  When performing the soft integral, one can perform the $\ell^0$ integral using contours first as well.  Explicit evaluation yields
\begin{align}
\mathcal{I}^{\chi}_h &= - \frac{i}{16\s\pi^2} \frac{\Gamma(1+\epsilon)}{M^2} \left(\frac{4 \pi \mu^2}{M^2}\right)^\epsilon \frac{(\Gamma(-\epsilon))^2}{\Gamma(1-2\epsilon)} \notag\\[6pt]
\mathcal{I}^{\chi}_c &=+ \frac{i}{16\s\pi^2} \frac{\Gamma(1+\epsilon)}{M^2} \left(\frac{4 \pi \mu^2}{m_\chi^2}\right)^\epsilon \left(\frac{n\cdot p}{\nu}\right)^{-\eta} \frac{\Gamma(-\epsilon)\Gamma(-\eta)}{\Gamma(1-\epsilon-\eta)} \notag\\[6pt]
\mathcal{I}^{\chi}_{\bar{c}} &=+ \frac{i}{16\s\pi^2} \frac{\Gamma(1+\epsilon)}{M^2} \left(\frac{4 \pi \mu^2}{m_\chi^2}\right)^\epsilon  \left(\frac{\bar{n}\cdot \bar{p}}{\nu}\right)^{-\eta}  \frac{\Gamma(-\epsilon)\Gamma(-\eta)}{\Gamma(1-\epsilon-\eta)}\notag\\[6pt]
\mathcal{I}^{\chi}_s &=- \frac{i}{16\s\pi^2} \frac{\Gamma(1+\epsilon)}{M^2} \left(\frac{4 \pi \mu^2}{m_\chi^2}\right)^\epsilon \left(\frac{m_\chi}{\nu}\right)^{-\eta}  \frac{2^{1-\eta}}{\eta} \frac{\Gamma(1/2 - \eta/2)\Gamma(\epsilon+\eta/2)}{\Gamma(1/2)\Gamma(1+\epsilon)}  \,,
\label{eq:IchiEval}
\end{align}
which can be expanded for small $\eta$ and small $\epsilon$ to yield\footnote{The naive expansion of $\frac{\Gamma(1/2-\eta/2)}{\Gamma(1/2)} = 1-(1/2) \psi^{(0)}(1/2)$, where $\psi^{(n)}(z)$ is a polygamma function.  To expand this expression in a form that is more useful for our purposes, one can apply the ``Legendre duplication formula,''  which in the case at hand implies
\begin{align}
\frac{\Gamma(1/2-\eta/2)}{\Gamma(1/2)} = \frac{\Gamma(-\eta)}{2^{-\eta-1}\Gamma(-\eta/2)} = 1 + \frac{1}{2} (\gamma_E+2\log 2) \eta + \mathcal{O}(\eta^2)\,.
\label{eq:expan}
\end{align}
This then combines nicely with the other factors that appear to yield the expression for $\mathcal{I}_s^\chi$ in \cref{eq:IchiEvalExp}.}
\begin{align}
\mathcal{I}^{\chi}_h &= - \frac{i}{16\s\pi^2} \frac{1}{M^2}\left(\frac{1}{\epsilon^2}+\frac{1}{\epsilon }\log \frac{\mu ^2}{M^2}+\frac{1}{2} \log^2\frac{\mu ^2}{M^2}-\frac{\pi ^2}{12} +\mathcal{O}\big(\lambda^2\big)\right)  \notag\\[6pt]
\mathcal{I}^{\chi}_c &=- \frac{i}{16\s\pi^2} \frac{1}{M^2}\left(-\frac{1}{\eta \s \epsilon }-\frac{1}{\eta }\log \frac{\mu ^2}{m_\chi^2}+\frac{1}{\epsilon }\log \frac{n\cdot p}{\nu }+\log\frac{\mu ^2}{m_\chi^2} \log \frac{n \cdot p}{\nu}+\frac{\pi ^2}{6}+\mathcal{O}\big(\lambda^2\big) \right)\notag\\[6pt]
\mathcal{I}^{\chi}_{\bar{c}} &=- \frac{i}{16\s\pi^2} \frac{1}{M^2} \left(-\frac{1}{\eta \s \epsilon }-\frac{1}{\eta}\log \frac{\mu^2}{m_\chi^2}+\frac{1}{\epsilon}\log\frac{\bar{n}\cdot \bar{p}}{\nu }+\log\frac{\mu ^2}{m_\chi^2}\log\frac{\bar{n}\cdot \bar{p}}{\nu }+\frac{\pi ^2}{6} +\mathcal{O}\big(\lambda^2\big)\right)\notag\\[6pt]
\mathcal{I}^{\chi}_s &=- \frac{i}{16\s\pi^2} \frac{1}{M^2} \left(-\frac{1}{\epsilon ^2}+\frac{2}{\eta \s \epsilon }+\frac{2}{\eta }  \log\frac{\mu ^2}{m_\chi^2}-\frac{1}{\epsilon }\log\frac{\mu ^2}{\nu^2}\right.\notag\\[2pt]
&\left.\hspace{104pt}-\frac{1}{2} \log^2\frac{\mu ^2}{m_\chi^2}- \log\frac{\mu ^2}{m_\chi^2} \log\frac{m_\chi^2}{\nu^2}+\frac{\pi^2}{12} +\mathcal{O}\big(\lambda^2\big)\right) \,,
\label{eq:IchiEvalExp}
\end{align}
where the factor of $\gamma_E - \log 4\,\pi$ has been absorbed into the renormalization parameter $\mu^2$ as per the $\overline{\text{MS}}$ scheme.  Note the (anti-) collinear integral has a $1/(\epsilon\s\eta)$ double divergence, which traces the $\log \mu^2/m_\chi^2 \log n\cdot p/\nu$ double log, in contrast to the (anti-) collinear contributions to the massless Sudakov integral which include $1/\epsilon^2$, see~\cref{eq:MasslessSudakovRegionsEval}.  This is a manifestation of the fact that the rapidity regulator is designed to be sensitivity to direction.

Adding these regions together, we find
\begin{align}
\mathcal{I}^{\chi} = -\frac{i}{16\s\pi^2}\frac{1}{M^2}\bigg[\frac{1}{2} \log^2 \frac{M^2}{m_\chi^2} + \frac{\pi^2}{3} + \mathcal{O}\big(\lambda^2\big)\bigg]\, ,
\label{eq:MassiveSudakovIRSum}
\end{align}
where we have replaced $\bar{n}\cdot p = -n\cdot\bar{p} = M$.  We identify the factor $\log^2 M^2/m_\chi^2$ as the expected Sudakov double log.  The combined result is finite as it must have been since $\mathcal{I}^\chi$ does not have any divergences, and the precise form agrees with the expanded full theory integral in~\cref{eq:MassiveSudakovFullEval}.  Note that the all $\mu$ and $\nu$ dependence has canceled in this combination as it had to, since both types of divergences are an artifact of our choice to separate the total finite integral into regions.  

\vspace{5pt}\mybox{
\begin{itemize}
\item {\bf Exercise:}  Evaluate the integrals in \cref{eq:MassiveSudakovWithRapadityReg} to derive \cref{eq:IchiEval}.  You may find it useful to restore the ``$i\s 0$'' factors, and then for the soft (collinear) do the $\ell^0$ ($n\cdot \ell$) integral using contours first.  For the soft integral, it is useful to note that there are four poles, three of which are below the real axis; it is straightforward to evaluate this integral by closing a contour above the real axis.  For the collinear integrals, note that sending $p \leftrightarrow \bar{p}$ and $n\leftrightarrow\bar{n}$ within $\mathcal{I}_c^{\chi}$ returns $\mathcal{I}_{\bar{c}}^{\chi}$, so no additional evaluation is required.  We again emphasize that the collinear integral only has support in the range $-n\cdot p < n\cdot \ell < 0$, which can be determined by examining the behavior of the poles as one varies $n\cdot \ell$.
\end{itemize}}

We repeat again for emphasis that we can simply sum these integrals instead of applying the zero-bin corrected formula \cref{eq:ZeroBinProcedure} above.  This is justified by the discussion above \cref{eq:IchiEval}, where we explained how the rapidity regulator eliminated the zero-bin from the collinear integrals, while still allowing for a non-zero soft contribution.  We also see a novel feature appear if we add the three IR integrals $\mathcal{I}^\chi_c + \mathcal{I}^\chi_{\bar{c}} + \mathcal{I}^\chi_s$, namely there is a subleading $\log \mu^2/m_{\chi}^2$ term which is multiplied by a large $\log M^2/m_\chi^2$.  This is a ``rapidity logarithm'' and looking at the individual expressions in \cref{eq:IchiEvalExp}, we see that the rapidity scale $\nu$ separates the scales inside this log.  This allows one to write down a rapidity RG that governs the evolution of Wilson coefficients as a function of $\nu$, thereby summing these subleading logarithms~\cite{Chiu:2011qc, Chiu:2012ir}.  

This completes our discussion of the massive Sudakov integral -- for the rest of these lectures we will only concern ourselves with the massless Sudakov case.  Our focus for the next section will be to introduce an EFT of the soft and collinear modes that allows us to sum Sudakov logarithms.  Since we will only sum the leading double logarithm, we will ignore issues of rapidity renormalization for the rest of these lectures (other than a brief mention in \cref{sec:WilsonLines} below).  Next, we will set up a scalar version of SCET and will use it to derive a set of RGEs that sum the double logs appearing in \cref{eq:IMasslessSudakovCombined}.

\section{Toying Around with Soft Collinear Effective Theory}
\label{sec:ToySCET}
We have now explored the IR structure of field theory in the presence of light-like external interacting particles by analyzing various leading power regions of two example integrals.  In addition, this exposed the presence of potentially large double logarithms, whose argument is a function of only physical scales.  In this section, we will introduce an EFT -- a Soft Collinear Effective Theory (SCET) -- whose degrees of freedom capture the IR divergence structure of a wide class of \FT~processes.  As with all EFTs, it is critical to find the appropriate propagating states along with their transformation rules under the symmetries that persist to low energies.  Then by tracking power counting, we can organize the allowed operators.\footnote{For a recent discussion of power counting and higher power operators in SCET applied to the QCD observable thrust, see~\cite{Moult:2018jjd}.}  We will show how SCET allows us to match and run, thereby separating scales and systematically summing the Sudakov double log and its subleading counterparts.  

In this section, we will explore many features of SCET using a toy scalar model.\footnote{What follows was heavily inspired by~\cite{Becher:2014oda}, where scalar SCET was written down and many aspects of what appears below were explained.  They worked out the summation for scalar SCET in 6 dimensions, while here we will provide the RG analysis for a 4 dimensional example.  The summation of the Sudakov double logs for $\phi^3$ theory in 4 and 6 dimensions is also discussed in the earlier review of Sudakov form factors~\cite{Collins:1989bt}.}  In particular, we will show how one can sum the massless Sudakov double log that was derived in \cref{eq:IMasslessSudakovCombined}.  Some differences will emerge when lifting these techniques to gauge theory, as will be discussed \cref{sec:RealSCET} below.

Our first task is to discuss how the \FT~IR logs of interest can be recast as UV logs for SCET.  This is critical to understanding how RG techniques (a UV phenomena) can be applied to sum the Sudakov double log (that results from dynamics in the IR).

\subsection{Mapping IR Logs to UV Logs}
There are many non-trivial differences between SCET and the Lorentz preserving EFTs we have studied so far.  As we were reminded in \cref{sec:RGE}, RG techniques emerge due to UV divergences, whose regulation introduces a spurious scale $\mu$.  It is the requirement that physical observables be $\mu$-independent, which yields the RGEs.  Therefore, the first question we must address is how it is logically possible to sum IR logs using an RG formalism.  This will yield an interesting insight into the EFT approach since the principles discussed here were also at work above.  Since a full understanding of this subtlety was not really required there, we chose to postpone the discussion until now.

The use of EFTs to sum logs relies on the fact that the IR of the two descriptions is equivalent.  One way of thinking about this is to zoom in on a given scale.  From this vantage point, all other scales can be organize as either contributing to UV or IR divergences.  We absorb the UV dependence into Wilson coefficients through the renormalization procedure, and IR divergences must cancel by unitarity arguments since we are not resolving the IR by assumption.  As a result, when we take a log that depends on a ratio of two physical scales (generalized to include regulating scales) and pull it apart into two pieces, the IR log with respect to zooming in on a high scale is necessarily matched with a UV log with respect to zooming in on a low scale.  Again we emphasize that the UV theory does not have IR divergences, but it can generate the large logs we are trying to pull apart.  

Said another way, in the high energy limit, it should be a very good approximation to take all the light states to be massless, since we should not be sensitive to the specific features of the IR.  This manifests mathematically in the statement that the matching coefficients must be analytic as one takes the light scales to zero.  On the other hand, the IR description does rely on the detailed properties of the light modes, but is only sensitive to the UV through matching.  The matching procedure trades the IR divergences of the \FT~(which emerge when the light scales are taken to zero), to the UV divergences of the EFT.\footnote{This was the first to noticed in the study of summing the large logarithms that appear for cusped Wilson lines in gauge theory~\cite{Korchemsky:1987wg}.  See \cref{sec:CuspAnnDim} below for more discussion.}  From the EFT perspective, the logs of interest are now due to UV dynamics, so that we can apply RG techniques to run the theory to long distances.

To see how this story plays out mathematically, take a simple IR divergent integral\footnote{For a more detailed toy example, see Sec.~5.8 of~\cite{Manohar:2018aog}.}
\begin{align}
\int_{\Lambda_\text{IR}}^\infty \D x\, \frac{1}{x^2} = \frac{1}{\Lambda_\text{IR}}\,.
\end{align}
Our goal is to map this mock IR divergence to a UV divergence.  To do so, break the integral into two parts using a cutoff $\Lambda_\text{UV}$
\begin{align}
\int_{\Lambda_\text{IR}}^{\Lambda_\text{UV}} \D x\, \frac{1}{x^2} +\int_{\Lambda_\text{UV}}^\infty \D x\, \frac{1}{x^2}= \left(\frac{1}{\Lambda_\text{IR}} - \frac{1}{\Lambda_\text{UV}}\right) + \left(\frac{1}{\Lambda_\text{UV}}\right)\,.
\end{align}
We see that $\Lambda_\text{UV}$ tracks the IR divergence structure of the full integral.  This is exactly what we are doing when we match and run: we introduce a convenient intermediate cutoff -- the matching scale -- which allows us to peel apart the logarithm of interest into a UV and IR contribution.  Since the arguments of our logarithms must always be dimensionless, this procedure necessitates introducing a spurious scale -- the RG scale $\mu$ -- which can then be leveraged to derive RGEs.

Now that we have a general sense of how SCET can be used to sum IR logs, we turn to setting up the structure of the EFT.  Our focus is on summing the leading Sudakov double log, and as such we will specialize to the EFT for that process, although many of the features that appear in what follows generalize.

\subsection{Identifying the Modes}
Our \FT~process of interest is $\Phi \rightarrow \phi\,\phi$, which yields a massless Sudakov double log, see \cref{eq:intAgain}.  In order to write down the relevant EFT, we need to identify the degrees of freedom that model the soft and collinear IR limits of the \FT.  From a Wilsonian point of view, we want to integrate out the hard momentum shells of our \FT~field $\phi$, and leave behind two types of independent propagating modes, one to capture each of the two types of IR divergences.  For our $\Phi \rightarrow \phi\,\phi$ process, the physical picture takes the schematic form:
\begin{center}
\vspace{10pt}
\includegraphics[width=0.7\textwidth, valign=c]{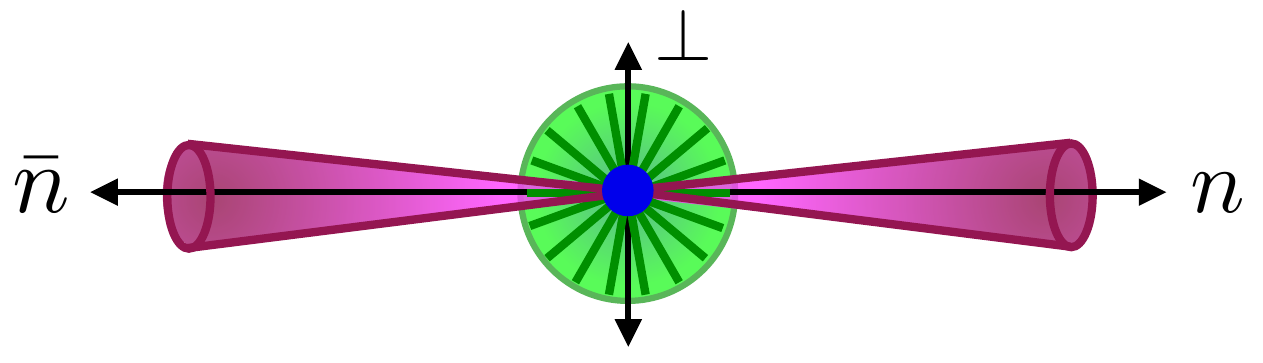}
\vspace{10pt}
\end{center}
where the blue dot in the center is our hard annihilation process (the decay of the heavy state $\Phi$), which will be modeled as a local interaction in the EFT, the purple cones are collinear and anti-collinear degrees of freedom, which are localized close to the $n^\mu$ and $\bar{n}^\mu$ directions respectively (where ``close'' is determined by our power counting parameter as we will see below), and the green lines represent isotropic (ultra)soft radiation, which has no preferred direction but whose momentum is restricted by our power counting parameter.  Depending on the physics one is interested in modeling, additional modes may be required.  Identifying the modes is non-trivial, and in fact the first attempts at an SCET-like theory~\cite{Dugan:1990de} missed the need for soft modes.   Here, we will largely focus on the massless Sudakov process, so we will restrict our discussion to the appropriate three modes.  Note that we will treat them as independent propagating states since $p^2$ scales homogeneously in power counting for each.

From a more mathematical point of view, what we are trying to do is develop a field theory whose diagrammatic structure will reproduce the sum of integrals we derived in our regions analysis above, \cref{eq:MasslessSudakovRegions}, and which of course generalizes to higher order in the loop and power counting expansions.  Unsurprisingly, this requires a propagating mode\footnote{There is an intriguing recent proposal for constructing SCET without performing this mode expansion~\cite{Goerke:2017ioi}.  Instead, they model the dynamics of each jet in a given process as a copy of QCD that is not able to resolve the detailed structure of the other jets in the event -- from the point of view of one of these QCD theories, the other jets are simply Wilson lines (see~\cref{sec:WilsonLines} and \cref{eq:WilsonLinesSCET} below).  Then the presence of ultrasoft modes and the related issues of zero-bin subtraction manifest as an ``overlap subtraction'' that avoids the double counting that occurs when shifting between the points of view of the different QCD copies.} whose associated momentum scales according to each of the on-shell regions that contribute:
\begin{align}
\big[\text{collinear}\big]\quad \quad \phi_c\,\,\s: &\quad  p_c\hspace{5.5pt} \sim \big(\lambda^2,1,\lambda\big) \notag\\[2pt]
\big[\text{anti-collinear}\big] \quad \quad \phi_{\bar{c}}\,\,\s:  &\quad p_{\bar{c}} \hspace{5.5pt}\sim \big(1,\lambda^2,\lambda\big)  \notag\\[2pt]
\big[\text{ultrasoft}\big] \quad\quad \phi_{us}: &\quad p_{us} \sim \big(\lambda^2,\lambda^2,\lambda^2\big)\,.
\label{eq:SCETmomPower}
\end{align}

In order to construct a Lagrangian for these modes, one can start from the \FT~and expand the fields in terms of the SCET modes as\footnote{This assumes we have already expanded the UV modes of $\phi$ and have subsequently integrated them out.}
\begin{align}
\phi(x) = \phi_c(x) + \phi_{\bar{c}}(x) + \phi_{us}(x)\,.
\label{eq:phiExpanded}
\end{align}
Using power counting to expand the structure of the interactions allows us to determine a hierarchy that organizes the allowed operators.  This relies on deriving the power counting for the SCET fields, which is the topic of the next section.

\subsection{Power Counting for SCET Fields}
In this section, we will derive the power counting for the propagating degrees of freedom in scalar SCET, as was first done for QCD in~\cite{Bauer:2002uv}.  In exact analogy with our relativistic example, see~\cref{sec:PowerCountingForFields} above, we enforce that the propagation of the dynamical fields is not power suppressed -- the power counting must be chosen so that the contribution to the action from the kinetic term scales as $\mathcal{O}(1)$:
\begin{align}
\int \D^4 x\, \frac{1}{2}\, \partial_\mu \phi_{c,\s\bar{c},\s us}(x)\, \partial^\mu \phi_{c,\s\bar{c},\s us}(x) \,\, \sim\,\, \mathcal{O}(1)\,.
\label{eq:scalingKinTermRequirement}
\end{align}
First, we assume collinear scaling for the momentum of our field.  Then we can infer the required scaling of $x$ by enforcing that the exponent of the Fourier transform kernel is scaleless:
\begin{align}
x\cdot p_c = \frac{1}{2}\, n\cdot x\, \bar{n}\cdot p_c + \frac{1}{2}\,\bar{n}\cdot x\, n\cdot p_c + x_\perp\cdot p_{c,\s\perp} \,\,\sim\,\, \mathcal{O}(1)\,,
\label{eq:FTkernel}
\end{align}
which is satisfied if
\begin{align}
\hspace{-10pt}n\cdot x \,\,\sim\,\,  1\qquad \qquad \bar{n} \cdot x \,\,\sim\,\,  \frac{1}{\lambda^2}\qquad\qquad x_\perp \sim \frac{1}{\lambda} \,,
\label{eq:collinearScalingPosition}
\end{align}
such that
\begin{align}
 \D^4 x = \big(\D n\cdot x\big)\big(\D \bar{n}\cdot x\big) \big(\D^2 x_\perp\big) \,\,\sim\,\, \frac{1}{\lambda^4}\qquad\qquad \big[\text{collinear}\big]\,.
\label{eq:collinearScalingPositionD4x}
\end{align}
Since the collinear virtuality is $\partial^2 \sim p_c^2 \sim \lambda^2$, consistency with \cref{eq:scalingKinTermRequirement} requires $\phi_c(x) \sim \lambda$.  Simply swapping $n \leftrightarrow \bar{n}$ in the previous derivation, we can deduce that $\D^4 x \sim \lambda^{-4}$ for an anti-collinear mode, implying that $\phi_{\bar{c}}(x) \sim \lambda$ as well.  

Finally, we perform the same analysis for the ultrasoft mode.  Since it scales homogeneously in spacetime, each component of $x \sim \lambda^{-2}$ so that 
\begin{align}
\D^4 x \,\,\sim\,\, \frac{1}{\lambda^{8}} \qquad\qquad \big[\text{ultrasoft}\big]\,,
\label{eq:ultrasoftScalingPositionD4x}
\end{align}
implying that $\phi_{us}(x) \sim \lambda^2$. To summarize
\begin{align}
\phi_c(x) \,\,\sim\,\, \lambda\qquad\qquad \phi_{\bar{c}}(x) \,\,\sim\,\, \lambda\qquad\qquad \phi_{us}(x) \,\,\sim\,\, \lambda^2\,.
\end{align}

When we perform this analysis for gauge fields below, it will be more convenient to extract the power counting of the fields from the Feynman propagator directly.  Therefore, we will quickly demonstrate how to find the power counting from this point of view.  The time ordered two-point function for the free field is given by (see~\cref{eq:twoPtFn} above)
\begin{align}
\vev{0\big|\s T \s \phi_0(x) \s\phi_0(y)\big|0} = \int \frac{\D^4 p}{(2\s\pi)^4} \frac{i}{p^2 - m^2 + i0} \,e^{i\s p\s\cdot(y-x)}\,.
\end{align}
If the momentum is collinear, $\D^4 p_c \sim \lambda^4$ and $p_c^2\sim \lambda^2$, implying that the fields must scale as $\phi_c(x) \sim \lambda$ for consistency.  Similarly, if the momentum is ultrasoft, then $\D^4 p_{us} \sim \lambda^8$ and $p_{us}^4 \sim \lambda^2$, and so $\phi_{us}(x) \sim \lambda^2$.

In the next section we will use the naive expansion in \cref{eq:phiExpanded}, augmented by the field power counting we have just determined, to derive the interaction structure of SCET from our \FT~Lagrangian.

\subsection{Interactions in Position Space: the Multipole Expansion}
\label{sec:SCETIntPosSpace}
Our power counting choice implies that summing the two point function into the propagator is self-consistent.  We can follow the same logic to infer the power counting for the interactions and local operators.  First, we will show how to power count interactions in position space.  This will expose a subtlety that arrises when considering collinear-ultrasoft interactions.  In the next section, we will show how to account for the same effects directly in momentum space, which has the benefit that the origin of the resulting Feynman rules will be more transparent.

The first step is to determine what interactions are allowed within the confines of our EFT.  In particular, our EFT fields can exchange momentum amongst themselves, and so we must ensure that these exchanges do not change the nature of a field.  For example, a collinear field could in principle transfer a large amount of momentum to an ultrasoft field, violating the assumed ultrasoft momentum scaling in \cref{eq:SCETmomPower}.  

Starting with the \FT~in position space, we expand in terms of our SCET fields at the matching scale:
\begin{align}
\mathcal{L}_\text{Int} \supset \frac{a}{3!}\, \phi^3 &= \frac{a}{3!}\,\big(\phi_c + \phi_{\bar{c}} + \phi_\text{us}\big)^3 \notag \\[7pt]
&= \frac{a}{3!}\Big( \phi_c^3 + \phi_{\bar{c}}^3 + \phi_{us}^3 + 3\, \phi_c^2\, \phi_{\bar{c}} + 3\, \phi_c\, \phi_{\bar{c}}^2 \notag\\[2pt]
&\hspace{35pt}+ 3\,  \phi_c^2\, \phi_{us}  + 3\,  \phi_c\, \phi_{us}^2 + 3\,  \phi_{\bar{c}}^2 \,\phi_{us}  + 3\,  \phi_{\bar{c}}\, \phi_{us}^2 + 6\, \phi_c\, \phi_{\bar{c}}\, \phi_{us}\Big)\,.
\end{align}
Note that the coefficients of each term are only equal to $a$ at the high scale, and running within the EFT will cause these to deviate from each other -- each interaction must be given its own independent coefficient when we write the final form of the EFT below, see~\cref{eq:LIntScalarSCET}.

Power counting can be used to check if the momentum sum at the vertex is consistent with the assumed momentum scalings within the EFT.  This is done by comparing the power counting of each momentum component for all the states that are involved in the interaction.  The rule is that there should be two (or more) contributions in each momentum component at the highest order in $\lambda$.  For example, if $p_1 \sim \big(1,\lambda^2, \lambda\big)$ and $p_2 \sim \big(\lambda^2,\lambda^2,\lambda^2\big)$, then $p_1 + p_2 \sim \big(1,\lambda^2, \lambda)$, which is clearly non-zero.  The interpretation is that momentum conservation is not possible in this case.  On the other hand, if $p_1 \sim \big(1,\lambda^2, \lambda\big)$ and $p_2 \sim \big(1,\lambda^2, \lambda\big)$, then $p_1 + p_2 \sim \big(0,0,0)$, which is simply a propagating collinear line.  

We start by checking the self-interaction terms, which yield
\begin{align}
\phi_c^3: \begin{array}{rl}p_1&\sim \big(\lambda^2, 1,\lambda\big) \\[2pt] p_2 &\sim \big(\lambda^2, 1,\lambda\big) \\[2pt] p_3 &\sim \big(\lambda^2, 1,\lambda\big)\\[5pt] \sum p_i&\sim\,\, (0,0,0)\end{array} \,;
\quad
\phi_{\bar{c}}^3:\begin{array}{r}p_1\sim \big(1, \lambda^2, \lambda\big) \\[2pt] p_2\sim \big(1,\lambda^2,\lambda\big) \\[2pt] p_3\sim \big(1,\lambda^2,\lambda\big) \\[5pt] \sum p_i\sim\,\, (0,0,0) \end{array}\,;
\quad
\phi_{us}^3:\begin{array}{rl} p_1&\sim \big(\lambda^2, \lambda^2,\lambda^2\big) \\[2pt] p_2 & \sim \big(\lambda^2, \lambda^2,\lambda^2\big) \\[2pt] p_3 & \sim \big(\lambda^2, \lambda^2,\lambda^2\big) \\[5pt] \sum p_i & \sim\,\, (0,0,0) \end{array}\,.
\end{align}
So we see that SCET fields inherit cubic self interactions.  Next, we can check interactions between collinear and ultrasoft:\footnote{Note that one of the consequences of breaking Lorentz invariance by working in an explicit frame is that crossing symmetry no longer holds.  So technically, one should specify which momenta are incoming/outgoing when performing this kind of analysis.  For example, $p_c \rightarrow p_c + p_{us}$ is allowed, while $p_c+p_c \rightarrow p_{us}$ violates EFT momentum conservation.}
\begin{align}
\phi_c^2\, \phi_{us}:\,\, \begin{array}{rl}p_1 &\sim \big(\lambda^2, 1,\lambda\big) \\[2pt] p_2 &\sim \big(\lambda^2, 1,\lambda\big) \\[2pt] p_3 &\sim \big(\lambda^2, \lambda^2,\lambda^2\big)\end{array}\,\,\,\, \sim\,\,\,\,\,\, \begin{array}{rl}p_1&\sim\big(\lambda^2, 1,\lambda\big) \\[2pt] p_2&\sim \big(\lambda^2, 1,\lambda\big)\end{array} \,\,\,\, \sim\,\,\,\,\, \sum p_i \sim (0,0,0) \,,
\end{align}
and the same holds for $\phi_{\bar{c}}^2\, \phi_{us}$.  In the second step, we used the fact that the ultrasoft momentum does not change the power counting of the collinear fields, so we expanded it away.  These interactions are allowed within the EFT, and the interpretation is that the direction of a collinear particle is unchanged when it interacts with an ultrasoft field.

Next, we can check some interactions that violate momentum conservation: 
\begin{align}
\phi_c^2\, \phi_{\bar{c}}:\,\, \begin{array}{rl} p_1 &\sim \big(\lambda^2, 1,\lambda\big) \\[2pt] p_2 &\sim \big( \lambda^2, 1,\lambda\big) \\[2pt] p_3 &\sim \big(1, \lambda^2, \lambda\big)\end{array}\,\,\, \sim\,\,\,\, \big(1,0,0\big)\,\,\,\, \cancel{\sim}\,\,\,\, (0,0,0)\,.
\end{align}
A similar analysis shows that $\phi_c\, \phi_{\bar{c}}^2$ and $\phi_c\, \phi_{\bar{c}}\,\phi_{us}$ also violate momentum conservation.  This conclusion is intuitive -- an anti-collinear field has a large component of momentum that has the capacity to change the direction of the collinear field outside the range modeled within the EFT.  Therefore, this interaction should not be included.  We conclude that our EFT interaction Lagrangian only includes terms schematically of the form
\begin{align}
\mathcal{L}_\text{Int}^\text{SCET} &= \frac{1}{3!}\Big(a_{c}\, \phi_c^3 + a_{\bar{c}}\, \phi_{\bar{c}}^3 + a_{us^3}\,\phi_{us}^3 + 3\, \as \, \phi_c^2\, \phi_{us}  + 3\,  \asb \, \phi_{\bar{c}}^2\, \phi_{us}\Big)\,.
\end{align}

Consistently power counting the interactions requires one additional step, as was first developed in~\cite{Beneke:2002ph, Beneke:2002ni}.  In position space, we begin with a multipole expansion for the field $\phi(x)$.  For example, expanding about the  $\bar{n}\cdot x$ direction,
\begin{align}
\hspace{-8pt}\phi(x) = \phi\big(\bar{n}\cdot x\big) + x_\perp \cdot \partial_\perp \phi\big(\bar{n}\cdot x\big) + \frac{1}{2}\,n \cdot x \,\,\bar{n}\cdot \partial \phi\big(\bar{n}\cdot x\big) + \frac{1}{2} \big(x_\perp \cdot \partial_\perp\big)^2 \phi\big(\bar{n}\cdot x\big) + \cdots\,.
\label{eq:multipoleExpand}
\end{align}
This will be relevant for interactions involving a collinear line that points in the $n^\mu$ direction.

We want to explore the implications for the $\phi_c^2\,\phi_{us}$ interaction.  Given the multipole expansion of the collinear field $\phi_c$, we can power count each term  using the scalings  in \cref{eq:collinearScalingPosition} that are relevant when position is conjugate to collinear momentum:
\begin{align}
\phi_{c}\big(\bar{n}\cdot x\big) &\,\,\sim\,\, \lambda \notag\\[2pt]
x_\perp \cdot \partial_\perp \phi_{c}\big(\bar{n}\cdot x\big) &\,\,\sim\,\, \frac{1}{\lambda}\times \lambda\times \lambda \,\,=\,\, \lambda \notag\\[2pt]
\frac{1}{2}\,n \cdot x\,\, \bar{n}\cdot \partial \phi_{c}\big(\bar{n}\cdot x\big) &\,\,\sim\,\, 1\time 1\times \lambda \,\,=\,\, \lambda \notag\\[2pt]
\frac{1}{2} \big(x_\perp \cdot \partial_\perp\big)^2 \phi_{c}(\bar{n}\cdot x) &\,\,\sim\,\, \left(\frac{1}{\lambda}\times \lambda\right)^2\times \lambda \,\,=\,\, \lambda\,,
\end{align}
so we see that all the multipole expanded components have the same scaling, and therefore we should sum the expanded field back into the full field $\phi_c(x)$, which is what appears in the interaction Lagrangian below.\footnote{This conclusion could have been anticipated by noting that the collinear scaling for position was derived by requiring $x\cdot p_c \sim \mathcal{O}(1)$, and we are applying spatial derivatives to the collinear field.}

Next, we specialize to the ultrasoft field.  We again power count using the scalings given in \cref{eq:collinearScalingPosition} that are relevant when position is conjugate to collinear momentum, since the collinear field dominates the momentum flowing through the vertex:
\begin{align}
\phi_{us}\big(\bar{n}\cdot x\big) &\,\,\sim\,\, \lambda^2 \notag\\[2pt] 
x_\perp \cdot \partial_\perp \phi_{us}\big(\bar{n}\cdot x\big) &\,\,\sim\,\, \frac{1}{\lambda}\times \lambda^2\times \lambda^2 \,\,=\,\, \lambda^3 \notag\\[2pt]
\frac{1}{2}\,n \cdot x\,\, \bar{n}\cdot \partial \phi_{us}\big(\bar{n}\cdot x\big) &\,\,\sim\,\, 1 \times \lambda^2\times \lambda^2 \,\,=\,\, \lambda^4 \notag\\[2pt]
\frac{1}{2} \big(x_\perp \cdot \partial_\perp\big)^2 \phi_{us}\big(\bar{n}\cdot x\big) &\,\,\sim\,\, \left(\frac{1}{\lambda}\times \lambda^2\right)^2\times \lambda^2 \,\,=\,\, \lambda^4\,.
\end{align}
This implies that when the ultrasoft field is multipole expanded about the collinear direction, it organizes itself as a hierarchal expansion in $\lambda$.  In particular, the leading order contribution is given by $\phi_{us}(\bar{n}\cdot x)$, and so we should drop the higher order terms when constructing the leading power Lagrangian.  Similarly, expanding $\phi_{us}$ around the anti-collinear direction shows that only $\phi_{us}(n\cdot x)$ contributes at leading power.   There is an intuitive way to understand what this multipole analysis is accomplishing.  The ultrasoft fields are restricted to only have long wavelength fluctuations, and as such they are unable to resolve any of the transverse structure of the collinear sector(s).  All they can see is the collinear particle's world line, so that is the only nontrivial position dependence that can appear in the interactions involving the ultrasoft fields.

To summarize, expanding to leading power and keeping track of the arguments of the fields yields the resulting interactions:
\begin{align}
\mathcal{L}^\text{SCET}_\text{Int} &= \frac{1}{3!}\Big(a_{c}\, \phi_c^3(x) + a_{\bar{c}}\, \phi_{\bar{c}}^3(x) + a_{us^3}\,\phi_{us}^3(x)\notag\\
& \hspace{35pt}+ 3 \, \as\,\phi_c^2(x)\, \phi_{us}(\bar{n}\cdot x)  + 3\,\asb\,  \phi_{\bar{c}}^2(x)\, \phi_{us}(n\cdot x)\Big)\,,
\label{eq:LIntScalarSCET}
\end{align}
which includes both super-leading power interactions, $\phi_c^3$, $\phi_{\bar{c}}^3$, and\footnote{Note that $\phi_{us}^3\sim \lambda^6$ is super-leading power since in this case one should use the scaling of $\int \D^4 x \sim 1/\lambda^8$ appropriate for ultrasoft, see~\cref{eq:ultrasoftScalingPositionD4x}.} $\phi_{us}^3$, and leading power interactions $\phi_c^2\,\phi_{us}$ and $\phi_{\bar{c}}^2\,\phi_{us}$.  We will see a non-trivial interplay between the power counting of these interactions against that of the local operators when we sum the Sudakov double log below.  

This position space multipole expanded result ensures that only the $n\cdot p_{us}$ components of the ultrasoft momenta are added to collinear momenta in momentum space interactions.  This appears as an apparent non-conservation of momentum when Fourier transforming the interaction.  Since we are ultimately interested in momentum space Feynman rules, the next section is devoted to the so-called label formalism, which is an alternative way to capture the impact of the multipole expansion.

\subsection{Interactions in Momentum Space: the Label Formalism}
\label{sec:SCETIntMomSpace}
In this section, we will derive the form of momentum space SCET interactions more directly, utilizing the so-called label formalism~\cite{Bauer:2000yr, Bauer:2001yt}.  This approach allows one to implement the multipole expansion by applying power counting to the momenta that flow through the Feynman diagrams directly by separating the ``large'' component of momentum from the ``small'' component.  The hard background process determines the large component, while the fluctuations within the EFT are governed by the small component, see \emph{e.g.}~Fig.~5 of~\cite{iain_notes} for an illustration of this separation. 

We begin with the Fourier transform of the EFT field
\begin{align}
\tilde{\phi}(p) = \int \D^4 x \, e^{i \s p\cdot x}\, \hat{\phi}(x)\,,
\end{align}
where $\hat{\phi}(x)$ is the position space EFT field, and $\tilde{\phi}(p)$ is the momentum space EFT field.

For concreteness, we will focus on the interaction between two collinear fields and an ultrasoft field.  This example is interesting since it is non-trivially modified by power counting as we saw in the previous section.  Furthermore, we will need it for our massless Sudakov example below.  Given a momentum $p$, we separate it into a large ``label'' momentum $p_L$ and a small ``residual'' momentum $p_r$:
\begin{align}
p^\mu &= p_{L}^\mu +  p_{r}^\mu \,.
\end{align}
The scaling of the residual momentum is always chosen to be the softest scale.  For our example, this is $p_r \sim \big(\lambda^2,\lambda^2,\lambda^2\big)$.   This can then be used to derive the scaling for the label momentum that is appropriate for a given interaction.  Taking the example $\phi_c^2\,\phi_{us}$, the label and residual momentum scalings for a collinear field are
\begin{align}
p_{c,L} & \sim (0,1,\lambda)\notag\\[3pt]
p_{c,r} & \sim \big(\lambda^2,\lambda^2,\lambda^2\big)\,,
\label{eq:pLabelpResScaling}
\end{align}
while for an ultrasoft field, 
\begin{align}
p_{us,L} & \sim (0,0,0)\notag\\[3pt]
p_{us,r} & \sim \big(\lambda^2,\lambda^2,\lambda^2\big)\,.
\label{eq:pLabelpResScalingus}
\end{align}
Then an integral over collinear momentum, including an contributions from both the large and residual components, is given by 
\begin{align}
 \int \D^4 p_c \,\,\longrightarrow\,\, \sum_{p_{c,L} \neq 0} \int \D^4 p_{c,r}\,,
\end{align}
where we have excluded the bin where the label momentum goes to zero, since this corresponds to the ultrasoft region and we want to avoid double counting.\footnote{This is in principle how we would like to formulate the EFT.  In practice, when extending the integration ranges to the full loop momentum space (so that we can simply apply dim reg), one must be careful not to double count any regions, see the discussion of zero-bin subtraction in~\cref{sec:ZeroBin}.}  An integral over ultrasoft momentum is simply given by
\begin{align}
 \int \D^4 p_{us}\,\, \longrightarrow\,\,  \int \D^4 p_{us,r}\,.
\end{align}

We can use these definitions to trade our field's explicit label dependence for a sum over many fields that are now labeled by their large momentum $p_L$.  This can be done using a field redefinition:
\begin{align}
\hat{\phi}(x) = \sum_{p_L\neq 0} e^{-i\s p_L \cdot x} \int \frac{\D^4  p_r}{(2\s\pi)^4} \,e^{-i\s p_r \cdot x}\,\tilde{\phi}_{p_L}(p_r) = \sum_{p_L\neq 0} e^{-i\s p_L \cdot x}\, \phi_{p_L}(x)\,,
\label{eq:labelFieldRedef}
\end{align}
which decomposes our momentum space field as promised, since now our degree of freedom only depends on the residual fluctuations within the EFT.  

In order to access the label of a field, it is useful to define the label momentum operator $\mathcal{P}_\mu$, whose action is
\begin{align}
\mathcal{P}_\mu\, \phi_{p_{L}}(x) \equiv p_{L,\s \mu}\, \phi_{p_L}(x)\,.
\label{eq:labelPaction}
\end{align}
To see the implications of \cref{eq:labelFieldRedef}, we can Fourier transform a labeled field.  Noting that it is $p_r$ that appears in the transform, we have
\begin{align}
\phi_{p_L}(x) = \int \frac{\D^4 p_r}{(2\s\pi)^4}\, e^{-i\s  p_r \cdot x}\, \tilde{\phi}_{p_L}(p_r)\,.
\label{eq:labeledPhiFT}
\end{align}
This implies that label and residual momentum are separately conserved:
\begin{align}
\int \D^4 x\, e^{i\s (p_L -q_L)\cdot x}\, e^{i\s (p_r - q_r)\cdot x} = (2\s\pi)^4\, \delta_{p_L,q_L}\, \delta^4(p_r - q_r)\,.
\end{align}
Conservation of label momentum has a variety of consequences: 
\begin{enumerate}[label=\roman*)]
\item Emitting ultrasoft modes from (anti-)collinear modes leaves the (anti-)collinear labels unchanged.
\item Interactions among collinear fields changes labels.  
\item Collinear and ultrasoft interactions preserve the direction $n$; only a hard scattering process (which takes us outside the EFT description) can change the direction $n$.
\end{enumerate}

Now the point of this formalism should be clear. The labels provide a self consistent way to power count momenta that flow through EFT Feynman diagrams.  Taking the collinear field (see \cref{eq:pLabelpResScaling} for the scalings) as an example, we see that \cref{eq:labeledPhiFT} implies that 
\begin{align}
n\cdot \mathcal{P}\s \phi_{c,\s p_L}(x) &\,\,\sim\,\, 0 \notag\\[3pt]
\bar{n}\cdot \mathcal{P}\s \phi_{c,\s p_L}(x) &\,\,\sim\,\, \lambda^0\, \phi_{c,\s p_L}(x) \notag\\[3pt]
 \mathcal{P}_{\perp\s,\s\mu}\s \phi_{c,\s p_L}(x) &\,\,\sim\,\, \lambda\,\phi_{c,\s p_L}(x) \notag\\[3pt]
\partial_\mu\s \phi_{c,\s p_L}(x) &\,\,\sim\,\, \lambda^2 \, \phi_{c,\s p_L}(x)\,.
\end{align}
This is how the label formalism encodes power counting of derivatives within the momentum space version of the EFT.

Now that we understand the mechanics, it is useful to simplify the notation. Utilizing the $\mathcal{P}$ operator, we can now rewrite
\begin{align}
\hat{\phi}(x) = \sum_{p_L\neq 0} e^{-i\s p_L \cdot x}\, \phi_{p_L}(x) = e^{-i\s \mathcal{P}\cdot x} \sum_{p_L\neq 0} \phi_{p_L}(x) \equiv e^{-i\s \mathcal{P}\cdot x}\, \phi(x)\,,
\end{align}
where in the last step we have defined $\phi(x) = \sum \phi_{p_L}(x)$.  In other words, the $\mathcal{P}$ operator allows us to take the label dependence outside of the sum.  One consequence of this manipulation is that products of fields behave simply as
\begin{align}
\hat{\phi}(x) \, \hat{\phi}(x) = e^{i\s \mathcal{P}\cdot x} \big(\phi(x)\, \phi(x)\big)\,,
\end{align}
where the label operator acts on both fields within the parenthesis.  Conservation of label momentum is encoded in this overall phase factor, which has no impact on the squared amplitude.  Hence, we will drop the labels from here forward since there is no reason to make the labels explicit -- their only impact on the physics is to determine which components of momentum flow through an EFT vertex.  

Finally, we return to our EFT interaction of interest $\phi_c^2\, \phi_{us}$.  By separating into label and residual momentum using \cref{eq:pLabelpResScaling} and \cref{eq:pLabelpResScalingus}, we see that the collinear label is unchanged by the ultrasoft exchange.  Therefore, the EFT vertex is
\begin{align}
\includegraphics[width=0.45\textwidth, valign=c]{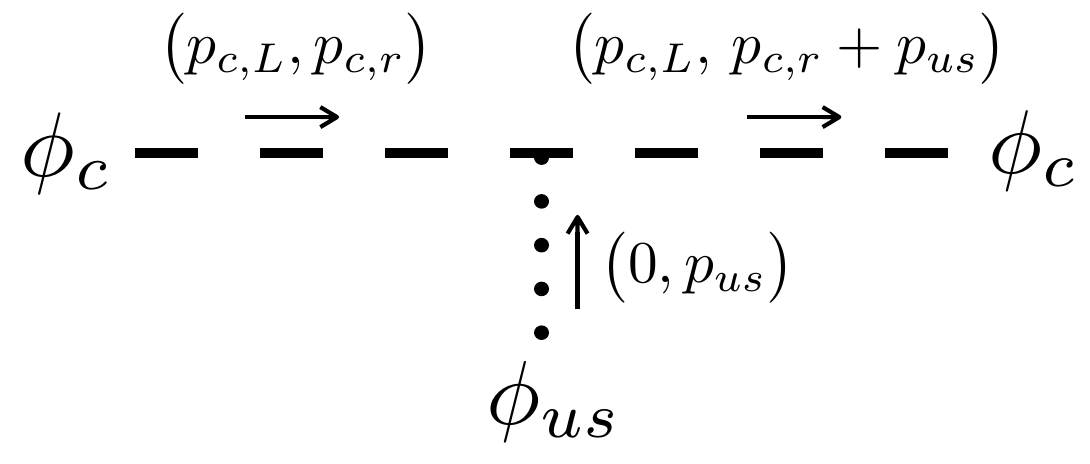} \,\,\,,
\end{align}
where the momentum flow is labeled explicitly.  There is an analogous derivation for the interaction $\phi_{\bar{c}}^2\, \phi_{us}$.  Practically, this allows us to track the residual momentum flow through the propagators that connect to this vertex.  We will make this explicit when we compute the one-loop diagrams for the Sudakov process in \cref{sec:ResummingSudakov} below.

Now that we understand the SCET degrees of freedom and their interactions, we turn to a discussion of the local operator structure relevant for our massless Sudakov example.  These contact operators encode the physics at the hard scale and provide a mechanism for injecting large momentum into the EFT.

\subsection{Local Operators and the Sudakov Process}
\label{sec:LocalOpsScalarSCET}
We need a portal that can inject a large source of energy and momentum into our EFT.  This can by modeled through the inclusion of what are usually referred to as local operators or contact operators: products of EFT fields and external current sources evaluated at a single (local) spacetime point.  Physically, a local operator is an interaction whose detailed structure cannot be resolved by the low energy degrees of freedom.  Each local operator has a coupling, \emph{i.e.}, the Wilson coefficient, whose size is determined by a matching calculation at the hard scale.  Then summation of large logs is accomplished by RG evolving these coefficients from the UV to the IR.  This will separate scales in exact analogy with the examples studied above.  Although much of what follows will be specific to our goal of summing the massless Sudakov logs, we will emphasize the general lessons as they appear.

We can determine the local operator structure and hierarchy within the EFT using symmetries and power counting.  Note that a collinear field can be multiplied by a factor of $(\bar{n}\cdot \partial)^{\s j}$ for any choice of $j$ since this derivative power counts as $\mathcal{O}(1)$.  In position space, this implies that the tower of operators can be summed into a translation along the collinear direction via the Taylor expansion:
\begin{align}
 \sum_{j=0}^{\infty} \frac{t^j}{j!} \big(\bar{n}\cdot \partial\big)^j \phi_c(x) = \phi_c\big(x+t\s\bar{n}\big)\,.
\end{align}
Therefore, our ``local'' operators in the EFT should be built out of these extended fields, and we must include an integral over the affine parameter $t$ when constructing operators.  We will see this manifest in momentum space through the momentum dependence of our Feynman rules, which is straightforward to derive with a matching calculation.

For our Sudakov process, we want a current that generates two hard back-to-back particles, \emph{i.e.}, it couples to a collinear and an anti-collinear field.  A simple way to model this process is to imagine that there is a heavy scalar $\Phi$ that decays into our light scalars $\phi$.  We can then perform a matching calculation starting with our \FT~interaction
\begin{align}
\mathcal{L}_\text{Int}^\textsc{Full} = \frac{b}{2}\, \Phi\, \phi^2  \qquad \Longleftrightarrow \qquad \includegraphics[width=0.15\textwidth, valign=c]{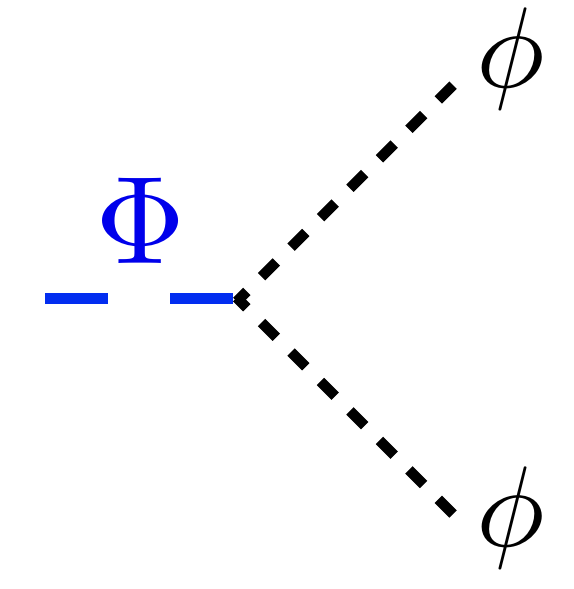} \,.
\label{eq:PhiphiphiDecayFull}
\end{align}
Matching this onto the EFT, including the convolution over the affine parameters, yields\footnote{Here we first encounter the fact that Wilson coefficients can also be functions of kinematics.  We are introducing the notation $\mathbb{C}$ to specify that it includes kinematic information.  This will be contrast with our use of $C$ below, which is the non-kinematic part of the Wilson coefficient.}
\begin{align}
\mathcal{L}_\text{Local} \supset \int \D t\, \D \bar{t}\, \mathbb{C}_2(\mu, t, \bar{t}\,\big)\, J(x)\, \phi_c\big(x + t\s \bar{n}\big)\, \phi_{\bar{c}}\big(x + \bar{t}\s n\big) \qquad \Longrightarrow \quad  \includegraphics[width=0.15\textwidth, valign=c]{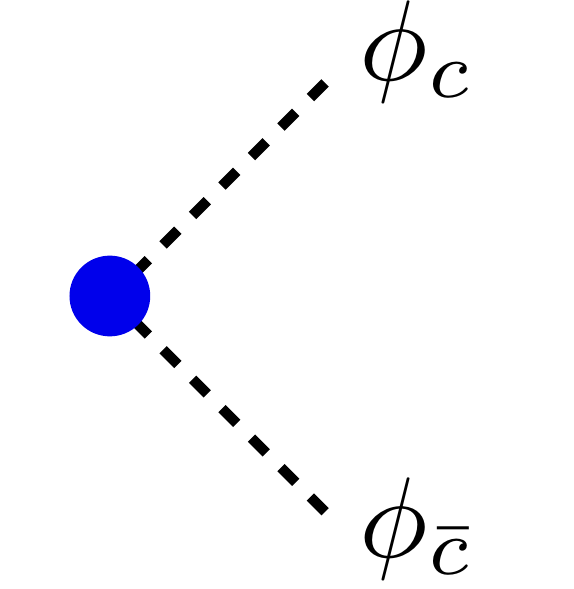}.
\label{eq:LLocalC2}
\end{align}
where $\mathbb{C}_2(\mu, t, \bar{t}\,\big)$ is the Wilson coefficient, which is explicitly a function of the RG scale $\mu$ and additionally captures the possible dependence on $t$, $\bar{t}$ (which will manifest as non-trivial momentum dependence in the Wilson coefficient),  and $J(x)$ is a static background source, whose role is to inject $M$ of energy into the light system (the blue dot in the Feynman rule).  

Given our process, we can match the \FT~given in \cref{eq:PhiphiphiDecayFull} onto the two particle local operator in the momentum space EFT given in~\cref{eq:LLocalC2}.  This requires the straightforward comparison of tree-level amplitudes:
\begin{align}
\includegraphics[width=0.15\textwidth, valign=c]{Figures/SudakovFullccBar.pdf} \quad = \quad\hspace{10pt} \includegraphics[width=0.15\textwidth, valign=c]{Figures/SudakovOpccBar.pdf} \qquad \Longrightarrow \qquad \mathbb{C}_2\big(\mu_M\big)=  b\big(\mu_M\big) = C_2\big(\mu_M\big)\,,
\label{eq:EFTOpccBarMatching}
\end{align}
where the identification between $\mathbb{C}_2$ and $C_2$ is trivial since there is no kinematic dependence.

Next, we compute one-loop corrections to the matching coefficient within the EFT.  This will require a pair of higher power local operators:
\begin{align}
\mathcal{L}_\text{Local} &\supset \int \D t_1\, \int \D t_2\, \int \D \bar{t}\, \mathbb{C}_3\big(\mu,t_1,t_2, \bar{t}\,\big)\, J(x)\, \phi_c\big(x + t_1\s \bar{n}\big)\,\phi_c\big(x + t_2\s \bar{n}\big)\, \phi_{\bar{c}}\big(x + \bar{t}\s n\big) \notag\\[5pt]
 &+\int \D t\, \int \D \bar{t}_1\, \int \D \bar{t}_2\, \overline{\mathbb{C}}_3\big(\mu,t,\bar{t}_1, \bar{t}_2\big)\, J(x)\, \phi_c\big(x + t \s \bar{n}\big)\,\phi_{\bar{c}}\big(x + \bar{t}_1 \s n\big)\, \phi_{\bar{c}}\big(x + \bar{t}_2 \s n\big) \notag\\[5pt]
&\qquad\qquad \Longrightarrow \qquad \includegraphics[width=0.15\textwidth, valign=c]{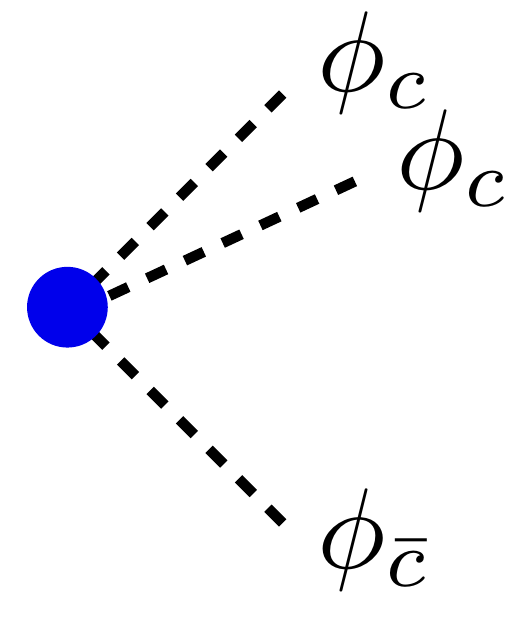} \qquad + \qquad \includegraphics[width=0.15\textwidth, valign=c]{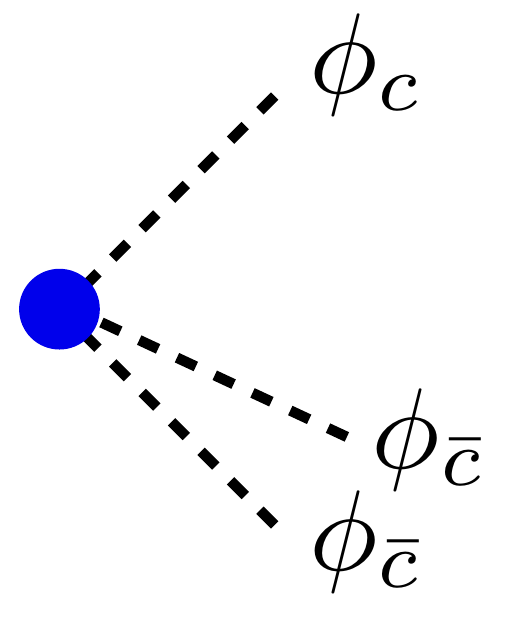} \,,
\label{eq:LLocalC3}
\end{align}
where, similar to \cref{eq:LLocalC2}, $\mathbb{C}_3$ and $\overline{\mathbb{C}}_3$ are Wilson coefficients, which can also include momentum dependence.\footnote{The position space form of these operators is given in Eq.~(3.31) of~\cite{Becher:2014oda};  since we will be able to extract the momentum space form of $C_3$ directly, we will not include the position space formulas here.}

From here forward, we will work in momentum space.  The impact of the $t$ integrals on $\mathbb{C}_3$ will yield an expansion in terms of inverse derivatives in the ``large'' momentum direction.  There is a connection to Wilson lines that will be made in \cref{sec:RealSCET} below, where we discuss how to promote $\phi$ to a gauge boson.  Note that one of the idiosyncrasies of scalar SCET (in four dimensions) is that our coupling $\phi_c^3$ is super-leading power.  This will imply that we will need a term in our local operator expansion that is higher order in power counting to capture the full physics of the massless Sudakov process at one-loop order.  Then the power counting of the diagram involving an insertion of this sub-leading power operator will be reduced by the super-leading power counting of $\phi_c^3$, see \cref{sec:ResummingSudakov} where this is made explicit.   This is in contrast with the QCD case where the gauge coupling is marginal.  Furthermore, when we move to QCD, we will find that an analogous analysis will lead us to include Wilson lines in our local operators.  Said another way, one feature that makes our toy theory non-QCD like is that the coupling expansion is tied to the power expansion.

To derive $\mathbb{C}_3$, we can again match at tree-level.  We power expand and keep only leading power contributions.
\begin{align}
 \includegraphics[width=0.15\textwidth, valign=c]{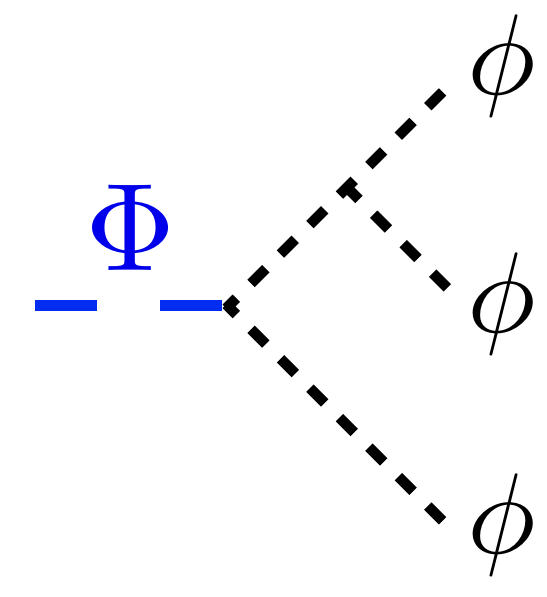} \quad+ \quad \includegraphics[width=0.15\textwidth, valign=c]{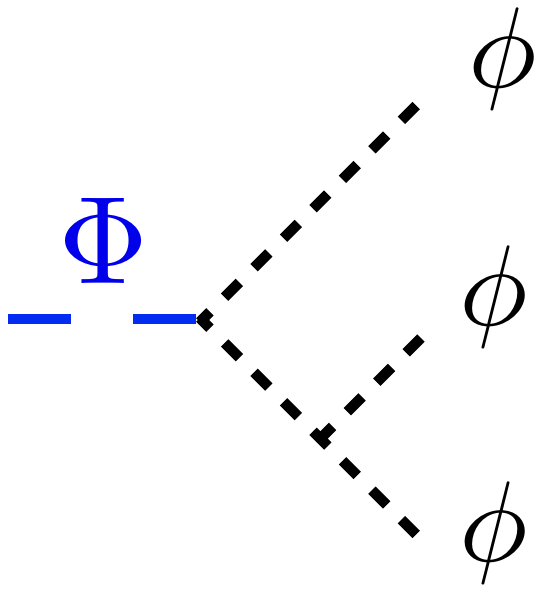} \quad = \quad\includegraphics[width=0.13\textwidth, valign=c]{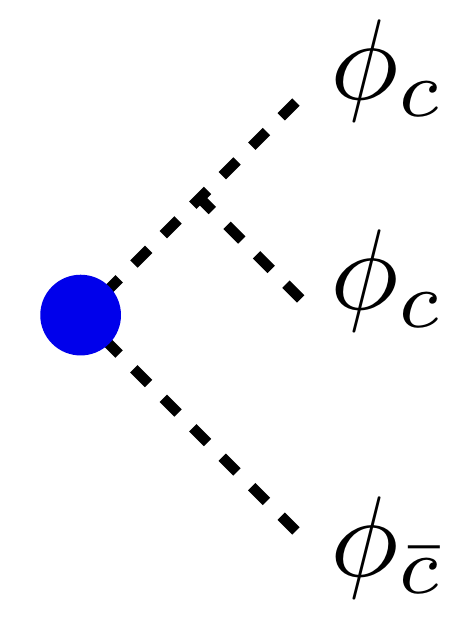} \quad +\quad \includegraphics[width=0.14\textwidth, valign=c]{Figures/SudakovOpcccBar.pdf}\,.
\end{align} 
Then noting that
\begin{align}
 \includegraphics[width=0.15\textwidth, valign=c]{Figures/SudakovFullcccBar.pdf} \quad = \quad \includegraphics[width=0.13\textwidth, valign=c]{Figures/SudakovDiagcccBar.pdf} 
 \label{eq:SCETCollinearMatchingEqualDiagrams}\,\,,
\end{align}
we can simply compute the matching coefficient by equating
\begin{align}
\includegraphics[width=0.15\textwidth, valign=c]{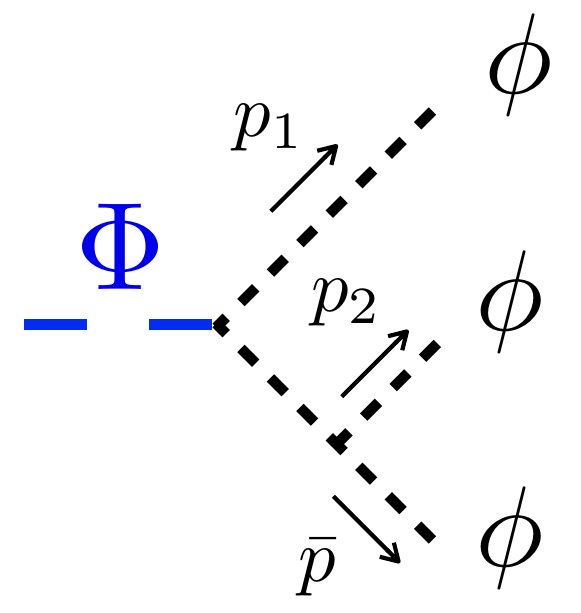}  = \includegraphics[width=0.14\textwidth, valign=c]{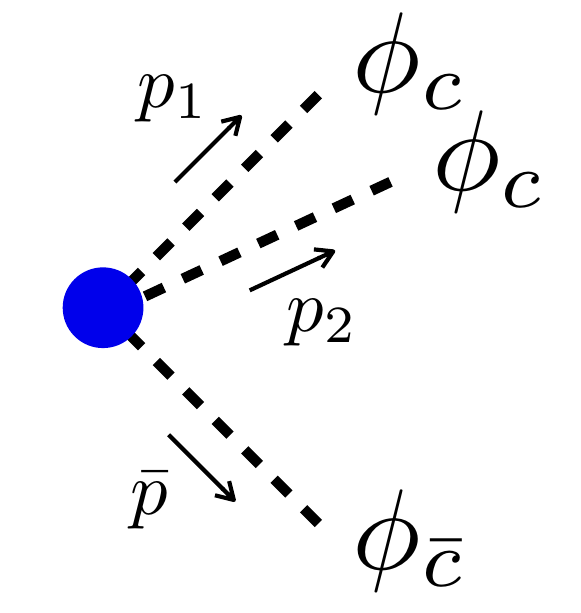} \quad\Longrightarrow\quad a\,b\, \frac{1}{(p_2+\bar{p})^2} = a\,b\, \frac{1}{\bar{n}\cdot p_2\,n\cdot \bar{p}} = \mathbb{C}_3 \,.
\label{eq:EFTOpcccBarMatching}
\end{align}
So we see that our the Wilson coefficient at the high scale is 
\begin{align}
\mathbb{C}_3\big(\mu_M\big) = a\big(\mu_M\big)\,b\big(\mu_M\big)\,\frac{1}{\bar{n}\cdot p_2\,n\cdot \bar{p}} = C_3\big(\mu_M\big)\,\frac{1}{\bar{n}\cdot p_2\,n\cdot \bar{p}} \,,
\label{eq:C3Match}
\end{align}
at the high scale where we match the \FT~to the EFT, and we have defined $C_3$ as the non-kinematic part of the Wilson coefficient.  

Next, we can perform a similar matching calculation for the diagram with a single collinear and two anti-collinear emissions:
\begin{align}
\includegraphics[width=0.15\textwidth, valign=c]{Figures/SudakovFullccBarcBar.pdf} \quad +\quad \includegraphics[width=0.15\textwidth, valign=c]{Figures/SudakovFullcccBar.pdf} \quad =\quad \includegraphics[width=0.13\textwidth, valign=c]{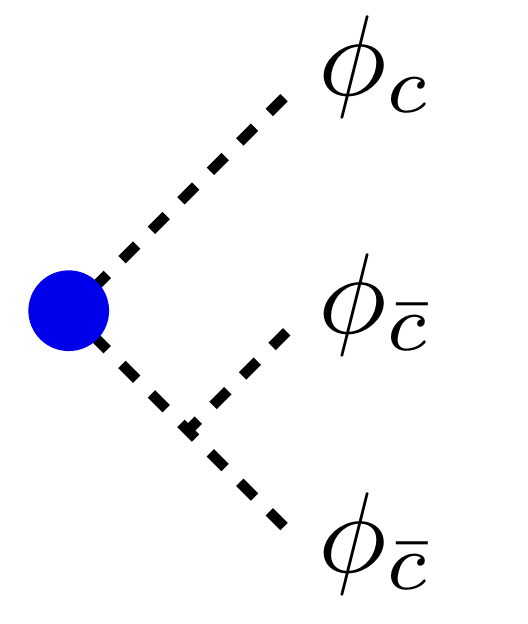} \quad + \quad\includegraphics[width=0.14\textwidth, valign=c]{Figures/SudakovOpccBarcBar.pdf}\,\,.
\end{align} 
Following the same argument as in \cref{eq:SCETCollinearMatchingEqualDiagrams}, we have
\begin{align}
\includegraphics[width=0.15\textwidth, valign=c]{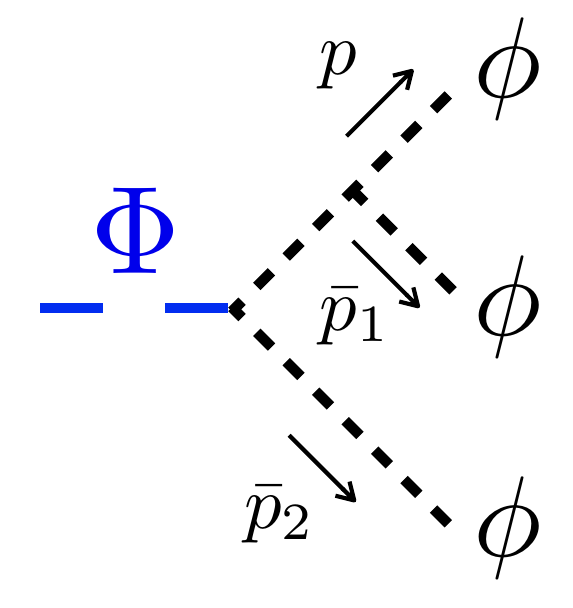}  = \includegraphics[width=0.14\textwidth, valign=c]{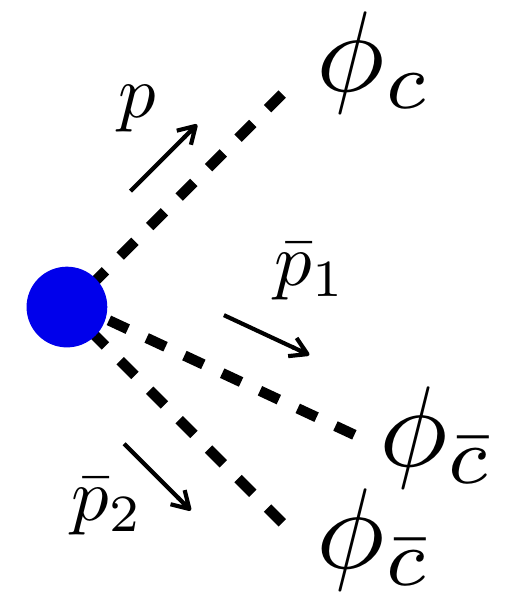} \quad \Longrightarrow \quad a\,b \frac{1}{(p+\bar{p}_1)^2} = a\,b \frac{1}{\bar{n}\cdot p\, n\cdot \bar{p}_1} = \overline{C}_3 \,,
\label{eq:EFTOpccBarcBarMatching}
\end{align}
where again matching yields
\begin{align}
\overline{\mathbb{C}}_3\big(\mu_M\big) = a\big(\mu_M\big)\,b\big(\mu_M\big)\,\frac{1}{\bar{n}\cdot p\,n\cdot \bar{p}_1} = \overline{C}_3\big(\mu_M\big)\,\frac{1}{\bar{n}\cdot p\,n\cdot \bar{p}_1} \, .  
\end{align}

These are independent local operators, so we will run their Wilson coefficients separately.  This is another difference from SCET with gauge bosons, where these coefficients will be related to each other though the enforcement of gauge invariance.  Practically, this will manifest as the appearance of Wilson lines in the QCD local operators.  Then we will interpret the Wilson lines as summing the tower of operators generated by the QCD analogy to \cref{eq:EFTOpcccBarMatching} and \cref{eq:EFTOpccBarcBarMatching}.  The implication is that the QCD versions of $\mathbb{C}_2$, $\mathbb{C}_3$, $\overline{\mathbb{C}}_3$, \dots~should be treated as the same Wilson coefficient, see \cref{eq:WilsonLinesSCET}.

The presence of inverse derivatives in the local interaction might appear strange, especially since inverse derivatives are often a harbinger of non-locality.  However, within the EFT, the components of the derivatives that appear in these denominators are ``large'' in that they power count as $\mathcal{O}(1)$.  This implies that one can not access the non-local nature of these interactions within the EFT, and as such our interpretation that these objects make a contribution to the set of SCET local operators is self-consistent.  Next, we will use these local operators to model the IR of the Sudakov process, and run the Wilson coefficients to sum the large double log.

\subsection{Summing Sudakov Logs with Scalar SCET}
\label{sec:ResummingSudakov}
Now that we have the relevant Feynman rules for the scalar SCET that will model the massless Sudakov double log, we can construct the one-loop diagrams, and use them to derive the RGE for the Wilson coefficient $C_2$.  The relevant scales in the problem are schematically 
\begin{align}
\label{eq:RGPathScalarSCET}
\includegraphics[width=0.45\textwidth, valign=c]{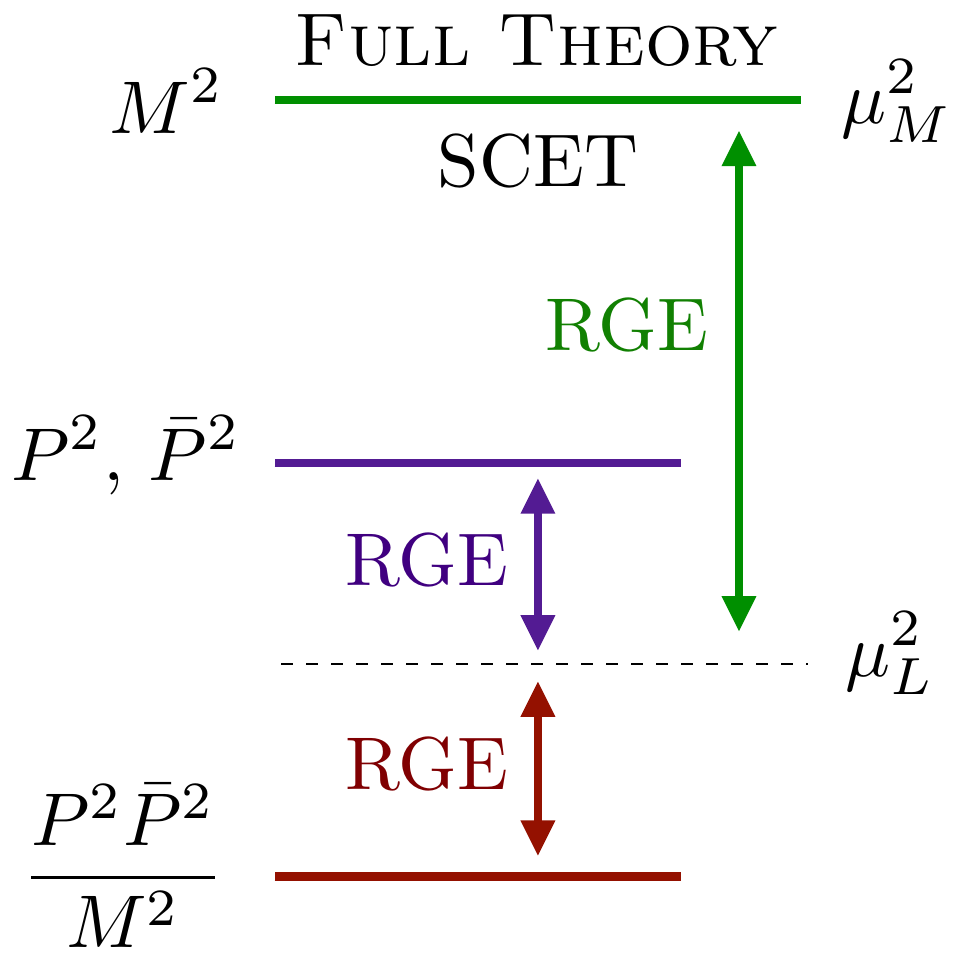}\\[-15pt]
\notag
\end{align}
As anticipated when we performed our regions analysis in \cref{sec:RegionsMasslessSudakov}, there will be four contributions: hard with canonical scale $M$, (anti-)collinear with canonical scale $P^2$ \big($\bar{P}^2$\big), and ultrasoft with canonical scale $P^2\,\bar{P}^2/M^2$.  SCET will allow us to separate scales at $\mu_M$, to absorb the hard scale into a Wilson coefficient through matching, and then sum the logarithms within the EFT by running them to a common scale $\mu_L$.  For consistency, we will only run $C_2$, since it is the Wilson coefficient for the leading power local operator that generates our two light-like directions.  The operators with Wilson coefficients $\mathbb{C}_3$ and $\overline{\mathbb{C}}_3$ are higher power, so their running begins as the next loop order.   Along the way, we will carefully track power counting, to justify that we have derived the leading power contributions to our RGE.  We will need a subset of the interactions derived above in \cref{eq:LIntScalarSCET}:
\begin{align}
\mathcal{L}_\text{EFT}^\text{Int} \supset \,\frac{\ac}{3!}\,\phi_c^3 + \,\frac{\acb}{3!}\,\phi_{\bar{c}}^3 + \frac{\as}{2}\,\phi_c^2\,\phi_{us}+ \frac{\asb}{2}\,\phi_{\bar{c}}^2\,\phi_{us}  \,,
\label{eq:SCETLInt}
\end{align}
where $\ac = \acb = \as = \asb = a$ at the matching scale, and we have not included the $\phi_{us}^3$ interaction since it does not play a role in our one-loop calculation.  Two classes of diagrams contribute at one loop, a pair where a (anti-)collinear particle runs in the loop (see \cref{eq:SCETcollinearloop} and \cref{eq:SCETanticollinearloop}), and one where an ultrasoft particle is exchanged between the two light-like lines (see \cref{eq:softDiag}).

Before we calculate, we should understand how power counting determines which diagrams contribute to the scalar Sudakov process at one loop.  We will choose to assign a power counting to the current $J$, although this is not typically done since all that matters is power counting relative to the lowest order diagram that reproduces the process of interest.\footnote{Obviously, one does not want the source to power count at O(1), since its purpose is to simply model a hard energy injection into the EFT.  One way of thinking about what we are doing here is that we are modeling an exclusive matrix element utilizing the Lagrangian of the EFT degrees of freedom.  The source $J$ cannot play a non-trivial role since it is external to the EFT.   Any conclusions you might try to draw from the power counting of $J$ are irrelevant to our goal of modeling the Sudakov process.}  First, note that our local operator given in \cref{eq:LLocalC2} is schematically of the form
\begin{align}
\int \D^4 x\, J\,\phi_c\, \phi_{\bar{c}} \,\,\sim\,\, \lambda_J \times \frac{1}{\lambda^{2}}\,,
\label{eq:EFTcouplingscalingsC2}
\end{align}
where we have used that $\D^4 x \sim \lambda^{-4}$ for collinear fields, and $\lambda_J$ is the power counting associated with the current.  Our goal is to sum corrections to this tree-level operator.  We know from the full theory calculation leading to \cref{eq:IMasslessSudakovCombined}, that the one-loop corrections will yield a double logarithm $\log^2 \lambda$ at leading power.  This is in contrast with our heavy-light example above in \cref{eq:SepScalesHLlog}, where our full theory yielded a power suppressed correction of the form $\lambda \,\log \lambda$.  We should therefore choose the power counting of $J$ such that this operator power counts as $\sim \mathcal{O}(1)$.  Concretely, this implies that $\lambda_J \sim \lambda^2$, which then feeds into the determination for the power counting of the local operators in \cref{eq:LLocalC2} and \cref{eq:LLocalC3} that will contribute to our process of interest:
\begin{align}
\int \D^4 x\, J\,\phi_c\, \phi_{\bar{c}} \,\,\sim\,\,  \mathcal{O}(1)  \quad \qquad \int \D^4 x\,J\,\phi_c^2\, \phi_{\bar{c}} \,\,\sim\,\, \lambda\quad \qquad \int \D^4 x\,J\,\phi_c\, \phi^2_{\bar{c}}\,\, \sim\,\,\lambda\,,
\label{eq:EFTcouplingscalings}
\end{align}
where again we have used that $\D^4 x \sim \lambda^{-4}$ for collinear fields and $\lambda_J \sim \lambda^2$.

We also need the power counting for the interactions in \cref{eq:SCETLInt}:
\begin{align}
\int \D^4 x\, \phi_c^3 \,\,&\sim\,\, \frac{1}{\lambda} \,\hspace{69pt} \int \D^4 x\, \phi_{\bar{c}}^3 \,\,\sim\,\, \frac{1}{\lambda} \,\notag\\[8pt]
 \int \D^4 x\, \phi_c^2 \,\phi_{us} \,\,&\sim\,\,  \mathcal{O}(1) \,\qquad\quad \int \D^4 x\, \phi_{\bar{c}}^2 \,\phi_{us} \,\,\sim\,\,  \mathcal{O}(1)\,,
\label{eq:EFTcouplingscalings2}
\end{align}
where we again use $\D^4 x \sim \lambda^{-4}$ as relevant for collinear momenta, since these dominate the Fourier transform kernel in \cref{eq:FTkernel}.  The operators that scale as $\lambda^{-1}$ are super-leading power.  Therefore, sub-leading power local operators multiplied by a super-leading power interaction can contribute at leading power.  Working with super-renormalizable couplings in an EFT is analogous to working in a relativistic theory where one can construct a dimensionless operator from the combination of $a^2/M^2$, a diagram that involves two insertions of a mass dimensionful couplings $a$, and an insertion of a higher dimension operator suppressed by $1/M^2$.

Now our task is to find one-loop contributions to \cref{eq:LLocalC2}, which scale as $\mathcal{O}(1)$.  Note since $a$ carries unit mass dimension, the appropriate compensating factors of $1/M$ must (and will) be generated by a combination of the matching coefficients and the loops themselves.  Computing these loops, extracting the RGE equations, and solving them is the topic of the rest of this section.
  
\subsubsection*{The Collinear Diagrams}
 First, we will analyze the diagram with a collinear mode running in the loop.  Using the Feynman rules and power counting of the previous section, the leading contribution is\footnote{Naively, one might expect a symmetry factor of $1/2$ for this diagram.  However, the SCET $\phi_c^2\,\phi_{\bar{c}}$ vertex treats the two $\phi_c$ states differently, so this accounts for its absence.}
\begin{align}
\includegraphics[width=0.15\textwidth, valign=c]{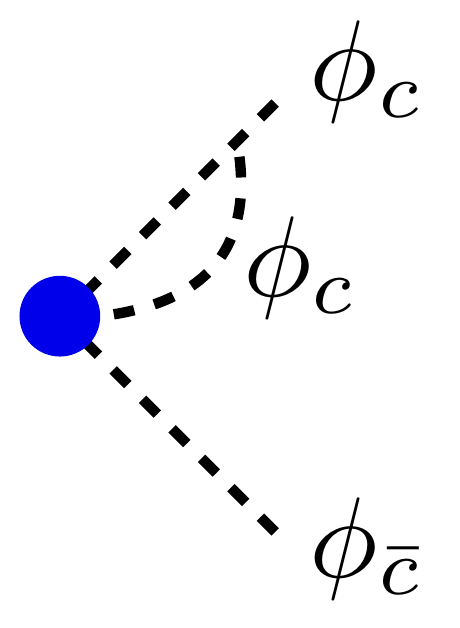} &=\,\,  \int \frac{\D^4 \ell}{(2\s\pi)^4} \,\mathbb{C}_3(\mu,n\cdot \ell,\bar{n}\cdot\ell) \,\ac(\mu)\, \frac{1}{\ell^2} \,\frac{1}{(\ell+p)^2}\notag\\[8pt]
& =\,\,  \frac{C_3(\mu)\,\ac(\mu)}{ n\cdot \bar{p}} \int \frac{\D^4 \ell}{(2\s\pi)^4} \,\frac{1}{\bar{n}\cdot \ell}\, \frac{1}{\ell^2} \,\frac{1}{(\ell+p)^2}\,,
\label{eq:SCETcollinearloop}
\end{align}
where in the second step we have used the matching coefficient derived in \cref{eq:C3Match}.  Note that the two propagator factors are from the loop, while the additional kinematic denominator comes from the Wilson coefficient.  This diagram is built out of an insertion of the local operator $\phi_c^2\,\phi_{\bar{c}}$ and the interaction $\phi_c^3$.  The scalings in \cref{eq:EFTcouplingscalings} and \cref{eq:EFTcouplingscalings2} imply that the diagram scales as $\mathcal{O}(1)$.

Next, we can compute the diagram where an anti-collinear mode runs in the loop:
\begin{align}
\includegraphics[width=0.15\textwidth, valign=c]{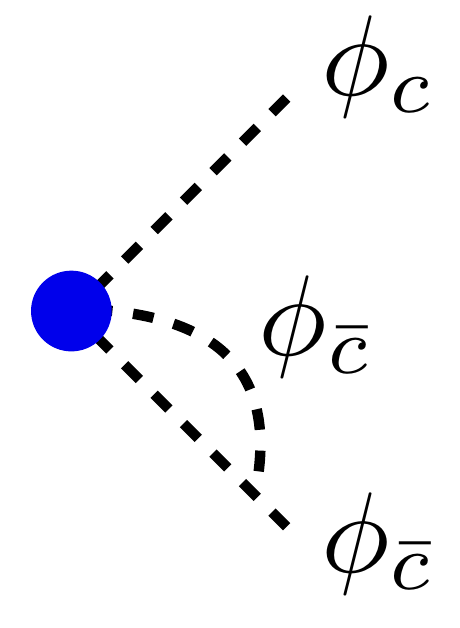} = \,\, \frac{\overline{C}_3(\mu) \, \acb(\mu )}{\bar{n}\cdot p} \int \frac{\D^4 \ell}{(2\s\pi)^4}\, \frac{1}{ n\cdot \ell}\, \frac{1}{\ell^2}\, \frac{1}{(\ell+\bar{p})^2}\,.
\label{eq:SCETanticollinearloop}
\end{align}
Similar to \cref{eq:SCETcollinearloop}, this contribution scales as $\mathcal{O}(1)$.

Although we are renormalizing $C_2$, these diagrams are proportional to $C_3$ and $\overline{C}_3$, implying operator mixing.  As already stated, we can self-consistently ignore the running of $C_3$ and $\overline{C}_3$ at this loop order since they are high power operators.  

\subsubsection*{The Ultrasoft Diagram}
Now we turn to the remaining SCET diagram at leading power and one loop, which involves an ultrasoft exchange.  To get the correct factors for the soft sector diagram,\footnote{In gauge theory, these considerations are streamlined through the use of Wilson lines, see \cref{sec:SoftWilsonFact}.  Wilson lines including scalars show up in supersymmetric theories, see \emph{e.g.}~\cite{Maldacena:1998im}.} we need to be careful about momentum labels.  We separate label and residual momentum power counting as above, see \cref{eq:pLabelpResScaling} and \cref{eq:pLabelpResScalingus}:
\begin{align}
p &= (p_{L}, p_{r}) \sim \Big(\big(0,1,\lambda\big), \big(\lambda^2,\lambda^2,\lambda^2\big)\Big) \notag\\[6pt]
\bar{p} &= \big(\bar{p}_L, \bar{p}_r\big) = \Big(\big(1,0,\lambda\big), \big(\lambda^2,\lambda^2,\lambda^2\big)\Big) \notag\\[6pt]
p_{us} &= \big(0, p_{us,r}\big) = \Big(0, \big(\lambda^2,\lambda^2,\lambda^2\big)\Big)\,.
\end{align}
Then the momentum flowing through the collinear and anti-collinear propagators respectively is given by the inverse of the $\Box$ operator appropriately power counted for our setting
\begin{align}
\Box_{c+us}^{-1} &= \big(\mathcal{P}_c + p_{r} + p_{us}\big)^{-2} = \big(\mathcal{P}_{c,\perp}^2 + \bar{n}\cdot \mathcal{P}_c\, n\cdot p_{us,r}\big)^{-1} + \mathcal{O}\big(\lambda^3\big) \notag \\[10pt]
\Box_{\bar{c}+us}^{-1} &= \big(\mathcal{P}_{\bar{c}} + \bar{p}_{r} + p_{us}\big)^{-2} = \big(\mathcal{P}_{\bar{c},\perp}^2 + n\cdot \mathcal{P}_{\bar{c}} \,\bar{n}\cdot p_{us,r}\big)^{-1} + \mathcal{O}\big(\lambda^3\big)\notag \\[10pt]
\Box_{us}^{-1} &= p_{us}^{-2}\,.
\end{align}
Then we can apply this to the propagators appearing in our loop diagram with the identification $p_{us} = \ell$.  We point the collinear momentum $p$ out of the diagram, while $\bar{p}$ points in as above, and we take $p^2 = \bar{p}^2 = 0$, which implies that the $\mathcal{P}_{c,\perp}^2 =0$.  The ultrasoft propagator simply implies a factor of $1/\ell^2$.  The resultant ultrasoft exchange EFT diagram is then
\begin{align}
\includegraphics[width=0.15\textwidth, valign=c]{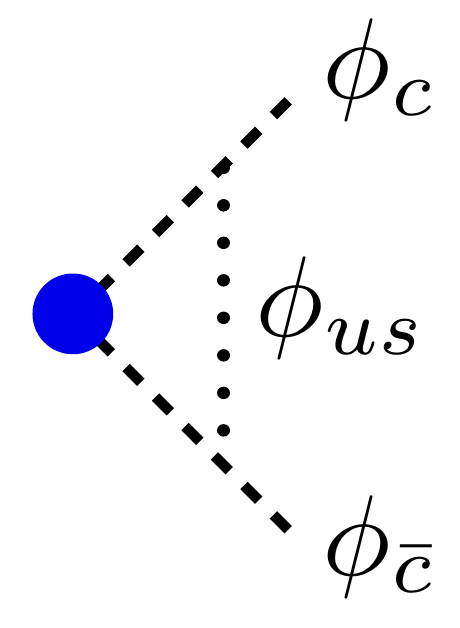} 
&= \,\,C_2\, \as\,\asb \int \frac{\D^4 \ell}{(2\s\pi)^4}\, \frac{1}{\ell^2}\,\frac{1}{\bar{n}\cdot p\, n\cdot \ell}\,\frac{1}{ n\cdot \bar{p}\, \bar{n}\cdot \ell}\,.
\label{eq:softDiag}
\end{align}
This diagram is built out of an insertion of the local operator $\phi_c\,\phi_{\bar{c}}$ and the interactions $\phi_c^2\,\phi_{us}$ and $\phi_{\bar{c}}^2\,\phi_{us}$.  Using the scalings in \cref{eq:EFTcouplingscalings} and \cref{eq:EFTcouplingscalings2}, we find that this contribution power counts as $\mathcal{O}(1)$, and so it contributes at the same order as the (anti-)collinear diagrams.  This completes the relevant leading power one-loop diagrams required to sum the Sudakov double log.  Deriving the RGEs is the topic of the next section.

\subsubsection*{Summation}
Now we have the one-loop leading power diagrams within scalar SCET that model the IR of the Sudakov process, \cref{eq:SCETcollinearloop}, \cref{eq:SCETanticollinearloop}, and \cref{eq:softDiag}.  Note that these loops are all IR divergent (in fact they are scaleless when $p^2 = \bar{p}^2 = 0$).  Therefore, we must regulate the IR to extract the UV divergence.  We might as well regulate them using same procedure as in \cref{eq:MasslessSudakovRegionsEval}, where we took $p$ and $\bar{p}$ slightly off-shell.  As mentioned before, one can think of this physically as a two jet final state, with non-vanishing jet masses, whose scale is set by $-p^2 = P^2\neq 0$ and $-\bar{p}^2 = \bar{P}^2\neq 0$.  So we will add IR mass regulators by hand to these integrals.
  
Before we renormalize, we note that the \FT~wave function renormalization vanishes
\begin{align}
\includegraphics[width=0.25\textwidth, valign=c]{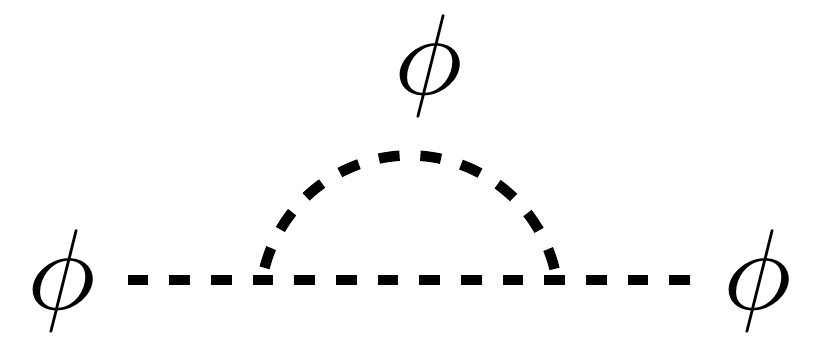} = \frac{1}{2}\,a^2 \int \frac{\D^4 \ell}{(2\s\pi)^4} \frac{1}{\ell^2}\frac{1}{(p+\ell)^2} = \frac{i\s a^2}{16\s\pi^2}\left(\frac{1}{\epsilon} + \log\frac{\mu^2}{p^2} + 1\right) \,,
\end{align}
where the $1/2$ is a symmetry factor.  Since this is not proportional to $p^2$, it yields mass renormalization.\footnote{Note that this comment regarding mass renormalization is technically not quite right since we are regulating the IR with $p^2$.  If $p^2 = 0$, then the particle is on-shell you get scaleless integral that vanishes.  When using dim reg, a contribution to the mass of $\phi$ requires a non-zero mass.}  Exactly the same argument holds in the EFT.  Therefore, one-loop renormalization only requires subtracting the divergences from vertex corrections.

\vspace{5pt}\mybox{
\begin{itemize}
\item {\bf Exercise:}  Convince yourself that the one-loop wave function renormalization for the collinear scalar vanishes in SCET.  There is a possible contribution to consider from both a (anti-)collinear loop and a soft loop.
\end{itemize}}

Next, we define counter terms $C_2^0 = Z_2\,C_2$, see \cref{eq:DefCounterTerms} above.  Then we can use $Z_2$ to compute the anomalous dimension for the $C_2$ Wilson coefficient within the EFT.  We do this by summing the tree plus collinear, anti-collinear, and ultrasoft diagrams, 
\begin{align}
&\includegraphics[width=0.13\textwidth, valign=c]{Figures/SudakovOpccBar.pdf} \quad +\quad  \includegraphics[width=0.1\textwidth, valign=c]{Figures/SudakovIntEFTcccBar.pdf} \quad +\quad  \includegraphics[width=0.1\textwidth, valign=c]{Figures/SudakovIntEFTccBarcBar.pdf} \quad +\quad \includegraphics[width=0.1\textwidth, valign=c]{Figures/SudakovIntEFTccBarus.pdf} \quad +\quad  \text{counterterm} \notag\\[10pt]
&\hspace{70pt} =  C_2\,\left[1+\left(\ac\,\frac{C_3}{C_2}\,\mathcal{I}_c + \acb\,\frac{\overline{C}_3}{C_2}\,\mathcal{I}_{\bar{c}} +\as \,\asb\,C_2\, \mathcal{I}_{us}\right) - \big(Z_2-1 \big)\right] \,,
\end{align}
whose evaluation is given in \cref{eq:MasslessSudakovRegionsEval}.  Solving for the $\overline{\text{MS}}$ counter term yields
\begin{align}
Z_2 -1
 =&\, -\frac{1}{8\s\pi^2}\frac{1}{M^2}\bigg[\, \ac\,\frac{C_3}{C_2}\left(-\frac{1}{\epsilon^2} - \frac{1}{\epsilon} \log \frac{\mu^2}{P^2}\right)+\acb\,\frac{\overline{C}_3}{C_2}\left(-\frac{1}{\epsilon^2} - \frac{1}{\epsilon} \log \frac{\mu^2}{\bar{P}^2}\right)\notag \\[8pt]
&\hspace{90pt}+ \as\,\asb \left(\frac{1}{\epsilon^2} + \frac{1}{\epsilon} \log \frac{\mu^2\,M^2}{P^2\,\bar{P}^2} \right)\bigg]\,. 
\label{eq:ZCSCET}
\end{align}

We compute the anomalous dimensions for $C_2$ following the same procedure as in \cref{eq:DeriveRGEC4} and \cref{eq:RGEOneLoopC4C6}, but keeping track of the fact that in this case $Z_2$ explicitly depends on $\log\mu^2$.   Start with
\begin{align}
\gamma = -\frac{1}{Z_2}\, \frac{\D Z_2}{\D\log \mu^2}\,.
\end{align}
Although we are not going to include the running for any of the $a$'s (these contribute to the running of $C_2$ at higher order), $C_3$, or $\overline{C}_3$ (these are power suppressed as mentioned above), we need to be cognizant of the fact that in $d = 4-2\s\epsilon$ dimensions they develop tree-level $\mu$ dependence to maintain consistent mass dimension.  Following the exact same steps that led to \cref{eq:GammaC4Classical}, we find the leading order expression
\begin{align}
& \frac{\D C_2}{\D \log \mu^2} = 0\qquad \frac{\D C_3}{\D \log \mu^2} = -\frac{\epsilon}{2}\, C_3\qquad \frac{\D \overline{C}_3}{\D \log \mu^2} = -\frac{\epsilon}{2}\, \overline{C}_3  \notag\\[8pt]
&\hspace{50pt}\frac{\D a_{c,\,\bar{c},\,us,\,\overline{us}}}{\D \log \mu^2} = -\frac{\epsilon}{2}\, a_{c,\,\bar{c},\,us,\,\overline{us}}\qquad \big[\text{leading order}\big] \,.
\end{align}
We can then iterate our solution as we did to derive \cref{eq:DDLogMuC4} above, which gives us a general form of the one-loop anomalous dimensions\footnote{For simplicity we will compute the total anomalous dimension directly, which clearly includes the effects of operator mixing since $\gamma$ depends on both $C_2$ and $C_3$.}
\begin{align}
\gamma &= \lim_{\epsilon\rightarrow 0}\frac{1}{Z_2} \left(\frac{\epsilon}{2}\, \ac \frac{\partial Z_2}{\partial \ac} +\frac{\epsilon}{2}\, \acb \frac{\partial Z_2}{\partial \acb} +\frac{\epsilon}{2}\, \as \frac{\partial Z_2}{\partial \as} +\frac{\epsilon}{2}\, \asb \frac{\partial Z_2}{\partial \asb} \right. \notag\\[7pt]
 &\left. \hspace{80pt}+\frac{\epsilon}{2}\, C_3 \frac{\partial Z_2}{\partial C_3} +\frac{\epsilon}{2}\, \overline{C}_3 \frac{\partial Z_2}{\partial \overline{C}_3}  - \frac{\partial Z_2}{\partial \log \mu^2}\right) \quad \big[\text{one loop}\big]\,.
\end{align}
Plugging in the explicit leading order expression for $Z_2$ from \cref{eq:ZCSCET}, we have
\begin{align}
\gamma = \frac{1}{8\s \pi^2}\frac{1}{M^2}\left( - \as\,\asb\,\log\frac{\mu^2\,M^2}{P^2\,\bar{P}^2}+  \ac\,\frac{C_3}{C_2} \,\log\frac{\mu^2}{P^2} + \acb\,\frac{\overline{C}_3}{C_2} \,\log\frac{\mu^2}{\bar{P}^2}\right)\,.
\end{align}

Finally, we can use the general form given in \cref{eq:RGEgeneral} above, to write down the RGE equation:
\begin{align}
\frac{\D C_2}{\D \log \mu^2}  &= \frac{1}{8\s \pi^2}\frac{1}{M^2}\left( - \as\,\asb \, C_2 \,\log\frac{\mu^2\,M^2}{P^2\,\bar{P}^2}+ \ac\,C_3 \,\log\frac{\mu^2}{P^2} + \acb\,\overline{C}_3 \,\log\frac{\mu^2}{\bar{P}^2}\right)\,,
\label{eq:Cresummed}
\end{align}
which can be solved to sum the Sudakov double log!  

The explicit full solution is not particularly illuminating, so we will not write it here, in contrast with the beautiful result provided for gauge theory in \cref{eq:SCETResummed} below.  The difference comes from the fact that $C_3$ does not run at this loop order, while in the gauge theory case, the analogous Wilson coefficient does run.

There is plenty we can still learn by studying the leading order expanded solution to our RGE in \cref{eq:Cresummed}.  Specifically, the algebra is non-trivial due to the presence of squared logarithms, so it is an instructive exercise to see explicitly how all the moving parts work together.  To this end, we solve this RGE keeping only the leading log squared terms:\footnote{This expression has been simplified using the fact that 
$$
\log^2 H-\log^2 L - 2\, \log X\, \log \frac{H}{L} = \log^2 \frac{H}{L} - 2\,\log \frac{X}{L} \log \frac{H}{L}\,.
$$
}   
\begin{align}
\!\!C_2\big(\mu_L\big)_\text{Expanded} =&\, C_2\big(\mu_M\big) + \frac{1}{16\s\pi^2}\bigg[
\Big(\as\,\asb\s C_2 -\ac\s C_3 -\acb\s \overline{C}_3 \Big)\log^2\frac{\mu_M^2}{\mu_L^2} \notag\\[10pt]
&\hspace{45pt}- 2\s\bigg(\as\,\asb \s C_2 \log\frac{P^2\,\bar{P}^2}{M^2\,\mu_L^2} -\ac\s C_3 \log \frac{P^2}{\mu_L^2}- \acb\s \overline{C}_3 \log \frac{\bar{P}^2}{\mu_L^2} \bigg) \log \frac{\mu_M^2}{\mu_L^2}\bigg] \notag\\[5pt]
&\hspace{310pt} \text{\big[LL\big]}\,,
 \label{eq:SCETSolRGEUnsimplified}
\end{align}
where all the scale dependent terms in the brackets are evaluated at the matching scale $\mu_M$.  
\vspace{5pt}\mybox{\begin{itemize}
\item \textbf{Exercise:} Derive \cref{eq:SCETSolRGEUnsimplified} from \cref{eq:Cresummed}.
\end{itemize}}
Using our tree-level matching relations at $\mu_M$, namely $\ac = \acb = \as = \asb = a$ and $C_3 =\overline{C}_3 = a\,C_2$, we can simplify \cref{eq:SCETSolRGEUnsimplified}, yielding 
\begin{align}
C_2\big(\mu_L\big)_\text{Expanded} =&\, C_2\big(\mu_M\big)\left[1 -\frac{1}{16\s\pi^2}\frac{a^2}{M^2}\left(\log^2\frac{\mu_M^2}{\mu_L^2} - 2 \log\frac{\mu_M^2}{\mu_L^2} \log\frac{M^2}{\mu_L^2}\right)\right]\qquad\text{\big[LL\big]}\,.
\label{eq:scalarSCETLLSol}
\end{align}

Next we can compute the hard matching one-loop correction at scale $\mu_M$ 
\begin{align}
\includegraphics[width=0.12\textwidth, valign=c]{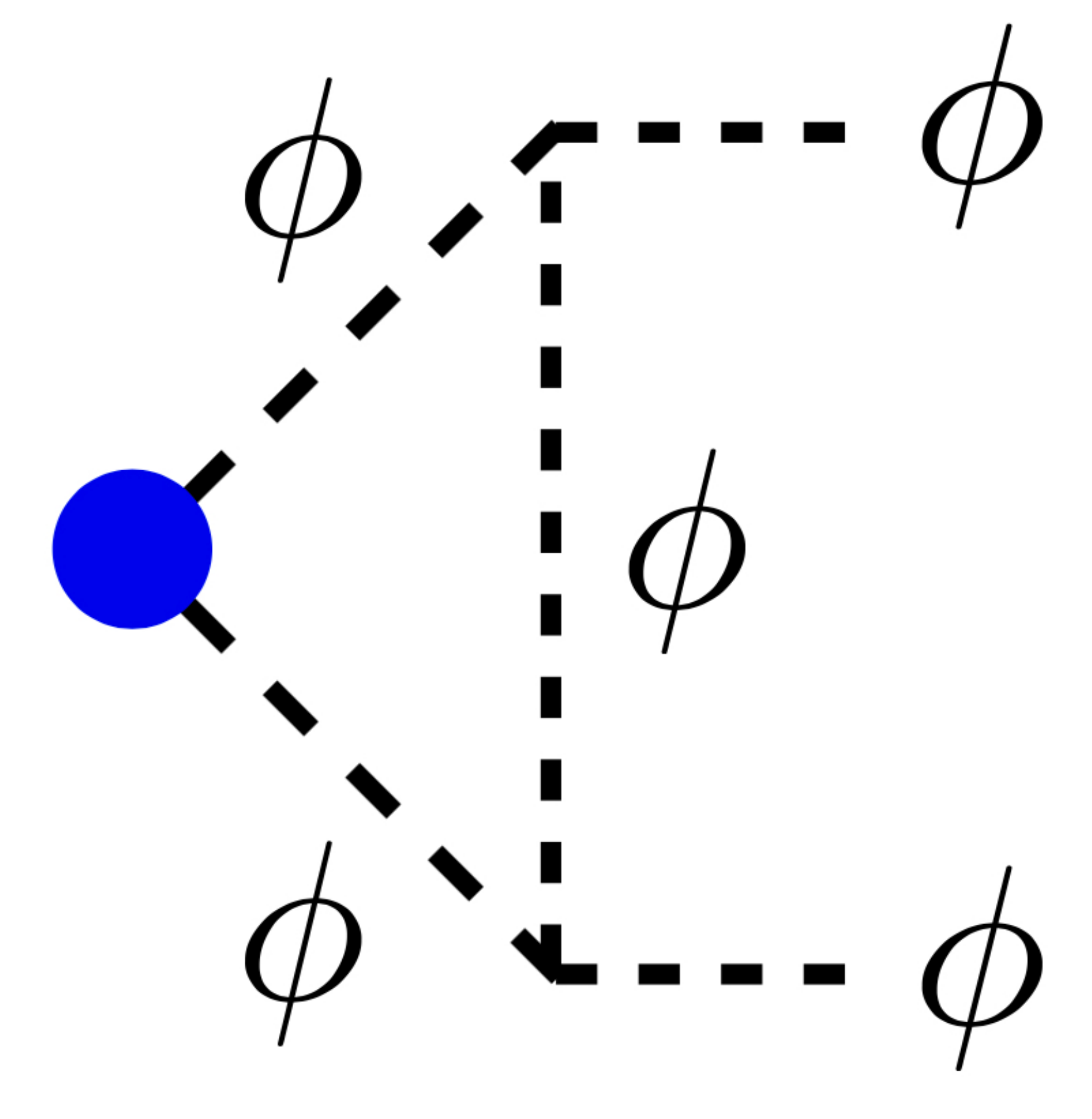} - \left( \includegraphics[width=0.1\textwidth, valign=c]{Figures/SudakovIntEFTcccBar.pdf} + \includegraphics[width=0.1\textwidth, valign=c]{Figures/SudakovIntEFTccBarcBar.pdf} +\includegraphics[width=0.1\textwidth, valign=c]{Figures/SudakovIntEFTccBarus.pdf} \right)\, =\,\, \mathcal{I}_h\,,
\end{align}
where $\mathcal{I}_h$ is the hard integral given above in \cref{eq:MasslessSudakovRegions}, and its evaluation is given in \cref{eq:MasslessSudakovRegionsEval}.  Extracting the leading log, we see that 
\begin{align}
C_2\big(\mu_M\big)_\text{Match} = b\,\frac{1}{32\s\pi^2}\frac{a^2}{M^2}\log^2\frac{\mu_M^2}{M^2}\,.
\end{align}
Then we can use this as a high scale boundary condition for our RGE.  Plugging this into the expanded solution given in \cref{eq:scalarSCETLLSol} to run to the low scale derives the expanded Wilson coefficient $C(\mu_L)$:
\begin{align}
C_2\big(\mu_L\big)_\text{Expanded} &= C_2\big(\mu_M\big)\left(1+\frac{1}{16\s\pi^2}\frac{a^2}{M^2}\log^2\frac{\mu_M^2}{M^2}\right)\notag\\[5pt]
&\hspace{20pt}\times\left[1 -\frac{1}{16\s\pi^2}\frac{a^2}{M^2}\left(\log^2\frac{\mu_M^2}{\mu_L^2} - 2 \log\frac{\mu_M^2}{\mu_L^2} \log\frac{M^2}{\mu_L^2}\right)\right]\notag\\[9pt]
&= C_2\big(\mu_M\big) \left(1 +  \frac{1}{16\s\pi^2}\frac{a^2}{M^2}\left(\log^2\frac{\mu_M^2}{M^2}- \left[\log^2\frac{\mu_M^2}{\mu_L^2} - 2 \log\frac{\mu_M^2}{\mu_L^2} \log\frac{M^2}{\mu_L^2}\right]\right) \right) \notag \\[9pt]
&= C_2\big(\mu_M\big) \left(1 + \frac{1}{16\s\pi^2}\frac{a^2}{M^2}\log^2\frac{M^2}{\mu_L^2} + \cdots\right) \qquad \text{\big[LL + NLO$_M$\big]}\,, 
\end{align}
where the first parenthesis of the first line is from the tree and one-loop hard scale matching, and the term in brackets is from solving the RGE.  This shows how our matching and running approach yields the low scale result for the Wilson coefficient.

Finally, to derive the full LL + NLO expanded result, we can compute the one-loop corrections at the low scale (using our $P^2$ and $\bar{P}^2$ IR regulators) to leading log order.  But this is just the corrections from $\mathcal{I}_c + \mathcal{I}_{\bar{c}} + \mathcal{I}_{us}$ evaluated at $\mu_L$, which gives us the low scale one-loop correction
\begin{align}
C_2(\mu_L) &=  C_2 \left(1 + \frac{1}{16\s\pi^2}\frac{a^2}{M^2}\left(\log^2\frac{M^2}{\mu_L^2}  -\log^2 \frac{\mu_L^2}{P^2} - \log^2 \frac{\mu_L^2}{\bar{P}^2} + \frac{1}{2} \log^2 \frac{\mu_L^2\,M^2}{P^2\,\bar{P}^2}\right)\cdots\right) \notag\\[7pt]
 &= b \left(1+  \frac{1}{16\s\pi^2}\frac{a^2}{M^2}\left(\log\frac{M^2}{P^2}\log\frac{M^2}{\bar{P}^2} \right)\right) \qquad \text{\big[NLO\big]}\,,
\label{eq:CLlowscale}
\end{align}
exactly the form we expected.

So to summarize, our expanded leading log summed massless scalar Sudakov including the high scale and low scale threshold corrections is 
\begin{align}
\hspace{-9pt}C_2\big(\mu_L\big)_\text{Expanded} &=  b\, \left[1 -\frac{1}{16\s\pi^2}\frac{a^2}{M^2}\left(\log^2\frac{\mu_M^2}{\mu_L^2} - 2 \log\frac{\mu_M^2}{\mu_L^2} \log\frac{M^2}{\mu_L^2}\right)\right] \notag \\[7pt]
&\hspace{10pt} \times \left[1 + \frac{1}{16\s\pi^2}\frac{a^2}{M^2}\left(\log^2\frac{M^2}{\mu_M^2}  -\log^2 \frac{\mu_L^2}{P^2} - \log^2 \frac{\mu_L^2}{\bar{P}^2} + \frac{1}{2} \log^2 \frac{\mu_L^2\,M^2}{P^2\,\bar{P}^2}\right)\right] \notag\\[5pt]
&\hspace{280pt} \text{\big[LL + NLO\big]}\,,
\end{align}
where we have dropped the finite terms.  This demonstrates that the separation of scales has been achieved: the first line includes the double log terms that are summed by the full solution to the RGE in \cref{eq:Cresummed}, and the fixed order logs are written in the second line.

\begin{itemize}
\item \textbf{Exercise}:  Note that while we have achieved our goal of summing the Sudakov double log, we have not actually completed our task of making all the logs small.  In particular, if we make the scale choice $\mu_M^2 \sim M^2$ and $\mu_L^2 \sim P^2, \bar{P}^2$, one fixed order log remains large due to the simultaneous presence of (anti-)collinear and ultrasoft modes.  To make all the logs small requires an extra step of matching at the collinear scale $P^2$ onto a theory with only the ultrasoft modes.  This will have the effect of introducing yet another RG scale such that it can be chosen to make the ultrasoft fixed order log small.  Your exercise is to perform this additional matching explicitly to minimize all the leading logs.  Note that we will come at this from a slightly different angle below in gauge theory by proving a factorization theorem that implements this additional matching automatically.
\end{itemize}

Note that our RGE was complicated by the fact that $C_2$ runs at one-loop, while $C_3$ does not.  This can be tracked to the fact that the power counting of the local operators in \cref{eq:EFTcouplingscalings}, which in turn results from the power counting for the scalars.  If we wanted to perform an NLL summation, we would need the set of two loop diagrams, and an operator of the schematic form $C_4\,\phi_c^3\,\phi_{\bar{c}}$ (and others) would contribute.  This is due to the interplay of our super-leading power interaction and the inherent non-renormalizability of our EFT expansion; at each order in perturbation theory higher power operators are required for consistency.  As we will show in what follows, this issue does not arise for gauge bosons, where all relevant interactions and local operators power count as $\mathcal{O}(1)$.  Furthermore, we will see that the gauge theory factorizes into a set of functions that have their own independent RGEs.  Highlighting these differences, and showing how the LL summation is performed for QCD are the subject of the next section.

\section{SCET in the Real World}
\label{sec:RealSCET}
Building on the technology developed so far, this section will highlight some new features that emerge when working with SCET for gauge theories with fermions, \emph{i.e.}, QCD.  Obviously, the two main complications that take us beyond our toy scalar theory are the fact that now particles carry spin, and that we have to keep track of charge conservation/gauge invariance.  We will explain the soft and collinear expansion for gauge bosons, we will show how to take the collinear limit of fermions, we will discuss the connections between collinear Wilson lines and local operators, and between soft Wilson lines and factorization.  All of this physics is explained in more detail elsewhere (including many applications and examples), \emph{e.g.}~\cite{Becher:2014oda, Becher:2018gno, iain_notes} for some reviews.  See~\cref{tech:Conventions} for some conventions that we will need for the first time here.

\subsection{Soft and Collinear Gauge Bosons}
We start with a non-Abelian SU$(N)$ gauge theory and assume that the only modes which contribute to our process of interest are collinear, anti-collinear, and ultrasoft.  In analogy with \cref{eq:phiExpanded}, we expand our gauge field into these separate modes as
\begin{align}
A^\mu(x) = A^\mu_c(x) + A^\mu_{\bar{c}}(x) + A^\mu_{us}(x) \,,
\end{align}
with the understanding that each field is restricted have the momentum scaling in \cref{eq:scaling}.  These expressions implicitly assume that if the theory is non-Abelian, then the fields are matrices, $A_\mu \equiv A_\mu^a\, T^a$.  

\subsubsection*{Power Counting}

Next, we can derive the scalings for the gauge fields.  If our modes are going to behave as propagating spin one degrees of freedom in the IR, then they must each have a notion of gauge invariance within the EFT.  Therefore, the SCET gauge boson kinetic terms will take the form $F_{\mu\nu}^2$, and each will have a Feynman propagator that takes the standard form (in $R_\xi$-gauge):
\begin{align}
\vev{0\big| T A_0^\mu(x) A_0^\nu(x) \big| 0} &= \int \frac{\D^4 p}{(2\s \pi)^4}\, \frac{-i\,\,\,\,}{p^2 + i0} \,\bigg[\s g^{\mu\nu} - (1-\xi)\,\frac{p^{\mu}\,p^{\nu}}{p^2}\bigg] \notag\\[5pt]
&= \int \frac{\D^4 p}{(2\s \pi)^4}\, \frac{-i\,\,\,\,}{p^4 + i0} \,\Big[\s p^2\,g^{\mu\nu} - (1-\xi)\,p^{\mu}\,p^{\nu}\Big]\,,
\end{align}
where $A_0^\mu$ is a free field.  Noting that $\D^4 p/p^4 \sim \mathcal{O}(1)$, the term in brackets in the second line exactly tracks the power counting of the combination $A^\mu\,A^\nu$, so we will use this to determine the power counting for the different components of $A^\mu$.  

First, we will infer the power counting for collinear gauge bosons $A_c^\mu$, where collinear points in the $n^\mu$ direction.  Recall that collinear momentum has virtuality $p_c^2 \sim \lambda^2$.  Taking the $\mu$ and $\nu$ indices to be in the $\perp$ direction, we find that schematically the term in brackets scales as $\lambda^2 g_{\perp} - (1-\xi)\,\lambda^2$.  This tells us that $A_{c}^{\mu_\perp} \sim \lambda$.  Next, we project $A^\mu$ onto $n$ and $\bar{n}$.  Recall that $n_\mu \, n_\nu\,g^{\mu\nu} = \bar{n}_\mu \, \bar{n}_\nu\,g^{\mu\nu} = 0$, see \cref{eq:gmunuLC}.  Then $n\cdot A_c\,n\cdot A_c \sim (1-\xi)\,\big(n\cdot p_c\big)^2 \sim \lambda^4$, and so $n\cdot A_c \sim \lambda^2$.  Similarly,  $\bar{n}\cdot A_c\,\bar{n}\cdot A_c \sim (1-\xi)\,\big(\bar{n}\cdot p_c\big)^2 \sim \mathcal{O}(1)$, and so $\bar{n}\cdot A_c \sim \mathcal{O}(1)$.  Putting this all together, we see that the power counting for the components of the collinear gauge boson scales in the same way as the components of its momentum.  One can check that the implications for the scaling of the mixed projection $n\cdot A\,\bar{n}\cdot A$ is self consistent.\footnote{One should be wary of the fact that these scalings are not gauge invariant (note the presence of $\xi$ in expressions), and additionally we are neglecting any discussion of ghosts, which can generically appear depending on the gauge choice.}

The scaling for the anti-collinear gauge field $A_{\bar{c}}^\mu$ follows from identical reasoning, with $n \leftrightarrow \bar{n}$.  The ultrasoft gauge field $A_{us}^\mu$ should not have a preferred direction since soft fields are homogeneous.  It is easy to check that all projections yield a scaling $A_{us}^\mu \sim \lambda^2$.  To summarize\footnote{A quick derivation uses that the scaling of the gauge boson had to track its momentum if there was going to be any consistent notion of a covariant derivative $D_\mu = \partial_\mu - i\s g\,A_\mu$ with which one could build $F_{\mu\nu} = (i/g) \big[D_\mu,D_\nu\big]$.  However, note the momentum of a charged field is not gauge invariant, so again we emphasize that care should be taken when interpreting these scalings.}
\begin{align}
A_c^\mu \,\,\sim\,\, \big(\lambda^2,1,\lambda\big) \qquad\qquad A_{\bar{c}}^\mu \,\,\sim\,\, \big(1,\lambda^2,\lambda\big) \qquad\qquad A_{us}^\mu \,\,\sim\,\, \big(\lambda^2,\lambda^2,\lambda^2\big)\,.
\end{align}

\vspace{5pt}\mybox{\begin{itemize}
\item \textbf{Exercise:} Convince yourself that the scaling of the components of the gauge boson should exactly follow the scaling of the components of its momentum.
\end{itemize}}

\subsubsection*{Interactions}

Now we have the tools we need to take the \FT~non-Abelian gauge boson Lagrangian and expand it by power counting in $\lambda$~\cite{Bauer:2000ew,Bauer:2000yr,Chay:2002vy,Beneke:2002ph}.  Gauge fixing and ghosts are treated in the standard way.  However, the power counting tells us what interactions are allowed by momentum conservation within our SCET theory, in exact analogy with \cref{sec:SCETIntPosSpace} and \cref{sec:SCETIntMomSpace} above.  For instance, we should multipole expand and power count our interactions as above, which tells us schematically that three-point interactions take the form (see \cref{eq:LIntScalarSCET} above):
\begin{align}
\mathcal{L}_\text{int} \supset g\,A_{us}\big(\bar{n}\cdot x \big)\, A_c\big(x\big)\, A_c\big(x\big)\,,
\end{align}
so that when working in momentum space, we must be careful to include label dependence and power counting for soft interactions with collinear gauge bosons.

There is one new feature due to the fact that the different components of the gauge field have different scalings.  The only component of the ultrasoft gauge boson that is not power suppressed with respect to the corresponding component of the collinear gauge boson is $n\cdot A_{us}\sim \lambda^2$ compared to $n\cdot A_c \sim \lambda^2$.  This tells us that when computing interactions, we should replace 
\begin{align}
A^\mu(x) \quad \longrightarrow \quad \Big( n\cdot A^\mu_c(x) + n\cdot A_{us}\big(\bar{n}\cdot x\big) \Big) \frac{\bar{n}^\mu}{2} + \bar{n}\cdot A_c(x) \frac{n^\mu}{2} + A_{c}^{\mu_\perp}\,.
\label{eq:SoftCollinearGaugeBosonPowerCounted}
\end{align}
Then a Feynman rule for interacting two hard gauge bosons with a soft gauge boson takes the form
\begin{align}
\includegraphics[width=0.25\textwidth, valign=c]{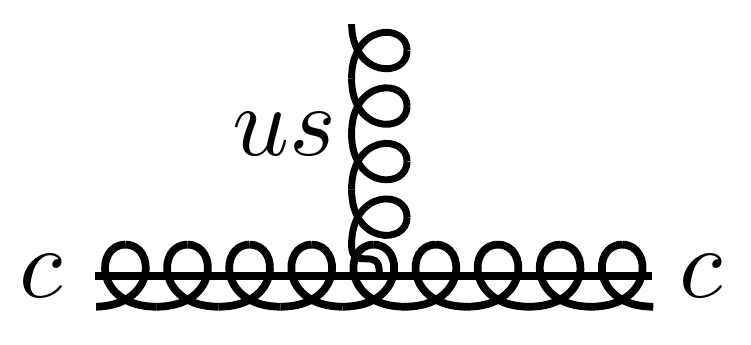} = g\,f^{abc}\,\bar{n}\cdot p\, n^\mu g^{\nu\rho}\,,
\label{eq:GaugeTripleVertex}
\end{align}
where we have drawn lines through the collinear gauge bosons as is standard practice, and the momentum $p^\mu$ is for the collinear gauge boson associated with either Lorentz index $\nu$ or $\rho$.  The ultrasoft gauge boson has a Lorentz index $\mu$, \emph{i.e.}, it must be contracted with $n^\mu$ as follows from \cref{eq:SoftCollinearGaugeBosonPowerCounted}.
Interactions between anti-collinear and ultrasoft work similarly with the replacement $c \leftrightarrow \bar{c}$ and $n\leftrightarrow \bar{n}$.  The full set of Feynman rules can be found in~\cite{iain_notes}.

\subsubsection*{Gauge Transformations}

In order to determine the gauge transformations of the SCET fields, we can multipole expand the gauge transformation in analogy with \cref{eq:multipoleExpand}, see \cref{tech:Conventions} for a review of the relevant conventions.  This will imply that each sector has its own independent gauge transformations:
\begin{align}
U_c(x) &= \exp\big(i\s \alpha_c^a(x)\,T^a\big)\notag\\[5pt]
U_{us}(x) &= \exp\big(i\s \alpha_{us}^a(x)\,T^a\big)\,,
\label{eq:UcUus}
\end{align}
and similar for anti-collinear.  If these transformations are going to consistently shift the appropriate gauge fields, they must have a scaling
\begin{align}
n\cdot \partial\, \alpha_c^a(x) & \,\,\sim\,\, \lambda^2\,\alpha_c^a(x) \notag\\[4pt]
\bar{n}\cdot \partial\, \alpha_c^a(x) & \,\,\sim\,\, \lambda^0\,\alpha_c^a(x) \notag\\[4pt]
 \partial_{\perp\s,\s\mu}\s \alpha_c^a(x) & \,\,\sim\,\, \lambda\,\alpha_c^a(x) \notag\\[4pt]
 \partial_\mu\s \alpha_{us}^a(x) & \,\,\sim\,\, \lambda^2\,\alpha_{us}^a(x) \,.
\end{align}
Then using these scalings, we can power expand the gauge invariance of the \FT~to separate soft and collinear transformations in the EFT, yielding
\begin{align}
\label{eq:SCETgaugebosonGaugeTrans}
{\rm c}: 
&\quad 
\begin{array}{l}
A_{c}^\mu \,\,\longrightarrow \,\, U_{c} \, A_{c}^\mu\, U_{c}^\dagger -  \frac{i}{g}\, \left(\left[ \partial^\mu - i\s g\, \frac{\bar{n}^\mu}{2}\,n\cdot A_{us}(\bar{n}\cdot x) \right]U_{c}\right) \,U_{c}^\dag 
\\[5pt]
A_{us}^\mu  \,\,\longrightarrow \,\, A_{us}^\mu
\end{array}
 \notag\\
\hspace{40pt}&\hspace{320pt} \,,\\
s: &\quad 
\begin{array}{l}
A_{\rm c}^\mu  \,\,\longrightarrow \,\, U_{us} (\bar{n}\cdot x)\, A_{c}^\mu\, U_{us}^\dagger(\bar{n}\cdot x) 
\\[5pt]
A_{us}^\mu  \,\,\longrightarrow \,\, U_{us}\,  A_{us}^\mu\, U_{us}^\dagger + \frac{i}{g}\,\big(\partial^\mu U_{us}\big)\, U_{us}^\dagger
\end{array}\notag
\end{align} 
where the anti-collinear gauge field does not transform under collinear transformations, and the anti-collinear transformations take the same form as in \cref{eq:SCETgaugebosonGaugeTrans} with the replacement $n \leftrightarrow \bar{n}$.  

Note that there is a global transformation under which the collinear and ultrasoft sectors transform equivalently.  We can choose to identify this global transformation with the ultrasoft local transformation.  Then in order to avoid double counting, we must set the boundary condition 
\begin{align}
U_c(n\cdot x \rightarrow -\infty) = 1\,.
\label{eq:UcBC}
\end{align}  
This is a symmetry of the EFT and holds to all powers.\footnote{One way to derive these gauge transformations is to use the background field method, see \emph{e.g.}~\cite{Abbott:1981ke} for a review.  The collinear modes see the softs as slowly varying background fields, which allows one to justify separating the gauge transformations and fixing the collinear transformation at infinity.}

\subsubsection*{Reparametrization Invariance}
Finally, we note that RPI for the gauge boson is straightforward to derive since $A^\mu$ is a Lorentz vector, and as such the total gauge field is RPI invariant.  The discussion in \cref{sec:LightCone} provided the RPI transformations for the vectors $n^\mu$ and $\bar{n}^\mu$.  Therefore, if one wishes to check RPI involving components of the gauge boson, all that must be tracked is the transformation of $n$ or $\bar{n}$.  For example,
\begin{align}
n\cdot A_c \quad \longrightarrow\quad n\cdot A_c + \Delta^\perp \cdot A_c\,,
\end{align}
under an RPI-I transformation.

Now that we have the basic gauge boson SCET building blocks, we will introduce the concept of a Wilson line.  This is an object that is very convenient to work with when one encounters charged particle lines in spacetime, as is typically the case when SCET is relevant.  As we will show below in \cref{eq:WilsonLinesSCET}, there are independent Wilson lines for the collinear and ultrasoft gauge bosons.  Collinear Wilson lines play a key role in the structure of the building blocks for local operators, while the ultrasoft Wilson lines can be used to show soft-collinear factorization at the level of the Lagrangian.  This motivates the inclusion of the physics described in the next \Primer.  First we provide a discussion of the general theory of Wilson lines.  Then we review the soft limit of gauge theory and the emergence of the universal eikonal factor.

\scenario{Wilson Lines and Eikonalization}
\addcontentsline{toc}{subsection}{\color{colorTech}{Primer~\thescenario.} Wilson Lines and Eikonalization}
\label{sec:WilsonLines}
In this \Primer, we develop two advanced topics in gauge theory.  First, we will introduce the notion of a Wilson line, and will discuss its role in maintaining the gauge invariance in situations where one has an extended object charged under a gauge group.  Then we will discuss the soft limit, and explain what it means for a gauge theory to eikonalize.  This will provide the background for understanding the ultrasoft Wilson lines in SCET.  For a more complete treatment, see \emph{e.g.}~\cite{Schwartz:2013pla}.

\subsubsection*{Wilson Lines in Gauge Theory}
A Wilson line tracks the gauge dependence along a trajectory in spacetime.  For example, if we have a U$(1)$ theory with a charged scalar, then the difference 
\begin{align}
\phi(y) - \phi(x) = e^{i\s \alpha(y)}\,\phi(y) - e^{i\s \alpha(x)}\,\phi(x)
\label{eq:gaugeAmbiguity}
\end{align}
is not gauge invariant.  To write a difference that respects gauge covariance, we define a Wilson line
\begin{align}
W(x,y) = \exp\left(i\s e\int_y^x A_\mu(s)\, \D s^\mu\right)\,.
\end{align}
It has the gauge transformation property
\begin{align}
W(x,y) \quad \longrightarrow \quad e^{i\s\alpha(x)}\,W(x,y)\,e^{-i\s\alpha(y)}\,,
\end{align}
which is straightforward to derive from the gauge transformation defined in \cref{eq:YMgaugeTrans} above.  

Using the Wilson line, we can define a gauge covariant notion of the difference between fields evaluated at two points:
\begin{align}
W(x,y)\,\phi(y) - \phi(x) \quad \longrightarrow \quad & e^{i\s\alpha(x)}\,W(x,y)\,e^{-i\s\alpha(y)}\,e^{i\s\alpha(y)}\,\phi(y) - e^{i\s \alpha(x)} \phi(x)\notag \\
& = e^{i\s \alpha(x)}\Big[W(x,y)\,\phi(y)-\phi(x) \Big]\,,
\end{align}
which eliminates the ambiguity in \cref{eq:gaugeAmbiguity}.  A difference of charged product of fields at different spacetime points is only gauge covariant if the fields are connected by Wilson lines.  

The non-Abelian extension requires path ordering for the generators since they do not commute:
\begin{align}
W = \textbf{P} \left\{\exp\left(i\s g \int_y^x A_{\mu}^a(s)\,T^a\,\D s^\mu\right)\right\}\,,
\end{align}
where $\textbf{P}$ is the path ordering symbol and $W$ is now a matrix.  Then the non-Abelian Wilson line transforms as
\begin{align}
W(x,y) \quad \longrightarrow \quad U(x)\,W(x,y)\,U(y)^\dag\,,
\label{eq:WilsonLineGaugeTrans}
\end{align}
where $U(x)$ is the gauge transformation defined in \cref{eq:UxGaugeT}.

\subsubsection*{Soft Gauge Bosons and the Eikonal Factor}
Before returning to SCET, we will briefly discuss the interactions of charged particles with soft gauge bosons.  Although our emphasis here will be on the universality of these interactions, there is also a beautiful connection one can make with charge conservation as was first shown in~\cite{Weinberg:1964ew}.  The idea of an Eikonal interaction will appear in the soft sector of SCET as we will see below.  For more details, see \emph{e.g.}~Sec.~9.5 of~\cite{Schwartz:2013pla}.

We begin with a process (illustrated by the pink blob) that involves at least one external negatively charged scalar $\phi$ with mass $m$, labeled by $j$, with momentum $p_j$:
\begin{align}
i\s \mathcal{A}_0\big(p_j\big) = \includegraphics[width=0.15\textwidth, valign=c]{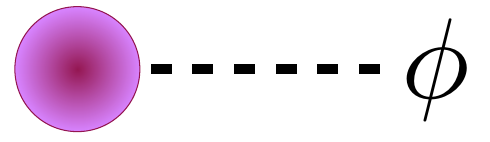}\,.
\end{align}
Then we can attach a photon to this external line, carrying away momentum $q$, and keeping $p_j$ fixed, which yields
\vspace{-10pt}
\begin{align}
i\s \mathcal{A}_j\big(p_j\big) = \includegraphics[width=0.15\textwidth, valign=c]{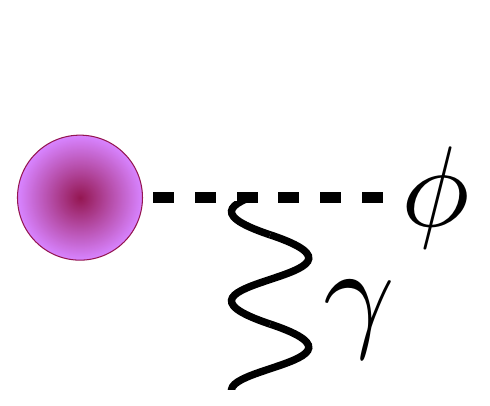} = e\,\frac{p_j^\mu + \Big(p_j^\mu - q^\mu \Big)}{\big(p_j - q\big)^2-m^2} \,\epsilon_\mu(q)\,\mathcal{A}_0\big(p_j -q\big)\,,
\end{align}
where $\epsilon_\mu(q)$ is the polarization vector for the photon.  This can be simplified by using the on-shell conditions $p_j^2 = m^2$, $q^2 = 0$, and $\epsilon\cdot q = 0$:
\begin{align}
\mathcal{A}_j\big(p_j,q\big) = -e\, \frac{p_j\cdot \epsilon}{p_j \cdot q} \, \mathcal{A}_0\big(p_j -q\big) \simeq -e\, \frac{p_j\cdot \epsilon}{p_j \cdot q} \, \mathcal{A}_0\big(p_j\big)\,,
\end{align}
where in the last step we took the soft limit for the photon momenta, \emph{i.e.}, $\big|q \cdot p_j\big| \ll \big| p_k\cdot p_m \big|$, for all external momenta $p_k$, $p_m$.  

Attaching the photon to a charged loop suffers extra suppression since those propagators are off-shell, so deriving the leading soft-photon amplitude only requires considering interactions with the external charged legs.  This yields
\begin{align}
\mathcal{A} \simeq e \, \mathcal{A}_0 \,\Bigg[\,\sum_{j\,\in\, \text{in}} Q_j \,\frac{p_j\cdot \epsilon}{p_j\cdot q} - \sum_{j\,\in\, \text{out}} \, Q_j\,\frac{p_j\cdot \epsilon}{p_j\cdot q}\, \Bigg]\,,
\label{eq:EikonalFactor}
\end{align}
where $Q_j$ is the charge of the $j^\text{th}$ state, and the sum is taken over all the external charge particles in the diagram.

This is an amazing result -- given an amplitude $\mathcal{A}_0$, it is trivial to then compute a new amplitude $\mathcal{A}$ that includes one additional soft photon by simply summing over $\mathcal{A}_0$ multiplied by an ``eikonal factor'' $p\cdot \epsilon/ p\cdot q$ for each charged external line.  Moreover, the gauge boson only knows about the charge of the lines it is interacting with since the eikonal factor is independent of the mass, spin, or any other properties of the external lines, \emph{e.g.}~if the line is associated with a fundamental or composite object.  We will use this universality to our advantage in SCET, where we will see that modeling the physics associated with the soft EFT gauge bosons relies on Wilson lines which are constructed to sum an infinite number of eikonal interactions.

Before we discuss the details of soft Wilson lines, we will work out the detailed properties of collinear Wilson lines, and will emphasize their role in building the local operators in SCET.

\subsection{Collinear Wilson Lines and Local Operators with Gauge Bosons}
\label{eq:WilsonLinesSCET}
The goal of this section is explain the role of collinear Wilson lines in SCET.  We will avoid the minor additional complications that arise for non-Abelian gauge theory by working with a $\text{U}(1)$ gauge theory.  Our model contains a heavy charged scalar $\Phi$ of mass $M$ with momentum $p = \big(M, \vec{0}\,\big)$ that can then decay to a light massless charged scalar $\phi$ and a photon $\gamma$ using an insertion of a scalar current $J = \Phi^\dag\, \phi$ (for brevity, we have set the Wilson coefficient for this heavy-light current insertion to unity):
\vspace{-10pt}
\begin{align}
 \includegraphics[width=0.2\textwidth, valign=c]{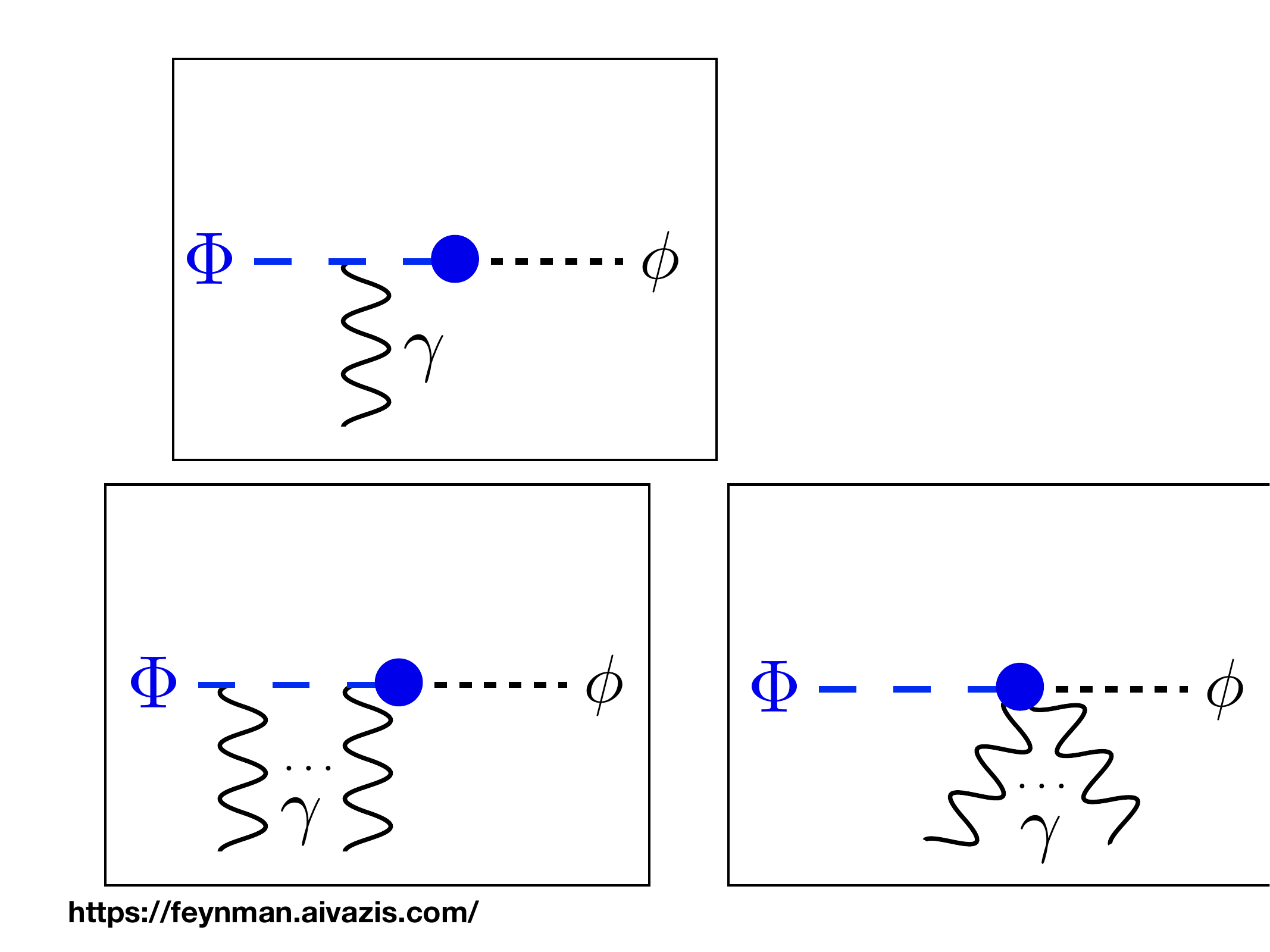} &= i\s e\, \big(2\,p_\mu + k_\mu\big) \,\epsilon_k^\mu\, \frac{i}{(p+k)^2 -M^2} = -e\, \frac{2\, p\cdot \epsilon_k}{p^2 + 2\,p\cdot k - M^2} \notag \\
 &= -e\, \frac{n\cdot p\, \bar{n}\cdot \epsilon_k}{n\cdot p\,\bar{n}\cdot k} + \mathcal{O}(\lambda) =-e\, \frac{\bar{n}\cdot \epsilon_k}{\bar{n}\cdot k} + \mathcal{O}(\lambda)\,,
\end{align}
where we note that the current insertion can absorb momentum, which allows the limit with $\phi$ and $\gamma$ collinear to be physical.  In the second line, we have defined $\lambda \sim k/M$ and power counted using the scaling for both collinear momentum and the collinear gauge field -- critically, the factor $\bar{n}\cdot A_n/\bar{n}\cdot k$ power counts as $\mathcal{O}(1)$ when $k \sim n$.   

Next, we note that we can attach $m$ collinear gauge bosons to our heavy line to yield
\vspace{-10pt}
\begin{align}
\includegraphics[width=0.2\textwidth, valign=c]{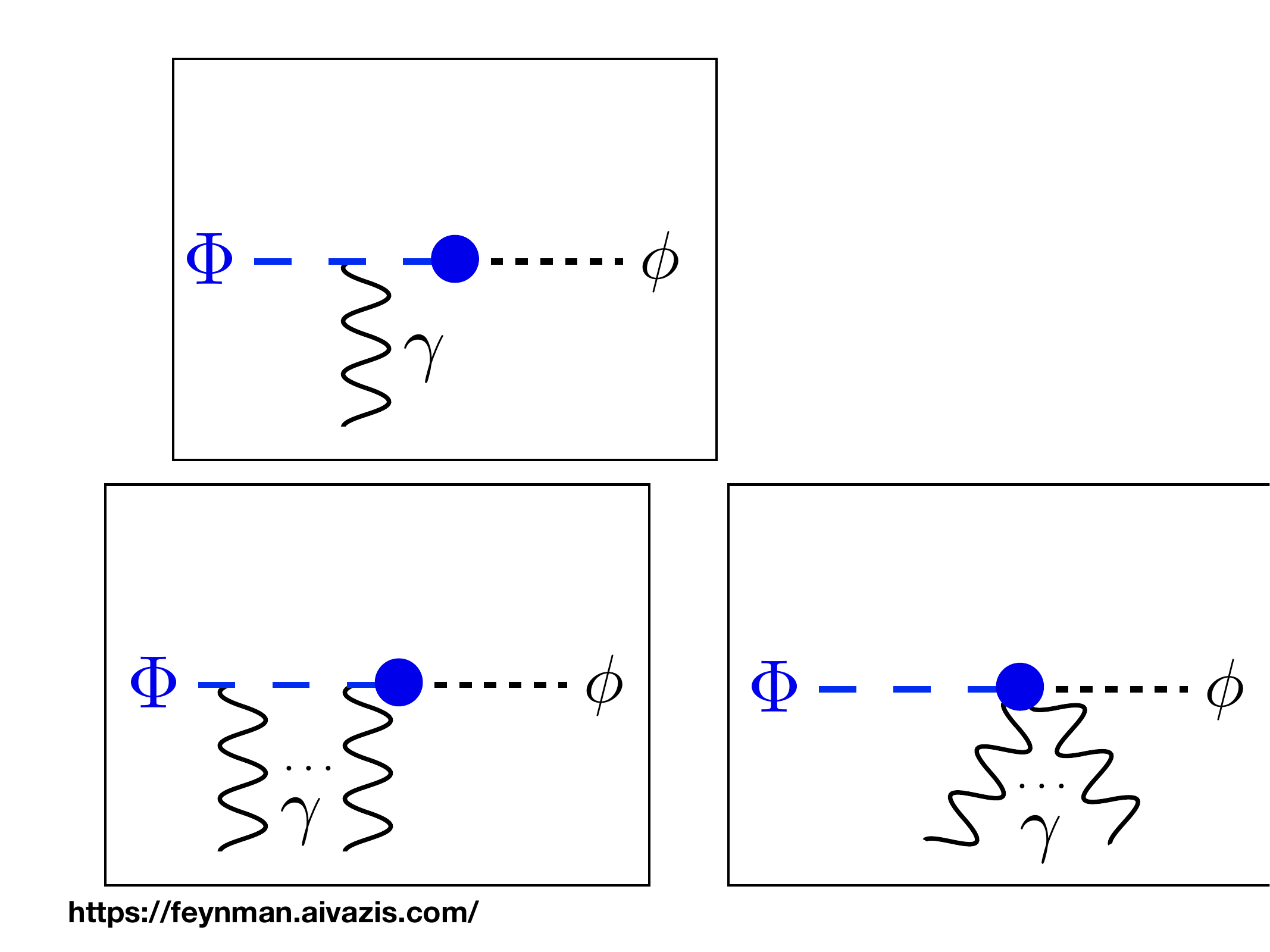} + \text{perms} &= \frac{-e\big(2\,p+ k_1\big)\cdot \epsilon_{k_1} }{\big(p+k_1\big)^2 - M^2}\,\, \cdots\,\,  \frac{-e\big(2\,p+ \sum k_m \big)\cdot \epsilon_{k_m}}{\big(p+\sum k_m \big)^2 - M^2} + \text{perms}  \notag \\
&= \left(-e\, \frac{\bar{n}\cdot \epsilon_{k_1}}{\bar{n}\cdot k_1}\right)  \,\,\cdots \,\, \left(-e\, \frac{\bar{n}\cdot \epsilon_{k_m}}{\bar{n}\cdot \left(\sum k_m \right)}\right) + \text{perms} + \mathcal{O}(\lambda)\,,
\end{align}
where ``perms'' refers to all the possible permutations of the ways to attach the photon lines to the heavy line.  Again, each of these factors scales as $\mathcal{O}(1)$.  

The key insight is to realize that there is a natural object one can write down, which accounts for the potentially infinite number of emissions, thereby generating an operator whose structure is 
\vspace{-10pt}
\begin{align}
i\s \mathcal{A} = \includegraphics[width=0.2\textwidth, valign=c]{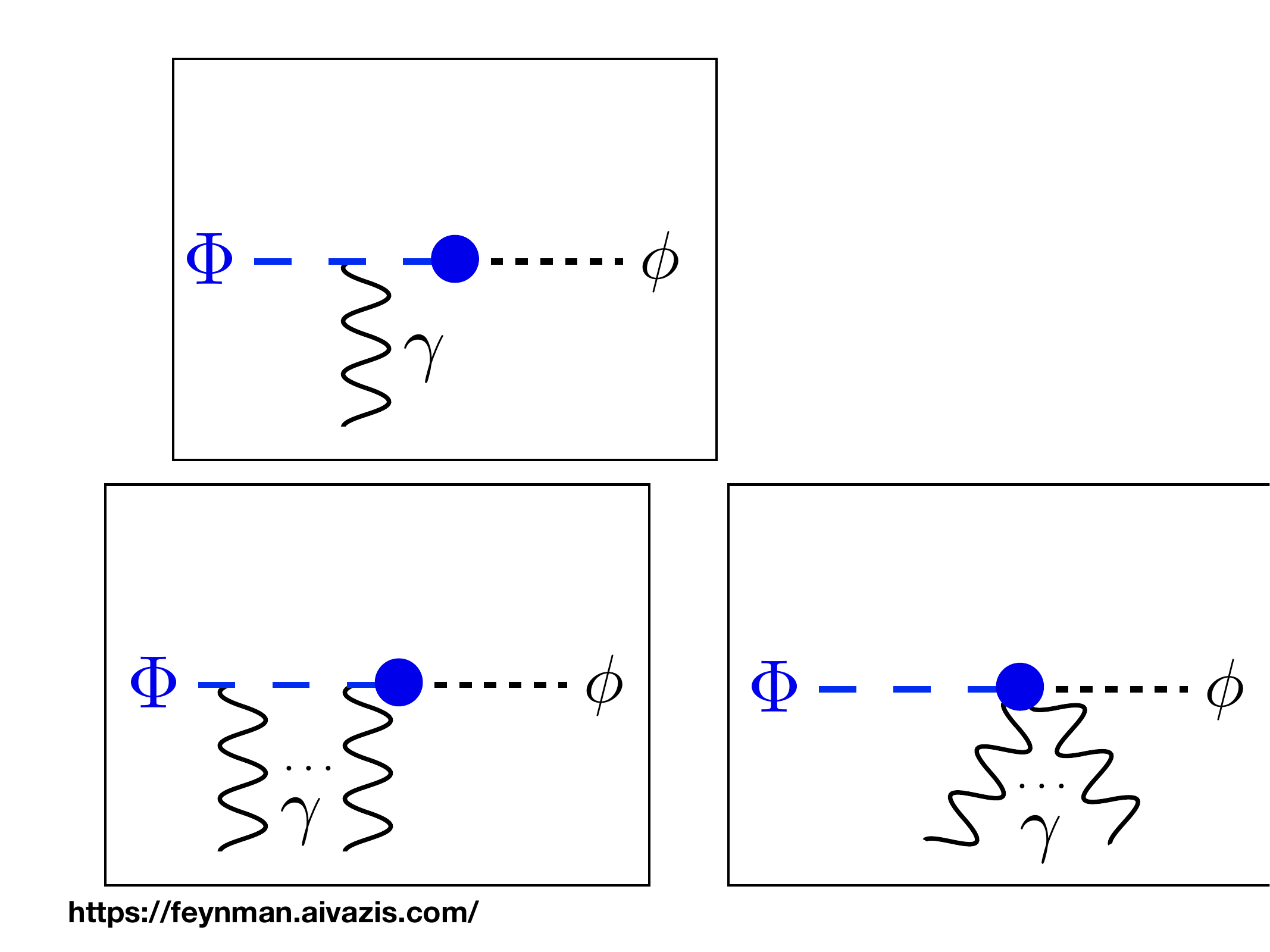} = \left(-e\, \frac{\bar{n}\cdot \epsilon_{k_1}}{\bar{n}\cdot k_1}\right)  \,\,\cdots \,\, \left(-e\, \frac{\bar{n}\cdot \epsilon_{k_m}}{\bar{n}\cdot \left(\sum k_m \right)}\right) + \text{perms} + \mathcal{O}(\lambda)\,.
\end{align}
This object is the collinear Wilson line, associated with the light-like direction $n$:
\begin{align}
W_c = \sum_m \sum_\text{perms} \frac{(-e)^m}{m!} \left(\frac{\bar{n}\cdot A_c(k_1) \cdots \bar{n}\cdot A_c(k_m)}{\big[\bar{n}\cdot k_1\big] \cdots \big[\bar{n} \cdot \sum k_m\big]}\right)\,,
\label{eq:WcMomSpace}
\end{align}
which can be Fourier transformed to position space\footnote{We have chosen explicitly to point our affine parameter along the $\bar{n}$ direction.  However, all that is required is to introduce a reference null vector $r$ that points away from $n$, such that $n \cdot r \sim \mathcal{O}(1)$.}
\begin{align}
W_c(0,-\infty) = \textbf{P} \left\{\exp\left(i\s g \int_{-\infty}^{\,0} \D s \, \bar{n}\cdot A_c\big(\bar{n}\s s\big)\right)\right\} = W_c(0)\,,
\end{align}
where again $\textbf{P}$ is the path ordering symbol, which is required to correctly lift this derivation to theories with non-Abelian gauge bosons.\footnote{We have switched notation for our gauge coupling from $e \rightarrow g$ to emphasize that all of these formulae hold for SU$(N)$ gauge theories.}  We have introduced the shorthand notation $W_c(0)$ because the Wilson line is integrated along a collinear line corresponding to one of our light-like charged states, so one end is always fixed at $-\infty$ and the other end terminates at the local interaction point.

Before discussing the detailed properties of these collinear Wilson lines, we note the relation to the scalar theory example above.  In \cref{sec:LocalOpsScalarSCET}, we saw a similar phenomena occur.  However, we paid a power counting price for each insertion.  This is related to the fact that our local operator expansion was not homogenous in power counting, which was compensated for by our super-renormalizable interaction.  For gauge theory, the fact that the factor from a collinear emission power counts as $\mathcal{O}(1)$ begs us to sum these vertices, and yields one of the crucial differences between scalars and gauge bosons.  
 
From the momentum representation of the collinear Wilson line in \cref{eq:WcMomSpace}, it is clear that
\begin{align}
\bar{n}\cdot \mathcal{P} \, W_c = \bar{n}\cdot \mathcal{P} \left(-g\,\frac{\bar{n}\cdot A_c(k)}{\bar{n}\cdot k} + \cdots\right) = -g\,\frac{\bar{n}\cdot k}{\bar{n}\cdot k} \bar{n}\cdot A_c(k) + \cdots = - \bar{n}\cdot A_c(k)\,W_c\,,
\end{align}
where we have used \cref{eq:labelPaction} for the action of $\mathcal{P}$.  We can rewrite this in the form of the Wilson line ``equation of motion''
\begin{align}
i\s \bar{n}\cdot D_c\,W_c \equiv \bar{n}_\mu\big(\mathcal{P}^\mu+ g\s A_c^\mu\big)\,W_c = 0\,.
\end{align}
Using similar manipulations, we can check the action of $i\s \bar{n}\cdot D_c$ on a product involving $W_c$ and an arbitrary operator $O$:
\begin{align}
i\s \bar{n}\cdot D_c\,\big(W_c\,O\big) &= \bar{n}_\mu\big(\mathcal{P}^\mu+ g\s A_c^\mu\big)\,\Big(W_c\,O\Big) \notag \\[8pt]
&= \Big(\bar{n}_\mu\big(\mathcal{P}^\mu+ g\s A_c^\mu\big)\,W_c\Big)\,O + W_c\,\big(\bar{n}\cdot \mathcal{P} O\big) = W_c\,\big(\bar{n}\cdot \mathcal{P} O\big)\,,
\end{align}
which can be expressed as an operator equation
\begin{align}
i\s \bar{n}\cdot D_c\,W_c = W_c \,\bar{n}\cdot \mathcal{P}\,.
\end{align}
Then using this result along with the fact that $W_c^\dag\,W_c = 1$, we have
\begin{align}
i\s \bar{n}\cdot D_c = W_c\, \bar{n}\cdot \mathcal{P}\,W_c^\dag \qquad\qquad\qquad \bar{n}\cdot\mathcal{P} = W_c^\dag\, i\s \bar{n}\cdot D_c \,W_c\,,
\end{align}
which can be inverted to derive 
\begin{align}
\frac{1}{i\s \bar{n}\cdot D_c} = W_c\,  \frac{1}{\bar{n}\cdot\mathcal{P}}\,W_c^\dag \qquad\qquad\qquad \frac{1}{\bar{n}\cdot\mathcal{P}} = W_c^\dag\,\frac{1}{i\s \bar{n}\cdot D_c} \,W_c\,.
\end{align}
These identities allow one to exchange collinear covariant derivatives for Wilson lines, and will be useful for re-expressing the canonical couplings between collinear gauge bosons and collinear charged fermions below.

Finally, we note the gauge transformation properties of collinear Wilson lines, which can be inferred from the general Yang-Mills gauge transformation for Wilson lines given in \cref{eq:WilsonLineGaugeTrans}, appropriately modified for the collinear and ultrasoft transformations given in \cref{eq:SCETgaugebosonGaugeTrans}:
\begin{align}
c:&\quad W_c(x) \,\,\longrightarrow \,\, U_c(x)\,W_c(x)\notag\\[5pt]
us:&\quad W_c(x) \,\,\longrightarrow \,\, U_{us}(x)\,W_c(x)\,U_{us}^\dag(x)\,,
\label{eq:WcGaugeTrans}
\end{align}
where for the collinear Wilson line we have used the fact that we fixed the gauge transformation matrix at $x\rightarrow - \infty$ to unity, see~\cref{eq:UcBC}, while for the ultrasoft Wilson line we have used the fact that the multipole expansion implies that only the ultrasoft field at $x$ contributes.  Said another way, the collinear Wilson line appears as a local operator from the point of view of the ultrasoft radiation.

Finally, we note that using the label momentum operator allows us to write a compact expression for the collinear Wilson line
\begin{align}
W_c(x) = \left[\,\sum_\text{perms}\exp\left(-\frac{g}{\bar{n}\cdot\mathcal{P}}\,\bar{n}\cdot A_c(x)\right)\right]\,,
\end{align}
where $1/\bar{n}\cdot\mathcal{P}$ acts on all fields to the right, yielding the denominator structure in \cref{eq:WcMomSpace}.  If rapidity regulation is required, one can use this form to incorporate the regulator introduced in \cref{sec:RegionsMassiveSudakov} as an operator expression~\cite{Chiu:2011qc, Chiu:2012ir}:
\begin{align}
W_c(x) = \left[\,\sum_\text{perms}\exp\left(-\frac{g}{\bar{n}\cdot\mathcal{P}}\, \frac{\big|\bar{n}\cdot \mathcal{P}\big|^{-\eta}}{\nu^{-\eta}}\,\bar{n}\cdot A_c(x)\right)\right]\,,
\label{eq:WcRapidityReg}
\end{align}
along with an analogous modification of the soft Wilson line given below in \cref{eq:SoftWilsonLine}.

\subsubsection*{Local Operators with Gauge Bosons}
One of the key applications of the collinear Wilson line is that it allows us to write down a novel object that can be used to build local operators involving external gauge bosons.  We introduce the combination~\cite{Bauer:2000yr, Bauer:2001ct}
\begin{align}
\mathcal{B}_{c\s\perp}^\mu(x) = \frac{1}{g}\,W_c^\dag(x)\, i D_{c\s\perp}^\mu\, W_c(x)\,,
\end{align}
where the covariant derivative only acts on the Wilson line to the right.  This operator is gauge invariant under the collinear gauge transformations in~\cref{eq:WcGaugeTrans}, and can be interpreted as a physical collinear gauge boson emitted from the hard process.  This connection is reinforced by Taylor expanding:\footnote{Often $\mathcal{B}_{c\s\perp}^\mu$ is defined in a slightly more complicated way to eliminate the $\mathcal{P}_{c\s\perp}^\mu$ term, see \emph{e.g.}~\cite{iain_notes}.}
\begin{align}
\mathcal{B}_{c\s\perp}^\mu = \frac{1}{g}\,\mathcal{P}_{c\s\perp}^\mu + A_{c\s\perp}^\mu - \frac{k_\perp^\mu}{\bar{n}\cdot k}\,\bar{n}\cdot A_{c,\s k} + \cdots\,,
\label{eq:BcOriginal}
\end{align} 
where the subscript $k$ refers to the label momentum.  We see that this operator includes the physical perpendicular components of the gauge boson as the first term, followed by a tower of $k_\perp^\mu\times(\bar{n}\cdot A_{c,\s k}/\bar{n}\cdot k)^m$ emissions (where $m$ is an integer), all of which scale as $\sim \lambda$, inherited from the scaling of $k_\perp^\mu$.  This object makes it entirely straightforward to model collinear gauge bosons that are emitted from the hard vertex, and can be used as a building block for local operators.

\subsection{Soft Wilson Lines and Factorization}
\label{sec:SoftWilsonFact}
In \cref{eq:EikonalFactor}, we derived the eikonal factor, which is a universal coupling between a soft gauge boson and a charged external line.  These factors can be summed in a straightforward way into a Wilson line:
\begin{align}
Y_n(x) = \sum_{m=0}^\infty \sum_\text{Perms} \frac{(-g)^m\,n\cdot A\big(k_1\big)\,\cdots\,n\cdot A\big(k_m\big)}{n\cdot k_1\,n\cdot \big(k_1+k_2\big)\, \cdots\, n\cdot \big(\sum_j k_j\big)}\,,
\label{eq:SoftWilsonLineMomSpace}
\end{align}
where propagators come with $+i0$.  The soft Wilson line $Y_n(x)$ is defined with respect to the $n^\mu$ direction -- the only information that the soft gauge bosons know about the collinear sector is the direction and the charge/representation.  Note that the propagators in \cref{eq:SoftWilsonLineMomSpace} all have virtuality $\sim \lambda^2$.  This implies that we should interpret the soft Wilson lines as living within SCET.  This can be contrasted with the collinear Wilson lines in \cref{eq:WcMomSpace}, whose propagators have virtuality of $\mathcal{O}(1)$, and are therefore interpreted as resulting from matching at the hard scale.

The soft Wilson line can be Fourier transformed to position space:
\begin{align}
Y_n(x) = \textbf{P}\s \exp\left[ i\s g\int_{-\infty}^{\,0}\D s\, n\cdot A_{us}(x+n\s s)\right]\,,
\label{eq:SoftWilsonLine}
\end{align}
where $s$ is an affine parameter that tracks the collinear direction.  There are additional subtleties which must be kept track of depending on if the external lines are incoming or outgoing~\cite{Chay:2004zn, Arnesen:2005nk}, see \emph{e.g.}~\cite{iain_notes} for a discussion.  If required, the Wilson line can be modified to include a rapidity regulator~\cite{Chiu:2011qc, Chiu:2012ir}, similar to \cref{eq:WcRapidityReg} above.

The soft Wilson lines satisfy a variety of useful identities.  First, note that it is clear from the definition that 
\begin{align}
Y_n(x)^\dag \,Y_n(x) = Y_n(x) \,Y_n^\dag(x) =  1\,.
\label{eq:YdagY}
\end{align}
The soft Wilson lines projected along the collinear direction satisfies an ``equation of motion''
\begin{align}
i\s n\cdot D_{us}\,Y_n\big(\bar{n}\cdot x\big) = 0 \,,
\label{eq:SoftWilsonLineEOM}
\end{align}
where
\begin{align}
i\s n\cdot D_{us} =  \frac{\partial}{\partial \bar{n}\cdot x} -i\s g\s n\cdot A_{us}(\bar{n}\cdot x)\,,
\end{align}
which is straightforward to derive using the light-cone decomposition of the derivative \cref{eq:DecomposeDer}.  Finally, we note that 
\begin{align}
\partial_\mu\,Y_n(\bar{n}\cdot x) = \bar{n}_\mu\, \frac{\partial}{\partial \bar{n}\cdot x}\,Y_n(\bar{n}\cdot x) \,,
\label{eq:DerivYn}
\end{align}
where again we have used \cref{eq:DecomposeDer}.  These three properties will be crucial for factorizing the SCET Lagrangian, which is the subject we turn to next.

\subsubsection*{Factorizing the Gauge Interactions with a Field Redefinition}
We now have all the required tools to factorize the leading power SCET Lagrangian, as was first done in~\cite{Bauer:2002nz}.  The strategy is to use a ``decoupling field redefinition'' (also referred to as a BPS field redefinition) involving the ultrasoft Wilson line in \cref{eq:SoftWilsonLine}.  The result is a derivation of independent Lagrangians for the collinear, anti-collinear, and ultrasoft sectors.  We will show how this works for the non-Abelian interactions between the gauge bosons, and will leave the analogous derivation for fermions as an exercise in \cref{sec:CollinearFermions} below.

Factorization has a long tradition in field theory, beginning with the famous result proving that the parton distribution functions can be factorized for the Drell-Yan process using \FT~QCD Feynman diagrams.\footnote{It is worth emphasizing that the rigorous demonstration that the parton distribution functions factorize from the hard process is somewhat different from the factorization of the hard process into hard/jet/soft functions as is done in SCET.}  For a review of the traditional approach to proving factorization, see \emph{e.g.}~the review~\cite{Collins:1989gx} and the book~\cite{Collins:2011zzd}.

To explore how factorization is derived in SCET, we begin with the collinear Yang-Mills field strength tensor, defined by the commutator of covariant derivatives:
\begin{align}
F_{c}^{\mu\nu} = \frac{i}{g}\Big[D^\mu,D^\nu\Big]\,,
\end{align}
which should be interpreted as a matrix expression, see \cref{eq:FmunuFromDD}, and where 
\begin{align}
D^\mu &= \frac{\bar{n}^\mu}{2}\, n\cdot D + \frac{n^\mu}{2}\,\bar{n}\cdot D_c + D_{c,\s\perp}^\mu \,,
\label{eq:DmuCollinear}
\end{align}
with
\begin{align}
i\s n\cdot D &= i\s n\cdot \partial + g\s n\cdot A_c(x) + g\s n\cdot A_{us}(\bar{n}\cdot x)\notag\\[2pt]
i\s \bar{n}\cdot D_c &= \bar{n}\cdot \mathcal{P} + g\s \bar{n}\cdot A_c(x) \notag\\[2pt]
i\s D_{c,\s\perp}^\mu &= \mathcal{P}_\perp^\mu + g\s A_{c,\s \perp}^\mu(x) \,.
\end{align}
Then the collinear Yang-Mills Lagrangian
\begin{align}
\mathcal{L}_c = -\frac{1}{4}\,\text{Tr}\Big[F_{c,\s\mu\nu}\,F_c^{\mu\nu}\Big]
\end{align}
includes the interaction between collinear and ultrasoft given in~\cref{eq:GaugeTripleVertex}.  

The decoupling field redefinition for the collinear gauge boson is
\begin{align}
A_c^\mu(x) = Y_n(\bar{n}\cdot x)\,A_c^{(0)\s\mu}(x)\,Y_n^\dag(\bar{n}\cdot x)\,,
\label{eq:Azero}
\end{align}
where $A_c^{(0)\s\mu}(x)$ will be the collinear gauge boson that appears in the factorized Lagrangian.  The superscript $(0)$ makes explicit that this is the leading power field, and emphasizes that the factorization is only being derived to leading power.

Our task is to compute the collinear Yang-Mills Lagrangian after performing this field redefinition, so we need the redefined versions of the covariant derivatives.  Two of the components are trivial
\begin{align}
i\s \bar{n}\cdot D_c &= Y_n(\bar{n}\cdot x)\,i\s \bar{n}\cdot D_c^{(0)}\,Y_n^\dag(\bar{n}\cdot x) \notag\\[5pt]
i\s D_{c,\s\perp} &= Y_c(\bar{n}\cdot x)\,i\s D_{c,\s\perp}^{(0)\s\mu}\,Y_n^\dag(\bar{n}\cdot x)\,,
\label{eq:Dczero}
\end{align}
where we have used \cref{eq:DerivYn} to move $Y_n(\bar{n}\cdot x)$ through $\bar{n}\cdot \partial$ and $\partial_\perp^\mu$, and the $(0)$ superscript on the covariant derivative means replace $A_c^\mu \rightarrow A_c^{(0)\s\mu}$.  The non-trivial component is
\begin{align}
i\s n\cdot D\,O(x) &=\Big[i\s n\cdot \partial + g\s n\cdot \big(Y_n \,A_c^{(0)}(x)\,Y_n^\dag \big) + g\s n\cdot A_{us}(\bar{n}\cdot x)\Big]O(x) \notag \\[8pt]
&= \Big[\big(i\s n\cdot \partial + g\s n\cdot A_{us}(\bar{n}\cdot x)\big)Y_n\,Y_n^\dag + g\s n\cdot \big(Y_n \,A_c^{(0)}(x)\,Y_n^\dag \big)\Big]O(x) \notag\\[8pt]
&= \Big[\big(i\s n\cdot \partial + g\s n\cdot A_{us}(\bar{n}\cdot x)\big)Y_n\Big]\,\big(Y_n^\dag\,O(x)\big) \notag\\[4pt]
&\hspace{20pt} + Y_n\,\Big[i\s n\cdot \partial \,\big(Y_n^\dag\,O(x)\big)\Big] + Y_n\big(g\s n\cdot A_c^{(0)}(x)\,Y_n^\dag\,O(x) \big) \notag\\[8pt]
&= Y_n\,\Big[i\s n\cdot \partial+g\s n\cdot A_c^{(0)}(x)\Big]\,Y_n^\dag\,O(x) = Y_n\,i\s n\cdot D_c^{(0)}\,Y_n^\dag\,O(x)\,,
\end{align}
where $O(x)$ is an arbitrary operator, we have written $Y_n = Y_n(\bar{n}\cdot x)$ for brevity, in the second line we used \cref{eq:YdagY}, in going from the third to the fourth line we have used \cref{eq:SoftWilsonLineEOM}, and the final equality implicitly defines $n\cdot D_c^{(0)}$.

Putting this all together, we see that 
\begin{align}
D^\mu = Y_n(\bar{n}\cdot x)\,D_c^{(0)\,\mu}\,Y_n^\dag(\bar{n}\cdot x)\,,
\end{align}
which only depends on the ultrasoft gauge boson through the Wilson lines $Y_n$.  Then the Yang-Mills field strength for $A_c^{(0)}$ is 
\begin{align}
F_c^{\mu\nu} = Y_n(\bar{n}\cdot x)\,\bigg(\s \frac{i}{g}\s\Big[D^{(0)\,\mu},D^{(0)\,\nu}\Big]\bigg)\, Y_n^\dag(\bar{n}\cdot x) = Y_n(\bar{n}\cdot x)\, F_{c}^{\s(0)\,\mu\nu}\, Y_n^\dag(\bar{n}\cdot x)\,,
\end{align}
so that
\begin{align}
\mathcal{L}^{(0)}_c &= -\frac{1}{4}\, \text{Tr}\Big[Y_n(\bar{n}\cdot x)\, F_{c,\s\mu\nu}^{\s (0)}\, Y_n^\dag(\bar{n}\cdot x)\,Y_n(\bar{n}\cdot x)\, F_{c}^{\s (0)\,\mu\nu}\, Y_n^\dag(\bar{n}\cdot x)\Big] \notag\\[7pt]
&= -\frac{1}{4}\, \text{Tr}\Big[F_{c,\s\mu\nu}^{\s (0)}\, F_{c}^{\s (0)\,\mu\nu}\Big]\,,
\end{align}
where in the last step we used the cyclic property of the trace to eliminate the ultrasoft Wilson lines.  Obviously, the exact same steps can be taken to eliminate the ultrasoft gauge boson dependence from the anti-collinear Lagrangian yielding $\mathcal{L}^{(0)}_{\bar{c}}$, by making the appropriate substitutions, \emph{e.g.}~$Y_n(\bar{n}\cdot x) \rightarrow Y_{\bar{n}}(n\cdot x)$.

Next, we note that the ultrasoft gauge boson Lagrangian is given by 
\begin{align}
\mathcal{L}_{us}^{(0)} = -\frac{1}{4}\, F_{us,\s \mu \nu}\,F^{\mu\nu}_{us}\,,
\end{align}
where
\begin{align}
F_{us}^{\mu\nu} &= \frac{i}{g}\Big[D_{us}^\mu,D_{us}^\nu\Big]\notag \\[5pt]
i\s D_{us}^\mu &= i\s \partial^\mu + g\s A_{us}^\mu(x)\,,
\end{align}
which is just a standard Yang-Mills Lagrangian, including self-interactions among the soft gauge bosons, but which critically does not include interactions with any collinear states.

Therefore, our Lagrangian takes the factorized form\footnote{It is important to note that the true Lagrangian does not actually factorize due to the presence of the Glauber modes briefly mentioned in the exercise in \cref{sec:RegionsMasslessSudakov} above.  A true demonstration of factorization requires justifying that the Glauber contributions can be neglected (as was first done by for QCD~\cite{Collins:1989gx}).  This in turn allows one to ignore the part of the Hilbert space where the Glauber modes live, allowing us to say that the Hilbert space factorizes.   In particular, there is often a cancelation among the Glauber diagrams that allows them to be absorbed into the soft sector. For an operator level treatment in SCET, see~\cite{Rothstein:2016bsq}, and in particular their Eq.~(8.1) writes the version of this Lagrangian that includes Glauber effects.  Another place where Glauber modes are important is in the demonstration that the hard matching coefficient for back-to-back quark production when including additional spectator external states is independent of the choice of which external states to use~\cite{Bauer:2010cc}.}
\begin{align}
\mathcal{L}_\text{YM}^{(0)} = \mathcal{L}^{(0)}_c + \mathcal{L}^{(0)}_{\bar{c}} + \mathcal{L}^{(0)}_{us}\,,
\end{align}
implying that our Hilbert space for the gauge bosons factorizes:
\begin{align}
\big|X \big\rangle = \big|X_c \big\rangle \otimes \big|X_{\bar{c}}\big\rangle\otimes \big|X_{us} \big\rangle\,,
\end{align}
where $\big|X \big\rangle$ is a state in the \FT~Hilbert space where all modes above the hard scale have been integrated out, \emph{i.e.}, in the regime where the EFT is expected to hold.  It is now expressed in terms of a direct product of states in independent EFT Hilbert spaces (to leading power).

The decoupling field redefinition led to the factorization of the interactions between collinear and ultrasoft fields.  However, there is another implication of this field redefinition, which is that now ultrasoft fields will appear in the local operator structure.  For example, our gauge boson local operator building block in \cref{eq:BcOriginal} is redefined following \cref{eq:Dczero}:
\begin{align}
\mathcal{B}_{c\s\perp}^\mu(x) = Y_n(\bar{n}\cdot x)\,\mathcal{B}_{c\s\perp}^{(0)\s\mu}(x)\,Y_n(\bar{n}\cdot x)^\dag\,.
\end{align}

To explore this further, we will take a simple example.  Consider a \FT~process of a singlet initial state producing a pair of gauge boson final states, \emph{e.g} a heavy scalar boson (this could be the Higgs boson in the Standard Model) decaying to a pair of gluons through an interaction $\mathcal{L}_\textsc{Full} \supset C\, \Phi\,F_{\mu\nu}\,F^{\mu\nu}$, where $C$ is a constant.  Then we can model this process in the EFT~\cite{Ahrens:2008nc} 
\begin{align}
F_{\mu\nu}\,F^{\mu\nu} \longrightarrow \,\, \mathcal{B}_{c\s\perp\s\mu}\,\mathcal{B}^{\mu}_{\bar{c}\s\perp} = Y_n(\bar{n}\cdot x)\,\mathcal{B}_{c\s\perp\s\mu}^{(0)}(x)\,Y_n(\bar{n}\cdot x)^\dag\,Y_{\bar{n}}(n\cdot x)\,\mathcal{B}_{\bar{c}\s\perp}^{(0)\s\mu}(x)\,Y_{\bar{n}}(n\cdot x)^\dag \,,
\end{align}
where we have suppressed the color contractions and the matching between the \FT~and SCET is leading order, see \emph{e.g.}~\cite{Ahrens:2008nc} for more detailed expressions.   We want to compute the expectation value of this object in the presence of the collinear and anti-collinear hard lines.  Therefore, we should take a matrix element using the incoming state $|A_{\bar{c}}(\bar{p})\rangle$ and the outgoing state $\langle A_{c}(p)|$:
\begin{align}
&\vev{A_{c}(p)\left|Y_n(\bar{n}\cdot x)\,\mathcal{B}_{c\s\perp\s\mu}^{(0)}(x)\,Y_n(\bar{n}\cdot x)^\dag\,Y_{\bar{n}}(n\cdot x)\,\mathcal{B}_{\bar{c}\s\perp}^{(0)\s\mu}(x)\,Y_{\bar{n}}(n\cdot x)^\dag\right|A_{\bar{c}}(\bar{p})}\notag\\[8pt]
&\hspace{10pt}=\tensor[_{us}]{\vev{0\left|Y_n(\bar{n}\cdot x)\,Y_n(\bar{n}\cdot x)^\dag\,Y_{\bar{n}}(n\cdot x)\,\,Y_{\bar{n}}(n\cdot x)^\dag\right|0}}{_{us}}\notag\\[5pt]
&\hspace{35pt}\times \vev{A_{\bar{c}}(\bar{p})\left|\mathcal{B}_{c\s\perp\s\mu}^{(0)}(x)\right|0}_c \times \tensor[_{\bar{c}}]{\vev{0\left|\mathcal{B}_{\bar{c}\s\perp}^{(0)\s\mu}(x)\right|A_{\bar{c}}(\bar{p})}}{_{}}\,.
\end{align}
Our task is reduced to computing a ``hard function'' $\mathcal{H}(M,\mu)$, a ``collinear jet function'' $\mathcal{J}_c\big(P^2,\mu\big)$, an ``anti-collinear jet function'' $\mathcal{J}_{\bar{c}}\big(\bar{P}^2,\mu\big)$, and an ``ultrasoft function'' $\mathcal{S}\big(P^2\,\bar{P}^2/M,\mu\big)$.  The hard function is determined by a matching calculation, and the other three functions have operator definitions within SCET.  Diagrammatically, up to one-loop order, we have jet functions
\begin{align}
\mathcal{J}_c\big(P^2,\mu\big) &= \vev{A_{\bar{c}}(\bar{p})\left|\mathcal{B}_{c\s\perp\s\mu}^{(0)}(x)\right|0}_c = \includegraphics[width=0.17\textwidth, valign=b]{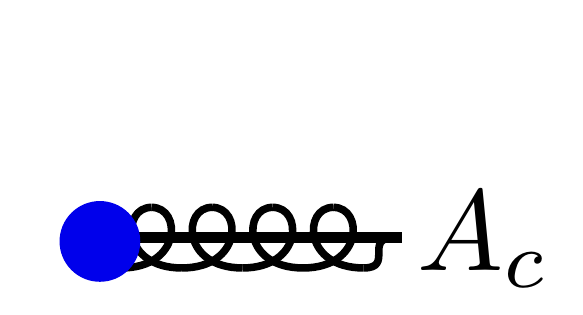} + \includegraphics[width=0.18\textwidth, valign=b]{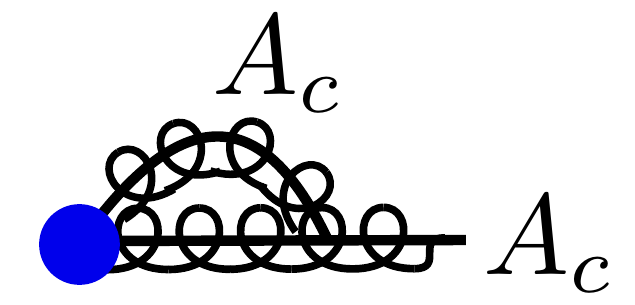} + \cdots \notag\\[5pt]
\mathcal{J}_{\bar{c}}\big(\bar{P}^2,\mu\big) &= \tensor[_{\bar{c}}]{\vev{0\left|\mathcal{B}_{\bar{c}\s\perp}^{(0)\s\mu}(x)\right|A_{\bar{c}}(\bar{p})}}{_{}} = \includegraphics[width=0.17\textwidth, valign=b]{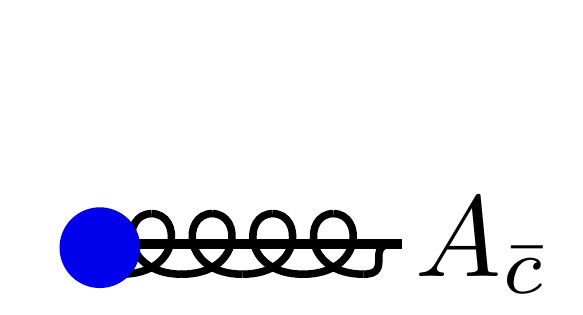} + \includegraphics[width=0.18\textwidth, valign=b]{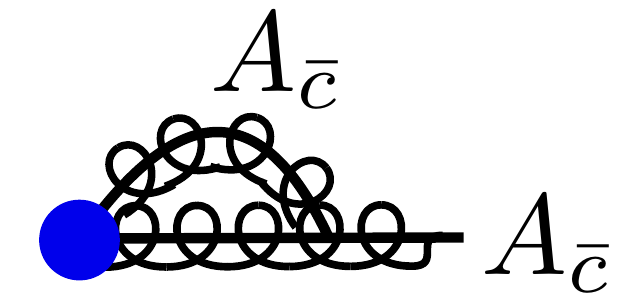} + \cdots \,,
\label{eq:DefineJFns}
\end{align}
where the top line is for collinear and the bottom line is for anti-collinear, and an ultrasoft function\footnote{This expression makes it seem like we should draw four Wilson lines in our diagram.  However, note that the $Y_n$ is in the fundamental representation, while the gauge boson is in the adjoint.  Then one can interpret a product of two $Y_n$ Wilson lines as modeling an adjoint Wilson line.}
\begin{align}
\mathcal{S}\big(P^2\,\bar{P}^2/M,\mu\big) &= \tensor[_{us}]{\vev{0\left|Y_n(\bar{n}\cdot x)\,Y_n(\bar{n}\cdot x)^\dag\,Y_{\bar{n}}(n\cdot x)\,\,Y_{\bar{n}}(n\cdot x)^\dag\right|0}}{_{us}} \notag\\[7pt]
&= \mathbb{1} + \includegraphics[width=0.12\textwidth, valign=c]{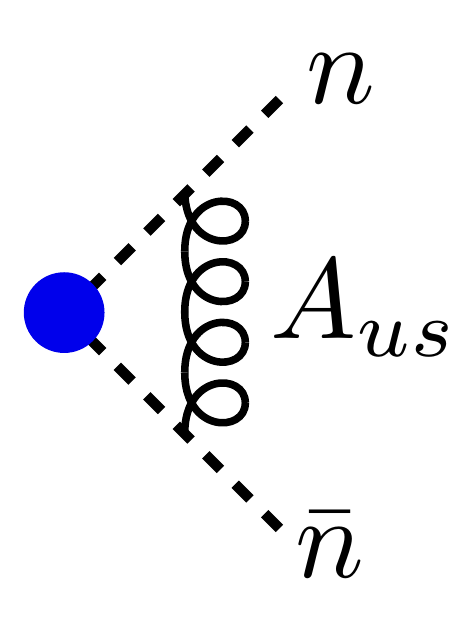} + \cdots \,,
\label{eq:DefineSFns}
\end{align}
such that our final summed observable is given by the product of these four functions, run to the common renormalization scale $\mu$.  This factorization can be illustrated schematically as (see \emph{e.g.}~\cite{Becher:2014oda, Becher:2018gno})
\begin{align}
\includegraphics[width=0.15\textwidth, valign=c]{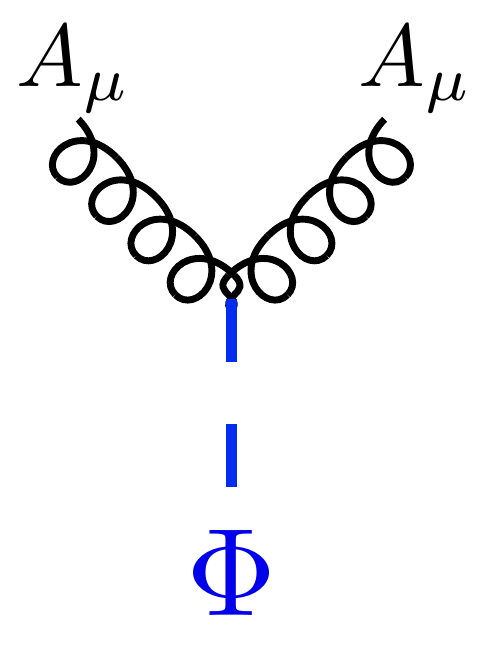} \qquad \boldsymbol{\longrightarrow} \qquad \includegraphics[width=0.3\textwidth, valign=c]{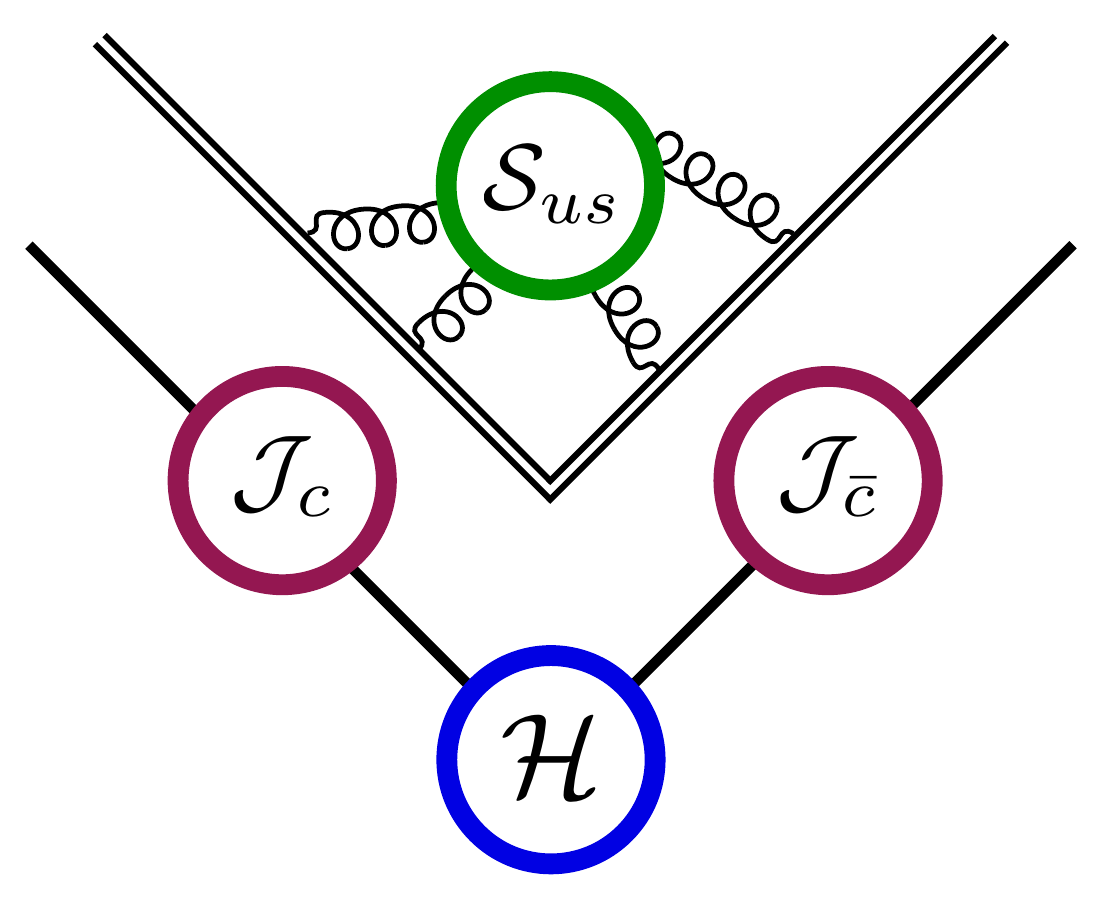} + \mathcal{O}\big(\lambda^2\big)\,.
\label{eq:FactorizeFig}
\end{align}

To summarize, we have shown how to decompose the gauge boson into collinear and ultrasoft modes.  Then calculations for gauge theories in the IR are recast in terms of hard, jet, and soft functions that can be computed independently, run to a common scale, and then multiplied together to yield the RG improved amplitude of interest.  Moreover, the jet and soft functions are universal, which implies that can be used to construct predictions for a wide variety of observables.  This procedure separates scales and sums potentially large Sudakov double logarithms.  Furthermore, it provides us with a systematic procedure for summing logs of arbitrary high order, and for computing sub-leading power corrections as an expansion in $\lambda$.

Below, we will provide an explicit demonstration of how the Sudakov double logs can be exponentiated using the vector current process built out of charged fermions.  To this end, we must develop the technology for treating fermions in the collinear limit, which is the topic of the next section.

\subsection{Collinear Fermions}
\label{sec:CollinearFermions}
In this section, we will explain how to model collinear fermions in SCET, using two-component notation.  Our starting point is a \FT~left-handed Weyl fermion $u$, which has a momentum scaling of $p\sim\big(\lambda^2,1,\lambda\big)$.  The first step is to find a set of projection operators that allow us to separate our collinear fermion into the degrees of freedom whose helicity is aligned with the $n^\mu$ direction $u_{c,\s n}$ from those whose helicity points in the opposite direction $u_{c,\s\bar{n}}$.  

The Lagrangian for $u$ is
\begin{align}
\mathcal{L} = i\s u^\dag(x) \,\bar{\sigma}\cdot \partial \,u(x)\,,
\label{eq:LFreeFermion}
\end{align}
which admits the standard plane-wave solution $u(x^\mu) =  \int \textrm{d}^3 \,p\,\, x(p)\,\exp(-i\s p^\mu x_\mu)$.  In \cref{eq:LFreeFermion} and many expressions that follow, the spinor indices are implicitly contracted, following the two component conventions of~\cite{Dreiner:2008tw}.

In order to see the connection between the projection operators and helicity, we can boost the free-theory solution for $u(x)$ along the light-cone $n^\mu$, which points in the $\hat z$ direction:\footnote{\footnotesize For example, see Sec.~3.3 of \cite{Peskin:1995ev}.}   
\begin{align}
x(p)\big|_n &=\left[\sqrt{E+p_z}\left(\frac{1-\sigma^3}{2}\right) + \sqrt{E-p_z}\left(\frac{1+\sigma^3}{2}\right)\right] \bar{\sigma}^0 \,\xi  \sim \left[ \begin{array}{c}
\lambda  \\
\mathcal{O}(1) \end{array} \right] \,,
\label{eq:uCollinearLimit}
\end{align}
where $\xi$ is a two-component spinor, we have inserted the appropriate factor of $\bar{\sigma}_0$ to make the spinor index structure consistent, and for the last step we have used the collinear scalings for the momentum, $\bar{n} \cdot p \sim \mathcal{O}(1)$ and $n \cdot p \sim \lambda^2$.  We see that the upper component of $x(p)|_n$ is power suppressed, and the collinear fermion is given by the lower component of $x(p)|_n$.  This is allows us to separate the \FT~fermion into a leading and subleading part.

The following combinations of Pauli matrices
\begin{align}
\label{eq:SigmaNotation}
&\hspace{-10pt}\left(\frac{\ns}{2}\right)_{\alpha \dot{\alpha}} =  \frac{1}{2}\big(\sigma^0 - \sigma^3\big)_{\alpha \dot{\alpha}}  = 
\left[ \begin{array}{cc}
0 & 0  \\
0 & 1  \end{array} \right]_{\alpha \dot{\alpha}} \hspace{25pt} \left(\frac{\nbsb}{2} \right)^{\dot{\alpha}\alpha} =    \frac{1}{2}\big(\bs_0 - \bs_3\big)^{\dot{\alpha}\alpha}  = 
\left[ \begin{array}{cc}
0 & 0  \\
0 & 1  \end{array} \right]^{ \dot{\alpha}\alpha} \notag\\[10 pt]
&\hspace{-10pt}\left(\frac{\nbs}{2}\right)_{\alpha \dot{\alpha}}  = \frac{1}{2}\big(\sigma^0 + \sigma^3\big)_{\alpha \dot{\alpha}} = 
\left[\begin{array}{cc}
1 & 0  \\
0 & 0  \end{array} \right]_{\alpha \dot{\alpha}}  \hspace{25pt} \left(\frac{\nsb}{2}\right)^{\dot{\alpha}\alpha}  =  \frac{1}{2}\big(\bs_0 + \bs_3\big)^{\dot{\alpha}\alpha} = 
\left[\begin{array}{cc}
1 & 0  \\
0 & 0  \end{array} \right]^{ \dot{\alpha} \alpha} \,,
\end{align}
can be used to infer projections operators $P_n$ and $P_{\bar{n}}$,
\begin{align}
u = \left(P_n + P_{\bar{n}}\right) u = u_{c,\s n} + u_{c,\s \bar{n}}\,.
\end{align}
Comparing to \cref{eq:uCollinearLimit}, we can derive the explicit forms
\begin{align}
\begin{array}{ll}
P_{n} \, u_{c,\s n} = \frac{n\cdot \sigma}{2}\frac{\bar n\cdot \bar \sigma}{2}\,u_{c,\s n} = u_{c,\s n} \qquad\qquad\quad\quad &P_{\bar{n}}\,u_{c,\s \bar{n}} =  \frac{\nbs}{2}\frac{\nsb}{2}\,u_{c,\s \bar{n}} = u_{c,\s \bar{n}}  \\[5pt]
P_{\bar{n}} \, u_{c,\s n} = 0 &P_n \,u_{c,\s \bar{n}}  = 0\,.
\end{array}
\label{eq:uProjections}
\end{align}
These operators project out half the helicity states, namely $u_{c,n,1} = 0$ and $u_{c,\bar{n},2} = 0$; the two component collinear/anti-collinear projection operators are equivalent to the chiral projection operators.

We can expose a few interesting features by analyzing the free fermion Lagrangian.  Starting with \cref{eq:LFreeFermion}, expanding out $u = u_{c,\s n} + u_{c,\s \bar{n}}$, and expressing it in components yields
\begin{align}
-i\s\mathcal{L}_u &= u_{c, n,\dot{2}}^\dagger\, n \cdot \partial\s u_{c,n,2}  +  u_{c,\bar{n}.\dot{1}}^\dagger \bar{n}\cdot \partial \s u_{c,\bar{n},1}\notag\\[4pt]
&\hspace{15pt}+ u_{c,\bar{n},\dot{1}}^\dagger \left( \bar{\sigma} \cdot \partial_\perp \right)^{\dot{1}2} \, u_{c,n,2} + u_{c,n,\dot{2}}^\dagger \left(\bar{\sigma} \cdot \partial_\perp \right)^{\dot{2}1} \, u_{c,\bar{n},1}\,,
\label{eq:fullThyLfermionComponent}
\end{align}
where we have used $(\bar{\sigma} \cdot \partial)^{\dot{2}2} = \bar{n} \cdot \partial$ and related identities to simplify this expression.

We can derive power counting for the fermion as above.  Assuming collinear scaling implies $\text{d}^4 x \sim \lambda^{-4}$, so that $\mathcal{L}_u \sim \lambda^4$ to yield an unsuppressed action.  This fixes
\begin{align}
u_{c,\s n} \,\,\sim\,\, \lambda\qquad \text{and} \qquad u_{c,\s \bar{n}} \,\,\sim\,\, \lambda^2\,. 
\end{align}

Next, we must identify a time component, which we take to be $n\cdot \partial$ by convention.  Then the expanded \FT~Lagrangian, \cref{eq:fullThyLfermionComponent}, implies that $u_{c,\s n}$ is a propagating degree of freedom (by identifying that there is a term with a time derivative acting on $u_{c,\s n}$), while $u_{c,\s \bar{n}}$ is not.  It will be convenient to write the Lagrangian exclusively in terms of the propagating mode, so we integrate out the anti-collinear modes by solving for its classical equation of motion:
\begin{align}
u_{c,\s \bar{n}} =   -\frac{\nbs}{2}  \frac{1}{\bar{n} \cdot \partial}\, \big(\bar{\sigma} \cdot \partial_\perp\big) \,u_{c,\s n} \,.
\label{eq:anticollEOM}
\end{align}
Then the collinear and anti-collinear fermion modes can both be expressed in terms of the propagating mode 
\begin{align}
& u_{c,\s n} =  \left[ \begin{array}{c}
0 \\
u_{c,n,2} \end{array} \right] 
& u_{c,\s \bar{n}} =  \left[ \begin{array}{c}
\frac{(-\partial_{\perp,1} + i\s \partial_{\perp,2})}{\nbp}\,u_{c,n,2} \\
0 \end{array} \right] \,.
\label{eq:matrixfermions}
\end{align}
Plugging \cref{eq:anticollEOM} into the \cref{eq:fullThyLfermionComponent} yields the leading power collinear Lagrangian for a free Weyl fermion:
\begin{align}
{\cal L}_{u_{c,\s n}} &=  u_{c,\s n}^\dagger \left(i\s n\cdot \partial - \frac{\partial_\perp^{\s 2}}{i\s\bar n \cdot \partial}  \right) \frac{\bar n\cdot \bar \sigma}{2}\,u_{c,\s n} \,.
\label{eq:Lfree}
\end{align}
Then the propagator for a collinear fermion can be extracted by inverting the free momentum space Lagrangian: 
\begin{equation}
\raisebox{-0.5\height}{\includegraphics[width=1.5cm]{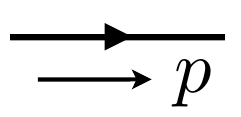}} =\,i\s\frac{n\cdot \sigma}{2}\,\frac{\bar n\cdot p}{(n\cdot p)(\bar n \cdot p) + p_\perp^2}\,.
\end{equation}
This is how we model propagating collinear fermions in SCET.

\subsubsection*{RPI Transformations of Fermions}
Before introducing any interactions for our collinear fermion, we will take the opportunity to discuss its RPI properties, with an emphasis on the connection to Lorentz invariance as discussed in \cref{sec:LightCone} above.  For brevity, we will drop the $c$ and $n$ subscripts on $u_{c,n,\alpha}$ for this subsection, and identify the propagating degree of freedom with $u_2$ as we did previously.  We will use $\beta$ ($\theta$) as a boost (rotation) parameter, associated with the $K$ ($J$) generators.  This discussion follows Appendix of~\cite{Cohen:2018qvn}, and more detail can be found there.

We begin with RPI-III, which is just a boost as generated by $K_3$:
\begin{align}
u_\alpha \quad \xrightarrow[\text{Boost}]{\makebox[1.0cm]{}}\quad  e^{-\beta_3 \sigma_3 /2}\, u_\alpha\,.
\end{align}
Due to the structure of $\sigma_3$, we see that 
\begin{align}
u_1 \quad  &\xrightarrow[\text{RPI-III}]{\makebox[1.0cm]{}} \quad  e^{- \beta_3/2}\, u_1 \notag\\[7pt]
 u_2 \quad & \xrightarrow[\text{RPI-III}]{\makebox[1.0cm]{}} \quad  e^{\beta_3/2}\, u_2\,,
\end{align}
where we identify the boost parameter $\beta_3$ with the $\alpha$ parameter of RPI-III.  So we see that the two components of the fermion transform oppositely with respect to RPI-III.

The derivations for the other RPI generators is complicated by the fact that RPI-I and RPI-II are a linear combination of boosts and rotations, see \cref{eq:LCgenbreaking}.  Using the Lorentz transformation properties of fermions, we can evaluate the action of the two RPI-I generators:
\begin{align}
u_\alpha \quad \xrightarrow[\text{Lorentz}]{\makebox[1.0cm]{}}\quad  \left(1 + \frac{i}{2}\,\theta_j\,\sigma_j - \frac{1}{2}\,\beta_j\,\sigma_j\right)\, u_\alpha\,.
\label{eq:uAlphaLorentz}
\end{align}
Then for example, we can identify one of the RPI-I transformations by taking the combination $K_1-J_2$, \emph{i.e.}, by setting $\beta_1 = -\theta_2$ with all other $\theta_j$ and $\beta_j$ equal to zero, which yields
\begin{align}
u_\alpha \quad \xrightarrow[\text{RPI-I}]{\makebox[1.0cm]{}}\quad \left(1 - \beta_1 \left(\begin{array}{cc} 0 & 1 \\ 0 & 0 \end{array}\right)\right) \,u_\alpha\,,
\end{align}
where we can identify $\beta_1$ with one of the RPI-I transformation parameters.  Then this makes it clear that RPI-I acts to rotate in the helicity component that has been integrated out:
\begin{align}
u_{c,\s n} = \left[ \begin{array}{c}
0 \\
u_2  \end{array} \right] \qquad  \xrightarrow[\text{RPI-I}]{\makebox[1.0cm]{}}    \qquad
\left[ \begin{array}{c}
\frac{1}{2}\,(  \Delta_\perp \cdot \sigma)_{1 \dot{2}} \,u_2  \\
u_2 \end{array} \right] \,,
\end{align}
where now we have expressed the two RPI-I transformation directions using a more covariant notation.

Analogous steps can be performed to derive the RPI-II transformations, which effectively acts as a rescaling of the fermion fields $u_{c,\s n}$:
\begin{align}
u_{c, \s n} = \left[ \begin{array}{c}
0 \\
u_2  \end{array} \right]   \qquad \xrightarrow[\text{RPI-II}]{\makebox[1.0cm]{}}    \qquad
\Bigg[ \begin{array}{c}
0  \\
\big( 1 +  \frac{1}{2}(\epsilon_\perp \cdot \sigma  )_{2\dot{1}} \,\frac{1}{ \bar{n}\cdot \partial}\, (\partial_{\perp} \cdot \bar{\sigma} )^{\dot{1}2}   \big) u_2 \end{array} \Bigg] \,.
\end{align}
\vspace{5pt}\mybox{\begin{itemize}
\item \textbf{Exercise:} Derive the RPI-I and RPI-II transformations for the fermion using the relations to boosts and rotations given in \cref{eq:LCgenbreaking} and the general form of the Lorentz transformation for a left handed fermion in \cref{eq:uAlphaLorentz}.
\end{itemize}}

\noindent Then one can use these RPI transformation properties to constrain the allowed form of the SCET Lagrangian.

\vspace{5pt}\mybox{\begin{itemize}
\item \textbf{Exercise:} Show that the kinetic term for a collinear fermion~\cref{eq:Lfree} is invariant under the three RPI transformations.  In doing so, one must be careful to track the RPI transformations of the various projected components of the derivatives.
\end{itemize}}
In the next section, we will charge our fermions and develop the technology to build local operators out of these interacting objects.

\subsubsection*{Collinear Fermions Coupled to Gauge Bosons}
Now that we have a propagating collinear fermion, and have understood its collinear spacetime transformation properties, we can give it a charge and couple it to ultrasoft and collinear gauge bosons by promoting $\partial_\mu \rightarrow D_\mu$ using \cref{eq:DmuCollinear}.  Under the collinear and ultrasoft gauge transformations given in~\cref{eq:UcUus}, 
\begin{align}
u_{c,\s n} &\,\,\longrightarrow \,\, U_c(x)\, u_{c,\s n}\notag\\[5pt]
u_{c,\s n} &\,\,\longrightarrow \,\, U_{us}(\bar{n}\cdot x)\, u_{c,\s n}\,,
\end{align}
where the argument of $U_{us}$ has been truncated to leading power.

We are interested in building local operators using the charged fermion fields.  This will allow us to model processes with light-like charged fermions as external legs.  Therefore, it is useful to define a gauge invariant collinear fermionic building block:
\begin{align}
\chi_{c}(x) = W_c^\dag(x)\,u_{c,\s n}\,.
\end{align}
Next, we note that the covariant derivative includes $A_{us}$ so our $\chi_c$ fermion has non-zero interaction with the ultrasoft gauge boson.  However, we can use the same strategy as in \cref{sec:SoftWilsonFact} above to redefine our field and eliminate this interaction.  Specifically, the field redefinition
\begin{align}
\chi_{c,\s n}(x) = Y_n(\bar{n}\cdot x)\,\chi_{c,\s n}^{(0)}(x)\,,
\label{eq:uRedefSoft}
\end{align}
yields an interacting Lagrangian for the collinear fermion that is independent of the ultrasoft gauge boson.
\vspace{5pt}\mybox{\begin{itemize}
\item \textbf{Exercise:} Following the same logic as in \cref{sec:SoftWilsonFact} above, show that the field redefinition in \cref{eq:uRedefSoft} eliminates the coupling between the collinear fermion and the ultrasoft gauge boson at leading power.
\end{itemize}}
This provides us with everything we need to model processes involving charged light-like fermions as external states.  In the next section, we will quote the formulas one can derive for the summation of the fermionic vector current by applying the technology we have developed in this section thus far.

\subsection{Sudakov Summation and the Cusp Anomalous Dimension}
\label{sec:CuspAnnDim}
In this section, we will provide the key results used to sum double Sudakov logs for a process with a pair of back-to-back charged fermionic final states.  This follows~\cite{Becher:2018gno}, where more details are provided.  Here we choose to highlight the final answer to emphasize new features of working with real SCET as opposed to the scalar toy SCET above.

We begin with a \FT~vector current process, which injects $M$ worth of energy into the system with only the light degrees of freedom:
\begin{align}
J_V^\mu = u_L^\dag\, \bar{\sigma}^\mu\,u_L + u_R \,\sigma^\mu\,u_R^\dag\,,
\end{align}
where $u_{L}$ $\big(u_R\big)$ is a left (right) handed Weyl fermion charged in the fundamental of an SU$(N)$ gauge group.  This current maps onto the SCET local operator
\begin{align}
&\chi_{c,L}^\dag(x)\, \bar{\sigma}^\mu \chi_{c,L}(x) + \chi_{c,R}(x)\, \sigma^\mu\, \chi^\dag_{c,R}(x) = Y_n(\bar{n}\cdot x)\Big(\chi_{c,L}^{(0)\s\dag}(x)\, \bar{\sigma}^\mu \chi^{(0)}_{c,L}(x) + \chi_{c,R}^{(0)}(x)\, \sigma^\mu\, \chi^{(0)\s\dag}_{c,R}(x)\Big)\notag\\[7pt]
&\hspace{10pt}= Y_n(\bar{n}\cdot x)\Big(u_{c,L}^{(0)\s\dag}(x)\, W_c(x)\,\bar{\sigma}^\mu \,W^\dag_c(x)\,u^{(0)}_{c,L}(x) + u_{c,R}^{(0)}(x) \,W_c^\dag(x) \, \sigma^\mu\,W_c(x)\,u^{(0)\s\dag}_{c,R}(x)\Big)\,.
\end{align}
Following the logic that led us to \cref{eq:FactorizeFig}, we know that our matrix element will take the factorized form 
\begin{align}
\mathcal{A}_V = \mathcal{H}\big(M^2,\mu^2\big)\,\mathcal{J}_c\big(P^2,\mu^2\big)\,\mathcal{J}_{\bar{c}}\big(\bar{P}^2,\mu^2\big)\,\mathcal{S}\big(P^2\,\bar{P}^2/M^2,\mu^2\big)\,.
\label{eq:AVfactorized}
\end{align}

Before writing down the explicit form of the RGEs, it is worth highlighting a few features.  Given \cref{eq:AVfactorized}, the anomalous dimensions for each of the functions $F$ take the form
\begin{align}
\gamma^\text{tot}_F = \gamma_\text{cusp}\,L_F + \gamma_F\,,
\label{eq:gammaFTot}
\end{align}
where $L_F$ is the logarithm whose argument depends on the scale associated with $F$, and $\gamma_\text{cusp}$ is a universal factor that we will describe in what follows.  This is the most general form we can have if we are going to realize scale separation.  Recall that the \FT~result must be independent of the RG scale $\mu$, since this is an artifact of matching onto SCET.  In particular, the fact that the anomalous dimensions are linear in the logarithm and all proportional to the same coefficient $\gamma_\text{cusp}$ implies that they can recombine to sum to zero and eliminate the $\mu$ dependence.  This not only provides a constraint on the logarithmic contribution (in that they must all be proportional to the same factor), but also implies that the non-log dependent part of the anomalous dimensions sum to zero:
\begin{align}
\gamma_{\mathcal{H}}(\alpha) - \gamma_{\mathcal{J}_c}(\alpha) - \gamma_{\mathcal{J}_{\bar{c}}}(\alpha) + \gamma_{\mathcal{S}}(\alpha) = 0\,.
\end{align}

Next, we note that the common factor  $\gamma_{\text{cusp}}(\alpha)$ has a beautiful physical interpretation as a  ``cusp anomalous dimension''~\cite{Polyakov:1980ca, Brandt:1981kf}.  This nomenclature is based on the observation that the anomalous dimension for the soft function is equivalent to a divergence which emerges from straightening a Wilson loop that is cusped, \emph{i.e.}, a change in its direction from $n$ to $\bar{n}$ at some spacetime point.  This is exactly the physical situation we are computing in \cref{eq:DefineJFns} and \cref{eq:DefineSFns}.  Then the required relation between the logarithmic terms in the anomalous dimensions implies that $\gamma_{\text{cusp}}(\alpha)$ appears in all four anomalous dimensions.\footnote{A cusped Wilson line additionally has been shown to encode the universal form of the QCD splitting function at large momentum fraction~\cite{Korchemsky:1992xv}.  Finally, to emphasize the universality of $\gamma_{\text{cusp}}$, we mention an intriguing connection to higher spin operators in CFTs, whose anomalous dimensions can be related to the Sudakov factor~\cite{Alday:2007mf}.}

We can write the general solution to the RGE equation
\begin{align}
\frac{\D}{\D \log\mu^2}\,F\big(\mu) = \gamma_F\big(\mu\big) \, F\big(\mu\big)\,,
\end{align}
which is solved by 
\begin{align}
F\big(\mu_L\big) = U_F\big(\mu_F,\mu_L\big) \, F\big(\mu_F\big)\,,
\end{align}
where $U_F$ is known as the ``evolution kernel'' and encodes the solution to the RGEs, see \emph{e.g.}~\cite{Ellis:2010rwa} for a general form of $U_F$.  Note that one must also consistently include the running of the gauge coupling $\alpha$ when working at a particular N$^k$LL order.  Both $\gamma_\text{cusp}$ and $\gamma_F$ can be computed as expansions in $\alpha$ within SCET.

The arguments for the universality of the eikonal factor, which is used to derive the soft Wilson line, imply that the cusp anomalous dimension is also universal in that it only depends on the charge of the states that define the collinear direction.  Since it is so ubiquitous and determines the structure of the log dependent term in the anomalous dimensions, this object has been studied extensively, \emph{e.g.}~the first two loop calculation for QCD was done in~\cite{Korchemsky:1987wg}: 
\begin{align}
\gamma_\text{cusp}(\alpha) = \frac{\alpha}{4\s\pi}\,4\,C_F + \left(\frac{\alpha}{4\s\pi}\right)^2\left(C_A\,C_F\left(\frac{67}{9}-\frac{\pi^2}{3}\right)- \frac{20}{9}\,n_f\,T_f\,C_F\right)\,,
\end{align}
where $C_A = N$ and $C_F = (N^2-1)/(2\,N)$ are the quadratic Casimir operators for SU$(N)$ adjoints and fundamental fermion fields respectively, $n_f$ is the number of charged fermion flavors, and $T_f = 1/2$ for the fundamental representation.

For our vector current process, we can now write the general form of the RGEs:
\begin{align}
\frac{\D}{\D \log \mu_{\mathcal{H}}^2}\, \mathcal{H} \big(M^2,\mu_{\mathcal{H}}^2\Big) &=\hspace{9pt} \frac{1}{2}\left[\gamma_\text{cusp}(\alpha) \log\frac{M^2}{\mu_{\mathcal{H}}^2} + \gamma_{\mathcal{H}}(\alpha)\right] \,\mathcal{H}\big(M^2,\mu_{\mathcal{H}}^2\big) \notag\\[10pt]
\frac{\D}{\D \log \mu_{\mathcal{J}_c}^2}\, \mathcal{J}_c \big(P^2,\mu_{\mathcal{J}_c}^2\Big) &= - \frac{1}{2}\left[\gamma_\text{cusp}(\alpha) \log\frac{P^2}{\mu_{\mathcal{J}_c}^2} + \gamma_{\mathcal{J}_c}(\alpha)\right] \,\mathcal{J}_c\big(P^2,\mu_{\mathcal{J}_c}^2\big) \notag\\[10pt]
\frac{\D}{\D \log \mu_{\mathcal{J}_{\bar{c}}}^2}\, \mathcal{J}_{\bar{c}} \big(\bar{P}^2,\mu_{\mathcal{J}_{\bar{c}}}^2\Big) &= - \frac{1}{2}\left[\gamma_\text{cusp}(\alpha) \log\frac{P^2}{\mu_{\mathcal{J}_{\bar{c}}}^2} + \gamma_{\mathcal{J}_{\bar{c}}}(\alpha)\right] \,\mathcal{J}_{\bar{c}}\big(\bar{P}^2,\mu_{\mathcal{J}_{\bar{c}}}^2\big) \notag\\[10pt]
\frac{\D}{\D \log \mu_{\mathcal{S}}^2}\, \mathcal{S} \left(\frac{P^2\,\bar{P}^2}{M^2},\mu_{\mathcal{S}}^2\right) &=\hspace{9pt} \frac{1}{2}\left[\gamma_\text{cusp}(\alpha) \log\frac{P^2\,\bar{P}^2}{M^2\,\mu_{\mathcal{S}}^2} + \gamma_{\mathcal{S}}(\alpha)\right] \,\mathcal{S}\left(\frac{P^2\,\bar{P}^2}{M^2},\mu_{\mathcal{S}}^2\right) \,,
\label{eq:RGEvectorCurrent}
\end{align}
where $\alpha = g^2/(4\s\pi)$ is the fine-structure constant for the SU$(N)$ theory.  Integrating \cref{eq:RGEvectorCurrent} sums the large Sudakov double logs that emerge for the vector current at one-loop, and can be used to systematically sum logs to arbitrary order.  Note that we must evolve all four functions to a common scale $\mu_L$.  However, they each will depend on an initial $\mu$, and as such this initial choice should be made to minimize the logarithms contained within the anomalous dimensions, such that the RG evolution is kept under good perturbative control.  In our simple example, the optimal RG path is clearly~\cite{Becher:2018gno}
\begin{align}
\includegraphics[width=0.8\textwidth, valign=c]{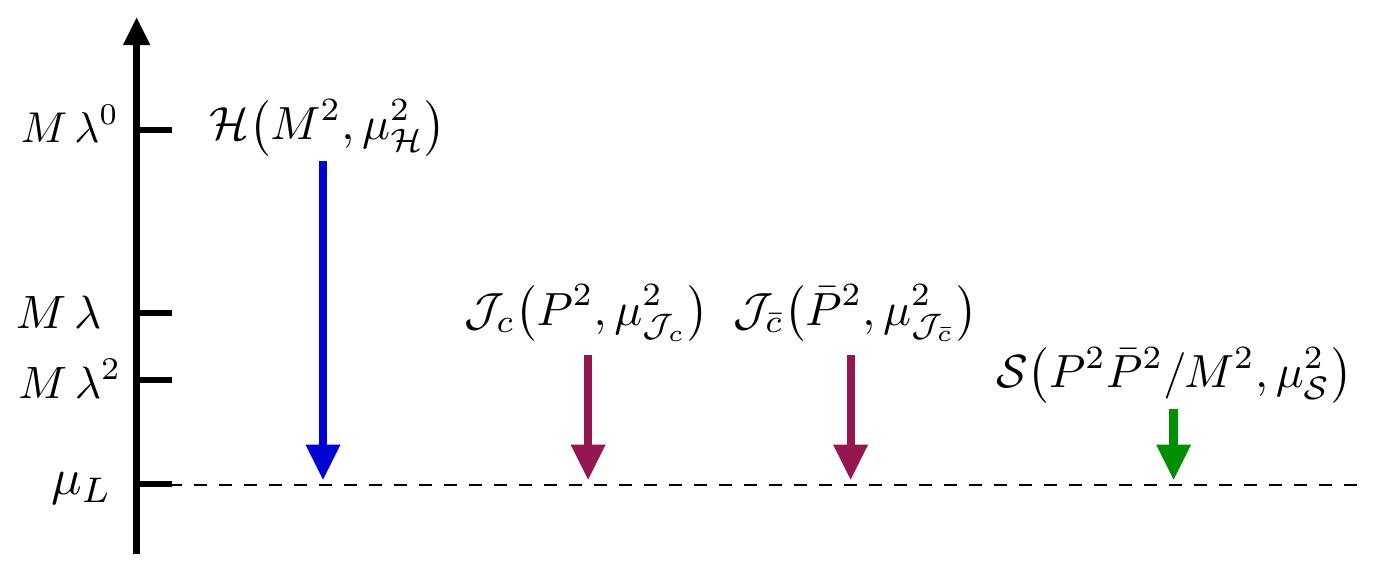}
\end{align}
Note the difference with the scalar SCET case above, see \cref{eq:RGPathScalarSCET}, where we did not factorize our matrix element into separate functions, which implied that we only had a single RG scale $\mu_L$.  We could have used a similar approach here, and just computed the contributions to the vector current process all together.  However, working with factorized individual functions has the benefit that one can analyze them independently when computing perturbative corrections.

For concreteness, we will sum the leading Sudakov logarithm.  This implies that we can treat $\gamma_\text{cusp}$ as a constant, and set the subleading anomalous dimensions to zero: $\gamma_{\mathcal{H}} = \gamma_{\mathcal{J}_c} = \gamma_{\mathcal{J}_{\bar{c}}} = \gamma_{\mathcal{S}} = 0$.  In this approximation, it is trivial to solve the RGEs in \cref{eq:RGEvectorCurrent}:
\begin{align}
\mathcal{H} \Big(M^2,\mu_L^2\Big) &= \mathcal{H} \Big(M^2,\mu_{\mathcal{H}}^2\Big) \, \exp\left[-\frac{\gamma_{\text{cusp}}}{4}\left(\log^2\frac{\mu_{\mathcal{H}}^2}{\mu_L^2}-2\,\log\frac{\mu_{\mathcal{H}}^2}{\mu_L^2}\,\log\frac{\mu_{\mathcal{H}}^2}{M^2}\right)\right] \notag\\[10pt]
\mathcal{J}_c \Big(P^2,\mu_L^2\Big) &= \mathcal{J}_c \Big(P^2,\mu_{\mathcal{J}_c}^2\Big)\,\exp\left[\frac{\gamma_{\text{cusp}}}{4}\left(\log^2\frac{\mu_{\mathcal{J}_c}^2}{\mu_L^2}-2\,\log\frac{\mu_{\mathcal{J}_c}^2}{\mu_L^2}\,\log\frac{\mu_{\mathcal{J}_c}^2}{P^2}\right)\right] \notag\\[10pt]
\mathcal{J}_{\bar{c}} \Big(\bar{P}^2,\mu_L^2\Big) &= \mathcal{J}_{\bar{c}} \Big(\bar{P}^2,\mu_{\mathcal{J}_{\bar{c}}}^2\Big)\,\exp\left[\frac{\gamma_{\text{cusp}}}{4}\left(\log^2\frac{\mu_{\mathcal{J}_{\bar{c}}}^2}{\mu_L^2}-2\,\log\frac{\mu_{\mathcal{J}_{\bar{c}}}^2}{\mu_L^2}\,\log\frac{\mu_{\mathcal{J}_{\bar{c}}}^2}{\bar{P}^2}\right)\right]  \notag\\[10pt]
\mathcal{S} \left(\frac{P^2 \bar{P}^2}{M^2},\mu_L^2\right) &= \mathcal{S} \left(\frac{P^2 \bar{P}^2}{M^2},\mu_{\mathcal{S}}^2\right)\,\exp\left[-\frac{\gamma_{\text{cusp}}}{4}\left(\log^2\frac{\mu_{\mathcal{S}}^2}{\mu_L^2}-2\,\log\frac{\mu_{\mathcal{S}}^2}{\mu_L^2}\,\log\frac{\mu_{\mathcal{S}}^2 M^2}{P^2\bar{P}^2}\right)\right]\,.
\label{eq:SCETResummed}
\end{align}
This shows that we have separated scales and exponentiated the Sudakov double log! 

This summation procedure is systematically improvable by computing higher $\alpha$ corrections to the anomalous dimension.  Following the notation in \cref{eq:gammaFTot}, this expansion schematically takes the form
\begin{align}
\big(\gamma_F^\text{tot}\big)^{[\text{LL}]} &\sim \left(\frac{\alpha}{4\s\pi}\right)\, \gamma_\text{cusp}^{(1)}\,L_F\notag\\[8pt]
\big(\gamma_F^\text{tot}\big)^{[\text{NLL}]} &\sim \Big[\left(\frac{\alpha}{4\s\pi}\right)\, \gamma_\text{cusp}^{(1)} + \left(\frac{\alpha}{4\s\pi}\right)^2\, \gamma_\text{cusp}^{(2)} \Big]L_F + \left(\frac{\alpha}{4\s\pi}\right)\, \gamma_F^{(1)}\notag\\[8pt]
\big(\gamma_F^\text{tot}\big)^{[\text{NNLL}]} &\sim \Big[\left(\frac{\alpha}{4\s\pi}\right)\, \gamma_\text{cusp}^{(1)} + \left(\frac{\alpha}{4\s\pi}\right)^2\, \gamma_\text{cusp}^{(2)}+  \left(\frac{\alpha}{4\s\pi}\right)^3\, \gamma_\text{cusp}^{(3)}\Big]L_F\notag\\[5pt]
&\hspace{25pt} + \Big[\left(\frac{\alpha}{4\s\pi}\right)\, \gamma_F^{(1)} + \left(\frac{\alpha}{4\s\pi}\right)^2\, \gamma_F^{(2)}\Big]\,,
\end{align}
where $L_F$ is the relevant log for the function $F$, and $\gamma_\text{cusp}^{(j)}$ $\big(\gamma_F^{(j)}\big)$ is coefficient of the $j^\text{th}$ order cusp (non-cusp) anomalous dimension term with the $\alpha/(4\s\pi)$ dependence removed.  Then if we ignore the running of $\alpha$, the evolution kernel takes the form
\begin{align}
U_F^{[\text{LL}]} & \sim \exp\Big(\left(\frac{\alpha}{4\s\pi}\right)\, \gamma_\text{cusp}^{(1)}\,L_F^2\Big) =1 + \left(\frac{\alpha}{4\s\pi}\right)  \, \gamma_\text{cusp}^{(1)}\,  L_F^2+ \frac{1}{2}\, \left(\frac{\alpha}{4\s\pi}\right)^2\,  \big(\gamma_\text{cusp}^{(1)}\big)^2 L_F^4 \notag\\[6pt]
& \hspace{22pt}+ \frac{1}{6}\, \left(\frac{\alpha}{4\s\pi}\right)^3\,\big(\gamma_\text{cusp}^{(1)}\big)^3\, L_F^6 + \cdots \notag\\[8pt]
\hspace{-10pt}U_F^{[\text{NLL}]} & \sim \exp\Big[\Big(\left(\frac{\alpha}{4\s\pi}\right)\, \gamma_\text{cusp}^{(1)} + \left(\frac{\alpha}{4\s\pi}\right)^2\, \gamma_\text{cusp}^{(2)} \big)L_F^2 + \left(\frac{\alpha}{4\s\pi}\right)\, \gamma_F^{(1)}\,L_F\Big)\Big] \notag\\[6pt]
&= 1 + \left(\frac{\alpha}{4\s\pi}\right)  \left[ \gamma_\text{cusp}^{(1)}\, L_F^2+ \gamma_F^{(1)}\, L_F\right] \notag\\[6pt]
& \hspace{22pt}+ \frac{1}{2}\,\left(\frac{\alpha}{4\s\pi}\right)^2 \left[\big( \gamma_\text{cusp}^{(1)}\big)^2\, L_F^4+2\, \gamma_\text{cusp}^{(1)}\, \gamma_F^{(1)}\, L_F^3+2\, \gamma_\text{cusp}^{(2)}\, L_F^2+\big(\gamma_F^{(1)}\big)^2\, L_F^2\right] \notag\\[6pt]
& \hspace{22pt} +\frac{1}{6} \, \left(\frac{\alpha}{4\s\pi}\right)^3 \left[\big( \gamma_\text{cusp}^{(1)}\big)^3\, L_F^6+3\, \big( \gamma_\text{cusp}^{(1)}\big)^2\, \gamma_F^{(1)}\, L_F^5+6\, \big( \gamma_\text{cusp}^{(1)}\big)\, \gamma_\text{cusp}^{(2)}\, L_F^4\right.\notag\\[6pt]
&\left. \hspace{100pt}+3\, \big( \gamma_\text{cusp}^{(1)}\big) \big(\gamma_F^{(1)}\big)^2\, L_F^4+6\, \gamma_\text{cusp}^{(2)}\, \gamma_F^{(1)}\, L_F^3+\big(\gamma_F^{(1)}\big)^3\, L_F^3\right] + \cdots\,.
\label{eq:UFexpanded}
\end{align}
So we can identify a pattern\footnote{For a more sophisticated treatment of the interpretation of what logs are summed at what order, see~\cite{Almeida:2014uva}.  In particular, they explore how to make direct comparisons between summed QCD and SCET results, providing insights into both approaches.}
\begin{align}
\includegraphics[width=0.9\textwidth, valign=c]{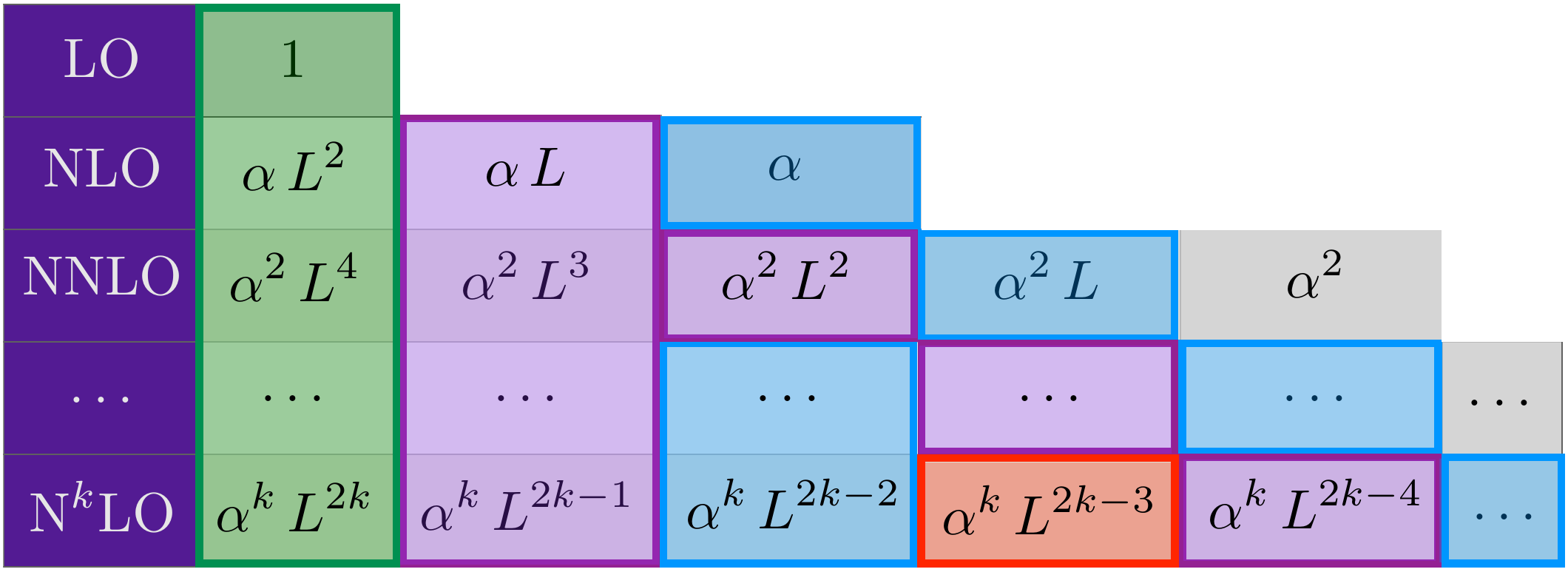}
\end{align}
where as in \cref{sec:RGE} above, the green shaded region corresponds to LL summation, the purple is NLL, the blue NNLL, and the red schematically contributes at N$^k$LL.  Note that the pattern is non-trivial, due to the fact that the anomalous dimension now includes two independent expansions in $\alpha$, the cusp anomalous dimension $\gamma_\text{cusp}$ and the non-cusp anomalous dimension $\gamma_F$.  Note again, we are neglecting the running of $\alpha$, which further complicates the expansions in \cref{eq:UFexpanded}.  This shows how SCET can be used to systematically sum the large logarithms due to IR soft and collinear divergences to arbitrary precision.

\subsection{A Few Remaining Concepts}
\label{sec:RemainingConcepts}
There are a few important additional ideas/techniques that often show up in SCET (and perturbative QCD) calculations, but were not required for our simple example processes explained here.  Therefore, we will conclude our discussion of SCET technology with a brief mention of a few of the most important ones.

SCET was invented to calculate the partial width for the heavy quark decay $B \rightarrow X_s\,\gamma$ (here $X_s$ is a ``jet'' that includes a strange quark) including the summation of large Sudakov logs~\cite{Bauer:2000ew}, see the discussion below in \cref{sec:Conc} for additional physics context.  This first calculation computed the partial width including the impact of this summation; we expect that the reader who has made it this far will now have the tools to read this paper.  Later it was understood how to compute the \emph{spectrum} of the outgoing photon~\cite{Bauer:2003pi, Bosch:2004th}.  This relies on the inclusion of a ``measurement function,'' where the unitarity cuts (see the discussion around \cref{eq:DaveExample_RCut}) are augmented by an additional constraint on the states crossing the cut.  For the $B \rightarrow X_s\,\gamma$ example, one would interpret the symbol ``\,\,\includegraphics[width=0.015\textwidth, valign=c]{Figures/CutSymbol.pdf}\,\,'' crossing a photon line as the replacement
\begin{align}
\frac{1}{p^2-m^2 + i 0} \quad\longrightarrow\quad -2\s i\s\pi\,\delta\big(p^2-m^2\big)\,\theta\big(p^0\big)\,\delta\big(E_\gamma - E_\gamma^\text{measured}\big)\,,
\end{align}
where $E_\gamma$ is the dynamical energy of the photon that flows through the diagram, and $E_\gamma^\text{measured}$ is the energy of the photon fixed by the observable.  This  yields a computation of the observable $\D \Gamma_B/\D E_\text{measured}$.  This technique is useful since it keeps careful track of various factors that appear when cutting all the possible channels of the relevant diagrams.  Furthermore, this measurement function can be given an operator definition, which allows one to apply power counting rules to the measurement function, thereby ensuring that the EFT models the impact of the measurement consistently.  Note that the region of validity where SCET is appropriate is in the limit $E_\text{measured} \rightarrow m_b/2$, known as the ``endpoint.''  When one is interested in an inclusive observable, then the operator product expansion likely provides a better approximation.

When one is computing a complicated observable, especially if it is differential in some variable as in the situation we were just discussing, one often finds that the resulting spectrum is a generalized function or distribution~\cite{WikiDist}, see~\cite{Gelfand:105396} for a book on generalized functions.  For example, the resulting endpoint differential spectrum for the $B$ meson decay~\cite{Bauer:2003pi, Bosch:2004th} contains on a factor of $\delta\big(m_b/2 - E_\gamma^\text{measured})$, and also depends on ``plus functions.''  A plus function arises from the expansion of the dimensionally regulated factor
\begin{align}
\frac{1}{(1-x)^{1+\epsilon}} = -\frac{1}{\epsilon}\delta(1-x) + \frac{1}{[1-x]_+}-\epsilon\left[\frac{\log(1-z)}{1-z}\right]_+ + \cdots\,,
\end{align}
where the plus function is defined as
\begin{align}
\int_0^1 \D x\, \frac{f(x)}{(1-x)_+} &= \int_0^1 \D x\,\frac{f(x)-f(1)}{1-x} \,,
\end{align}
which implies that $1/[1-x]_+ \neq 1/(1-x)$.  For example, this factor appears in the DGLAP evolution equations that govern the RG scaling of the parton distribution functions~see \emph{e.g.} the relevant discussion in~ \cite{Schwartz:2013pla} and~\cite{Peskin:1995ev}.    One way to think about why these objects appear is that the RG sums logs but does not change the power law structure of the observable.  Therefore, if the calculation of an observable includes a factor like $1/(1-x)$ to some power, then then one encounters plus functions when applying dim reg and Taylor expanding about $x = 1$.

Note that integrated distributions are normal functions, and so it is often more straightforward to work with an integrated distribution, and then differentiate it to yield the spectrum.  For example, we could define a cumulant for the $B$ partial width as
\begin{align}
\Gamma_B^\text{cumulant}\big(E_\gamma^\text{measured}\big) = \int_{E_\gamma^\text{measured}}^{m_b/2}\,\D E\,\frac{\D \Gamma_B}{\D E}\,,
\end{align}
which is simply a function of the photon energy.

Another complication that can arise is that factorization theorems may require non-perturbative input.  The most famous example is the factorization of parton distribution functions, which schematically takes the form\footnote{One might complain that this formula is not simply a product of functions, and so it is not technically ``factorized.''  However, this is simply an artifact of the space one is choosing to work in, \emph{e.g.} this formula becomes a simple product of functions when Fourier transformed to position space.  Then it can be convenient to perform scale setting in the factorized space, and then transform back.} 
\begin{align}
\sigma = \int_0^1 \D x_1 \, \int_0^1 \D x_2\, f_a\big(x_1\big)\,f_b\big(x_2 \big) \times \hat{\sigma}\big(x_1\,P,x_2\,P,\cdots\big)\,,
\end{align}
where the $x_i$ are momentum fractions, $f_{a/b}(x_i)$ is a parton distribution function, and $\hat{\sigma}$ is a partonic cross section.  Although one can derive operator definitions and RG evolution equations for the parton distribution functions (the DGLAP equations mentioned previously), they must ultimately be determined by comparing to data.  Note that when performing an inclusive measurement, this factorization holds.  However, in more complicated cases with measurements, this simple factorization is modified and additional tools like SCET become relevant.  Schematically, this picture looks something like
\begin{center}
\includegraphics[width=0.7\textwidth, valign=c]{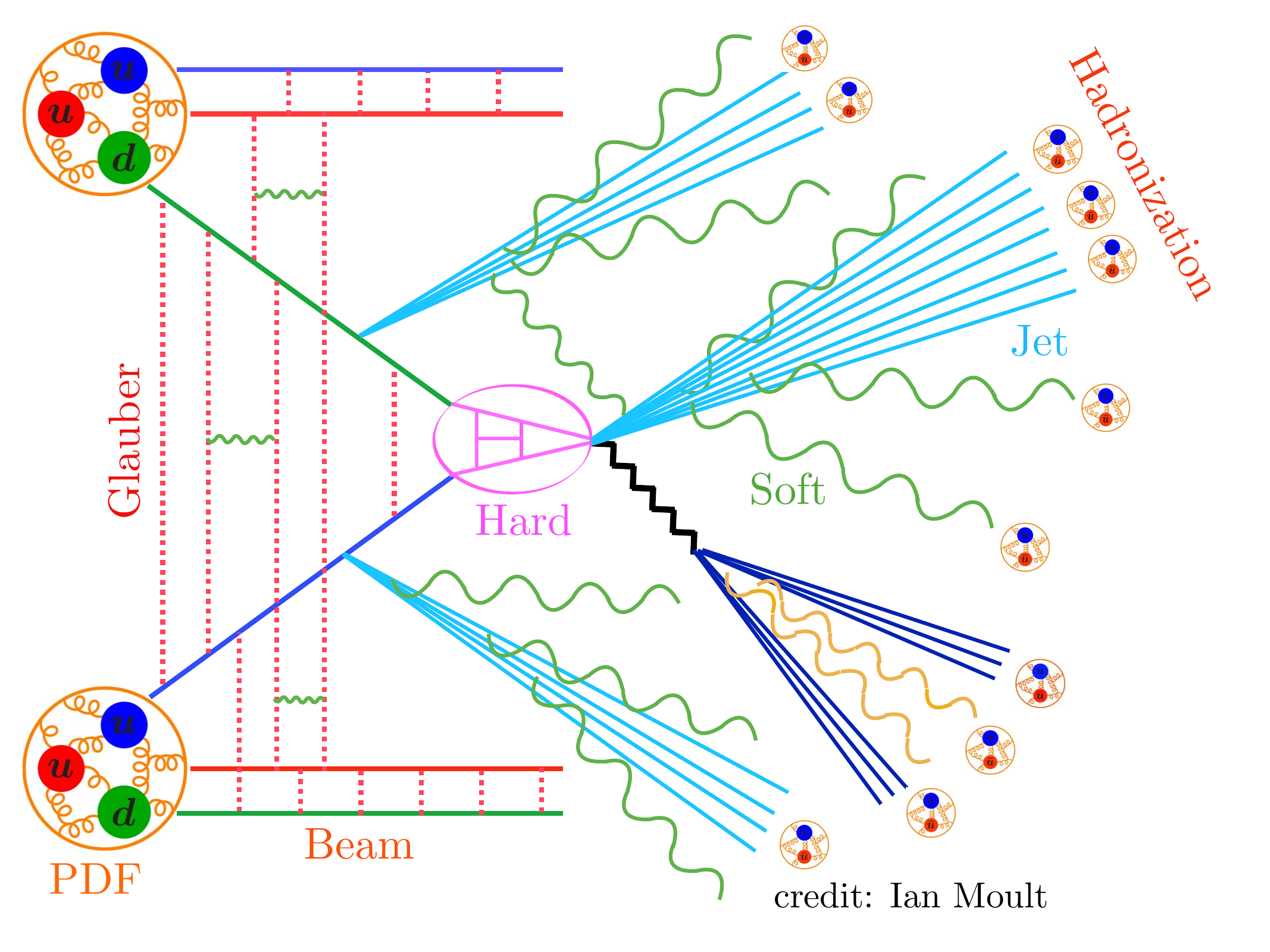}
\end{center}
where we have labeled a variety of effects (including the non-perturbative physics of parton distributions and hadronization) that must be taken into account if one wishes to compute a complicated realistic observable.

Finally, we note that the process dependence of SCET implies that one must be careful when expanding out the \FT~into EFT modes.  In particular, recent developments in additional factorizations, which can be relevant to multi-scale problems such as jet substructure~\cite{Bauer:2011uc, Larkoski:2014tva, Procura:2014cba, Larkoski:2015zka, Pietrulewicz:2016nwo}, and progress understanding the nature of factorization violation in SCET due to so-called Glauber modes~\cite{Rothstein:2016bsq}, makes it clear that there are a variety of modes that can in principle contribute (we have dropped an overall factor of $M$ for brevity):
\begin{align}
\text{hard:} & \quad p_h\hspace{3.5pt} \sim (1,1,1) \notag \\[3pt]
\text{collinear:} &\quad  p_c \hspace{4pt}\sim \big(\lambda^2,1,\lambda \big) \notag \\[3pt]
\text{anti-collinear:} &\quad  p_{\bar{c}} \hspace{4pt} \sim \big(1,\lambda^2,\lambda\big) \notag \\[3pt]
\text{soft:} &\quad  p_s \hspace{4pt}\sim \big(\lambda,\lambda,\lambda \big) \notag \\[3pt]
\text{ultra-soft:} &\quad  p_{us} \sim \big(\lambda^2,\lambda^2,\lambda^2 \big) \notag \\[3pt]
\text{soft-collinear:} &\quad  p_{c_s} \hspace{1pt}\sim (1-x)\,\big(1,\lambda^2,\lambda \big) \notag \\[3pt]
\text{anti-soft-collinear:} &\quad  p_{\bar{c}_s} \hspace{1pt}\sim (1-x)\,\big(\lambda^2,1, \lambda \big) \notag \\[3pt]
\text{non-relativistic:} &\quad  p_{nr} \sim \big(1,1,\lambda \big) \notag \\[3pt]
\text{Glauber:} &\quad  p_{\text{G}} \hspace{2pt}\sim \big(\lambda^2,\lambda^2,\lambda \big)\,,
\label{eq:AllTheModes}
\end{align}
where for the collinear-soft, we are assuming an additional scale (such as an endpoint measurement where the parameter $x \rightarrow 1$) implies the presence of an additional small parameter $(1-x)$, and the Glauber scaling can be generalized as discussed in~\cite{Feige:2014wja}.  This makes it clear that the IR of gauge theory can be quite complicated, and one must take care when matching a \FT~onto SCET.

This concludes our discussion of the technical aspects of SCET.  The final section closes these lectures by provides some much needed connections to physical examples, first for EFTs of Goldstone bosons, and then for SCET.

\newpage
\section{A Bit More Physics}
\label{sec:Conc}
If I have come close to accomplishing the goals I set for myself when concocting the topics covered in these lectures, you should now have a non-trivial level of comfort with the approach to separating scales where one matches a~\FT~onto an EFT in the UV, and subsequently sums logs by running the EFT Wilson coefficients to the IR.  The application of EFTs within particle physics is so ubiquitous, I made the hopeful assumption that the reader would find the motivation to read these lectures self evident.  However, it should also be obvious that some very important topics have been neglected.  Although you were forewarned, you might still be appalled at the lack of physics contained here.  To that end, I will take some time in this concluding section to emphasize some of the most important concepts that any reasonable Effective Field Theorist should have in her toolkit.  This discussion is additionally supplemented by a short EFT bestiary in \hyperlink{sec:EFTZoo}{Appendix A}.  Due to the enormity of what can be done with SCET, I also decided it would be useful to include some closing comments on the physics that has been explored with this framework as well.  I also will take this opportunity to remind you of the presence the annotated bibliography given in \hyperlink{sec:AnnBib}{Appendix B}, which I hope will provide some guidance for your future EFT endeavors.

\subsubsection*{(Goldstone) EFTs and Physics}
One intuitive jumping off point for a set of lectures on EFTs would be a discussion of Fermi theory and the weak interactions, and in fact many of the excellent lectures on EFTs that already exist do exactly that (see \hyperlink{sec:AnnBib}{Appendix B}).  This would be nicely followed by a discussion of applications to low energy QCD, since all on its own, the fact that the interactions of light mesons can be modeled in detail as a theory of Goldstone bosons, even though the \FT~description is strongly coupled, makes a deeply compelling case for the study of EFTs.  My early outlines of these lectures had prominently featured the EFT of Goldstones, and so as an attempt to assuage my guilt over the lack of physics contained in what you have in front of you, I will briefly describe the physics I would have reviewed if I had followed that approach.

A Goldstone boson is the massless (a condensed matter physicist might call this gapless) excitation associated with the spontaneous breaking of a global symmetry.\footnote{This assumes the breaking preserves Lorentz invariance.  For a treatment of Goldstone bosons in non-relativistic systems, see \emph{e.g.} \cite{Watanabe:2012hr, Watanabe:2014fva}.}   Since it is massless, it persists to the IR and should be included as a degree of freedom in a low energy EFT.  If the global symmetry is exact, then the EFT Lagrangian for the Goldstones is entirely determined by its particular symmetry breaking pattern, which can be derived by relying on the constraints imposed by the coset structure $G/H$, where $G$ is the full global symmetry group and $H$ is the subgroup of $G$ that remains unbroken when the spontaneous symmetry breaking is active~\cite{Coleman:1969sm, Callan:1969sn}.  Additionally, introducing explicit violations of the global symmetry in a controlled way yields masses for the now pseudo-Goldstone bosons, along with new interactions.

When applied to the Standard Model, this framework is known as the ``chiral Lagrangian'' or ``chiral perturbation theory'' since it is the manifestation of chiral symmetry breaking in the quark sector, see \emph{e.g.}~\cite{Leutwyler:1993iq, Ecker:1994gg, Scherer:2002tk, Pich:2018ltt} for some reviews.  The quark sector has an approximate $\text{U}(3)_L\times \text{U}(3)_R$ global symmetry since the up, down, and strange quark masses are small compared to the confinement scale of QCD.  Then the input assumption for chiral perturbation theory is that strong dynamics at the QCD scale break this symmetry down to the diagonal $\text{U}(3)_V$, and the light quark masses play the role of the small global symmetry breaking.  Identifying the pions and Kaons with the resulting pseudo-Goldstone bosons, yields a phenomenal phenomenological model of low energy QCD. One can derive the couplings of these Goldstones to nucleons, which yields a rich set of observables and gives a reasonable first approximation for nuclear binding effects.  It is additionally possible to gauge an unbroken global $\text{U}(1)$ symmetry, which induces couplings to the photon.  Furthermore, the anomalous nature of the global symmetry structure yields a calculable Wess-Zumino-Witten term, which leads to a variety of effects, \emph{e.g.}~it explains how the neutral pion can decay to a pair of photons.  There are a number of additional topics that could have been covered, \emph{e.g.}~naive dimensional analysis as an approach for estimating the size of operators, chiral perturbation theory at loop level, and the connection between extending the global symmetry to include a ``hidden local symmetry'' as a model for the $\rho$-meson.  The physics of light mesons also provides a nice segue to exploring the related meson phenomenology described by Heavy Quark Effective Theory~\cite{Isgur:1989vq, Grinstein:1990mj, Eichten:1989zv, Georgi:1990um, Neubert:1993mb, Manohar:2000dt}.  This framework allows one to treat the physics of heavy mesons in a controlled way by isolating the strongly coupled effects of QCD utilizing a factorization theorem.

There are a number of subjects that have compelling relations to the chiral Lagrangian.  One of my favorite examples is Seiberg duality~\cite{Seiberg:1994pq}, which serves as an incredibly rich playground for a calculable theory of mesons akin to the chiral Lagrangian.  Seiberg's brilliant insight was that supersymmetric QCD-like theories (labeled by a number of colors $N_c$ and a number of flavors $N_f$) can expose the physics of (supersymmetric) mesons in a completely controlled way, see \emph{e.g.}~\cite{Intriligator:1995au} for early lectures and~\cite{Terning:2006bq} for a comprehensive modern treatment.\footnote{One might argue that talking about Seiberg duality as an EFT is misleading since it neglects the fact there is no controlled expansion beyond leading order.  Specifically, the heavy mass limit for a quark only works in infinite or zero mass limit, \emph{i.e.}, Seiberg duality is a statement about the deep IR.  The two dual descriptions really are different theories that happen to flow in to the same IR description, see~\cite{deGouvea:1998ft} for an explicit demonstration of the consequences of this fact.} The duality is a mapping from a quark description that is strongly coupled in the IR to a theory of mesons that is weakly coupled in the IR.  This is a rich subject that works as a beautiful example of EFT reasoning.  It takes advantage of the powerful constraints imposed by supersymmetry (in particular that the superpotential is not renormalized), along with a concept for matching two theories that was not covered in these lectures known as ``anomaly matching''~\cite{tHooft:1979rat}.  This is the idea that if two theories have a chance of describing the same physics in the IR, they must manifest the same global symmetries.  Then one can perform current correlator calculations to check if any of these global symmetries are anomalous.  The presence of an anomaly is an all-orders statement, it cannot be altered by an RG flow.  Therefore, if any of the global symmetries are anomalous in one theory, the anomaly must also be present in the dual theory, which constrains the allowed particles and charges.  This provides a necessary condition that the two theories describe the same IR physics.  In this sense, one derives a non-perturbative matching between a quark and meson description of the same IR dynamics.\footnote{For the interested reader, I will highlight a few more features of Seiberg duality.  One of the key aspects of the construction relies on a non-perturbative instanton calculation for the case when $N_f = N_c -1$ that determines the precise form of the so-called ADS superpotential that is a consequence of strong dynamics~\cite{Affleck:1983mk, Affleck:1984xz}.  Then by relying on the physics of decoupling, one can introduce masses for the quarks, take them to be large and match this theory with a heavy quark onto an EFT that has one fewer flavor.  Various dynamical phases of supersymmetric gauge theories can be inferred, including an example theory that confines without breaking chiral symmetry, manifesting ``s-confinement''~\cite{Csaki:1996sm}.  Additionally, one can argue for the existence of the ``conformal window'' for theories that satisfy $3/2 < N_f/N_c < 3$~\cite{Seiberg:1994pq}.  The claim is that supersymmetric QCD models that satisfy this relation between $N_f$ and $N_c$ are conformal: the theories at the high edge of this window have an incalculable strongly coupled fixed point, while those at the bottom edge are an example of a weakly coupled ``Banks-Zaks''~\cite{Banks:1981nn} fixed point.  Finally, there are compelling connections to be made to a version of the chiral Lagrangian that incorporates the $\rho$ meson using the hidden local symmetry technique mentioned above~\cite{Komargodski:2010mc}.}

There is another incredible application of Goldstone EFTs that would have been exciting to highlight: the Effective Theory of Inflation~\cite{Cheung:2007st}, and see~\cite{Baumann:2009ds, Senatore:2016aui} for reviews.  This EFT is based on the realization that the physics of slow-roll inflation has a description in terms of a spontaneous breaking of the global Lorentz spacetime symmetry.  Specifically, the inflaton is identified as the Goldstone boson of spontaneously broken time-translation invariance, since the slowly evolving background can be interpreted as setting a clock.  It is possible to derive a universal Lagrangian for slow roll inflation as an EFT expansion, which allows one to infer the number of independent observables at a given order in power counting.  One can compute deviations from the Gaussian predictions within the EFT, which can be mapped onto interactions in the \FT.  This EFT sharpens the question of what aspects of UV models for inflation can be probed using observations.  There is another EFT relevant for cosmology, which is the EFT for large scale structure which allows one to connect Wilson coefficients directly to the results of $N$-body cosmological simulations~\cite{Baumann:2010tm, Carrasco:2012cv}.  This is an EFT to describe deviations from the perfect fluid which is a good approximation of the dark matter distribution until non-linear structure formation takes over.  Taken together, these formalisms demonstrate the relevance of EFT techniques to precision cosmology.

In an alternate timeline, the lectures you have before you would have introduced you to these topics.  For additional examples of EFT applications (including some overlap with what was just said here), see \hyperlink{sec:EFTZoo}{Appendix A}.

\subsubsection*{SCET and Physics}
Most of what you need to know to work with SCET was introduced in the main body of these lectures.  However, given the minimal connections to physics, it is worthwhile to briefly describe how this framework has been applied to real world processes.  I should emphasize that working with SCET has become a subfield all of its own, and a much more comprehensive discussion of applications is provided in~\cite{Becher:2014oda, Becher:2018gno}.

The framework of SCET was discovered as a way to sum logarithms that appear heavy meson decays, \emph{e.g.}~the $B$-meson decay channel $B\rightarrow X_s\,\gamma$, where $X_s$ is a ``jet'' that includes a strange quark, $B\rightarrow X_s\, e^+\,e^-$, and $B \rightarrow X_u\,\ell\,\bar{\nu}$, where $X_u$ is a ``jet'' that contains an up quark.   These partial widths can be computed using Heavy Quark Effective Theory, where the decay can be modeled using the operator product expansion, see \emph{e.g.}~\cite{Falk:1993dh, Kapustin:1995nr} for a calculation of $B\rightarrow X_s\,\gamma$.  These calculations manifest a kinematic Sudakov double log when one restricts the phase space of the final state, \emph{e.g.}~when the photon energy is near half the $B$ mass.  This motivated the invention of SCET~\cite{Bauer:2000ew}, which was used to sum the large logs and provide an RG improved prediction for the $B$ partial decay widths.  This was then followed up by refinements of the calculation and calculations of the spectrum near the endpoint $2\,E_\gamma/m_b$~\cite{Neubert:1993um, Bauer:2000ew, Bauer:2003pi, Neubert:2004dd, Becher:2006pu}, and sub-leading power contributions to $B$ decays were also computed~\cite{Beneke:2002ph}.  Around the same time, the SCET approach to deep inelastic scattering was pioneered in~\cite{Manohar:2003vb}. 

There have also been many applications to collider physics.  For example, SCET was used to sum QCD corrections to the thrust observable at an $e^+\,e^-$ collider, which allowed for a precision  $\alpha_s$ extraction from LEP data~\cite{Becher:2008cf}.  The thrust calculation has also recently been computed including sub-leading power corrections~\cite{Moult:2018jjd}, providing the first concrete calculation of a collider observable beyond leading order in SCET power counting.  SCET techniques have also been applied at hadron colliders.  The factorization of initial state jets from the parton distribution functions was explored in~\cite{Stewart:2009yx}.  The hadron collider production of the Higgs boson in the gluon fusion channel including summation for jet vetos was done in~\cite{Berger:2010xi, Stewart:2013faa}.  There has also been recent progress in taking a SCET description and factorizing it into a sub-SCET, which allows logarithms associated with jet substructure (\emph{e.g.}~logs of the jet radius parameter) to be summed~\cite{Bauer:2011uc, Larkoski:2014tva, Procura:2014cba, Larkoski:2015zka, Pietrulewicz:2016nwo}.  Finally, some interesting studies that turned SCET ideas on their head to make rigorous statements about perturbative QCD amplitudes was performed in~\cite{Becher:2009cu, Gardi:2009qi, Becher:2009qa, Becher:2009kw}.

Given the extent to which SCET (and for that matter so much of collider physics in general) depends on the factorization of observables, it is interesting to study the systematics of factorization violating effects.  One source of these effects are due to Glauber modes.  The proof that Glauber modes only contribute an overall phase to the matrix elements for the Drell-Yan process was a critical component of the original demonstration of the factorization for the parton distribution functions~\cite{Collins:1984kg}.  Not only are the presence of Glauber modes a potential harbinger of factorization violation, they are also a critical to include if one is modeling the forward scattering (or Regge) limit of a cross section when $|t| \ll s$.  SCET in the presence of Glauber modes has now been studied extensively~\cite{Rothstein:2016bsq}, including recent results computing the Glauber contribution from $t$-channel quark exchange~\cite{Moult:2017xpp}, and for electroweak factorization violating effects~\cite{Baumgart:2018ntv}.  There is another potential source of factorization violation, known as non-global logarithms~\cite{Dasgupta:2001sh}.  These can occur in the presence of a hierarchy of scales when radiation from one side of an event can non-trivially impact the kinematics of the other side.  The summation of this class of logarithms can be performed in the limit of a large number of colors using the so-called BMS equation~\cite{Banfi:2002hw}.  Non-global logarithms have additionally been cast in the language of SCET~\cite{Larkoski:2015zka, Larkoski:2016zzc}.  Using a conformal transformation, non-global logarithms can be related to the BFKL equation that governs the evolution of logarithms that appear in the Regge limit~\cite{Caron-Huot:2015bja}.  
 
I am compelled to close this discussion with examples of SCET applied to physics beyond the Standard Model, both because this is in close alignment with the topics covered at TASI 2018, and especially because this gives me the opportunity to unabashedly highlight some of my own work.  My personal entry into the world of SCET started with the realization that heavy wino annihilation to photons suffers from a large Sudakov log that has an $\mathcal{O}(1)$ impact on the NLO prediction~\cite{Hryczuk:2011vi}.  This large log is not due to kinematics, but instead emerges because the final state is restricted to contain a photon, which one is an allowed final state since electroweak symmetry is broken.  This causes an imprecise cancelation between the virtual and real emission diagrams, yielding a Sudakov double log.  This motivated myself and collaborators, along with two other groups, to apply SCET to sum the total annihilation rate~\cite{Baumgart:2014vma, Bauer:2014ula, Baumgart:2014saa, Ovanesyan:2014fwa, Baumgart:2015bpa, Ovanesyan:2016vkk}.  We then followed this up to compute the spectrum~\cite{Baumgart:2017nsr, Baumgart:2018yed}, which involves introducing a measurement function as discussed above, and relied on multi-stage factorization techniques to separate the multiple scales in the problem.  There have also been applications of SCET to general relativity in~\cite{Beneke:2012xa, Okui:2017all}, where the soft sector can be shown to eikonalize and factorizes in QCD, but the collinear sector does not manifest any divergences.  On another front, there have been recent studies applying SCET to a hypothetical post-discovery scenario where a new heavy beyond the Standard Model particle has been observed at the LHC or a future collider~\cite{Alte:2018nbn,Alte:2019iug}.  Finally, I have recently been involved in understanding the interplay of SCET and supersymmetry.  We now have understood how to formulate superspace in the collinear limit.  In~\cite{Cohen:2016jzp, Cohen:2016dcl}, we constructed theories in collinear superspace from the top down by ``integrating out half of superspace.''  Now, we have understood the bottom-up EFT rules for working with superfields in collinear superspace, with a particular emphasis on the action of RPI, and have uncovered some novel superfield objects that do not exist in the Lorentzian version of $\mathcal{N} = 1$ supersymmetry~\cite{Cohen:2018qvn}.

This discussion should make it pretty obvious that SCET has tremendous relevance for physics of the Standard Model and beyond.  There is no question that the framework of SCET will continue to be explored and applied to more phenomena.  
\vspace{5pt}\mybox{\begin{itemize}
\item \textbf{Exercise:} What did these lectures inspire you to go calculate?
\end{itemize}}

\section*{Acknowledgments}
\addcontentsline{toc}{section}{\hspace{14pt}Acknowledgments}
I am extremely grateful to my friends and collaborators who helped me in enumerable ways throughout this process.  X.~Lu was my first reader, and he was handed a rough early version of the notes.  He worked through every expression, catching tons of typos, and more importantly helped to clarify and streamline a variety of discussions, often realizing what I was trying to explain even when it was presented poorly.  The notes have a hope of getting most of the factors of two correct due to his diligent efforts.  Furthermore, he provided me with the discussion of operator counting (an area he has helped to pioneer) that is presented above.  My second reader was M.~Freytsis, who has one of the deepest understandings of the ideas presented here of anyone I know.  He was asked to read a more refined version, and provided countless insights, helping whip a number of conceptual problems into shape.  Furthermore, so much of my knowledge of EFTs comes from the many conversations we have had over the years.  He was also one of the key people I bounced ideas off of when writing the lectures that were presented to the TASI students, helping to shape their form into something (hopefully) clear, correct, and to the point.  Another of my most important sounding boards was I.~Moult.  He made his incredible knowledge of EFTs available to me, both when preparing for my lectures at TASI, and then as I developed these notes.  His explanations of many EFT concepts appear in the lectures you have before you.   I am additionally grateful for his permission to include his beautiful figure illustrating a real world QCD process.  D.~Soper is an inexhaustible wealth of knowledge regarding the IR of field theory and QCD.  He is always extremely generous with time and his willingness to help one make sense of it all.  He also provided many suggestions and clarifications throughout the draft, especially in \cref{sec:SoftCollinearDiv}.  M.~McCullough and N.~Craig have had innumerable conversations with me over the years regarding fine-tuning that have deeply shaped the way I think about naturalness.  I have done my best to channel their voices in my philosophical section on the hierarchy problem, and much of the discussion there was developed using feedback I received on the draft from M.~McCullough.  V.~Vaidya provided invaluable insights into understanding the power counting of scalar SCET.  G.~Elor has taught me most of what I know about RPI and collinear fermions.  M.~Solon gave me very useful insights into the EFT of inflation, large scale structure, and binary inspirals.  Y.~Kahn clued me in to just how interesting the Euler-Heisenberg is.  G.~Kribs, B.~Ostdiek, and S.~Chang put up with me barging into their offices or taking over lunch on at times a near daily basis to ask for help on factors of 2, especially when preparing to head to TASI, and also in writing up these notes.  In the same spirit, I owe the UO theory group tremendous gratitude for putting up with me as I lost myself in this process and was at times distracted or worse.  I am extremely grateful to all my collaborators over the years who have taught me so many of the insights and shared their intuition.  A very important thank you goes to the TASI 2018 students for a truly memorable lecturing experience, and for inspiring me to write these lectures.  These are for you first and foremost, and the thought of you studying them kept me going through the long hours it took to put this all together.  T.~Slatyer and T.~Plehn for asking me to do this in the first place.  I hope these notes live up to your expectations.  Additionally, T.~DeGrand deserves a mention for all he did to make TASI 2018 the amazing experience that it was for both students and faculty.  And finally, to my partner L.~Curto Serrano, for all her gracious support as I let myself be consumed by this process.  

For the updates in v2: I got a message from Michael Peskin during summer 2020 that he had been working through these lectures, and had many comments in \cref{sec:RegionsMasslessSudakov} through \cref{sec:RegionsMassiveSudakov}.  Not only did he catch a number of typos, but he also provided many suggestions and formulas which improved these sections dramatically.  I am additionally very grateful to Matthew Low and Akhil Premkumar for correcting various typos.  

TC is supported by the US Department of Energy under grant number DE-SC0011640.  

\newpage
\section*{Appendix~A.~The Effective Theory Zoo}
\addcontentsline{toc}{section}{\hspace{10pt} \textbf{Appendix~A.~The Effective Theory Zoo}}
\hypertarget{sec:EFTZoo}{}
There are far too many EFTs to list comprehensively.\footnote{To my knowledge, the best attempt can be found via the online course~\cite{iain_class}.}  However, in the interest of making the point that these techniques apply to many situations, I was compelled to provide this appendix.  I will briefly discuss some important EFTs, highlighting a few of their most salient features.  
\end{spacing}
\begin{spacing}{1.1}
\vspace{-10pt}
\begin{itemize}
\item \textbf{Fermi's Theory of the Weak Interactions}~\cite{Fermi:1934sk}: Perhaps the first EFT that a student of particle physics learns is Fermi theory, which describes the low energy limit of the weak interactions as invented by Fermi to model $\beta$ decay.  This EFT can be derived from the Standard Model by integrating out the $W$ boson, yielding a description of the weak interactions as contact operators between Standard Model fermions, \emph{e.g.}~$G_F\,\big(\bar{u}\,\gamma^\mu\,P_L\, s\big)\big(\bar{d}_k \,\gamma_\mu\,P_L\,u\big)$, where $u$ is the up quark, $d$ is the down quark, $s$ is the strange quark, $\gamma^\mu$ are the Dirac matrices, $P_L$ is the left-projection operator, and $G_F \sim g_2^2/m_W^2$ is the Fermi constant ($g_2$ is the weak coupling constant, and $m_W$ is the $W$-boson mass), which can be determined through a matching calculation.  This is effectively just a fancy real world version of the example we discussed above in \cref{sec:TreeMatching}.
\item \textbf{The Euler-Heisenberg Lagrangian}~\cite{Heisenberg:1935qt} (see \emph{e.g.}~\cite{Dunne:2012vv} for a review and the discussion in~\cite{Schwartz:2013pla}, and a recent experimental proposal that has a shot at measuring it for the first time~\cite{Eriksson:2004cz, King:2015tba, Bogorad:2019pbu}):  This is the EFT of electromagnetic field interactions, where the electron has been integrated out.  It describes photon-photon scattering when expanded in the weak field limit, with a coefficient $\sim \alpha^2/m_e^4$, where $\alpha$ is the electromagnetic fine-structure constant, and $m_e$ is the electron mass.  It is particularly interesting because this calculation holds to all orders in the electro-magnetic field, such that it can be extended to the strong field regime.  This allows one to describe effects like Schwinger pair production~\cite{Schwinger:1951nm}, when interpreting the term as arising from an instanton-like effect.  One recent extension has been to write down an EFT description for the propagation of charged particles in a magnetar magnetosphere, where both strong electric fields and a dilute plasma exist~\cite{Freytsis:2015qda}.  
\item \textbf{Chiral Perturbation Theory} (see \emph{e.g.}~\cite{Leutwyler:1993iq, Ecker:1994gg, Scherer:2002tk, Pich:2018ltt} for some reviews):  This is the EFT of the light mesons in QCD, which relies on identifying the pions and Kaons as Goldstone bosons of a flavor symmetry that is spontaneously broken by the confinement of QCD.  This is an example where matching is not useful, since the scale where one would attempt to perform a matching calculation would be $\mu \sim \Lambda_\text{QCD}$, where we do not have perturbative control of the theory.  However, this does not stand in the way of being able to use this EFT to extract a tremendous amount of physics by enforcing the symmetry structure, providing insight into the nature of low energy QCD and nuclear physics.  A discussion of this EFT was provided in \cref{sec:Conc} above.
\item \textbf{Heavy particle EFTs}~\cite{Isgur:1989vq, Grinstein:1990mj, Eichten:1989zv, Georgi:1990um} (also see the review~\cite{Neubert:1993mb} and book~\cite{Manohar:2000dt}): An EFT for heavy particles can be written down by modeling these states as classical static sources with small residual quantum fluctuations.  This implies a preferred frame, namely the rest frame of the heavy particle where $v^\mu = (1,0,0,0)$.  One can then write down a systematic operator expansion by expanding the momentum of the heavy state as $p^\mu = M\,v^\mu+ k^\mu$, where the small momentum $k^\mu$ provides a power counting parameter.  This approach has yielded many important quantitative successes when applied to heavy meson decays.
\item \textbf{EFTs for Non-relativistic Processes}~\cite{Caswell:1985ui, Luke:1999kz}, (also see~\cite{Rothstein:2003mp} for a review):  This can be seen as an extension of the heavy particle EFT, where a further matching is performed onto non-relativistic quantum mechanics so that perturbative bound state effects can be systematically treated.  This leads to a multi-scale problem since now $p \sim M\,v$ and $E \sim m\,v^2$ are both necessary power counting parameters.  This EFT has applications to quarkonia, can be used to calculate the Lamb shift in QED, and leads to the intriguing notion of a velocity RG.   A related approach has recently been applied for understanding black hole inspirals~\cite{Cheung:2018wkq}.  Another aspect of these EFTs is their application to scattering calculations.  This has been explored in the context of nucleon scattering in \emph{e.g.}~\cite{Lepage:1997cs, Kaplan:1998tg}.
\item \textbf{Soft Collinear Effective Theory}~\cite{Bauer:2000ew, Bauer:2000yr, Bauer:2001ct, Bauer:2001yt, Bauer:2002uv, Beneke:2002ph} (also see~\cite{Dugan:1990de} for an early attempt known as Large Energy Effective Theory):  Given that essentially half of these lecture notes are devoted to SCET, it had to be put in this list.  The basic idea is to develop an EFT for summing large Sudakov double logarithms.  This EFT is incredibly rich, and there are a tremendous number of applications, so if you are curious about it we encourage you to read the body of these lectures.
\item \textbf{The EFT of Inflation}~\cite{Cheung:2007st} (also see~\cite{Baumann:2009ds, Senatore:2016aui} for reviews):  There is an EFT description of inflation, where the inflaton is treated as a Goldstone boson that results from the spontaneous breaking of time-translation invariance due to the slowly varying vacuum energy that was present during inflation.  This EFT demonstrates the universal nature of inflation (in that the leading order Lagrangian only depends on the scale of inflation and the speed of sound), and allows a systematic organization of the observables that result from quantum fluctuations of the inflaton.
\item \textbf{The EFT of Gravitational Inspirals}~\cite{Goldberger:2004jt} (also see~\cite{Goldberger:2007hy, Porto:2016pyg} for a review):  This EFT is obviously relevant now that we live in a world where the observation of gravitational waves from binary inspirals has become routine.  This EFT organizes the post-Newtonian expansion of general relativity, where there are three relevant scales to keep track of:  the size of the inspiraling object, the orbital radius, and the wavelength of the gravitational radiation.
\item \textbf{The EFT of the Fermi Surface}~\cite{Benfatto:1990zz, Shankar, Weinberg:1993dw, Kapustin:2018dch}:  A demonstration of the fact that EFTs are a useful formalism to describe pseudo-particles in non-relativistic systems is the EFT that describes excitations of the Fermi surface.  This allows one to explore phases of the Fermi liquid using EFT techniques.
\end{itemize}
\end{spacing}
\begin{spacing}{1.3}
\noindent Since I have to stop somewhere, I might as well leave it at this, apologize to the reader who's favorite EFT did not make this list, and move on to my annotated bibliography.

\section*{Appendix~B.~A Brief Annotated Bibliography}
\addcontentsline{toc}{section}{\hspace{14pt}\textbf{Appendix~B.~A Brief Annotated Bibliography} \vspace{-5pt}}
\hypertarget{sec:AnnBib}{}
In this appendix, I will feature many of the resources that are available for learning about EFTs and advanced field theory techniques.  Furthermore, this will provide me with an opportunity to acknowledge many of the resources that were used heavily when constructing these lectures.  I have organized the references into three categories.  The first set are broadly EFT centric, and in particular provide many real world physical applications which are notably absent from what is presented in these lectures.  The second set are specific to SCET, and provide many details and applications that go beyond our goal of simply setting up the framework.  Finally, I highlight a few technical resources that one might find to be very useful when attempting to perform tricky integrals or deal with subtle aspects of dim reg and the RG.

\end{spacing}
\vspace{-10pt}
\begin{spacing}{1.1}
\begin{center}
\sc{A Variety of Perspectives on EFTs}
\end{center}
\vspace{-5pt}
\begin{itemize}
\item Aneesh~Manohar's 2017 Les Houches lectures~\cite{Manohar:2018aog} and 1995 Lake Louise Winter Institute lectures~\cite{Manohar:1995xr}:  The 2018 lectures provide details for many physical examples, both for determining the degrees of freedom and power counting schemes.  This is followed by a comprehensive introduction to matching and running.  In particular, the inspiration to study the heavy-light integral in \cref{eq:SepScalesHLlog} came from these notes.  The 1995 lectures include an extensive discussion of the EFT for the weak interactions at low energies at tree and loop level, a discussion of loop-level heavy particle decoupling, and an introduction to chiral perturbation theory, including coupling of heavy quarks to the Goldstones.
\item Matthias Neubert's 2017 Les Houches lectures~\cite{Neubert:2019mrz}:  These lectures cover renormalization theory for QED and QCD.  Multiple renormalization schemes are emphasized.  Then applications to EFTs and the notion of renormalizing a composite operator is discussed.  One of the goals of these notes is to serve as an introduction to the three 2017 Les Houches lectures highlighted here~\cite{Manohar:2018aog, Pich:2018ltt, Becher:2018gno}.
\item Antonio Pich's 2017 Les Houches lectures~\cite{Pich:2018ltt}:  These lectures provide a systematic introduction to EFTs of Goldstone bosons.  First the general formalism is reviewed.  This is followed by applications to chiral Perturbation theory, including a discussion of symmetry breaking effects, quantum anomalies, the large $N_c$ limit, and some phenomenological applications.  Then the Standard Model EFT is treated with an emphasis on the Goldstone nature of the Higgs boson.
\item David~B.~Kaplan's 2005 National Nuclear Physics Summer School lectures~\cite{Kaplan:2005es} and Institute for Nuclear Theory 1995 Summer School lectures~\cite{Kaplan:1995uv}:  The 2005 lectures cover many applications and techniques of EFTs, including a beautiful EFT driven discussion for why the sky is blue, chiral perturbation theory, nucleon EFTs, and color superconductivity.  The shorter 1995 lectures also provide many insights into the philosophy and framework of EFTs, along with a few examples. 
\item Ira~Rothstein's 2002 TASI lectures~\cite{Rothstein:2003mp}:  These notes provide a detailed introduction to the matching and running technology, followed by an introduction to EFTs for non-relativistic bound states (NRQCD), and an EFT approach to theories of large extra dimensions.  Strong emphasis is place on the details of the RG for these models.
\item Alexey~Petrov and Andrew~Blechman's book~\cite{Petrov:2016azi}:  This is a comprehensive introduction to EFTs.  After a discussion of techniques, many examples are presented including chiral perturbation theory, Heavy Quark Effective Theory, NRQCD, SCET, higher dimensional operators for the Standard Model, and EFTs of gravity.  They provide a nice organization of types of EFTs: those where it is possible to match onto a UV theory, and those where this is not possible but the symmetries of the EFT are enough to extract useful physics.
\item Iain~Stewart's 2013 edX course~\cite{iain_class}:  This is a full course on EFTs.  It is freely available online, and provides access to a remarkable set of resources including video lectures, homework assignments, and lecture notes.
\item Howard Georgi's 1993 review~\cite{Georgi:1994qn}:  This review emphasizes the philosophy of EFT techniques.  There are a purposefully minimal number of expressions, allowing plenty of space to emphasize the conceptual underpinnings of the EFT approach.
\item Aneesh~Manohar and Mark~Wise's book~\cite{Manohar:2000dt}:  This monograph provides a comprehensive introduction to Heavy Quark Effective Theory.  Many techniques and applications are presented, with an emphasis on the physics of heavy mesons in the Standard Model.
\item Matthew~Schwartz's book~\cite{Schwartz:2013pla}:  This book is referenced here for two reasons.  First, I have largely followed its conventions, so this is a useful resource if one is interested in double checking factors of 2 or $i$.  Furthermore, it includes introductions to the Euler-Heisenberg Lagrangian, Heavy Quark Effective Theory, SCET, and (of course) the Standard Model.
\item Witold Skiba's 2009 TASI lectures~\cite{Skiba:2010xn}:  These lectures begin with an explanation of EFT techniques, including a comprehensive discussion of the hierarchy problem.  This is followed by applications to the EFT approach to precision electroweak calculations.
\item Markus~Luty's 2004 TASI lectures~\cite{Luty:2005sn}: Although these are lectures on supersymmetry breaking, they open with an introduction to EFTs, emphasizing the hierarchy problem.
\item Joseph~Polchinski's 1992 TASI lectures~\cite{Polchinski:1992ed}:  These lectures provide a condensed matter focused point of view on EFTs.  The EFT of the low energy modes for a conductor are presented, with applications to Fermi liquids and high-$T_c$ superconductors.
\item Walter~Goldberger's 2006 Les Houches lectures~\cite{Goldberger:2007hy}:  Following a discussion of EFT techniques, these lectures provide an introduction to the EFT for the gravitational wave spectrum resulting from binary inspirals.
\item Kenneth~Wilson and John~Kogut's 1974 Physics Reports review~\cite{Wilson:1973jj}:  This reference is included here because it is a candidate for the first review of EFTs, co-authored by one of the inventors of the approach.  The emphasis here is on critical phenomena in statistical mechanics.
\end{itemize}
\begin{center}
\sc{Soft Collinear Effective Theory Reviews}
\end{center}
\begin{itemize}
\item Thomas Becher, Alessandro Broggio, and Andrea Ferroglia's book~\cite{Becher:2014oda} and Thomas Becher's 2017 Les Houches lectures~\cite{Becher:2018gno}:  The impact of this SCET book and lectures on the development of these notes cannot be overstated.  In particular, the notion of a scalar version of SCET was (to our knowledge) first introduced here, which is one of the main examples studied in what follows.  While they did explain and set up all of the general framework, these authors then chose to work out the details of a 6-dimensional version of the scalar SCET theory, which has some important differences with our 4-dimensional example.  Many explicit integrals have been taken from these resources.  Furthermore, the details of the QCD example presented below rely heavily on~\cite{Becher:2018gno}.  These resources additionally provide many detailed applications to QCD processes.
\item Iain~Stewart and Christian Bauer's lectures~\cite{iain_notes}:  These lecture notes provide one of the most comprehensive introductions to SCET on the market (even though they are currently unfinished), and the notes you have before you have been heavily influenced by them.  The first half is devoted to a detailed introduction setting up the EFT degrees of freedom, Lagrangian, and operator structure.  Then applications to QCD are emphasized, including $B$ meson decays, $e^+\,e^- \rightarrow 2\text{-jets}$, and more.
\end{itemize}
\begin{center}
\sc{Techniques and Integrals}
\end{center}
\begin{itemize}
\item Vladimir~Smirnov's books~\cite{Smirnov:2002pj} and~\cite{Smirnov:2012gma}:  These two books are very comprehensive resources for evaluating integrals that appear for the Feynman diagram expansions in EFTs.  They also provide an introduction to the method of regions approach to evaluating integrals that we will use below.  Specifically, ``Applied Asymptotic Expansions in Momenta and Masses''~\cite{Smirnov:2002pj} is entirely devoted to the systematics of regions analysis, while ``Analytic tools for Feynman integrals''~\cite{Smirnov:2012gma} provides insights into evaluating multi-loop integrals using modern techniques. 
\item John Collins's books~\cite{Collins:1984xc} and~\cite{Collins:2011zzd}:  These are two monographs on advanced topics in field theory.  The book ``Renormalization''~\cite{Collins:1984xc} provides a very comprehensive approach to the RG, including many of the technical details that underlie the formalism introduced above.  Highlights include a rigorous definition of dimensional regularization, a discussion of how to treat multi-loop integrals (including nested divergences), and a proof of the RG invariance of bare parameters.  The book ``Foundations of Perturbative QCD'' provides an extremely detailed introduction to factorization in QCD.
\item George Sterman's book~\cite{Sterman:1994ce}:  This introductory QFT text book has a comprehensive discussion of IR divergences and factorization in QCD.
\end{itemize}
\end{spacing}
\clearpage

\begin{spacing}{1.15}
\bibliographystyle{utphys}
\bibliography{TASI_EFT}
\end{spacing}
\end{document}